\documentclass[12pt]{article}
\pdfoutput=1

\usepackage{amssymb}
\usepackage{amsmath}
\usepackage{amsthm}
\usepackage{mathtools}
\usepackage[usenames,dvipsnames]{xcolor}
\usepackage{epsfig}
\usepackage{dcolumn}
\usepackage{upgreek}
\usepackage{setspace}
\usepackage{enumitem}
\usepackage{array,multirow,bigdelim,arydshln}
\usepackage{appendix}
\usepackage[export]{adjustbox}
\usepackage{xparse}
\usepackage[utf8]{inputenc}
\usepackage{microtype}
\usepackage{bm}
\usepackage{braket}
\usepackage{dsfont}
\usepackage{nccmath}
\usepackage{datetime}
\usepackage{bm}
\usepackage{multirow}
\usepackage{etoc}
\usepackage{empheq}
\usepackage[most]{tcolorbox}
\usepackage{transparent}
\usepackage{nccmath}
\usepackage{textgreek}

\newtcbox{\graybox}[1][]{%
	nobeforeafter, math upper, tcbox raise base,
	enhanced, colframe=white!95!black,
	colback=white!95!black, boxrule=0pt, arc=0pt,
	#1}

\usepackage{jheppub}

\definecolor{bg}{RGB}{63,63,63}
\usepackage[finalizecache=false,frozencache=true]{minted}
\setminted[mathematica]{frame=none,
	rulecolor=\color{white!80!black},
	fontsize=\small,
	numbers=right,
	numbersep=5pt,
	obeytabs=true,
	tabsize=4,
	breaklines=true,
	firstnumber=last,
	bgcolor=bg}

\usepackage[usenames,dvipsnames]{xcolor}
\usepackage{hyperref}
\hypersetup{
    pdfencoding=unicode,
	colorlinks=true,
	urlcolor=Maroon,
	linkcolor=RoyalBlue,
	citecolor=Maroon,
	pdftitle={Physics of the Analytic S-Matrix},
	pdfauthor={Sebastian Mizera},
	pdfdisplaydoctitle=true,
	pdfstartview=FitH,
	linktocpage=true
}
\DeclareMathAlphabet{\mathbbold}{U}{bbold}{m}{n}
\newcommand*{\boldone}{\mathbbold{1}}
\usepackage{physics}
\usepackage{lastpage}

\usepackage{tikz}
\usetikzlibrary{decorations.pathmorphing,decorations.pathreplacing,calligraphy,calc,snakes,cd}

\tikzset{
	photon/.style={
		decoration={snake, aspect=0.75, mirror, segment length=3mm, amplitude=1mm},
		decorate
	}
}

\allowdisplaybreaks
\graphicspath{{figures/}}
\numberwithin{figure}{section}
\DeclareMathOperator{\Disc}{Disc}

\DeclareMathOperator{\tanc}{tanc}

\def \be {\begin{equation}}
\def \ee {\end{equation}}
\def \nn {\nonumber}

\def \R {\mathbb{R}}
\def \D {\mathrm{D}}

\def \d {\mathrm{d}}
\def \eps {\varepsilon}

\def \leq {\leqslant}
\def \geq {\geqslant}

\def \Im {\mathrm{Im}}

\def \Cut {\mathrm{Cut}}

\def \e {\mathrm{e}}
\def \x {\mathbf{x}}
\def \p {\mathbf{p}}
\def \q {\mathbf{q}}
\def \Born {\mathrm{Born}}
\def \j {\mathrm{j}}
\def \t {\mathrm{t}}

\title{Physics of the Analytic S-Matrix}
\author{Sebastian Mizera}\emailAdd{smizera@ias.edu}
\affiliation{Institute for Advanced Study, Einstein Drive, Princeton, NJ 08540, USA}

\abstract{%
You might've heard about various mathematical properties of scattering amplitudes such as analyticity, sheets, branch cuts, discontinuities, etc. What does it all mean? In these lectures, we'll take a guided tour through simple scattering problems that will allow us to directly trace such properties back to physics. We'll learn how different analytic features of the S-matrix are really consequences of causality, locality of interactions, unitary propagation, and so on.

These notes are based on a series of lectures given in Spring 2023 at the Institute for Advanced Study in Princeton and the Higgs Centre School of Theoretical Physics in Edinburgh.
}

\begin{document}

\setcounter{tocdepth}{2}
\maketitle
\setcounter{page}{3}
\setcounter{section}{-1}

\setcounter{tocdepth}{3}
	
\pagebreak

\section{\label{sec:introduction}Introduction}

Scattering experiments are all around us. 
For example, by reading these lecture notes, you're currently participating in one: photons bouncing off the sheet of paper (or produced by the screen) are detected by your eyes and processed by the brain.
So in a sense, we've been understanding the world around us by interpreting scattering amplitudes since time immemorial.
Modern physics continues this exploration much more purposely and quantitatively through particle experiments, large and small. It therefore becomes important to understand how to infer physics from scattering amplitudes systematically.

As you might expect, scattering processes can become arbitrarily complicated and in general there's little hope for being able to compute them exactly. We typically resort to studying them in idealized situations, or in limits, such as perturbation theory. Nevertheless, it turns out that there are certain \emph{universal} properties that can be investigated without direct computations. Did the amplitude come from a causal process? Were the interactions local? Can we guarantee that probability was conserved? The goal of these lectures is to convince you that such features indeed leave universal signatures on scattering amplitudes, once treated as complex analytic functions. This observation dates back to around a century ago and is broadly known as the \emph{analytic S-matrix} program.
Remember about analyticity? It's the $\oint_{|z|=\eps} \frac{\d z}{z} = 2\pi i$ stuff.

There's no denying that this is an absolutely crazy concept. Somehow, particles traveling with complex momenta and scattering at imaginary angles seem to not only know about, but also tell us new things about \emph{real} physics. As we'll learn during the course of the lectures, there are a few mechanisms that allow us to establish direct connections between singularities of scattering amplitudes and physical principles, such as causality, locality, and unitarity. But despite involving high-brow concepts from complex analysis, fascination in this subject is far from being purely academic. Studying analyticity often leads to rather unexpected constraints on the S-matrix, including bounds on coupling constants and relations between amplitudes for different scattering processes. Indeed, the main motivation for pursuing this research program is to uncover underlying principles governing the S-matrix theory, which almost always turn out to push forward the state-of-the-art of computational techniques.

There's also no denying that the subject itself is extremely subtle, especially that it requires a certain level of mastery of complex analysis, which physicists normally study only at a superficial level.
When I was trying to learn this subject as a grad student, like many of others in the same situation, I had to plow through the classic but dated textbook on the topic, \emph{The Analytic S-Matrix} by Eden et al., which reflects how much confusion there was around the subject half a century ago, when it was written.%
\footnote{It might not entirely be a coincidence that many of the pioneers of the analytic S-matrix turned to seek higher powers. John Polkinghorne, for one, resigned from his professorship at the University of Cambridge to become an Anglican priest. Henry Stapp, on the other hand, ditched conventional physics to pursue his theory of quantum consciousness and free will. Look, I'm not saying that studying this subject is going to leave you searching for the purpose of life, but at least a certain amount of caution is advised.}
While there are other more up-to-date resources, some of which are mentioned below, I couldn't find one that explained the whole story clearly from scratch. The goal of these lecture notes is to introduce the subject of the analytic S-matrix from the physics perspective, the way I would've wanted to learn it back when I was a student.

One of the main challenges is that, for various technical reasons, not a whole lot is understood concretely about the analytic properties of scattering amplitudes in quantum field theory. In my opinion, a more fruitful path to learning the subject is to begin with way less sophisticated but more pedagogical theories, within which explicit computations can be made and analyticity can be traced back to its physical roots. They will, hopefully, leave little to no room for confusion.
This is going to be the underlying philosophy: we'll start with simple scattering problems, first even classical and spherically-symmetric, then we'll gradually branch out to quantum-mechanical problems in four dimensions, and finally reach relativistic quantum field theory. The idea is that in these simplified systems, we'll manage to understand everything through and through. This understanding will trickle into more involved problems, where technical challenges often multiply, but the basic intuition stays intact. The plan is to cover the following topics.
\begin{itemize}[leftmargin=*]
\item {\bfseries Lecture I: Causal transforms}, in which we'll encounter the basic link between classical causality and analyticity in simplified settings.
\item {\bfseries Lecture II: Finite-range scattering} will discuss how some of the analyticity statements need to be modified in quantum theory and establish the connection between locality and singularities.
\item {\bfseries Lecture III: Potential scattering}, where we're going to study more realistic processes and make connections between unitarity and various aspects of analyticity in perturbation theory.
\item {\bfseries Lecture IV: Maximal analyticity} is going to combine all the above knowledge to put the consequences of causality, locality, and unitarity into a single formula.
\item {\bfseries Lecture V: Quantum field theory} will cover a selection of more advanced topics illustrating the role analyticity plays in modern research, including crossing symmetry, dispersive bounds on effective field theories, and the S-matrix bootstrap.
\end{itemize}

Due to time limitations, we won't be able to dive into specialized topics that would normally be discussed in a modern course on scattering amplitudes, such as spinor-helicity variables or mathematical properties of Feynman integrals. Such topics don't lack great resources to learn from. We'll also avoid, as much as possible, various bits of formalism that might distract us from the physics. For example, I won't discuss Hilbert spaces or formal scattering theory at all. The main emphasis will be on illustrating all the key points on elementary and explicit examples, most of which you can plot and verify everything directly yourself.

\subsection{Organization}

These notes were originally based on a series of five lectures, two hours each (without the exercises). They were aimed at grad students. Prerequisites are rather minimal: it's enough to be familiar with quantum mechanics and basic concepts from complex analysis such as Cauchy's theorem. Lecture~\ref{sec:lecture5} requires some proficiency with quantum field theory at the level of tree-level Feynman diagrams.

However, the notes grew over time to contain much more material included for completeness. As a consequence, many parts can be skipped on first reading. Longer derivations will appear here and there, but only if they give some new insights into the results. In those situations, I'll warn you beforehand that some of the text can be jumped over, unless you're interested in the details.

Each lecture is followed by a series of programming exercises. It's really just an excuse to learn new things in an interactive way. Exercises can be solved in any language or software, but I will provide step-by-step solutions in \texttt{Mathematica 13.2}. Remember, it's not important to get all the nuances of the syntax right, as long as you follow the concepts and have fun doing so. If you're not a programming whiz, I'd encourage you to use artificial intelligence assistants or just copy-paste the code I'll give you.

\subsection{Useful resources}

There won't be any references. Most of the material has been known for decades, though maybe not always presented in the same way as here. Let me mention a few useful books on the subject that could serve as a complement to the lecture notes.

\pagebreak
\begin{itemize}[leftmargin=*]
	\item {\bf Scattering Theory: The Quantum Theory of Nonrelativistic Collisions} by Taylor. One of the two classic textbooks on potential scattering, which contains a short review of the analytic properties of the total and partial-wave amplitudes, among many other relevant topics. It's a great introduction to practical scattering theory, providing a lot of background for the later study of relativistic amplitudes.
	
	\item {\bf Scattering Theory of Waves and Particles} by Newton (no, not that one). The other popular textbook on potential scattering, which puts slightly more emphasis on the formal aspects. It contains a similar breadth of topics to Taylor, but I'd recommended it for more mathematically-inclined readers.
	
	\item {\bf Causality and Dispersion Relations} by Nussenzveig. An excellent book on analyticity of scattering amplitudes in non-relativistic theories. It goes into a lot of details and proves precise theorems on analytic properties for finite-range and Yukawa potentials. It's probably the closest in spirit to these lecture notes, in that it emphasizes the role of the physical principles such as causality, locality, and unitarity.
	
	\item {\bf The Theory of Complex Angular Momenta} by Gribov. This text focuses specifically on the analytic properties in the scattering angle and the complex angular momenta. It's a fast-paced introduction to Regge theory in quantum field theory relevant to hadron physics and beyond.
	
	\item {\bf Lectures on Topics in Quantum Mechanics} by Tong. Part of the legendary series of lecture notes. Chapter 6 contains a discussion of scattering amplitudes providing a great introduction to scattering theory for newcomers.
	
	\item {\bf Homology and Feynman integrals} by Hwa and Teplitz. A textbook on homological approaches to Feynman integrals in massive quantum field theories. It provides a pedagogical exposition of more sophisticated mathematical topics, such as spectral sequences, for physicists. 
	
	\item {\bf Graph theory and Feynman integrals} by Nakanishi. This book focuses on the analytic and graph-theoretic properties of Feynman integrals. It's a great resource on Schwinger-parametric representations of massive Feynman integrals and their discontinuity and singularity structure.
	
	\item {\bf Scattering in Quantum Field Theories: The Axiomatic and Constructive Approaches} by Iagolnitzer. A rare and thorough review of axiomatic approaches to relativistic scattering theory, including rigorous non-perturbative results on the analyticity domains of the S-matrix elements.
	
	\item {\bf The Analytic S-Matrix} by Eden, Landshoff, Olive, and Polkinghorne. Classic textbook on the topic of S-matrix analyticity. Written while the subject was still in its infancy, it focuses on plausibility arguments rather than bullet-proof results. It's riddled with mistakes that are difficult to spot unless you're already an expert, and hence should be read with great caution.
	
\end{itemize}

\subsection{Acknowledgments}
I would like to thank all the participants of the lecture courses given at the Institute for Advanced Study in Princeton and the Higgs Centre School of Theoretical Physics in Edinburgh. In particular, let me thank Nima Arkani-Hamed, Simon Caron-Huot, Carolina Figueiredo, Einan Gardi, Mathieu Giroux, Hofie Hannesdottir, Aaron Hillman, Alberto Nicolis, and Wayne Zhao for useful feedback.

This material is based upon work supported by the Sivian Fund and the U.S. Department of Energy, Office of Science, Office of High Energy Physics under Award Number DE-SC0009988.

\pagebreak
\section{Lecture I: Causal transforms}

In this lecture, we will learn about the basic link between analyticity, causality, and dispersive properties of media. Since we have to start simple, the discussion will remain entirely classical. We'll consider a primitive model for materials, the Lorentz oscillator model, which will be enough to illustrate the underlying relationship between causal properties in time and analyticity in frequency domain. We'll learn about dispersion relations, which provide a link between absorptive and dissipative properties of media. Towards the end, we'll apply the same mechanisms to simple models of spherical scattering. By demanding causality of wave propagation, we'll discover our first example of an amplitude and analyze its features.

\setcounter{tocdepth}{3}

\medskip
\noindent\rule{\textwidth}{.4pt}
\vspace{-2em}
\localtableofcontents
\noindent\rule{\textwidth}{.4pt}

\pagebreak

\subsection{Lorentz oscillator model}

Let us start with the Lorentz oscillator model, where an electron at the position $\mathbf{x}(t)$ is being harmonically bound to a nucleus at the origin with a natural frequency $\omega_0$ and a damping constant $\gamma > 0$. Once we place multiple such electron-nucleus systems next to each other,
\be
\begin{gathered}
	\begin{tikzpicture}
		\draw[snake=coil,segment length=3pt] [color=black][line width=0.5] (0,0) -- ++ (1,1);
		\filldraw[RoyalBlue] (0,0) circle (4pt) node[anchor=east]{\footnotesize nucleus\;\;};
		\filldraw[Maroon] (1,1) circle (2pt) node[anchor=west]{\footnotesize\, electron};
		\begin{scope}[rotate=108,scale=0.67,xshift=-30pt,yshift=-100pt]
			\draw[snake=coil,segment length=3pt] [color=black][line width=0.5] (0,0) -- ++ (1,1);
			\filldraw[RoyalBlue] (0,0) circle (4pt);
			\filldraw[Maroon] (1,1) circle (2pt);
		\end{scope}
		\begin{scope}[rotate=-149,scale=0.67,xshift=70pt,yshift=-70pt]
			\draw[snake=coil,segment length=3pt] [color=black][line width=0.5] (0,0) -- ++ (1,1);
			\filldraw[RoyalBlue] (0,0) circle (4pt);
			\filldraw[Maroon] (1,1) circle (2pt);
		\end{scope}
		\begin{scope}[rotate=207,scale=0.5,xshift=-200pt,yshift=30pt]
			\draw[snake=coil,segment length=3pt] [color=black][line width=0.5] (0,0) -- ++ (1,1);
			\filldraw[RoyalBlue] (0,0) circle (4pt);
			\filldraw[Maroon] (1,1) circle (2pt);
		\end{scope}
		\begin{scope}[rotate=300,scale=0.5,xshift=-90pt,yshift=-40pt]
			\draw[snake=coil,segment length=3pt] [color=black][line width=0.5] (0,0) -- ++ (1,1);
			\filldraw[RoyalBlue] (0,0) circle (4pt);
			\filldraw[Maroon] (1,1) circle (2pt);
		\end{scope}
		\begin{scope}[rotate=-30,scale=0.3,xshift=-300pt,yshift=-90pt]
			\draw[snake=coil,segment length=3pt] [color=black][line width=0.5] (0,0) -- ++ (1,1);
			\filldraw[RoyalBlue] (0,0) circle (4pt);
			\filldraw[Maroon] (1,1) circle (2pt);
		\end{scope}
	\end{tikzpicture}
\end{gathered}
\ee
the setup is a crude toy-model for studying collective excitations in a medium. 
We'll later relate it to absorption and dissipation phenomena. The lessons we'll learn about causal responses of this system will turn out to be rather universal and have a direct counterpart in scattering amplitudes. For the time being, we can just focus on an individual electron-nucleus system and think of it as a simple harmonic oscillator problem.

\subsubsection{Free motion}

In the absence of external forces, the motion of the electron is described by the equation
\be
\ddot{\mathbf{x}}  + 2\gamma \dot{\mathbf{x}} + \omega_0^2 \mathbf{x} = 0\, ,
\ee
where the dots denote time derivatives. 
After years of studying physics, you no doubt are an expert at solving simple harmonic oscillators, but let's recall the physics behind it nevertheless. Since it's a second-order equation, the most general solution is given by a superposition of two exponentials:
\be
\mathbf{x}(t) = \mathbf{c}_+ \e^{- i \omega_+ t} + \mathbf{c}_- \e^{- i \omega_- t}\, ,
\ee
where $\mathbf{c}_\pm$ are two constants determined by boundary conditions and $\omega_\pm$ are obtained by solving the quadratic equation
\be
\omega_\pm^2 +2 i \gamma \omega_\pm - \omega_0^2 = 0\, ,
\ee
which gives
\be\label{eq:omega-plus-minus}
\omega_\pm = \pm \sqrt{\omega_0^2 - \gamma^2} - i \gamma\, .
\ee
At this stage, let me call attention to the fact that both $\omega_\pm$ have a \emph{negative} imaginary part, as we'll see more carefully in a second. It's a consequence of the fact that $\gamma>0$, meaning that the system \emph{dissipates} energy (as opposed to gaining it). This fact will turn out to be extremely important later on.
In any case, putting everything together, we have
\be
\mathbf{x}(t) = \left( \mathbf{c}_+ \e^{-i\sqrt{\omega_0^2 - \gamma^2}t} + \mathbf{c}_- \e^{i\sqrt{\omega_0^2 - \gamma^2}t}\right) e^{-\gamma t}\, .
\ee
This is the standard form of a damped oscillation with relaxation time $\sim 1/\gamma$ or so. Now, it's time to do some scattering.

\subsubsection{Coupling to an electric field}

Let's say that we switch on the electric field $\mathbf{E}(t)$. The motion of the electron is described according to Newton's second law by the driven harmonic oscillator equation:
\be\label{eq:driven}
\ddot{\mathbf{x}}  + 2\gamma \dot{\mathbf{x}} + \omega_0^2 \mathbf{x} = -\frac{e}{m} \mathbf{E}\, ,
\ee
where $-e$ and $m$ are the electric charge and mass of the electron. Our goal is to understand how basic properties of the electric field can be read off from the response of the medium in the frequency domain. 

As usual, we need to find stationary solutions of \eqref{eq:driven} first. In other words, we apply a plane-wave electric field $\sim e^{-i\omega t}$ at a given frequency $\omega$ and look at solutions $\x(t)$ oscillating with the same frequency. This amounts to setting
\be
\mathbf{x}(t) = \tilde{\mathbf{x}}(\omega)\, \e^{-i\omega t},\qquad
\mathbf{E}(t) = \tilde{\mathbf{E}}(\omega)\, \e^{-i\omega t}\, .
\ee
Plugging these functions back into \eqref{eq:driven}, we find they satisfy
\be\label{eq:x-chi-E}
\tilde{\mathbf{x}}(\omega) = \tilde{G}(\omega) \tilde{\mathbf{E}}(\omega)
\ee
with
\begin{empheq}[box=\graybox]{equation}\label{eq:chi-tilde}
\tilde{G}(\omega) = \frac{e}{m}\frac{1}{(\omega - \omega_+)(\omega - \omega_-)}\, .
\end{empheq}
Here, $\tilde{G}(\omega)$ is the frequency-domain Green's function of the system, telling us how much kick the electron gets as a response to a plane-wave of a given frequency $\omega$. As you might've noticed, \eqref{eq:chi-tilde} is actually complex. Its phase measures the delay between the driving electric field and the response of the electron, also known as the \emph{phase shift}. As we'll see in due time, physically $\tilde{G}(\omega)$ measures the complex susceptibility of the system.

The two poles of $\tilde{G}(\omega)$ are the $\omega_\pm$ given in \eqref{eq:omega-plus-minus}. As we keep increasing $\gamma$, their positions in the complex $\omega$-plane move as follows:
\be
\includegraphics[valign=c]{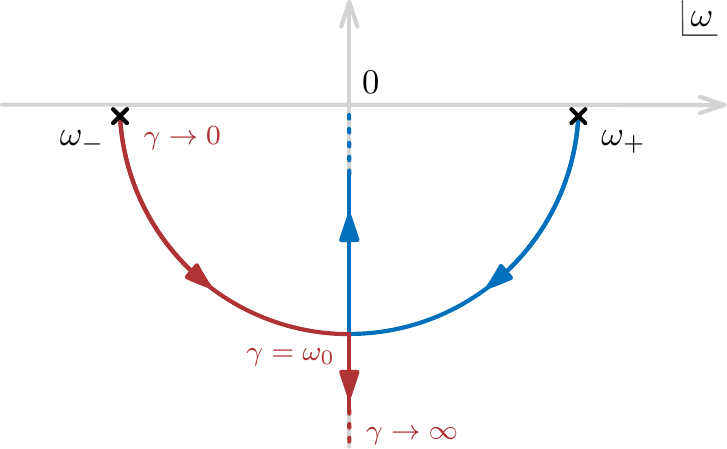}
\ee
Let us make some sanity checks. As we switch off the damping, $\gamma \to 0$, we have $\omega_\pm \to \pm \omega_0$, i.e., the response is the strongest near the natural frequency, as expected. On the other end, as $\gamma \to \infty$, the electron becomes very stiffly-bound to the nucleus and $\tilde{G}(\omega) \to 0$, which makes sense since external electric forces hardly influence the system.

\subsubsection{General solution}

Since the response of the system is linear, it's now easy to write down a general solution to \eqref{eq:driven} after some standard Fourier transform gymnastics.
An arbitrary electric field $\mathbf{E}(t)$ can be represented as a Fourier transform
\be
\mathbf{E}(t) = \frac{1}{2\pi} \int_{-\infty}^{\infty} \tilde{\mathbf{E}}(\omega)\, \e^{-i\omega t}\, \d \omega\, .
\ee
Using \eqref{eq:x-chi-E}, the solution of the equation of motion becomes
\begin{subequations}
\begin{align}
	\mathbf{x}(t) &= \frac{1}{2\pi} \int_{-\infty}^{\infty} \tilde{\mathbf{x}}(\omega)\, \e^{-i\omega t}\, \d \omega\\
	&= \frac{1}{2\pi} \int_{-\infty}^{\infty} \tilde{G}(\omega) \tilde{\mathbf{E}}(\omega)\, \e^{-i\omega t}\, \d \omega \, .
\end{align}
\end{subequations}
To tie everything together, let's re-express the right-hand side in terms of the $\mathbf{E}(t)$ in the time domain using its inverse Fourier transform:
\be
\mathbf{x}(t) = \int_{-\infty}^{\infty} G(t - t') \mathbf{E}(t')\, \d t'
\ee
with
\be\label{eq:chi}
G(t - t') = \frac{1}{2\pi} \int_{-\infty}^{\infty} \tilde{G}(\omega)\, \e^{-i \omega (t-t')} \d \omega \, .
\ee
In other words, the response $\mathbf{x}(t)$ to sending the electric field $\mathbf{E}(t)$ is given by convolving it with the temporal Green's function. To proceed further, we need to evaluate $G(t-t')$ explicitly.

\subsection{\label{sec:causality-analyticity}Causality implies analyticity}

Remember when I told you to pay attention to the imaginary part of $\omega_\pm$ being negative? Now, it's time to finally use this fact in evaluating \eqref{eq:chi}. This integral is actually not that difficult to compute if we use some contour deformations. Let's distinguish between two cases.

When $t < t'$, the integrand decays exponentially fast in the upper half-plane of $\omega$, since $\tilde{G}(\omega)$, as given in \eqref{eq:chi-tilde}, is only a rational function. This means we can add an integral over a large arc in the upper half-plane without changing the answer. The resulting contour looks something like this:
\be
\label{fig:omega-contour-UHP}
\includegraphics[valign=c]{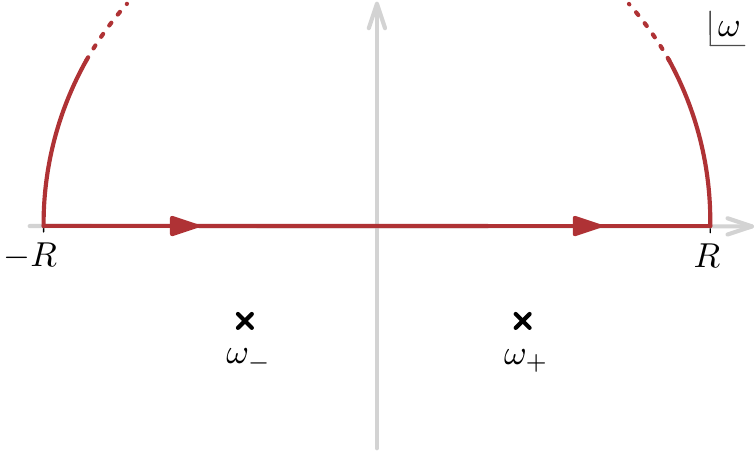}
\ee
where we take the limit as $R$ goes to infinity.
To make sure we haven't messed anything up, let's set $\omega = R \e^{i \phi}$ with $\phi \in (0,\pi)$ and look at the exponential in the integrand:
\be
\e^{-i \omega(t-t')} = \e^{iR  (t'-t) \cos \phi} \e^{-R (t'-t) \sin \phi}\, ,
\ee
where the second factor gives us the exponential suppression since both $t' - t$ and the sine are always positive. At this stage, we can simply deform the integration contour in \eqref{fig:omega-contour-UHP} until it shrinks to a point, proving that the integral vanishes when $t < t'$.

On the other hand, when $t > t'$, by similar arguments we can add a large arc in the lower half-plane, so that the resulting contour looks like this:
\be
\label{fig:omega-contour-LHP}
\includegraphics[valign=c]{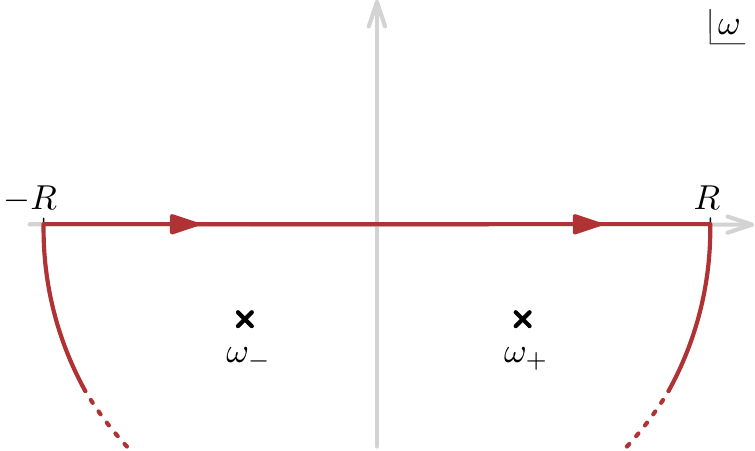}
\ee
In this case, we can once again deform the contour, but this time we are obstructed by the two poles at $\omega = \omega_\pm$ (let's for now ignore the case $\omega_0 = \gamma$ when the two poles are on top of each other). As a result, the integral evaluates to a sum of two residues as follows:
\begin{subequations}
\begin{align}
	G(t - t') &= -\frac{ie}{m} \sum_{s = \pm} \Res_{\omega = \omega_s} \frac{\e^{-i \omega (t-t')}}{(\omega - \omega_+)(\omega - \omega_-)} \\
	&= -\frac{ie}{m} \frac{\e^{-i \omega_+(t-t')} - \e^{-i \omega_-(t-t')}}{\omega_+ - \omega_-}
\end{align}
\end{subequations}
when $t > t'$.
Remember that there's an additional factor of $-2\pi i$ when converting from a small circular contour to a residue, with a minus sign due to the fact the circle was oriented clockwise. Computing the residue and putting everything together, we have the Green's function:
\be
G(t - t') = -\frac{e}{m} \frac{\sin\left[\sqrt{\omega_0^2 - \gamma^2}(t-t')\right]}{\sqrt{\omega_0^2 - \gamma^2}} e^{-\gamma (t-t')} \theta(t-t')\, ,
\ee
where $\theta$ is the Heaviside step function.
The answer describes the response in the position of the electron $\mathbf{x}(t)$ to the value of the electric field $\mathbf{E}(t')$ at the instant $t'$. The fact it vanishes for $t<t'$ is precisely the statement of \emph{causality} (often called \emph{primitive} causality condition): the cause has to precede the effect. At times $t>t'$, the electric field causes oscillation and its effect is attenuated after $\sim 1/\gamma$ or so. You can go ahead and repeat the exercise in the edge case $\omega_0 = \gamma$ with the same conclusions.

There's a basic lesson we've learned here. The fact that the Green's function $G(t - t')$ vanished for $t < t'$ was a direct consequence of the fact that its Fourier cousin $\tilde{G}(\omega)$ was \emph{analytic} in the upper half-plane of $\omega$, meaning that it didn't have any singularities there. Technically speaking, it was also important that $\tilde{G}(\omega)$ decayed sufficiently fast at infinity, so that we could add the arc in \eqref{fig:omega-contour-UHP} at no cost. The thing to appreciate here is that it's a completely general feature of \emph{any} system in which the input (in our case $\mathbf{E}(t')$) and the output (for us $\mathbf{x}(t)$) are related linearly through a time-translation invariant Green's function $G(t -t')$. Later on, we'll encounter a different incarnation of this phenomenon in scattering amplitudes.

The logic can be reversed to also show the other implication: causality implies analyticity of $\tilde{G}(\omega)$ in the upper half-plane. To see this, let us consider the inverse Fourier transform of \eqref{eq:chi}:
\be
\tilde{G}(\omega) = \int_{-\infty}^{\infty} G(\tau)\, \e^{i\omega \tau}\, \d \tau = \int_{0}^{\infty} G(\tau)\, \e^{i\omega \tau}\, \d \tau\, .
\ee
In the second equality we used the fact that $G(\tau)$ vanishes for $\tau < 0$. For any $\omega$ in the upper half-plane, the integral converges and hence defines an analytic function. This is because the exponential is
\be
\e^{i\omega \tau} = \e^{i \tau \Re \omega} \e^{- \tau\, \Im\, \omega} \, .
\ee
Therefore, as long as $G(\tau)$ doesn't grow too fast in the upper half-plane, causality implies $\tilde{G}(\omega)$ is analytic for $\Im\, \omega > 0$. Notice that, once again, we haven't used any details about the setup other than causality.

\subsection{\label{sec:dispersion-relations}Dispersion relations}

At this stage you might be questioning why any of this matters. After all, physics happens on the real axis, not in the complex plane of $\omega$. But the fact that $\tilde{G}(\omega)$ on the real axis is a boundary value of an analytic function (embodiment of causality) \emph{does} mean that it's rather special. In particular, it means that $\tilde{G}(\omega)$ can't just be any function whatsoever. It must satisfy certain constraints called \emph{dispersion relations}.

\subsubsection{\label{eq:dispersion-relations-derivation}Derivation}

The simplest way to arrive at dispersion relations is to consider the function
\be\label{eq:chi-tilde-omega}
\frac{1}{2\pi i} \frac{\tilde{G}(\omega')}{\omega' - \omega}\, .
\ee
We don't need to know anything about $\tilde{G}(\omega')$ other than the fact it's analytic in the upper half-plane together with the real axis.
Let's integrate it in $\omega'$ over the by-now familiar contour:
\be\label{fig:dispersion-contour}
\includegraphics[valign=c]{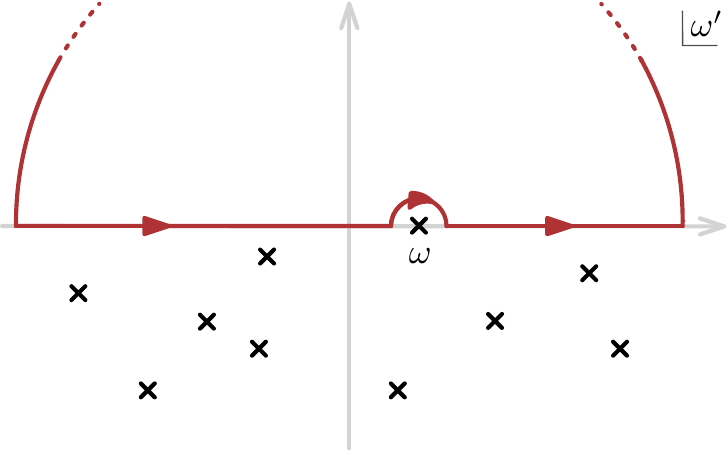}
\ee
We strategically placed $\omega$, which from this perspective is a constant, on the real axis of $\omega'$. We chose to route the contour with a small semi-circle around $\omega' = \omega$ sticking out upwards. For the time being, we will also demand that $\tilde{G}(\omega')$ decays at least as fast as $\sim 1/\omega'$ when $\omega' \to \infty$, so that the semi-circle at infinity doesn't contribute. We will revisit this assumption shortly.

Just as before, the contour can be closed up and hence \eqref{eq:chi-tilde-omega} integrated over \eqref{fig:dispersion-contour} gives zero. On the other hand, we can evaluate each part of the contour on its own, giving
\be\label{eq:dispersion-contour}
0 = -\frac{1}{2}\tilde{G}(\omega) + \frac{1}{2\pi i}\, \mathcal{P}\!\! \int_{-\infty}^{\infty} \frac{\tilde{G}(\omega')}{\omega' - \omega} \d \omega'\, .
\ee
The first term is $-\frac{1}{2}$ times the residue around $\omega' = \omega$. The half comes from the fact we're integrating over a semi-circle as opposed to a full circle and minus is due to its clockwise orientation. The second term is simply the remaining integration over the real axis. 
The symbol $\mathcal{P}$ denotes the \emph{Cauchy principal value}. It's a fancy way of saying that we integrate over the two half-lines $(-\infty,\omega-\eps)$ and $(\omega+\eps, \infty)$ and take $\eps \to 0$ at the end, i.e., the part of the contour \eqref{fig:dispersion-contour} parallel to the real axis. In practice, it amounts to using the Sokhotski--Plemelj\footnote{Believe it or not, but most of the questions I got after this lecture were about the pronunciation of \emph{lj} in Plemelj's name. Let me not leave you hanging. It's pronounced similarly to the English word ``million'' or saying ``call you''. Fun fact: Josip Plemelj was the first chancellor of the University of Ljubljana, which contains not one but two \emph{lj}'s!} identity
\be\label{eq:Sokhotski–Plemelj}
\mathcal{P}\frac{1}{z} = i\pi \delta(z) + \lim_{\eps\to 0^+}\frac{1}{z+i\eps}\, ,
\ee
where the first term is a Dirac delta function and the second displaces the pole into the lower half-plane. Graphically, it can be illustrated as follows:
\be
\includegraphics[scale=1,valign=c]{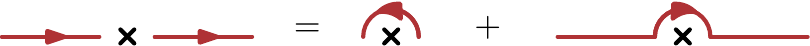}
\ee

The relative factor of $i$ between the two terms in \eqref{eq:dispersion-contour} might look innocent, but plays a crucial role. It's because taking the real and imaginary parts gives rise to two relations:
\begin{subequations}
\begin{empheq}[box=\graybox]{align}\label{eq:dispersion-relations}
	\Re \tilde{G}(\omega) &= \frac{1}{\pi}\, \mathcal{P}\!\! \int_{-\infty}^{\infty} \frac{\Im\, \tilde{G}(\omega')}{\omega' - \omega} \d \omega'\, ,\\ 
	\Im\, \tilde{G}(\omega) &= -\frac{1}{\pi}\, \mathcal{P}\!\! \int_{-\infty}^{\infty} \frac{\Re \tilde{G}(\omega')}{\omega' - \omega} \d \omega'\, .
\end{empheq}
\end{subequations}
They're called Kramers--Kronig or dispersion relations.

As a matter of fact, one can also show the either of these equations implies the other, and moreover that starting with either one, one can derive the causality condition $G(\tau) = 0$ for $\tau < 0$. In other words, dispersion relations are true if and only if the response of the system respects causality. If you'd like to look it up, all the statements are made rigorous by the \emph{Titchmarsh's theorem}. Any function $\tilde{G}(\omega)$ satisfying the above criteria is called a \emph{causal transform}.

\subsubsection{\label{sec:causal-transform-example}Example}

To get some practice with using dispersion relation, let's verify on an example that starting with any sufficiently-fast decaying $\Im\, \tilde{G}$, we can reconstruct a causal Green's function. Say we take
\be\label{eq:Im-chi-tilde-example}
\Im\, \tilde{G}(\omega) = -\frac{b}{(\omega-a)^2 + b^2}
\ee
for some real constants $a$ and $b>0$.
According to the equation \eqref{eq:dispersion-relations}, we can reconstruct the real part as follows:
\be
\Re \tilde{G}(\omega) = -\frac{b}{\pi}\, \mathcal{P}\!\! \int_{-\infty}^{\infty} \frac{\d \omega'}{(\omega' - \omega)[(\omega'-a)^2 + b^2]}\, .
\ee
Here, we can use the identity \eqref{eq:Sokhotski–Plemelj} around the pole at $\omega' = \omega$. The delta function term can be evaluated simply to give $-\frac{b}{\pi} \frac{i \pi}{(\omega-a)^2 + b^2}$, while the $i\eps$ term can be evaluated by residues. To be more precise we upgrade $\omega'$ to a complex variable and the pole is now at $\omega' = \omega - i\eps$, in addition to two others at $\omega' = a \pm i b$.
We can then close the contour, say in the upper half-plane and pick up the single residue at $\omega' = a + i b$. This finally gives
\be
\Re \tilde{G}(\omega) = -\frac{b}{\pi} \left( \frac{i\pi}{(\omega-a)^2 + b^2} + \frac{2\pi i}{2i b (a + i b - \omega)} \right) =  \frac{\omega-a}{(\omega-a)^2 + b^2}\, .
\ee
for real $\omega$.  As a cross-check, we see that the answer came out to be real, as a real part should be. Putting it together with \eqref{eq:Im-chi-tilde-example}, we reconstructed the whole Green's function:
\be
\tilde{G}(\omega) = \Re \tilde{G}(\omega) + i\, \Im\, \tilde{G}(\omega) = \frac{1}{\omega - a + i b}\, . 
\ee
The answer is indeed analytic in the upper half-plane: it only has a simple pole at $\omega = a - i b$. Note a posteriori that it was rather crucial that $b$ was chosen positive in the first place, otherwise the dispersion relations wouldn't have been valid.

We can finally go back to the time domain by performing the Fourier transform \eqref{eq:chi}. The details are very similar to the computation in Sec.~\ref{sec:causality-analyticity}: we can close the contour in the upper and lower half-planes depending on whether $\tau < 0$ or $\tau > 0$. The result is
\be
G(\tau) = - i \, \e^{- b \tau} \e^{- i a \tau} \theta(\tau)\, ,
\ee
with the Heaviside step function $\theta$ imposing causality.

\subsubsection{\label{sec:subtractions}Subtractions}

For completeness, let's come back to the small technicality: to close up the contour, we had to assume that $\tilde{G}(\omega')$ decays sufficiently fast at infinity. If this is not the case, e.g., $\tilde{G}(\omega')$ grows as $\sim \omega'^{k-1}$ for integer $k$, we simply modify the starting point \eqref{eq:chi-tilde-omega}. For instance, we can instead use
\be
\frac{1}{2\pi i}\frac{\tilde{G}(\omega')}{(\omega' - \omega) \prod_{i=1}^{k} (\omega' - \omega_i)}
\ee
with some constants $\omega_i$ called subtraction points. Repeating essentially the same steps, one arrives at dispersion relations with $k$ subtractions. We have basically traded the information about how the function $\tilde{G}(\omega')$ behaves at infinity for having to specify its value at the subtraction points. Alternatively, taking some of the $\omega_i$'s to be the same, we probe the information about the value of the function and its derivatives. In all the above examples, no subtractions were necessary, but they're often required in many physical setups.

\subsubsection{Physical interpretation}

At this stage, we can return back to the physics of the Lorentz oscillator model. Recall that the dipole moment of a single electron equals $-e \tilde{\mathbf{x}}(\omega)$. If we model our material by putting many such electron-nucleus system at density of $N$ per unit volume, the resulting polarization (dipole momentum per unit volume) is given by
\be
\mathbf{P}(\omega) = - N e \tilde{\mathbf{x}}(\omega) =  - N e \tilde{G}(\omega) \tilde{\mathbf{E}}(\omega) \, .
\ee
We'll assume that the medium is dilute enough, for example that it's a gas, so that we can ignore interactions between different electron-nucleus systems. I warned you it's a very crude model.
From electromagnetism classes you might remember that polarization can be also written in terms of the applied electric field:
\be
\mathbf{P}(\omega) = \epsilon_0 \chi_e(\omega) \tilde{\mathbf{E}}(\omega) \, ,
\ee
where $\epsilon_0$ is the permittivity of free space and the proportionality coefficient $\chi_e$ defines the \emph{electric susceptibility} of the medium. Comparing the two equations above, we find
\be
\chi_e(\omega) = -\frac{N e}{\epsilon_0} \tilde{G}(\omega).
\ee
So, as promised, $\tilde{G}$ measures the (complex) susceptibility up to a constant factor.

Finally, the complex refractive index $n(\omega)$ is related to susceptibility by
\be\label{eq:refractive-index}
\chi_e = n(\omega)^2 - 1 \approx 2 [n(\omega) - 1]\, .
\ee
Here, we expanded around $n(\omega) \approx 1$, which is the domain of validity of the Lorentz model that ignores interaction between individual electron-nucleus systems and hence holds only for dilute media.
The larger $\Re[n(\omega)]$, the less ``transparent'' the medium is to light, leading to light bending and lensing effects. Due to its $\omega$ dependence, different frequencies refract differently and allow the light to disperse, as you might remember from your lab classes or the Pink Floyd's album cover. On the other hand, $\Im[n(\omega)]$ is related to absorption of light in the medium. So the physical meaning of dispersion relations is to provide a direct link between dissipative and absorptive properties. It turns out that they are valid even without the simplifying approximation we used in \eqref{eq:refractive-index}.

As an example, let us consider a hypothetical medium that absorbs a specific frequency $\omega_0$ according to the Gaussian:
\be\label{eq:n-Gaussian}
n(\omega) = i \e^{-c(\omega - \omega_0)^2}\, ,
\ee
for some positive constant $c$. For real $\omega$, this function looks completely reasonable, but it actually blows up in the upper half-plane as $\sim i \e^{c\, \Im(\omega)^2}$, so it couldn't have originated from a causal medium! Another way to see it is recalling that a Fourier transform of a Gaussian \eqref{eq:n-Gaussian} is also a Gaussian in the time domain, which is certainly not causal.

In order to avoid violation of causality, absorption at $\omega_0$ must be accompanied by dissipation at other frequencies, in a coordinated way such that all the contributions interfere destructively to make $G(t-t')$ vanish for $t < t'$. Here's how a typical relation between absorption and dissipation looks like:
\be
\includegraphics[valign=c]{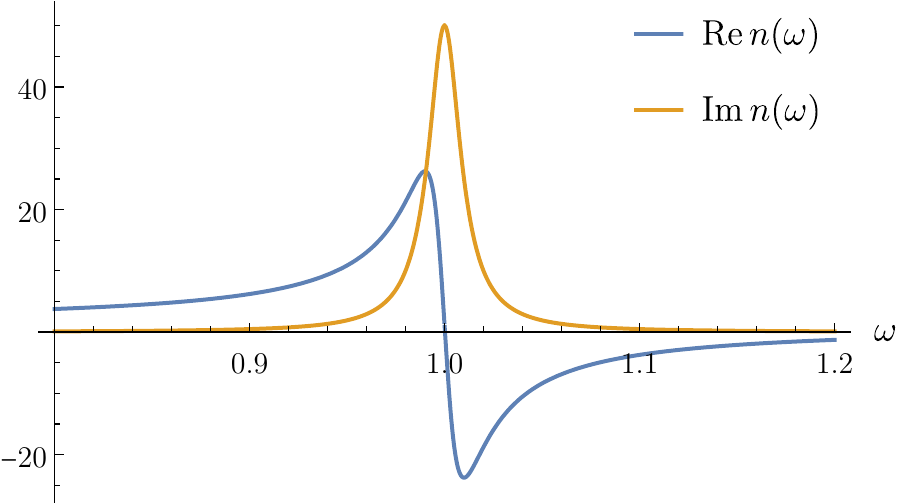}
\ee
We plotted \eqref{eq:chi-tilde} with $\frac{Ne^2}{2m \epsilon_0} = \omega_0 = 1$ and $\gamma = \frac{1}{100}$. Absorption peaked around $\omega = \omega_0$ (orange) has to be accompanied by dissipation at a range of frequencies (blue).

\subsection{\label{sec:spherical-wave}Spherical-wave scattering}

It would be a shame to finish this lecture without encountering an actual scattering amplitude. Let's consider an extremely simple model of classical spherical-wave scattering, which essentially reuses all the technology we already developed above, up to some variable relabeling. It will lead to arguably the simplest scattering amplitude you'll ever encounter.

\subsubsection{Wave equation}

Spherical waves $\psi(r,t)$ are simple because the only spatial coordinate entering is the radius $r$ and hence the dynamics is essentially one-dimensional. The scatterer, say a nucleus, is located near the origin. In vacuum $\psi$ satisfies the wave equation
\begin{align}\label{eq:wave-equation}
	\ddot{\psi} &= v^2\, \nabla^2 \psi\, ,
\end{align}
where $v$ is the speed. The Laplacian can be simplified to $\nabla^2 = \frac{\partial^2 }{\partial r^2} + \frac{2}{r} \frac{\partial}{\partial r}$ because of the spherical symmetry.
Recall that solutions corresponding to incoming waves traveling with momentum $p$ look like $\frac{1}{r} \e^{-i p (r + v t)}$ and outgoing ones are $\frac{1}{r} \e^{i p (r - v t)}$ (technically speaking, $p$ has the units of $1/[\mathrm{mass}]$, so it's the wave number, but we'll stick with calling it momentum). The intensity $|\psi|^2$ then decays as $\sim \frac{1}{r^2}$, which is the usual inverse-square law.

We're going to build the incoming and outgoing solutions by superimposing different momenta:
\begin{subequations}\label{eq:psi-in-out}
	\begin{align}
		\psi_{\mathrm{in}}(r,t) &= \frac{1}{r} \int_{-\infty}^{\infty} \tilde{\psi}_{\mathrm{in}}(p)\, \e^{-i p (r + v t)}\, \d p\, ,\\
		\psi_{\mathrm{out}}(r,t) &= \frac{1}{r} \int_{-\infty}^{\infty} \tilde{\psi}_{\mathrm{out}}(p)\, \e^{i p (r - v t)}\, \d p\, .
	\end{align}
\end{subequations}
The total wave $\psi(r,t)$ is a general solution given by the sum of the two contributions,
\be\label{eq:total-wave-function}
\psi(r,t) = \psi_{\mathrm{in}}(r,t) + \psi_{\mathrm{out}}(r,t)\, .
\ee
The fact that the solutions are real buys us additional constraints. To see this, let's compute the complex conjugate and simultaneously flip the integration variable $p \to -p$:
\be
\psi_{\mathrm{in}/\mathrm{out}}^\ast(r,t) = \frac{1}{r} \int_{-\infty}^{\infty} \tilde{\psi}_{\mathrm{in}/\mathrm{out}}^\ast(-p)\, \e^{\mp i p (r \pm v t)}\, \d p\, .
\ee
For this to equal \eqref{eq:psi-in-out} for an arbitrary choice of $\tilde{\psi}_{\mathrm{in}/\mathrm{out}}$, we need
\be\label{eq:psi-symmetry}
\tilde{\psi}_{\mathrm{in}/\mathrm{out}}(p) = \tilde{\psi}_{\mathrm{in}/\mathrm{out}}^\ast(-p)\, .
\ee
This constraint will turn out to be useful shortly.

The scattering amplitude is defined as the operation transforming the initial state $\tilde{\psi}_{\mathrm{in}}(p)$ into the final one $\tilde{\psi}_{\mathrm{out}}(p)$ and hence can be written as the ratio
\begin{empheq}[box=\graybox]{equation}\label{eq:S-0}
S_0(p) = -\frac{\tilde{\psi}_{\mathrm{out}}(p)}{\tilde{\psi}_{\mathrm{in}}(p)}\, .
\end{empheq}
As before, we assume that the interaction is linear.
The minus sign normalization is to make the scattering amplitude unity if no interactions happen. In that case, the wave equation \eqref{eq:wave-equation} is valid all the way to $r=0$ and we need $S_0(p) = 1$ for the total solution $\psi(r,t)$ to be regular as $r \to 0$. In other words, this is the choice making the $\frac{1}{r}$ poles cancel out between $\psi_{\mathrm{in}}(r,t)$ and $\psi_{\mathrm{out}}(r,t)$.

Additionally, as a consequence of \eqref{eq:psi-symmetry}, the scattering amplitude satisfies
\be\label{eq:s-wave-symmetry}
S_0(p) = S^\ast_0(-p)\, ,
\ee
called the symmetry relation.
Finally, the subscript ${}_0$ in $S_0(p)$ refers to the total angular momentum $j=0$ of the system. In the next lectures, we'll consider more general scattering amplitudes $S_j(p)$, which will share a lot of properties with the spherical one.

\subsubsection{Unitarity constraint}

We're going to assume that the total energy of the waves is conserved. In other words, the scatterer doesn't emit or absorb extra energy. We won't be able to avoid a little algebra exercise to compute the consequence of energy conservation, so let's get right to it. If you're just interested in the final result, you can skip to \eqref{eq:unitarity1}.

Recall that the energy flux through a sphere of radius $r$ is equal to $\Phi_r(t) = -4\pi r^2 \frac{\partial \psi}{\partial t} \frac{\partial \psi}{\partial r}$, which after plugging in \eqref{eq:total-wave-function} evaluates to
\begin{align}
	\Phi_r(t) \approx 4\pi v & \int_{-\infty}^{\infty} p \left[ \tilde{\psi}_{\mathrm{in}}(p)\, \e^{-ipr} + \tilde{\psi}_{\mathrm{out}}(p)\, \e^{ipr} \right] \e^{-ipvt}\, \d p\\
	&\times \int_{-\infty}^{\infty} p' \left[ \tilde{\psi}_{\mathrm{in}}(p')\, \e^{-ip'r} - \tilde{\psi}_{\mathrm{out}}(p')\, \e^{ip'r} \right] \e^{-ip'vt}\, \d p' \, .\nn
\end{align}
We dropped all terms subleading in $\frac{1}{r}$ since we're interested in the behavior at infinity. The statement of energy conservation is that $\Phi_r$ integrated over all time is zero. But the $t$-dependence above is very simple and we can just use
\be
\int_{-\infty}^{\infty} \e^{-i(p+p')vt} \,\d t = \frac{2\pi}{v} \delta( p + p')\, .
\ee
Using the delta function to localize $p' = -p$ and applying the symmetry relations \eqref{eq:psi-symmetry}, we're left with
\be
\int_{-\infty}^{\infty} \Phi_r(t)\, \d t \approx -8\pi^2 \int_{-\infty}^{\infty} p^2 \left[ |\tilde{\psi}_{\mathrm{in}}(p)|^2 - |\tilde{\psi}_{\mathrm{out}}(p)|^2 \right] \d p \stackrel{!}{=} 0\, .
\ee
The two contributions represent the total incoming and outgoing energy. Since the final equality must hold for all (square-integrable) choices of wave packets, it must be that the term in the brackets vanishes for every $p$. This immediately implies that the scattering amplitude satisfies
\begin{empheq}[box=\graybox]{equation}\label{eq:unitarity1}
|S_0(p)| = \frac{|\tilde{\psi}_{\mathrm{out}}(p)|}{|\tilde{\psi}_{\mathrm{in}}(p)|} = 1\, .
\end{empheq}
It's one of the central constraints we'll have at our disposal, called the \emph{unitarity condition}. The origin of this name will become clear in the next lecture. Together with \eqref{eq:s-wave-symmetry}, we also find $S_0(p)S_0(-p)=1$.

\subsubsection{Phase shift}

The equation \eqref{eq:unitarity1} simply means that the amplitude is a phase for real momenta $p$. Therefore, we can write it as
\be\label{eq:phase-shift}
S_0(p) = \e^{2i \eta_0(p)}\, ,
\ee
where $\eta_0(p)$ is a real function called the \emph{phase shift}. It characterizes the change of phase between the incoming and outgoing waves: the phase difference would've been $2pr$ in the absence of the scatterer and a non-trivial scattering shifts it to $2[pr + \eta_0(p)]$. The factor of $2$ is just a convention.
Moreover, the symmetry relation \eqref{eq:s-wave-symmetry} implies
\be
\eta_0(p) = -\eta_0(-p)\, ,
\ee
that is, the phase shift has to be an odd function of $p$. Later in Sec.~\ref{sec:time-delays}, we'll see that phase shifts are closely related to causality.

\subsubsection{\label{sec:simplest-amplitude}What's the simplest scattering amplitude?}

At this stage we know a fair bit about $S_0(p)$, even though we haven't even committed to a specific scattering process. We can play a little game and try to construct the simplest interacting scattering amplitude consistent with all the properties that appeared above. Most importantly, we found that $S_0(p)$ has to be a pure phase and that phase has to be odd in $p$. The simplest possibility is therefore
\begin{empheq}[box=\graybox]{equation}\label{eq:simplest-amplitude}
S_0^{\mathrm{HS}}(p) = \e^{-2ipR}\, .
\end{empheq}
For the exponent to be dimensionless, we also multiplied it by $R$ with units of length, which should be associated with a natural scale of the problem, for example the radius of the scatterer. There's nothing obviously wrong with \eqref{eq:simplest-amplitude}. Could it be an actual scattering amplitude?

It's actually not that difficult to set up a problem that leads to \eqref{eq:simplest-amplitude}, so let's just do it straight away. Consider an impenetrable or ``hard'' sphere of radius $R$. By definition, whichever wave packet we send in, the total wave function has to vanish at the surface $r=R$:
\be
\psi(R,t) = \frac{1}{R} \int_{-\infty}^{\infty} \tilde{\psi}_{\mathrm{in}}(p) \left[ \e^{-ipR} - S_0(p)\, \e^{ipR} \right] \e^{-i p v t}\, \d p \stackrel{!}{=} 0 \, .
\ee
for all times $t$. The only solution is \eqref{eq:simplest-amplitude}. Hence, it's the amplitude for scattering off of a hard sphere, which justified the superscript ${}^{\mathrm{HS}}$.

Notice one curious thing. The amplitude \eqref{eq:simplest-amplitude} is analytic in the upper half-plane, but it grows exponentially fast, as $\sim \e^{2 R\, \Im\, p}$.
On the other hand, in Sec.~\ref{sec:dispersion-relations} we've learned that causal transforms need to \emph{decay} at infinity for all the arguments to go through. Does it mean that the scattering amplitude isn't a causal transform? The answer is that indeed $S_0^{\mathrm{HS}}(p)$ itself is not, but it's closely related to one in such a way that it's still allowed to diverge in the upper half-plane. In fact, we'll see shortly that \eqref{eq:simplest-amplitude} puts a limit on this growth for arbitrarily-complicated scatterer confined within the radius $R$.

\subsubsection{\label{sec:high-energy-bound}High-energy bound}

Before doing a more detailed computation, let's pause to think about what causality could mean in this context. As with the harmonic oscillator, we're going to impose that effect can't precede the cause. More concretely, that the outgoing wave can't appear before the incoming wave has reached the surface of the scatterer at $r=R$. In other words, we demand that whenever
\be
\psi_{\mathrm{in}}(R,t) = \frac{1}{R} \int_{-\infty}^{\infty} \left[ \tilde{\psi}_{\mathrm{in}}(p) \e^{- ipR} \right] \e^{-i p v t}\, \d p = 0
\ee
for $t<0$, then also 
\be
\psi_{\mathrm{out}}(R,t) = -\frac{1}{R} \int_{-\infty}^{\infty} \left[ S_0(p)\tilde{\psi}_{\mathrm{in}}(p) \e^{ipR} \right] \e^{-i p v t}\, \d p = 0
\ee
for $t<0$.
It means that the terms in the square brackets are causal transforms, and in particular, that they are analytic in the upper half-plane of $p$. This fact also implies that their ratio
\be\label{eq:S0-normalized}
S_0(p)\, \e^{2i p R}
\ee
is analytic and bounded in the upper half-plane. (If the first square bracket had a zero, it would cancel out with the same zero of the second bracket, so \eqref{eq:S0-normalized} cannot have poles in the upper half-plane.) Certainly, \eqref{eq:simplest-amplitude} is an example of such a function (equal to $1$). It means that we can write down dispersion relations for \eqref{eq:S0-normalized}, which look similar to the ones we've seen earlier for susceptibility.

Let's now try to derive the explicit bound. We're going to look at $S_0(p)\tilde{\psi}_{\mathrm{in}}(p) \e^{ipR}$ with $p$ in the upper half-plane and write 
\be\label{eq:psi-out-bound}
S_0(p)\tilde{\psi}_{\mathrm{in}}(p) \e^{ipR} = \frac{1}{2\pi i} \int_{-\infty}^{\infty} \frac{S_0(p')\tilde{\psi}_{\mathrm{in}}(p') \e^{ip'R}}{p' - p} \, \d p'\, .
\ee
This equation is obtained using the same contour as in Sec.~\ref{eq:dispersion-relations-derivation}, but now $p$ is in the upper half-plane, so the term on the left-hand side doesn't have a $\frac{1}{2}$ and we don't have to worry about the principal value $\mathcal{P}$.

Up to now, the incoming wave $\tilde{\psi}_\mathrm{in}(p')$ has been pretty much arbitrary. At this stage we're going to make a special choice:
\be\label{eq:psi-in-choice}
\tilde{\psi}_{\mathrm{in}}(p')\e^{-ip'R} = \frac{1}{p' - p^\ast}\, .
\ee
From this perspective, we treat $p^\ast$ (complex conjugate of $p$) as a constant. It's an allowed choice because it satisfies all the causality requirements. As a matter of fact, this is precisely the example we studied earlier in Sec.~\ref{sec:causal-transform-example} with $\omega = p'$ and $a-ib = p^\ast$. It might not be entirely clear why this specific choice is interesting, but let's just run with it anyway. Plugging everything back into \eqref{eq:psi-out-bound} and rearranging a little, we get:
\be
S_0(p)\, \e^{2ipR} = \frac{\Im\, p}{\pi} \int_{-\infty}^{\infty} \frac{S_0(p')\, \e^{2ip' R}}{(p' - p)(p' - p^\ast)} \d p'\, .
\ee
To finish the derivation, we take the absolute value on both sides, resulting in
\be
|S_0(p)\, \e^{2ipR} | \leq \frac{\Im\, p}{\pi} \int_{-\infty}^{\infty} \frac{\d p'}{(p' - p)(p' - p^\ast)}\, .
\ee
Here, we used the fact that the absolute value of an integral is upper bounded by the integral of the absolute value.
Since $p'$ is real, we can also use the unitarity condition $|S_0(p')|=1$ and $|\e^{2i p' R}| = 1$. Finally, the remaining integral can be done by closing the contour up in either half-plane and gives $\frac{\pi}{\Im\, p}$, leading to the final result
\begin{empheq}[box=\graybox]{equation}\label{eq:complex-unitarity}
|S_0(p)\, \e^{2 i p R}| \leq 1 \, .
\end{empheq}
On the real $p$-axis, $|\e^{2 i p R}| = 1 $ and the resulting bound $|S_0(p)| \leq 1$ is indeed compatible with the real unitarity equation \eqref{eq:unitarity1}. In fact, \eqref{eq:complex-unitarity} should be viewed as the extension of unitarity to the upper half-plane. Using similar techniques, you can also derive that $S_0(p) = S_0^\ast(-p^\ast)$. It also hinged on causality, which was used as an ingredient in the derivation. In the next lecture we'll see that a similar constraint extends to higher angular momenta too.

We can now question whether \eqref{eq:psi-in-choice} was the most optimal choice and if perhaps a different one would give rise to a more stringent constraint. However, the fact we constructed an explicit example saturating \eqref{eq:complex-unitarity} means that the inequality must be optimal.

The result we proved is a version of the so-called Phragm\'en--Lindel\"of principle in complex analysis. Applied to our problem, it says that if an analytic function such as $S_0(p)\, \e^{2i p R}$ has its modulus bounded by $1$ on the real axis and by $\e^{\delta|p|}$ in the upper-half plane for some $\delta$, then in fact analyticity in the upper-half plane forces it to obey a much stronger bound \eqref{eq:complex-unitarity} for any $\Im\, p \geq 0$.

Before wrapping up this lecture, let me emphasize that the whole discussion so far has been classical. In the next lecture, we'll find that more general scattering amplitudes, even if coming from causal processes, don't have to be analytic in the upper half-plane. Actually, the fact that singularities can exist in the upper half-plane will be intimately tied to quantum mechanics.

\subsection{Exercises}

In this lecture, we've talked about the basic link between analyticity and physical properties such as causality. The purpose of the exercises will be to familiarize yourself with analytic functions a little more. As a concrete problem, let us imagine we're given scattering data for some real momenta $p$ and we'll try to reconstruct analytic properties in the complex plane.

In practice, we'll first just generate the data ourselves, add some noise, and then try to reconstruct analytic properties of the function we started from. First, let's define some ``random'' function $f(p)$:
\begin{minted}[firstnumber=1]{mathematica}
f[p_] := p/((p + 57I)*(p + 137I))
\end{minted}
We can then generate the data by sampling, say, $100$ points $p=1,2,\ldots,100$ and adding Gaussian noise. Let's start very gently and add only $0.1\%$ noise:
\begin{minted}{mathematica}
Noise[x_] := 1 + RandomVariate[NormalDistribution[]]*x;
data = Table[{p, Noise[0.001]*f[p]}, {p, 1, 100}]
\end{minted}
Later, you can start increasing the noise to $1\%, 2\%, 5\%$, etc. The function \texttt{Noise} adds a given percentage of Gaussian noise. The list \texttt{data} consists of pairs of $\{p, \hat{f}(p)\}$ for every sample point $p$, where $\hat{f}$ is a noisy version of $f$. Throughout the exercise, you can play with different choices of the initial function, number of samples, and the noise level. It's even more fun if you get \texttt{data} from a friend, try to reconstruct their $f(p)$ using techniques explained below, and then compare the results.

\subsubsection{Pad\'e approximants}

How do we go about reconstructing the analytic function if we only have access to a discrete number of (possibly noisy) data points? One extremely simple, but surprisingly effective, way is using \emph{Pad\'e approximants}. The idea just is to write a big ansatz for a function as a ratio of polynomials:
\be\label{eq:Pade-approximant}
g_{m,n}(p) = \frac{a_0 + a_1 p + a_2 p^2 + \ldots + a_m p^m}{b_0 + b_1 p + b_2 p^2 + \ldots + b_n p^n}\, .
\ee
The degrees of the two polynomials, $m$ and $n$, are our choice. Typically, it's desirable to use an ansatz that matches the asymptotic behavior we expect. Since the original function decays $f(p) \sim \frac{1}{p}$ (you should be able to read it off from \texttt{data}), it's wise to stick with $m = n-1$.

The goal is to find the coefficients $a_i$ and $b_i$ such that $g_{m,n}(p)$ best approximates our target function $f(p)$. Without loss of generality, we may fix one coefficient, e.g., $b_0 = 1$,  but numerics turns out to be better if we keep it unfixed. Let's write the ansatz as follows:
\begin{minted}{mathematica}
g[p_, m_, n_] := Sum[(Rea[i] + I*Ima[i])*p^i, {i, 0, m}]/
                 Sum[(Reb[i] + I*Imb[i])*p^i, {i, 0, n}];
\end{minted}
Here, we've split the undetermined variables $a_i$ and $b_i$ into their real and imaginary parts, which will help us in a second.
The next step is to find their optimal values. A reasonable choice would be to write a cost function by summing over all sample points $p$:
\be\label{eq:cost-function}
\sum_{p \in \texttt{data}} |\hat{f}(p) - g_{m,n}(p)|^2
\ee
and then minimize for it. This can be achieved with the following short snippet:
\begin{minted}{mathematica}
With[{m = 1, n = 2},
	solution = NMinimize[
		Sum[Abs[g[point[[1]], m, n] - point[[2]]]^2, {point, data}],
	Variables[g[1, m, n]]];
	g[p_] := g[p, m, n] /. solution[[2]];
]
\end{minted}
The function \texttt{NMinimize} takes as arguments the cost function \eqref{eq:cost-function} and the names of the real variables to solve for.
Its result \texttt{solution} is a pair consisting of the value of the minimum (measuring how good the fit is) and the optimal values of the parameters $a_i$ and $b_i$. The function \texttt{g[p]} is the resulting approximation.

\subsubsection{Reconstructing the function}

Try out a few different values of $m$, $n$, the level of the noise, number of samples, etc. You can scan over possible values and pick the ones that lead to the smallest value of the cost function (don't forget to normalize by the number of samples if they change).

One way to determine the quality of the reconstruction is whether the poles and zeros of $g_{m,n}(p)$ match approximately with those of $f(p)$. The way to determine them is
\begin{minted}{mathematica}
poles = NSolve[Denominator[g[p]] == 0, p]
zeros = NSolve[Numerator[g[p]] == 0, p]
\end{minted}
For example, using $(m,n) = (1,2)$, for my random sample I get two simple poles at $\approx -1.0 - 55.6 i$ and $\approx 5.4 - 137.2 i$ and a zero at $\approx 0.07 - 0.05 i$, which very roughly matches what we started with. As expected, increasing the number of samples and/or decreasing the noise helps with getting them more accurately.

Another idea is to compare residues of poles (or residues of $1/f(p)$ for the zeros) as follows:
\begin{minted}{mathematica}
<< NumericalCalculus`
Residue[f[p], {p, -137I}]
NResidue[g[p], {p, p /. poles[[2]]}]
\end{minted}
Can you come up with other ways of quantifying quality of the Pad\'e approximation?

The number of poles and zeros is built-in in the ansatz $g_{m,n}(p)$, so what happens to them when we increase $m$ and $n$? To visualize this situation, you can plot both $f(p)$ and $g_{m,n}(p)$ in the complex plane:
\begin{minted}{mathematica}
ComplexPlot3D[f[p], {p, -200 - 200I, 200 + 200I},
	        PlotRange -> {0, 0.1}, PlotPoints -> 100, Mesh -> Automatic]
\end{minted}
with an analogous routine for $\mathtt{g[p]}$.
The type of result you'd get for, say, $(m,n) = (3,4)$ is

\be
\!\!\includegraphics[scale=0.85,valign=c]{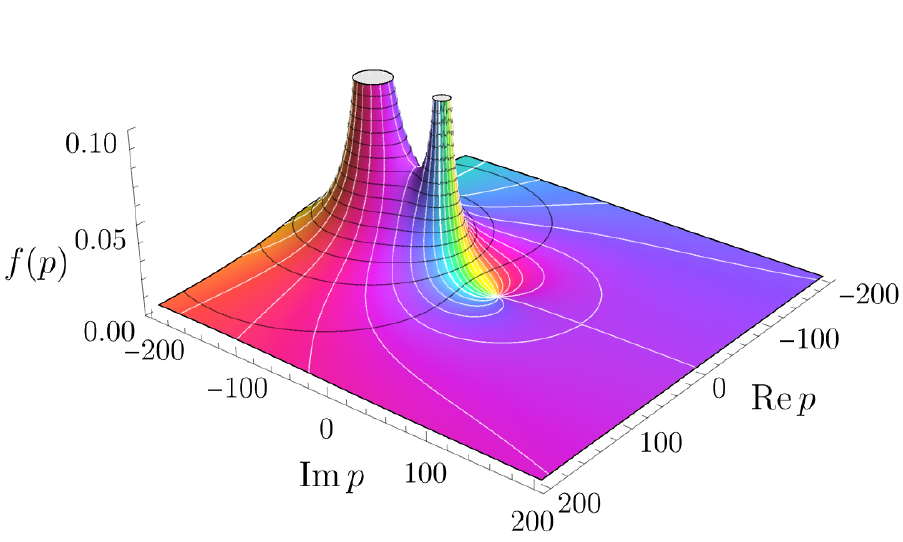}\;
\includegraphics[scale=0.85,valign=c]{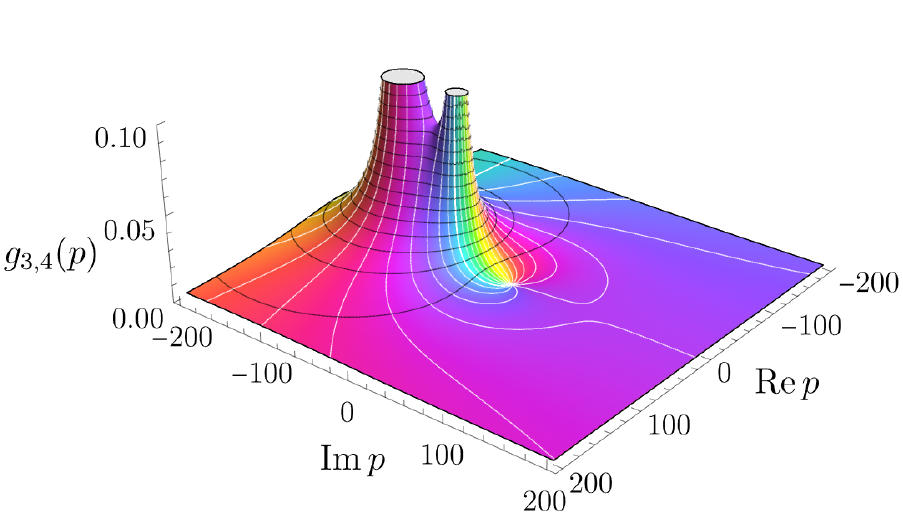}
\ee
The original $f(p)$ is on the left and $g_{3,4}(p)$ on the right. In this way of plotting complex functions, the darkness and height denotes the absolute value: poles are white and zeros are black, while the hue corresponds to the phase. As you can tell, poles on the right plot are roughly in the correct positions around $p = - 57i$ and $p = -137i$, even though the details of the function are slightly off compared to $f(p)$. There's however an accumulation of $2$ poles and $3$ zeros around $p \approx 0$. They nearly cancel out, giving back an approximate zero, as expected. That's why they're not even visible in the above plot. Appearance of pole-zero pairs that approximately cancel out, called \emph{Froissart doublets}, is a common feature of Pad\'e approximants.

Pad\'e approximants are designed to reconstruct rational functions approximations. What happens when you start with $f(p)$ featuring branch cuts? Experiment with a few examples.

\subsubsection{Unstability}

As the final exercise, let's ask if we could tell whether $f(p)$ is analytic in the upper half-plane just from the data on the real axis alone. For example, we can redo all the analysis above with $f(p) \to f(p) + \frac{1}{p-a}$, where $a$ is the position of a new pole. As you might expect by now, if $a$ was sufficiently close to the real axis in the range where the data is sampled, we can probably recover it approximately. However, if $a$ has a large imaginary part, it becomes more and more difficult to ``see it'' from the real axis. In general, we can't expect to recover all complex-analytic features from the real data. The best we can hope for is that we can reconstruct the complex neighborhood of the real axis. Fortunately, as we'll see in the next lecture, this is where a lot of physics hides. In the third set of exercises, we'll learn how to reconstruct the analytic properties near the real axis from experimental data.

This is the first sign of a general problem of \emph{unstability} of analytic continuation with respect to small changes. As a more extreme version, let's consider the deformation of the form
\be
f(p) \to f(p) + \epsilon\, \e^{-i p /\epsilon}
\ee
with small $\epsilon$. This modification leads to arbitrarily large changes of the analytic function in the upper half-plane, even though the data on the real axis hardly receives any corrections. In fact, the new function is not a causal transform because it blows up in the upper half-plane. If $\epsilon$ was small enough, we couldn't have possibly detected this fact from the real data alone.

We therefore learn an important lesson that reconstructing analyticity from the experimental data is in principle an impossible task, except for the neighborhood of the samples. Analyticity is a one-way street: as we've seen in the lecture, we can use it to make predictions about real physics, but the inverse problem is unstable.

\section{Lecture II: Finite-range scattering}

In the last lecture, we've learned about the connection between causality and analyticity in the momentum for classical $(1+1)$-dimensional problems. It's time to move on to scattering in $3+1$ dimensions in quantum mechanics. We're going to revisit the question of causality, which becomes more tricky in the presence of quantum effects.
Then, after reviewing what happens for higher angular momenta and plane-wave scattering, we'll make a connection between locality and analyticity.

\medskip
\noindent\rule{\textwidth}{.4pt}
\vspace{-2em}
\localtableofcontents
\noindent\rule{\textwidth}{.4pt}

\pagebreak

\subsection{\label{sec:barriers-wells}Spherical barriers and wells}

In this lecture we move on to quantum scattering problems. Instead of the wave equation, the wave function $\Psi(\x,t)$ is going to evolve according to the Schr\"odinger equation
\be
i \hbar \frac{\partial}{\partial t} \Psi(\mathbf{x},t) = H\, \Psi(\mathbf{x},t)\, ,
\ee
where $H$ is the Hamiltonian. We'll focus exclusively on time-independent Hamiltonians. In those cases, a wave function with energy $E$ can be written as
\be\label{eq:psi-x-t}
\Psi(\mathbf{x},t) = \e^{-i E t/\hbar}\, \psi(\mathbf{x})\, .
\ee
So the time-independent Schr\"odinger equation reads $H \psi(\x) = E \psi(\x)$.

Before diving deeper, let's for the time being stay within the realm of spherical cows and construct a quantum version of what we did in Sec.~\ref{sec:spherical-wave}. To this end, we'll consider a single particle of mass $m$ scattering off a spherically-symmetric potential $V(r)$. This setup is described by the Hamiltonian
\be
H(r) = -\frac{\hbar^2}{2m} \nabla^2 + V(r)\, .
\ee
Examples include the Coulomb or Yukawa potentials. However, it will be more instructive  to start with an even simpler case of the cutoff potential supported entirely within some finite radius $R$ of the origin:
\be\label{eq:potential-V}
V(r) = \begin{dcases}
	\frac{\hbar^2}{2m}U \;&\mathrm{for}\quad r<R\, ,\\
	0 \;&\mathrm{for}\quad r>R\, .
\end{dcases}
\ee
The height of the potential is characterized by the constant $U$, up to some prefactors we introduced to reduce the clutter later on. We'll consider cases with $U>0$ (potential barrier) and $U<0$ (potential well) in turn. The reason we use a finite-range potential is to emphasize the impact of \emph{locality} of interactions on analyticity. While the term ``locality'' can have different meanings depending on the context, here we simply mean that particles can influence each other significantly only at short distances. 

As before, we'll send in a spherical wave packet $\tilde{\psi}_{\mathrm{in}}(p)$ and read off the scattering amplitude from how it evolves into $\tilde{\psi}_{\mathrm{out}}(p)$ according to the Schr\"odinger equation. There's a crucial difference to the classical case, which is that wave packets evolve with the time dependence $\e^{-iEt}$ as in \eqref{eq:psi-x-t} instead of $\e^{-ipvt}$ (the difference is caused by the first derivative of time in the Schr\"odinger equations compared to the second derivative in the wave equation). Here, the energy $E$ and momentum $p$ are related by
\be\label{eq:E-p2}
E = \frac{\hbar^2 p^2}{2m} \geq 0\, .
\ee
In any case, the total wave function can be written as
\be\label{eq:total-wavefunction}
\Psi(r,t) = \frac{1}{r} \int_{-\infty}^{\infty} \tilde{\psi}_{\mathrm{in}}(p)\, \left[ \e^{-ipr} - S_0(p) \e^{ipr}\right] \e^{- iE t/\hbar}\, \d p\, ,
\ee
from which we can read-off the spherical scattering amplitude $S_0(p)$. Once again, it satisfied the unitarity condition $|S_0(p)|^2 = 1$, but this time it's a consequence of probability conservation. More precisely, for time-independent solutions the probability current is
\be
\mathbf{J}(\x) = \frac{1}{2i} \frac{\hbar}{m} \left( \Psi^\ast \nabla \Psi - \Psi \nabla \Psi^\ast  \right)\, .
\ee
Conservation of probability is governed by the continuity equation $\nabla \cdot \mathbf{J} = 0$. You can show that it plays a similar role to energy conservation in the classical case and ultimately leads to the unitarity constraint.

It's a bit of a hassle to carry the $\hbar$ around, so from now on we'll set $\hbar=1$ and only reinstate it where necessary.

\subsubsection{\label{sec:solving-Schrodinger}Solving the Schr\"odinger equation}

We're now ready to solve our first quantum scattering problem. Explicitly, the time-independent Schr\"odinger equation for the potential \eqref{eq:potential-V} is given by
\be
\left[ -\frac{\partial^2}{\partial r^2} - \frac{2}{r} \frac{\partial}{\partial r}  + U \theta(R-r) - p^2\right] \psi(r) = 0\, ,
\ee
where we plugged in the spherical Laplacian and $\theta$ is the Heaviside step function.

Solving this system is a small algebra problem. Inside the potential ($r<R$) we obtain the solution:
\begin{subequations}
	\begin{align}
		\psi(r) &= \frac{1}{r} \left[ A_1\, \e^{-i r \sqrt{p^2 - U}} + A_2\, \e^{i r \sqrt{p^2 - U}} \right]\\
		&= \frac{2i A_2}{r} \sin \left( r\sqrt{p^2 - U}\right)\, . \label{eq:psi-well-inside}
	\end{align}
\end{subequations}
In the second equality we demanded regularity at $r=0$, which allows us to set $A_1 = -A_2$ and leaving us with a single undetermined constant. Likewise, outside of the potential ($r>R$) we have the free solution
\begin{subequations}
	\begin{align}
		\psi(r) &= \frac{1}{r} \left[ B_1\, e^{-ipr} + B_2\, e^{ipr} \right]\\ &= \frac{B_1}{r} \left[  e^{-ipr} - S_0(p) e^{ipr} \right]\, .\label{eq:psi-well-outside}
	\end{align}
\end{subequations}
This is the asymptotic wave function, so after tucking on the time evolution, $\e^{-iEt}$, we should match it with the integrand in \eqref{eq:total-wave-function} for every momentum.
This straight away allowed us to identify the ratio $-B_2/B_1$ as the spherical scattering amplitude $S_0$. Therefore, $S_0(p)$ and $B_1$, remain as two undetermined constants from the perspective of $r$.

The two solutions \eqref{eq:psi-well-inside} and \eqref{eq:psi-well-outside} can be matched at the boundary $r=R$ by requiring that $\psi(R)$ and its derivative $\psi'(R)$ are continuous. This gives two equations for three total unknowns: $A_2, B_1, S_0$. But the overall normalization is not important to determine the scattering amplitude, so we can just solve for $B_1$ and $S_0$. After some unilluminating algebra, we get the final answer:
\begin{empheq}[box=\graybox]{equation}\label{eq:spherical-amplitude}
	S_0^{\mathrm{well}}(p) = \e^{-2 i p R} \frac{1 + i p R \tanc\left(R\sqrt{p^2 - U}\right)}{1 - i p R \tanc\left(R\sqrt{p^2 - U}\right)}\, ,
\end{empheq}
where $\tanc(x) = \frac{\tan(x)}{x}$. As a quick cross-check, you can verify that \eqref{eq:spherical-amplitude} is a pure phase for real $p$ (because the numerator and denominator are complex conjugates), so the unitarity condition $|S_0^{\mathrm{well}}(p)|=1$ is satisfied regardless of $R$ and $U$. You can also check that when we switch off the potential by sending either $U \to 0$ or $R \to 0$, $S_0^{\mathrm{well}}(p)$ approaches the trivial amplitude $1$.

\subsubsection{Complex poles}

The amplitude \eqref{eq:spherical-amplitude} is a meromorphic function, meaning that it only has isolated poles (one can show that this is a generic property for any potential supported within a finite radius). In fact, it has an infinite number of them. Any pole corresponds to the denominator of \eqref{eq:spherical-amplitude} vanishing so its position can be determined by solving
\be
1 = i p R \tanc \left(R \sqrt{p^2 - U}\right)\, .
\ee
There's no chance we can solve this equation exactly, but let's at least describe qualitatively what happens to the solutions as we tune the parameter $U$. The radius $R$ just sets the overall scale for the problem, so we can work in the units in which $R=1$.

There are two cases depending on whether we're dealing with a potential barrier or a well. In the first case, when $U>0$ and very small, the poles asymptote to $(k+\tfrac{1}{2})\pi - i\infty$ for any integer $k$. As we keep increasing $U$, they approach the real axis away from the origin, more or less like this:
\be
\includegraphics[valign=c]{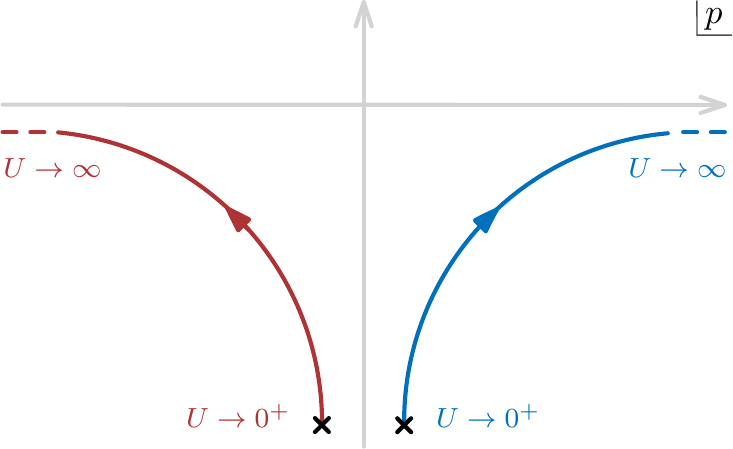}
\ee
Here, we only plotted two representative poles. The bottom line is that all of them are confined to the lower half-plane. In the limit as $U \to \infty$, all the poles escape to infinity and $S_0^{\mathrm{well}}(p)$ approaches the hard-sphere limit recovering the simplest amplitude from \eqref{eq:simplest-amplitude}.

The case of the potential well with $U<0$ is a bit more interesting. When $|U|$ is tiny, the poles asymptote to $k\pi - i\infty$ for integer $k$. As we increase the depth of the well $|U|$, they travel towards the imaginary axis until they meet in pairs at $p = -i$ when some critical depth $U_k < 0$ is reached. Then they shoot off to the upper and lower-half planes when the depth becomes larger. To summarize it all pictorially, a pair of poles moves like this:
\be
\includegraphics[valign=c]{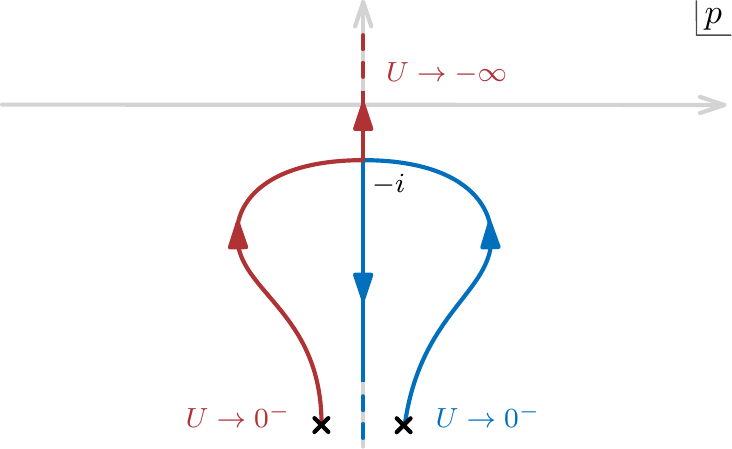}
\ee
The case $k=0$ is special, but it doesn't really matter for our purposes. The result of the analysis is that for any potential well, there will be a finite number of poles in the upper half-plane and an infinite number in the lower half-plane.

Since seeing is believing, let me plot the absolute value $| S_0^{\mathrm{well}}(p)\, \e^{2 i p R}|$ in the $p$-plane with $R=1$ and two values of $U$:
\be\label{eq:U-plots}
\includegraphics[width=0.43\textwidth,valign=c]{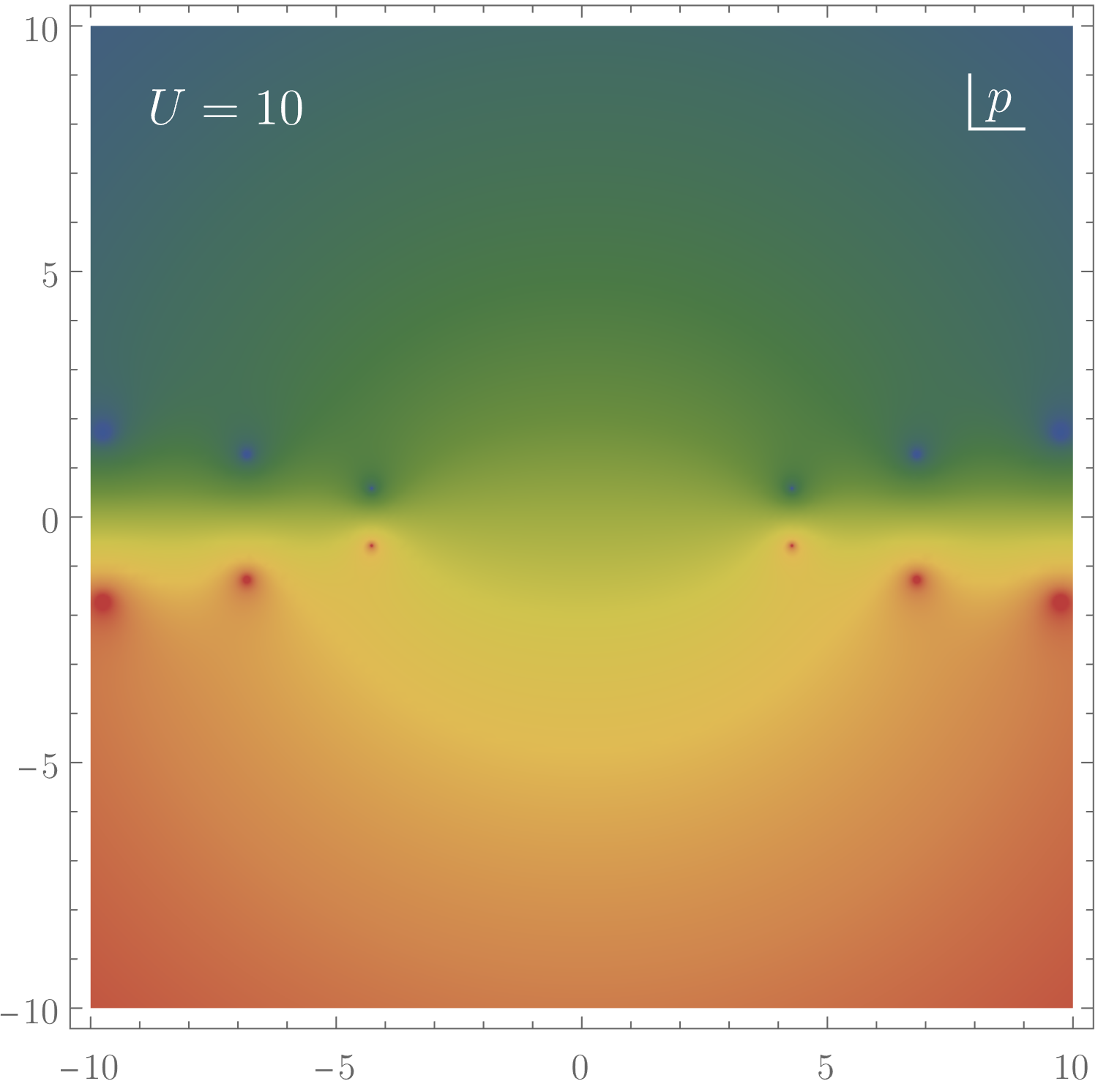}\quad
\includegraphics[width=0.43\textwidth,valign=c]{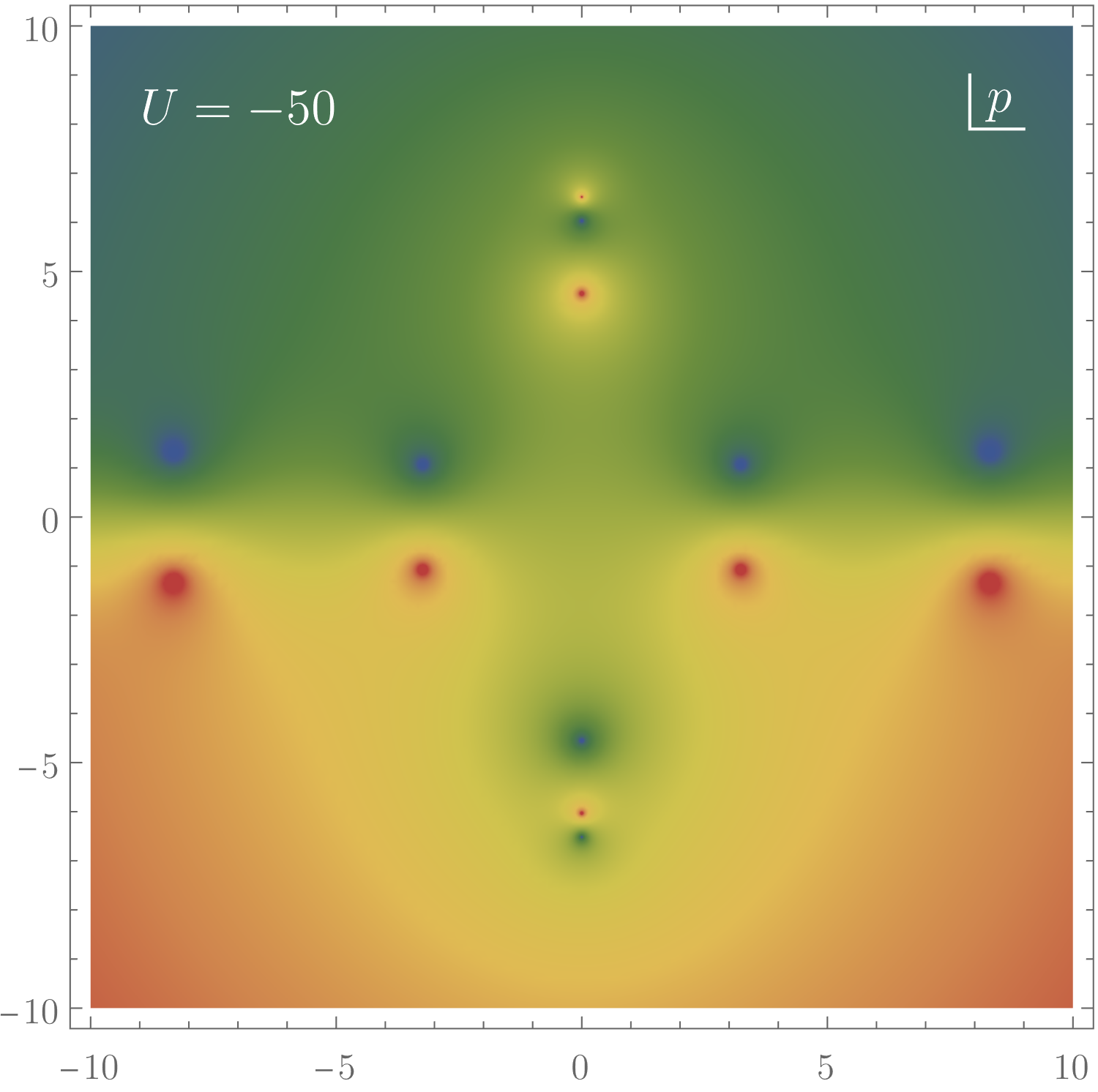}
\ee
Here, warmer colors correspond to larger values, and in particular the red dots are the poles and the blue ones are zeros. This might look a bit shocking after having spent the entire previous lecture showing how a classical amplitude can't have singularities in the upper half-plane if it comes from a causal process. What's going on?

\subsubsection{Total energy branch cut}

This is precisely where the difference between classical and quantum scattering comes in. Recall from Sec.~\ref{sec:barriers-wells} that quantum wave functions evolve in time as $\sim \e^{-i E t}$, while classical ones as $\sim \e^{-ipvt}$. Hence the quantum scattering amplitude naively is a causal transform with respect to the variable $E$, not $p$. We will come back to the question of causality shortly, but for the time being let's understand what happens when we commit to using $E$.

Fortunately, we don't have to do much additional work because $E$ is a square of $p$, as in \eqref{eq:E-p2}. It means that both $p$ and $-p$ map to the same energy $E = \frac{p^2}{2m}$, so there's simply too much information to contain in a single complex $E$-plane. Instead, $S_0(E)$ is defined on two complex planes connected by a branch cut. To get more intuition for it, let's take the second picture in \eqref{eq:U-plots} and see how it looks like in the energy variable:
\be\label{eq:first-second-sheet}
\includegraphics[width=0.43\textwidth,valign=c]{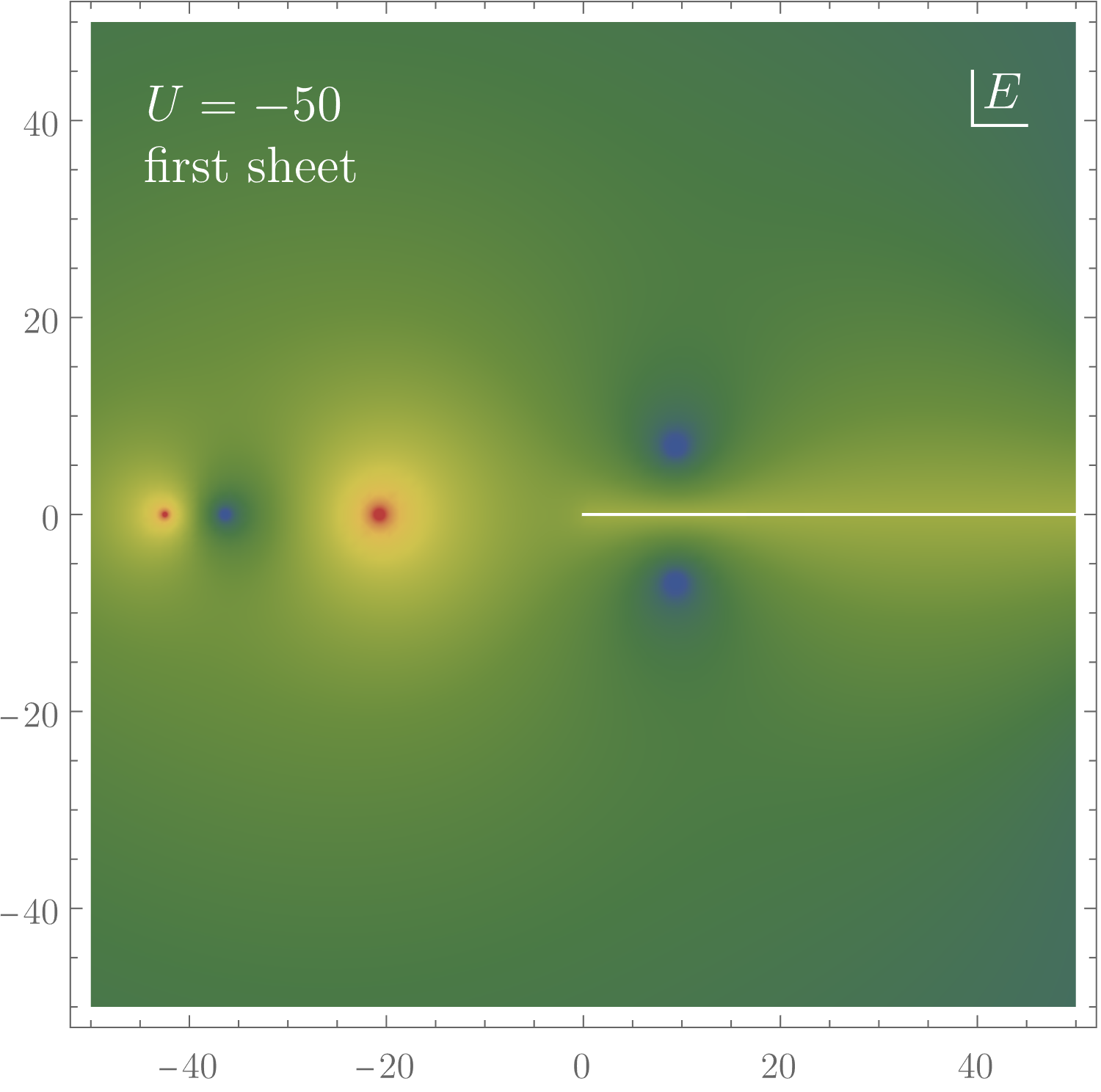}\quad
\includegraphics[width=0.43\textwidth,valign=c]{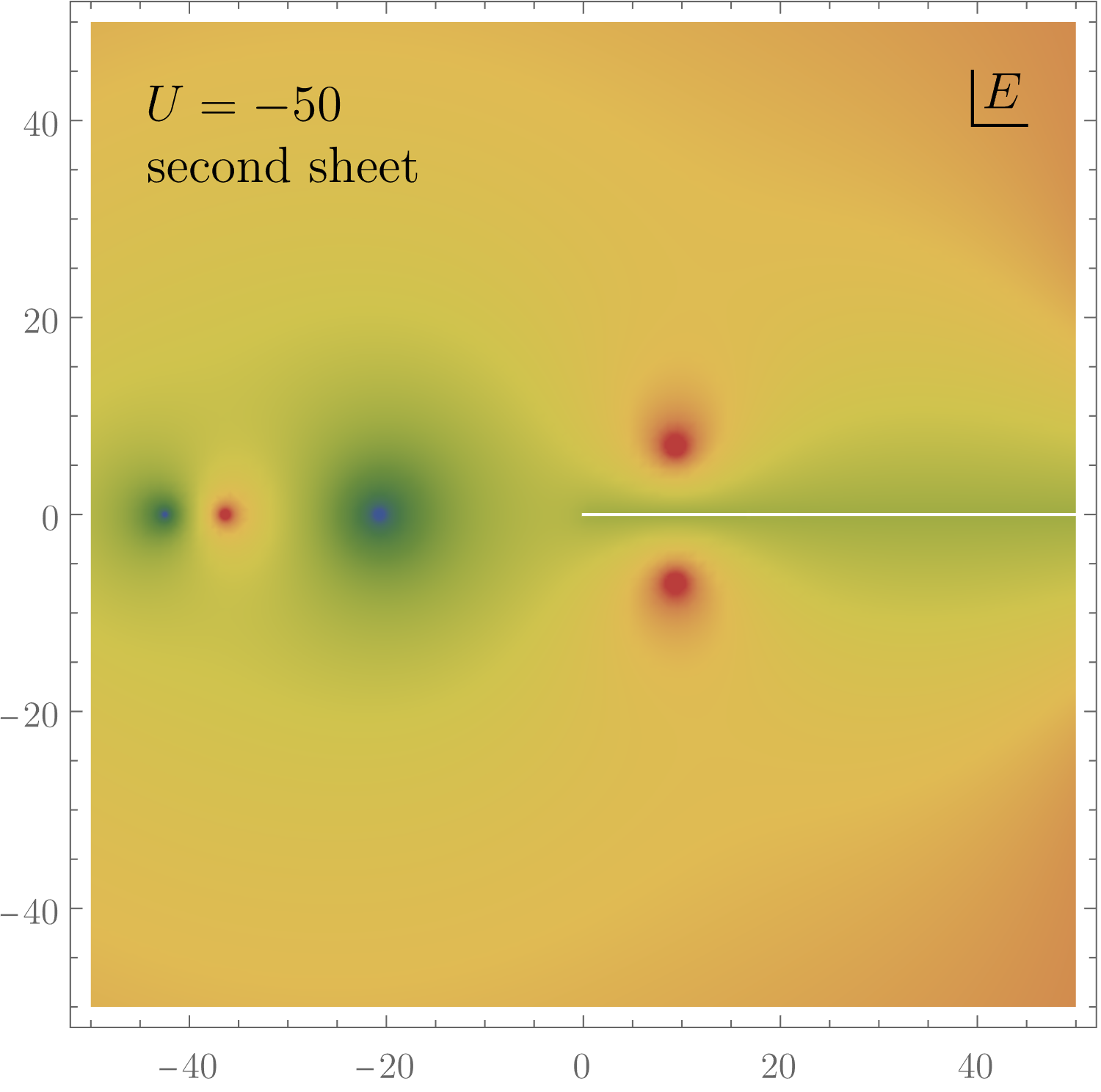}
\ee
For the purpose of plotting, we set $2m=1$.
Essentially, the upper half-plane in $p$ got stretched to the whole plane in $E$ on the first sheet, and likewise the lower half-plane in $p$ into the second sheet in $E$. Sometimes, they are called \emph{physical} and \emph{unphysical} sheet respectively. The two sheets are glued by a \emph{branch cut} at $E \geq 0$, such that approaching it from above on the first sheet, one reemerges from below on the second sheet and so on. You can identify some of the poles and zeros from the right-hand side plot of \eqref{eq:U-plots}: for example, the two poles in the upper half-plane in $p$ are now on the negative real $E$-axis on the first sheet. As promised, there are no poles in the upper half-plane of $E$ on the first sheet, but they do appear on the second sheet.

In practice, the expressions on the two sheets can be obtained by setting $p = \pm i \sqrt{-2mE}$ respectively. Notice that the amplitude \eqref{eq:spherical-amplitude} satisfies $S_0(p) S_0(-p) = 1$ even for complex $p$. Translated to the $E$ variables, this implies
\be\label{eq:S-E-unitarity}
S^{\mathrm{first}}_0(E)\, [S^{\mathrm{second}}_0(E)]^\ast = 1\, ,
\ee
which can be viewed as an analytic continuation of unitarity. It explains why the plots in \eqref{eq:first-second-sheet} look like color inverses of each other. In particular, \eqref{eq:S-E-unitarity} implies that a zero on the first sheet translates to a pole on the second and vice versa.

Note that we had to separate the two sheets by a branch cut, but placing it on the positive real axis was a \emph{choice}. We could've made any other choice without affecting the value of the amplitude. Branch cuts are just mathematical tools that we introduce after trying to cram too much information into a single complex plane. One can say that they are properties of physicists (who choose the variables) more than physics itself.
The specific choice made in \eqref{eq:first-second-sheet} becomes convenient later on. 

At this stage you can go ahead and write down dispersion relations for $S_0(p)$, or equivalently $S_0^{\mathrm{first}}(E)$, as long as you're careful about the extra poles that give additional contributions. We're going to come back to dispersion such relations much later in Sec.~\ref{sec:dispersion-once-again}, after we learn more about analyticity.

\subsubsection{\label{sec:revisiting-causality}Revisiting primitive causality}

Having briefly discussed what happens in the $E$ variable, let's return to the problem of formulating a causality condition. We already called attention to the fact that $E$ is the right variable to think about because it's Fourier-dual to the time $t$. However, the crucial difference to the classical case is that $E$ has to be positive (as opposed to $\omega$ or $p$ that could be either positive or negative). For example, the incoming wave function is
\be\label{eq:psi-in-E}
\Psi_{\mathrm{in}}(r,t) = \frac{1}{r} \int_{0}^{\infty} \left[ \tilde{\psi}_{\mathrm{in}}(E)\, \e^{-i pr} \right] \e^{-iEt}\, \d E\, ,
\ee
where the Jacobian from changing $p$ to $E$ was absorbed into the definition of $\tilde{\psi}_{\mathrm{in}}(E)$.

It's not difficult to see that the result of such a transform can never vanish in a finite time interval. The quickest way to see it is that the term in the square brackets in \eqref{eq:psi-in-E} is a causal transform times a step function $\theta(E)$. So its Fourier transform is a convolution of something with a sharp front proportional to $\theta(t)$ (Fourier transform of $\tilde{\psi}_{\mathrm{in}}(E) \,\e^{-ipr}$) with $\sim \frac{1}{t}$ (Fourier transform of $\theta(E)$). The resulting wave function has to spill out of $t>0$ at least a tiny bit:
\be
\includegraphics[scale=1,valign=c]{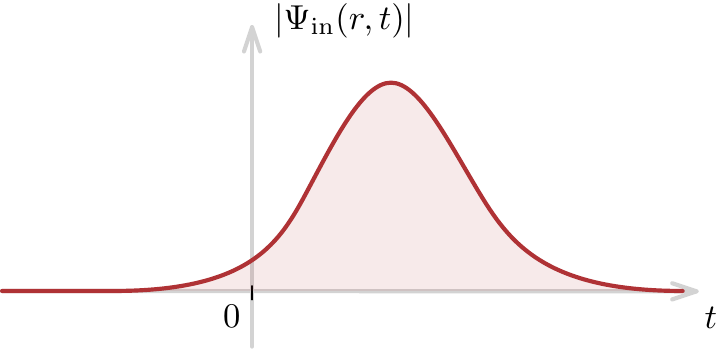}
\ee
Hence no matter how hard we try, we can't construct a wave packet that would probe primitive causality in the same way as in the classical case.

Physically, the problem is just a consequence of the Heisenberg uncertainty principle. Trying to construct a wave packet that switches on in an instant introduces arbitrarily-large energies, which in turn imply instantaneous spreading of the wave function. 
Nevertheless, the amount of acausality introduced by quantum effects can be made arbitrarily small and hence one might expect that there exists a modified set of arguments that would prove the required level of analyticity anyway.

Indeed, one can formulate other notions of what causality is supposed to mean in the quantum case. One of the most intuitive is Van Kampen causality, which is a smeared version of primitive causality. It translates to the statement that probability of finding the total wave packet outside of the scatterer can't exceed $1$ at any given time. We'll use a version of it in later on in Sec.~\ref{sec:time-delays} when we discuss time delays. One can show that this notion of causality leads to analyticity in the upper half-plane of $E$. Moreover, it allows us to prove that the amplitude is also analytic in the upper half-plane of $p$, except for a finite number of poles for attractive potentials. This is certainly in line with our observations above. The derivation is on the lengthier side and not entirely intuitive so we'll skip it.

\subsection{Bound states and resonances}

Let us now try to interpret physically the poles we encountered in the above examples. We formulated the problem as a scattering process, but ultimately everything amounted to solving the Schr\"odinger equation. So it's not too far-fetched to expect that the scattering amplitude also knows about more general solutions for a given potential.

Let's make a distinction between \emph{scattering states}, which look like waves and hence their spatial dependence goes as $\sim \e^{\pm i p r}$. On the other hand, we can have \emph{bound states} which are solutions localized in a region of space and otherwise decay as $\sim \e^{-\lambda r}$ for large $r$ with positive $\lambda$. You might know from quantum mechanics classes that such solutions are difficult to come by and usually occur for only for special (quantized) values of the energy. The cleanest way of distinguishing between the two classes is the sign of their energy: in the first case, it needs to be positive as in \eqref{eq:E-p2}, but in the second
\be
E = - \frac{\lambda^2}{2m} < 0 \, ,
\ee
which is always negative. Another difference is that scattering states aren't normalizable, but bound states are. In principle, we can also have ``runaway'' solutions to the Schr\"odinger equation that explode at infinity, which we'd usually dismiss as unphysical.

Let's say that the scattering amplitude has a simple complex pole at some $p=P$, for example
\be
S_0(p) \approx \frac{p - P^\ast}{p - P}\, .
\ee
Note that a pole doesn't violate any principle we discussed so far. For example, $| \frac{p - P^\ast}{p - P}| = 1$, which is consistent with unitarity. To isolate the contribution from this pole, we can deform the $p$ integration contour in the total wave function \eqref{eq:total-wavefunction} to enclose $p=P$:
\be\label{eq:psi-P}
\Psi_P(r,t) =  \frac{1}{r} \oint_{p = P} \tilde{\psi}_{\mathrm{in}}(p)\, \left[ \e^{-ipr} - \frac{p - P^\ast}{p - P} \e^{ipr}\right] \e^{- iE t}\, \d p\, .
\ee
Notice that the pole is precisely needed to select $\e^{iPr}$ over $\e^{-iPr}$. Ignoring the overall constants, the resulting wave function behaves as
\be\label{eq:psi-bound-state}
\Psi_P(r,t) \sim \e^{i (P r - \frac{P^2 t}{2m})}\, .
\ee
From here, we can read off the conditions for a bound state. We need that the wave function decays spatially and its energy is negative, which means
\be
\Im\, P > 0 \quad \mathrm{and}\quad P^2 < 0\, .
\ee
The second inequality implies $\Im P^2 = 2 (\Im P)(\Re P) = 0$, so $\Re P$ needs to vanish.
As expected, poles on the positive imaginary axis in $p$ (real negative axis on the physical sheet of $E$) correspond to bound state solutions with $P = i\lambda$.

The reason why a bound state gives rise to a singularity in $S_0(p)$ is needed precisely for a non-normalizable scattering state to become normalizable. In \eqref{eq:psi-P} it was responsible for filtering out the unwanted incoming wave contribution $\e^{-iPr} = \e^{\lambda r}$ that would spoil normalizability.

What about the poles in the lower half-plane of $p$? Let's write their positions as $P = P_0 - i \Gamma$ with positive $\Gamma$. The energy associated to a state with momentum $P$ is complex:
\be
E = \frac{P_0^2 - \Gamma^2}{2m}  -i \frac{P_0 \Gamma}{m}\, .
\ee
The temporal dependence of the wave function that it produces is $\Psi \sim \e^{-i E t}$, so for a solution that doesn't blow up in the infinite future, we need $\Im\, E < 0$, or equivalently $P_0 > 0$ by the above equation. These are the poles in the forth quadrant of $p$ (or the lower half-plane of $E$ on the unphysical sheet). They correspond \emph{resonances} or \emph{unstable states} that decay away with time. Their typical lifetime is $\sim \frac{m}{P_0 \Gamma}$. The quantity $\frac{2P_0 \Gamma}{m}$ is called the decay width of the state. For a physical interpretation we also need to demand that $\Gamma$ is much smaller than $P_0$ so that before decaying, the wave function looks like a state with a positive energy. It corresponds to the poles right below the positive $p$-axis (or right below the positive $E$-axis on the second sheet).

To summarize, poles in the complex $p$-plane can be labeled as follows:
\be
\includegraphics[scale=1.05,valign=c]{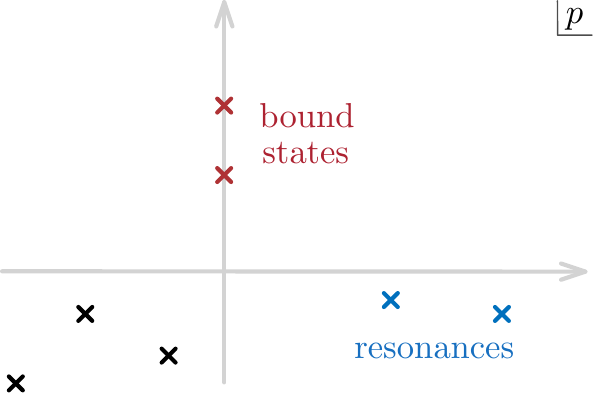}
\ee
All the other poles in the lower half-plane are also solutions of the Schr\"odinger equation, but become more and more difficult to interpret physically the farther away we get from the real axis. For example, the poles right below the negative real axis, sometimes referred to as \emph{capture states}, correspond to wave functions that increase in time exponentially.

\subsection{\label{sec:time-delays}Time delays}

As we've seen in Sec.~\ref{sec:revisiting-causality}, the notion of causality becomes more difficult to formulate in quantum theory, essentially because of the uncertainty principle. Our next goal would be to understand yet another physical formulation of causality. It states that the incoming wave has to first reach a given point in space before the scattered (outgoing) wave has a chance to respond to it.

\subsubsection{Stationary phase approximation}

We need a way to measure this time delay. One choice is to take the incoming wave to be a narrow wave packet $\tilde{\psi}_{\mathrm{in}}(E)$ centered around a specific energy $E_0$. In other words, we have
\be\label{eq:psi-stationary}
\Psi_{\mathrm{in}}(r,t) = \frac{1}{r} \int_{0}^{\infty} |\tilde{\psi}_{\mathrm{in}}(E)|\, \e^{i\alpha(E)}\, \e^{-i(pr + E t)}\, \d E\, ,
\ee
where we separated the absolute value of the wave packet $|\tilde{\psi}_{\mathrm{in}}(E)|$ from its phase $\e^{i\alpha(E)}$.

The integrand of \eqref{eq:psi-stationary} is an oscillating function. The biggest contribution arises from the energies for which its phase is stationary, which is where we have constructive interference. It's obtained by varying the phase with respect to the energy $E$ close to some $E=E_0$ where most of the wave is supported:
\be\label{eq:t-in}
\alpha'(E_0) -  \frac{m}{p_0}r - t_{\mathrm{in}} = 0\, .
\ee
At a given radius $r$, the narrow wave packet centered around $E_0$ arrives at the time $t_\mathrm{in}$ determined by the above equation. In other words, the wave packet travels with speed $v_0 = \frac{p_0}{m}$, where $E_0 = \frac{ p_0^2}{2m}$. We assume that $v_0 >0$ so the wave is really traveling inwards.

The strategy is now clear. We will assume that the outgoing wave hasn't spread out too much so that we can still talk about its well-defined center. Writing the spherical scattering amplitude as $S_0(E) = \e^{2i \eta_0(E)}$, we have
\be
\Psi_{\mathrm{out}}(r,t) = -\frac{1}{r} \int_{0}^{\infty} |\tilde{\psi}_{\mathrm{in}}(E)|\, \e^{i\alpha(E) +2i \eta_0(E)}\, \e^{-i(-pr + E t)}\, \d E\, .
\ee
We can apply the stationary phase method once again to get
\be\label{eq:t-out}
\alpha'(E_0) + 2 \eta_0'(E_0) + \frac{m}{p_0}r - t_{\mathrm{out}} = 0\, ,
\ee
which describes a wave packet traveling with opposite velocity $-v_0$, but shifted in time. 
The time delay $\Delta t_0$ is then obtained by comparing how much slower the outgoing wave packet emerges compared to the vacuum case:
\begin{subequations}
	\label{eq:time-delay}
	\begin{align}
		\Delta t_0(E_0) &= t_{\mathrm{out}} - t_{\mathrm{in}} - \frac{2m}{p_0} r\\
		&= 2 \eta_0'(E_0)   \, .
	\end{align}
\end{subequations}
Note that it depends on the specific energy $E_0$ at which we probe it. The subscript ${}_0$ denotes the fact we're looking at the $j=0$ time delay. As we'll see later on, higher angular momenta can have different delays.

\subsubsection{Simple examples}

Let's try to get some intuition for the time delay. In practice, it's easier to use the momentum derivative
\be\label{eq:eta-prime}
\eta_0'(p) = \frac{1}{2i} \frac{\partial \log S_0(p)}{\partial p} = \frac{p}{2m} \Delta t_0\, ,
\ee
which equals $\Delta t_0$ up to a positive constant.
In the absence of scatterer, $\eta_0(p) = 0$ and there's no time delay, as expected. The simplest non-trivial case is the hard sphere from Sec.~\ref{sec:simplest-amplitude}. It has $\eta_0^{\mathrm{HS}}(p) = -pR$ giving
\be\label{eq:eta-prime-hard-sphere}
\eta_0'^{\,\mathrm{HS}}(p) = - R\, 
\ee
or equivalently, $\Delta t_0^{\mathrm{HS}} = -\frac{2Rm}{p}$.
This is of course expected: bouncing off the surface at $r=R$, the wave skips the distance $2R$ that it would've traveled in free space (one $R$ on the way to the center and another on the way back to the surface). Traveling at the speed $\frac{p}{m}$ gives the above time advance.

One could expect that for a scatterer contained within radius $R$, \eqref{eq:eta-prime-hard-sphere} sets the largest time advance allowed by causality. This intuition would be correct classically. But quantum mechanically, we could expect corrections to at order of the wavelength $\sim \frac{\hbar}{E}$, where the classical intuition breaks down.

As a matter of fact, we can consider a simple generalization of the hard sphere amplitude, where
\be
S_0(p) = \e^{-2ipR}\, \frac{p - P^\ast}{p - P}\,
\ee
for some constant $P$. Plugging it into \eqref{eq:eta-prime} for real $p$ gives
\be\label{eq:eta-prime-pole}
\eta_0'(p) = - R - \frac{\Im\, P}{|p - P|^2}\, .
\ee
Hence we see that a simple complex pole changes the time delay. Bound state poles with $\Im\, P > 0$ make it more negative. The largest time advance at momentum $p$ is obtained for a pole at $P = i p$. Nevertheless, \eqref{eq:eta-prime-pole} still can't be arbitrarily negative. This motivates looking for a sharp lower bound.

\subsubsection{Wigner's causal inequality}

We're still lacking a precise lower bound on the time delay, so now we need to sit down and do a more precise calculation. The strategy is going to be estimating how much time $T_\mathrm{free}$ a given wave packet spends within the radius $r \leq R$ in free space and compare it with the same quantity in the presence of the scatterer, $T_\mathrm{scat}$. The difference of the two will be the expected time delay $\Delta t_0$. The fact that $T_\mathrm{scat}$ can't be negative is going to yield a lower bound on the time delay.

Making this statement quantitative will require a small technical computation. If you're just interested in the result, you can easily skip straight to \eqref{eq:Wigner-inequality}.

The probability density of finding a particle at time $t$ within the scatterer is
\be\label{eq:P-t}
4\pi \int_{0}^{R} r^2\, \d r\, |\Psi(r,t)|^2 = 1 - 4\pi \int_{R}^{\infty} r^2\, \d r\, |\Psi(r,t)|^2\, .
\ee
The volume element gave $4\pi$ from the angular part and the remaining $r^2\, \d r$ we ought to integrate over. Out of the two equivalent expressions, the latter (outside the scatterer) is the one we want to use because it can be related to the scattering amplitude. The time delay is the difference between \eqref{eq:P-t} and the same quantity evaluated with the free wave function $\Psi_{\mathrm{free}}$ (with no scatterer, so $S_0=1$) integrated over all times:
\be
\Delta t_0 = -4\pi \int_{-\infty}^{\infty} \d t \int_{R}^{\infty} r^2\, \d r\, \Big( |\Psi(r,t)|^2 -  |\Psi_{\mathrm{free}}(r,t)|^2 \Big)\, ,
\ee
Plugging in the definition \eqref{eq:total-wavefunction} with $S_0(E) = \e^{2i \eta_0(E)}$, it becomes
\begin{align}\label{eq:Delta-t-0}
	&\Delta t_0 = 4\pi \int_{-\infty}^{\infty}\d t \int_{R}^{\infty} \d r \int_{0}^{\infty} \d E \int_{0}^{\infty} \d E'\, \tilde{\psi}_{\mathrm{in}}(E)\, \tilde{\psi}^\ast_{\mathrm{in}}(E')\, \e^{-i(E-E')t}\\
	&\times \left[ \left(1 - \e^{2i (\eta_0(E) - \eta_0(E'))}\right) \e^{i(p-p')r} + \Big(\e^{2i \eta_0(E)}-1\Big) \e^{i(p+p')r} + \left(\e^{-2i \eta_0(E')}-1\right) \e^{-i(p+p')r} \right] .\nn
\end{align}
The first step is to integrate out $r$. For this purpose, we can use the identity
\be
\int_{R}^{\infty} \d r \, \e^{i (p - p') r} = \frac{i}{p - p'} \e^{i (p - p') R} = \frac{i}{2m} \frac{p + p'}{E - E'} \e^{i (p - p') R}
\ee
and analogous ones for the other terms. In the next step, we wish to integrate out $t$, which leads to the Dirac delta function $2\pi \delta(E - E')$. But we have to be a tiny bit careful with the first term in the square brackets in \eqref{eq:Delta-t-0}, which naively leads to a $0/0$ situation. To resolve it, let's expand the dangerous terms around $E = E'$ first:
\be
\left(1 - \e^{2i (\eta_0(E) - \eta_0(E'))}\right) \frac{p + p'}{E - E'} = -4i p\, \eta_0'(E) + {\cal O}(E - E')\, .
\ee
This is precisely how the derivative of the phase shift comes about in this derivation. Putting everything together gives
\be\label{eq:Delta-t-0-final}
\Delta t_0 =  \frac{16\pi^2}{m} \int_{0}^{\infty} |\tilde{\psi}_{\mathrm{in}}(E)|^2 \left( \eta_0'(E) + m\frac{\sin(2p R) - \sin[2(pR+\eta_0)]}{2p^2} \right) p\, \d E\, .
\ee
This is a more precise version of the relation \eqref{eq:time-delay}.

Recall that $\Delta t_0 = T_{\mathrm{scat}} - T_{\mathrm{free}}$, so if we can also compute $T_{\mathrm{free}}$, the constraint $T_{\mathrm{scat}} \geq 0$ will give as a lower bound on $\eta_0'(E)$. This computation follows the same lines as what we've just calculated above and gives:
\begin{subequations}
	\begin{align}
		T_{\mathrm{free}} &= 4\pi \int_{-\infty}^{\infty} \d t \int_{0}^{R} r^2\, \d r\, |\Psi_{\mathrm{free}}(r,t)|^2\\
		&= 16\pi^2 \int_{0}^{\infty} |\tilde{\psi}_{\mathrm{in}}(E)|^2 \left( \frac{R}{p} - \frac{ \sin(2pR)}{2p^2} \right) p\, \d E\, .
	\end{align}
\end{subequations}
The second term in the brackets cancels in the second term in \eqref{eq:Delta-t-0-final}. The inequality we obtain is
\be
T_{\mathrm{scat}} =  \frac{16\pi^2}{m} \int_{0}^{\infty} |\tilde{\psi}_{\mathrm{in}}(E)|^2 \left( \eta_0'(E) + \frac{mR}{p} - \frac{1}{4E} \sin[2(pR+\eta_0)] \right) p\, \d E \geq 0\, .
\ee
Since this inequality has to be true for an arbitrary choice of the wave packet $\tilde{\psi}_{\mathrm{in}}(E)$, the term in the bracket is non-negative. In addition, the sine function is lower-bounded by $-1$, so we get
\be
\hbar\, \eta_0'(E) \geq -\frac{m R}{p} - \frac{\hbar}{4E}\, .
\ee
This is the Wigner's \emph{causal inequality}. It might be more intuitive to express it in terms of the derivative with respect to the momentum instead of energy:
\begin{empheq}[box=\graybox]{equation}\label{eq:Wigner-inequality}
	\hbar\, \eta_0'(p) \geq - R - \frac{\hbar}{2p}\, .
\end{empheq}
We reinstated $\hbar$ to emphasize the difference between the two terms on the right-hand side.
The first term is the contribution we expected to see classically and corresponds to the time advance we'd get if the scattering wave simply bounced off the scatterer. The second term is the quantum correction. At high energies, it doesn't have a big effect since the incident wave hardly enters the scatterer, but it becomes more and more relevant at momenta of order the wavelength.

In the end, we have found that the scattering amplitude from \eqref{eq:eta-prime-pole} saturates the bound for a given fixed momentum $p_0$ if we set $P = ip_0$. Actually, you might ask what would be wrong with saturating the inequality \eqref{eq:Wigner-inequality} for all energies simultaneously. Integrating it would lead to the amplitude $S_0(p) = C \e^{-2ipR} p^{-i}$ and there doesn't exist a constant $C$ that would make it unitary across all momenta $p$.

\subsubsection{\label{sec:back}Back to the potential well and barrier}

To finish off the discussion on time delays, let's look back at the scattering amplitude \eqref{eq:spherical-amplitude} for the potential well and barrier. Here, we can compute explicitly
\be\label{eq:time-delay-U}
{\eta'^{\,\mathrm{well}}_0}(p) = \frac{UR}{p^2 - U} \frac{1- \tanc(R\sqrt{p^2 - U})}{1 + p^2 R^2 \tanc^2(R\sqrt{p^2 - U})}\, .
\ee
Let us plot this quantity for a few different values of $U$ in the units of $R=1$:
\be
\includegraphics[scale=1.1,valign=c]{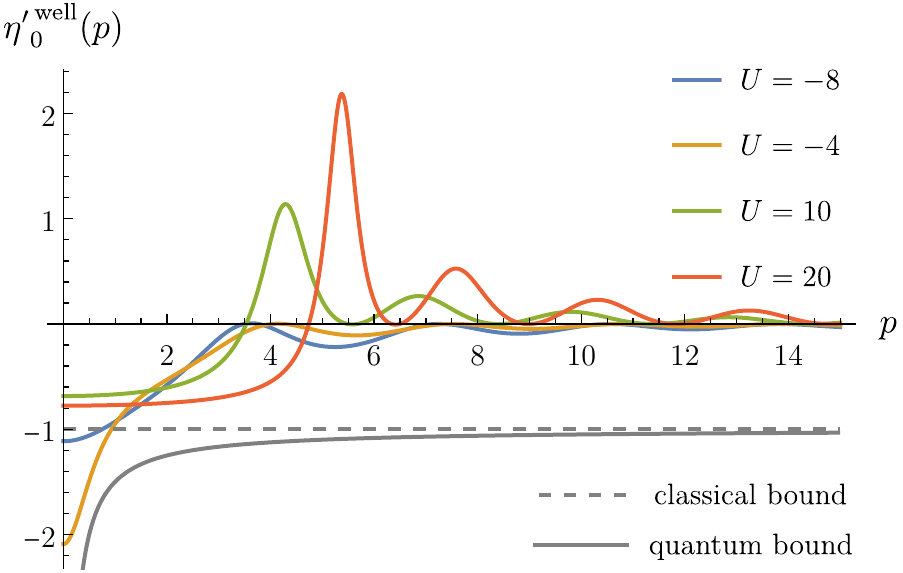}
\ee
The dashed gray line represents the classical bound $-R$. The lines for the potential well with $U<0$ both dip underneath it for small momenta, but still remains above the Wigner bound \eqref{eq:Wigner-inequality} indicated by the solid gray line. For potential barriers with $U>0$, the time delay remains above the classical bound. In both cases, at sufficiently high momenta $p$, the time delay decays to zero since scattering becomes less and less affected by the presence of the potential. At small $p$, the delay approaches a constant ${\eta'^{\,\mathrm{well}}_0}(0) = R [-1 + \tanc(R\sqrt{-U})]$. From here, we see that spherical wells can introduce arbitrarily large time delays or advances, but ${\eta'^{\,\mathrm{well}}_0}(0) \geq -R$ for spherical barriers. One can show that \eqref{eq:time-delay-U} respects Wigner's causal inequality for all energies.

\subsection{\label{sec:higher-angular-momenta}Higher angular momenta}

We've more or less exhausted the list of things we can learn from the spherical-wave scattering, so it's time to graduate to more realistic scenarios. A natural next step is to consider higher angular momenta $j$. The bottom line is that everything generalizes rather straightforwardly, but we need to involve some special functions to make it quantitative.

The generalizations of spherical incoming and outgoing waves for fixed momentum $p$ are 
\begin{subequations}\label{eq:psi-higher-j}
	\begin{align}
		\psi_{\mathrm{in}} = \frac{\e^{-ipr}}{r} &\quad\rightsquigarrow\quad -ip\, h_{j}^{(2)}(pr) P_j(\cos \theta)\,,\\
		\psi_{\mathrm{out}} = -S_0(p)\, \frac{\e^{ipr}}{r} &\quad\rightsquigarrow\quad -ip\, S_j(p)\, h_{j}^{(1)}(pr) P_j(\cos \theta)\, ,
	\end{align}
\end{subequations}
where the dependence on the radial and angular parts is factorized.
The factors $h_j^{(1,2)}$ are known as the spherical Hankel functions of the first and second kind. Explicitly, they are given by
\be\label{eq:spherical-Hankel}
h_j^{(1,2)}(pr) = \frac{\e^{\pm i p r}}{\pm i p r} \sum_{k=0}^{j} \frac{(\pm i)^{k-j}}{k!} \frac{(j+k)!}{(j-k)!} (2 p r)^{-k}\, .
\ee
The details are not terribly important. The main point is that from large distances away, $r\to \infty$, they are dominated by the $k=0$ term and hence look like spherical waves, up to a phase. So the total wave function $\psi = \psi_{\mathrm{in}} + \psi_{\mathrm{out}}$ looks asymptotically like
\be\label{eq:partial-wave-function}
\psi \sim \frac{(-i)^j}{r} \left[(-1)^{j} \e^{-ipr} - S_j(p)\,  \e^{ipr} \right] P_j(\cos \theta)\, .
\ee
In \eqref{eq:psi-higher-j}, the angular dependence is captured by the Legendre polynomial
\be\label{eq:Legendre-Rodrigues}
P_j(x) = \frac{1}{2^j j!} \frac{\partial^j}{\partial x^j} (x^2 - 1)^j\, .
\ee
They are special cases of spherical harmonics, equal to $Y_j^0(\theta,\phi)$ up to a constant, simplifying because our problem doesn't depend on the azimuthal angle $\phi$. For example, $P_0(x) = 1$, $P_1(x) = x$, $P_2(x) = (3x^2 - 1)/2$, and so on.

To summarize, the most general form of the wave packets scattering off a spherically-symmetric potential is given by
\be
\Psi_j (r,\theta,t) = P_j(\cos \theta) \int_0^{\infty} \tilde{\psi}_{\mathrm{in}}(E) \left[  h_j^{(2)}(pr) +  S_j(p)\, h_j^{(1)}(pr) \right] \e^{-iEt}\, \d E\, .
\ee
Let's call the radial part of the wave function, the term in the square brackets, $\psi_j(p,r)$ for fixed momentum $p$. It's also known as the partial wave function. In the literature, $j$ is also often referred to as the \emph{partial-wave spin}, but it shouldn't be confused with the intrinsic spin of particles. The Schr\"odinger equation for this quantity translates to
\be\label{eq:j-Schrodinger-equation}
\left[ -\nabla^2 + \frac{j(j+1)}{r^2} + U - p^2 \right] \psi_j(p,r) = 0\, ,
\ee
where as before we rescaled the potential with $V(r) = \frac{\hbar^2}{2m}U(r)$.
Compared to the scalar case, we have an extra term behaving as $\sim \frac{j(j+1)}{r^2}$, which essentially acts as a repulsive potential barrier at low radii. This is the centrifugal force making it difficult for the wave function to stay near the center when $j$ is large.

\subsubsection{Spherical well and barrier back again}

You can go ahead and solve the Schr\"odinger equation for the potential from \eqref{eq:potential-V} and compute the higher-$j$ scattering amplitudes. The derivation doesn't bring anything new to the table, so let me just quote the final answer:
\begin{empheq}[box=\graybox]{equation}\label{eq:S-j-spherical}
	S^{\mathrm{well}}_j(p) = -\frac{\beta\, \j_j(\alpha) {h_j'}^{(2)}(\beta) - \alpha\, \j_j'(\alpha) h_j^{(2)}(\beta)  }{\beta\, \j_j(\alpha) {h_j'}^{(1)}(\beta) - \alpha\, \j_j'(\alpha) h_j^{(1)}(\beta)  }\, ,
\end{empheq}
where $\alpha = \sqrt{p^2 - U} R$ and $\beta = pR$. Primes denote derivatives. Here, $\j_j = \frac{1}{2} (h_j^{(1)} + h_j^{(2)})$ is called the spherical Bessel function of the first kind. The expression might look intimidating, but after plugging in the definition \eqref{eq:spherical-Hankel}, it always evaluates to rational functions of $\alpha$ and $\beta$.

All lessons we've learned about $S_0(p)$ also hold for $S_j(p)$. In the above example, by solving for the positions of poles of \eqref{eq:S-j-spherical}, one can show that they only occur either in the lower-half plane of $p$ or on the real positive axis, just as in the $j=0$ case. Since $h^{(2)}_j$ is the complex conjugate of $h^{(1)}_j$, we see immediately that
\be\label{eq:S-j-unitarity}
|S_j(p)| = 1
\ee
for $S_j^{\mathrm{well}}(p)$ with real momenta $p$. This is a general result for finite-range potentials. As mentioned in Sec.~\ref{sec:revisiting-causality}, one can also derive results in the upper half-plane, if one is careful about a possible finite number of poles there. For example, it's known that for any potential supported within radius $R$, we can put the high-energy bound:
\be
| S_j(p) e^{2ipR}| \leq 1
\ee
asymptotically in the upper half-plane $|p| \to \infty$ with $\Im\, p > 0$. There are various other statements we won't have time to delve into.

This is around the time when it stops being immediately useful to stare at the expression \eqref{eq:S-j-spherical}, so to understand its meaning let's just plot it instead. A couple of examples are:
\be
\includegraphics[width=0.43\textwidth,valign=c]{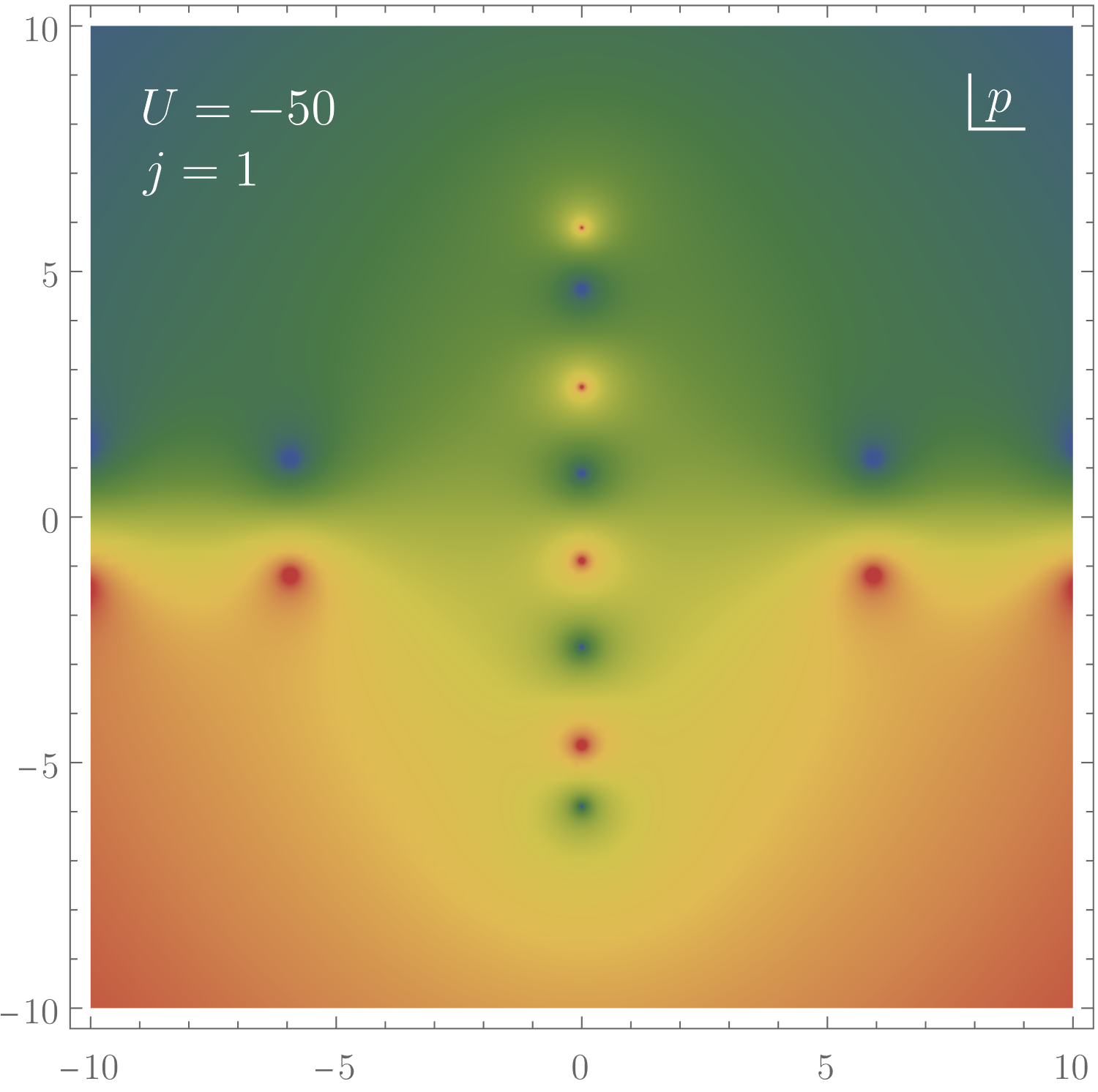}\quad
\includegraphics[width=0.43\textwidth,valign=c]{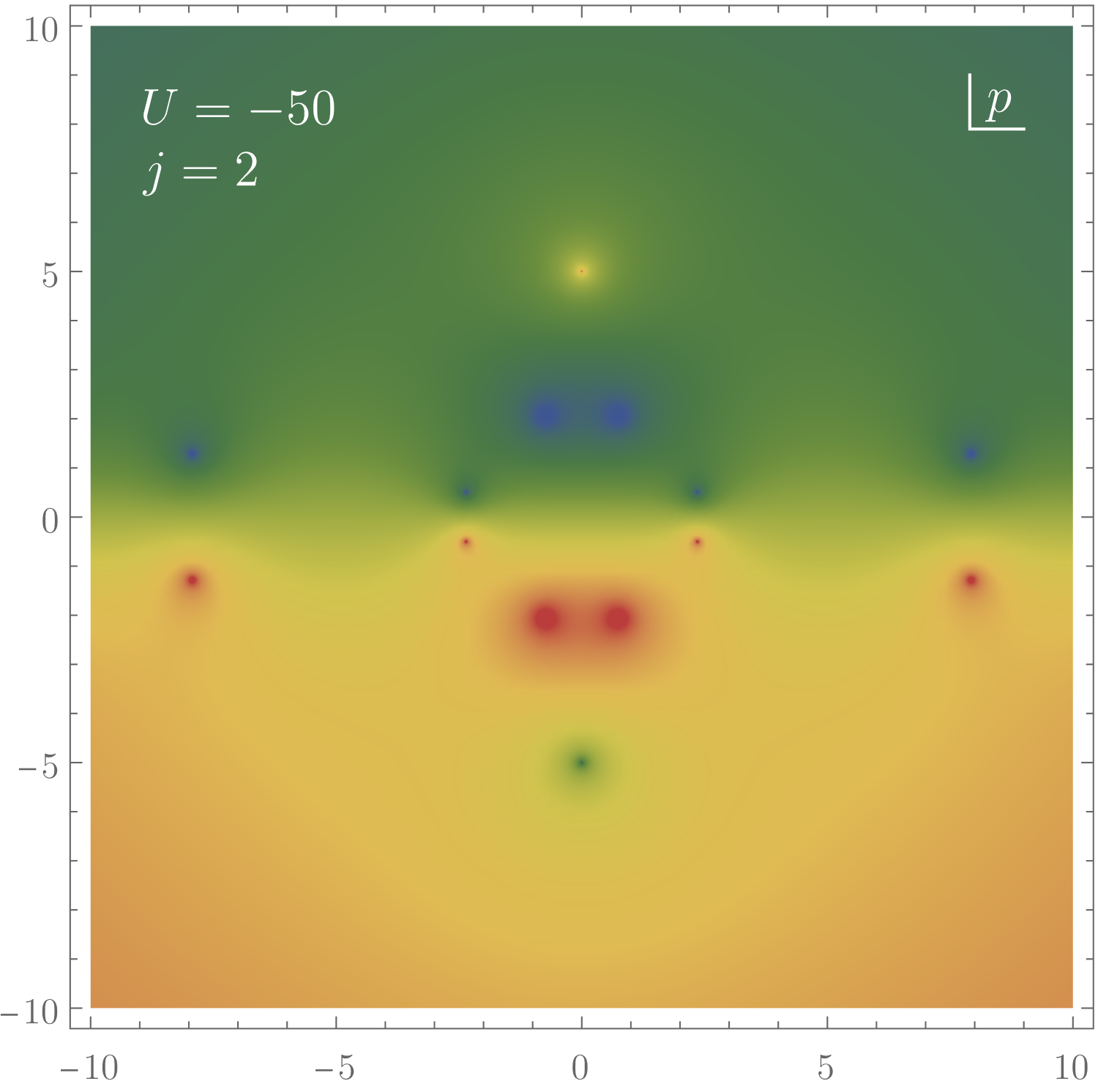}
\ee
As before, we plotted the absolute value $| S_j^{\mathrm{well}}(p)\, \e^{2 i p R}|$. Notice how the poles in the upper half-plane start disappearing as we increase $j$. This happens because of the centrifugal barrier we talked about, which makes it more and more difficult for bound states to be viable. Later in Sec.~\ref{sec:Regge-trajectories}, we'll find that bound states for different values of $j$ are actually all linked together.

To have another handle on the amplitude, let us take the limit $U \to \infty$, which correspond to the hard sphere scattering. In this case the terms proportional to $\alpha$ dominate and the expression simplifies to
\be
\lim_{U \to \infty} S_j^{\mathrm{well}}(p) = - \frac{h_j^{(2)}(pR)}{h_j^{(1)}(pR)}\, ,
\ee
which according to \eqref{eq:spherical-Hankel} is just $\e^{-2i p R}$ times a ratio of degree-$j$ polynomials in the momentum.

For future reference, it will also be convenient to introduce
\be\label{eq:partial-waves}
S_j(p) = 1 + 2i f_j(p)\, ,
\ee
which decomposes the amplitude into its interacting part $f_j(p)$ and free propagation. The factor of $2i$ is just a convention that will simplify some formulas later on. The $f_j$'s are often called \emph{partial wave amplitudes} and they share a lot of analytic features with $S_j$. In terms of higher-spin phase shift $\eta_j$, they can be expressed as $f_j = \frac{1}{2i}(\e^{2 i \eta_j}-1) = \e^{i \eta_j} \sin \eta_j$. Here's how they look like plotted as a function of $p$ with $U=10$, $R=1$ and different values of $j$:
\be\label{fig:TJ}
\includegraphics[scale=1.1,valign=c]{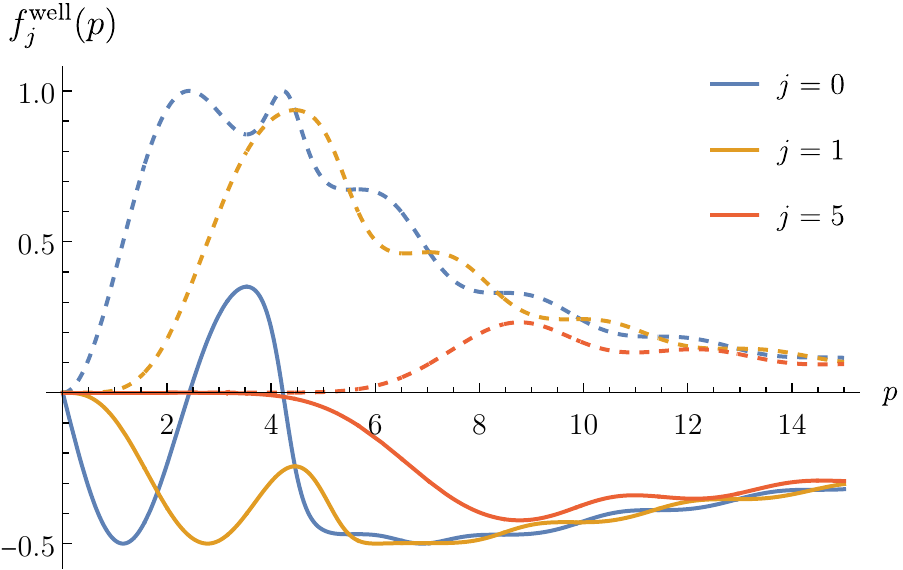}
\ee
The solid and dashed lines are the real and imaginary parts of $f_j^{\mathrm{well}}(p)$ respectively. Notice that the higher $j$, the slower the corresponding partial amplitude grows at low energies (in fact as $\sim p^{2j+1}$) and only gives significant contributions at higher momenta. We'll come back to this point shortly.

You might also notice that the imaginary part is always positive and bounded. This is not an accident. Expanding the unitarity constraint \eqref{eq:S-j-unitarity} in terms of $f_j$, we find
\be\label{eq:Im-fj}
\Im\, f_j(p) = |f_j(p)|^2\, .
\ee
for real $p$. The right-hand side is an absolute value, so it's obviously non-negative and unitarity also implies that it's bounded by $1$. From here we learn that $\Im f_j$ must satisfy
\be\label{eq:Im-fj-bound}
0 \leq \Im\, f_j(p) \leq 1
\ee
for all real momenta $p$.

\subsubsection{\label{sec:low-spin}Dominance of low partial-wave spin}

We can qualitatively understand the structure of peaks in \eqref{fig:TJ} as follows. A wave function with momentum $p$ interacting with the scatterer at radius $r$ contributes to the angular momentum approximately $j \sim p r$. Since the scatterer is compact, $r \leq R$, we'd expect partial waves with momentum $j$ to matter roughly when
\be\label{eq:p-j-R-inequality}
p \gtrsim \frac{j}{R}\, .
\ee
This is indeed the pattern we observe in \eqref{fig:TJ}. Increasing $j$ delays the peaks of the curves until \eqref{eq:p-j-R-inequality} is reached. To illustrate this point better, we plotted $f_{j=5}^{\mathrm{well}}(p)$, which is almost exactly zero until $p \approx 5$ where it switches on.

Another way to understand this result is by comparing the terms $\frac{j(j+1)}{r^2}$ and $p^2$ on the left-hand side of effective Schr\"odinger equation in \eqref{eq:j-Schrodinger-equation}. The angular momentum term acts as a potential barrier, such that small momenta $p \ll \frac{j}{R}$ don't really affect the process. It's only after \eqref{eq:p-j-R-inequality} is reached that non-trivial scattering can happen.

As a matter of fact, large spins $j$ are more than exponentially suppressed at any given momentum $p$. It's surprisingly tricky to show this directly from the representation \eqref{eq:S-j-spherical} of the spherical barrier/well amplitude. For our purposes, it will suffice to illustrate this point on a plot, for example $[\Re \log f_j^{\mathrm{well}}(p)]/j$ as a function of $j$ for any fixed $p$. Taking $U=10$ and $R=1$, we have
\be\label{eq:large-spin}
\includegraphics[valign=c,scale=1.1]{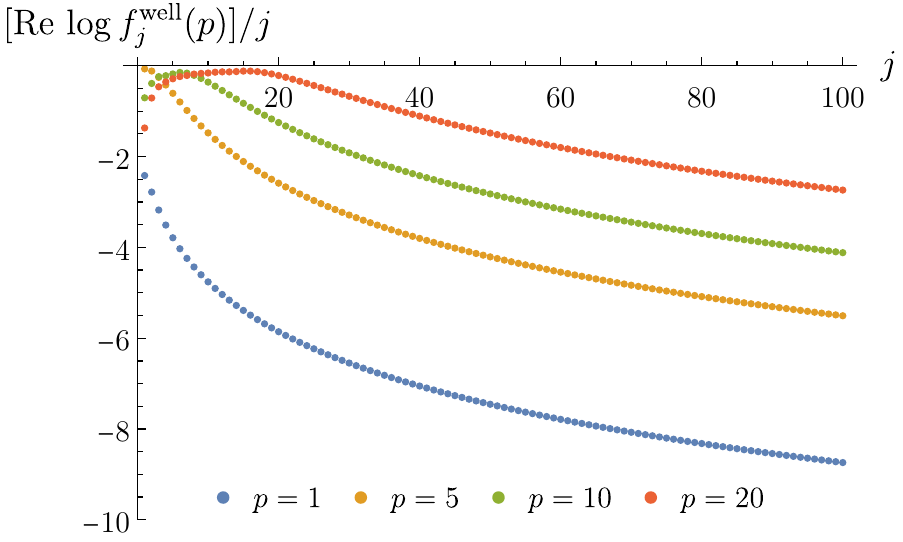}
\ee
A straight horizontal line on this plot would correspond to exponential suppression $\sim \e^{-c j}$ for some constant $c$. The fact that all the data points decay faster than that, indicates $f_j^{\mathrm{well}}$ decay faster than exponential for large $j$. As a matter of fact, it's possible to show that $f_j^{\mathrm{well}}$ decays asymptotically as $\e^{-2j \log j}$.

In general, one can show that if a scatterer is compact (contained within a finite region of space), higher partial waves are suppressed with the angular momentum more than exponentially. In the next section, we'll learn that this fact has a direct reflection in the analytic properties of scattering amplitudes. Note also that if you squint your eyes, the data points in \eqref{eq:large-spin} look like they align themselves along smooth curves. We'll see in Sec.~\ref{sec:Regge-theory} that, indeed, sometimes it becomes useful to think about $j$ as a continuous variable.

The physical principle at play here is that of \emph{locality}. To summarize, the fact that the scatterer has a finite size (interaction is local) leads to a faster-than-exponential suppression of high angular momenta.

\subsection{\label{sec:locality}Locality implies analyticity too}

So far, we've only looked at spherical potentials, so it was natural to consider fixed angular momentum $j$. But scattering problems are most commonly thought of in terms of particles with definite momenta, or equivalently, plane waves. We'll finish this lecture by connecting the discussion of spin to that of plane waves.

\subsubsection{Plane-wave scattering}

The total wave function will be written as a superposition of incoming and outgoing (scattered) waves: $\psi = \psi_{\mathrm{in}} + \psi_{\mathrm{out}}$.
As incoming states we'll take plane waves with a specific momentum $\p$:
\be\label{eq:psi-in}
\psi_{\mathrm{in}} = \e^{i \p \cdot \x} = \e^{i p r \cos\theta}\, ,
\ee
where $p = |\mathbf{p}|$, $r = |\x|$, and $\theta$ is the angle between $\p$ and $\x$.
As a result of a scattering process, we obtain the outgoing wave function. Let's say that as $r \to \infty$, it behaves as
\be\label{eq:psi-out}
\psi_{\mathrm{out}} \sim f(p,\theta) \frac{\e^{i p r}}{r}\, ,
\ee
where the coefficient $f(p,\theta)$ depending on the momentum, the angle, and details of the potential is the interacting part of the total scattering amplitude. To see how it's related to the partial amplitudes $f_j(p)$ we've seen earlier, let's decompose $\psi_{\mathrm{in}}$ and $f(p,\theta)$ into partial waves. A short computation tells us that asymptotically, the former behaves as
\be\label{eq:total-psi-in}
\psi_{\mathrm{in}} \sim -\frac{1}{2ipr} \sum_{j=0}^{\infty} (2j+1) \left[ (-1)^{j}\e^{-ipr} - \e^{ipr} \right] P_{j}(\cos \theta)\, .
\ee
So a plane wave from far away looks like an infinite superposition of incoming and outgoing partial waves, except for even $j$ we get a relative minus sign.

Similarly, decomposing $f(p,\theta)$ we get
\begin{empheq}[box=\graybox]{equation}\label{eq:partial-wave-expansion}
	f(p,\theta) = \frac{1}{p} \sum_{j=0}^{\infty} (2j+1) f_j(p) P_j(\cos \theta)\, .
\end{empheq}
We're going to claim that $f_j(p)$ are the same partial amplitudes we've just encountered in the previous section. To see this, let's look at the asymptotic form of the total wave function by combining \eqref{eq:psi-out} and \eqref{eq:total-psi-in}:
\be
\psi \sim -\frac{1}{2ipr} \sum_{j=0}^{\infty} (2j+1) \Big[ (-1)^j \e^{-ipr} - (1+2i f_j(p)) \e^{ipr} \Big] P_j(\cos \theta)\, .
\ee
The term in the square brackets is precisely the partial wave function from \eqref{eq:partial-wave-function} after substituting $S_j = 1 + 2i f_j$, so everything matches as promised. 

The formula \eqref{eq:partial-wave-expansion} is the partial-wave expansion, which teaches us how to compute the plane-wave scattering amplitude if we knew those for fixed angular momentum with any $j$. Using orthogonality of Legendre polynomials we can also go the other way:
\begin{empheq}[box=\graybox]{equation}\label{eq:total-to-partial}
	f_j(p) = \frac{p}{2} \int_{-1}^{1} f(p,\theta) P_j(\cos\theta) \,\d(\cos\theta)\, .
\end{empheq}
Here, we essentially traded the angular momentum $j$ for the scattering angle $\theta$. Now, we want to learn about analyticity properties of scattering amplitudes in this new variable.

\subsubsection{\label{sec:causal-transforms-angles}Causal transforms with angles}

Let's return back to causality once again, this time for the total scattering amplitude. We first reconsider the primitive causality argument from the previous lecture. We've already seen it isn't entirely correct in the quantum case, but it's intuitive and its essential points can be proven quantum-mechanically using a longer chain of arguments anyway.
In Section~\ref{sec:higher-angular-momenta}, we've encountered examples of $f_j(p)$ that can be exponentially large as $|p| \to \infty$ in the upper half-plane and go as $\sim \e^{-2i p R}$. The physical origin of this growth is the time advance of each partial wave reflected off the surface of the scatterer.

The situation with plane-wave scattering is completely different, because such a wave doesn't hit the scatterer all at once. In fact, a uniform time advance of the type $\sim \e^{-2i p R}$ would be in contradiction with causality, because it would imply that the information about the wave front hitting a single point at the surface at $r=R$ would instantaneously propagate throughout the whole surface. So causality tells us that the asymptotic bound on $f(p,\theta)$ in the upper half-plane has to be \emph{weaker} than that on the individual $f_j(p)$'s.

As in the case of the spherical wave, we can learn a lot by looking at classical wave fronts. Actually, almost all the steps are the same as for $S_0(p)$, except the time difference between the incoming and outgoing wave is different. We can understand it with the following picture:
\be\label{eq:reflection}
\includegraphics[valign=c,scale=1]{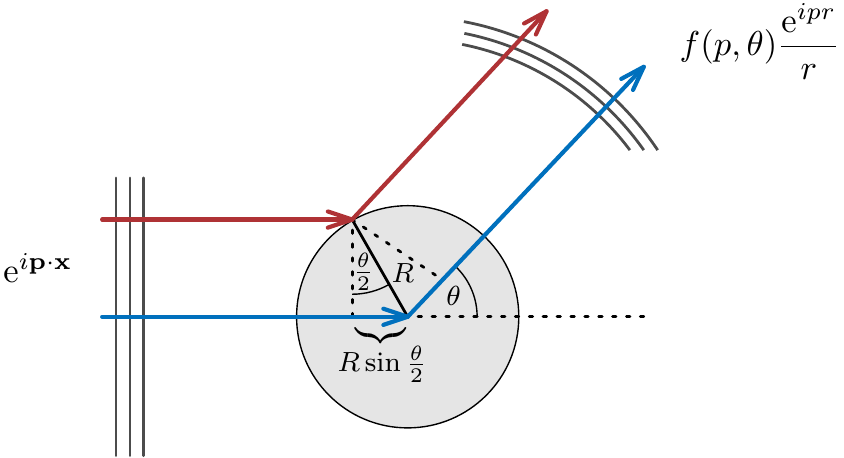}
\ee
Here, $\theta$ is the scattering angle. We see that in the worst-case scenario, the wave bounces off the surface along the red path, which shortens its trajectory by $2 R \sin \frac{\theta}{2}$ compared to the blue path.
This amounts to replacing $2R$ with $2R \sin \frac{\theta}{2}$ (and $S_0(p)$ with $f(p,\theta)$) in all the derivations in Sec.~\ref{sec:high-energy-bound}. We learn that the quantity
\be\label{eq:f-p-theta-rescaled}
f(p,\theta)\, \e^{2i p R \sin \frac{\theta}{2}}
\ee
is analytic in the upper half-plane of $p$ with real angles $\theta$. Moreover, using dispersion relations one can show it satisfies the bound
\be\label{eq:T-bound}
| f(p,\theta)\, \e^{2i p R \sin \frac{\theta}{2}} | \leq |f(\Re p, \theta)|\, .
\ee
We haven't really derived a bound on the right-hand side yet, but in practice it has to grow rather slowly so that the total cross-section doesn't grow too fast. We'll learn more about such bounds in later lectures.

Using a more careful quantum-mechanical treatment, one can show that \eqref{eq:f-p-theta-rescaled} is still analytic in the upper half-plane of $p$, except for a finite number of bound-state poles. The bound \eqref{eq:T-bound} also remains true asymptotically as $|p| \to \infty$ in the upper half-plane. Note that in the setup \eqref{eq:reflection}, by modeling the scattering process with a hard sphere, we assumed that the potential is repulsive ($U>0$). In contrast, classically, attractive potentials ($U<0$) can only give time delays. This can be simply illustrated by once again drawing trajectories of light rays, which in this situation can only lengthen their path compared to vacuum:
\be
\includegraphics[valign=c,scale=1]{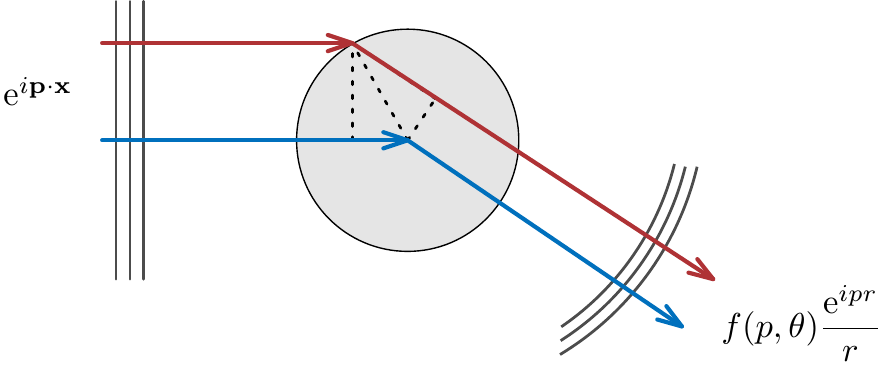}
\ee
However, as we've already seen on examples in Sec.~\ref{sec:time-delays}, at the quantum-mechanical level even attractive forces can lead to potential time advances.

Especially for very shallow angles $\theta$, all partial waves have to interfere destructively in order to knock down the exponential growth to a much more tame polynomial bound \eqref{eq:T-bound}. This tells us that the causality constraint on the plane-wave amplitude $f(p,\theta)$ contains \emph{more} information than causality of individual partial amplitudes $f_j(p)$ combined. It means that there have to be correlations between different $f_j$'s. In fact, we've already seen an example of this in Sec.~\ref{sec:low-spin}, where we found that $f_j$'s decay very quickly in $j$. We now turn to converting this property into yet another analyticity statement.

\subsubsection{Cauchy--Hadamard theorem}

At this stage we need to learn a tiny bit of math. It turns out that other than causal transforms, there's another way of proving analyticity. It goes as follows. Let's say we know enough about a function $f(z)$ at the origin that we can compute its Taylor series
\be
f(z) = \sum_{k=0}^{\infty} c_k\, z^k\, .
\ee
The question is which property of the coefficients $c_k$ could guarantee that $f(z)$ doesn't have singularities near the origin? Consider, for example, the function
\begin{subequations}\label{eq:1-za}
	\begin{align}
		\frac{1}{1-z/a} &= 1 + \frac{z}{a} + \left(\frac{z}{a} \right)^2 + \cdots \\
		&= \sum_{k=0}^{\infty} a^{-k} z^k\, .
	\end{align}
\end{subequations}
In this example $c_k = a^{-k}$ for some constant $a$. As we can read-off from the left-hand side, increasing $a$ allows us to extend the function $f(z)$ from the origin to the complex plane up to $|z| < |a|$ without encountering any singularities.

As a generalization of this statement, the Cauchy--Hadamard theorem says that as long as the coefficients $|c_k|$ die out as $\sim a^{-k}$ for large $k$, the function $f(z)$ is always analytic in the disk $|z| < |a|$, in which case $|a|$ is called the radius of convergence. If $|c_k|$ decay slower than exponentially, the series doesn't converge. If the decay is faster, the radius of convergence is infinity and the function can be extended to the whole complex plane.

In this example, the obstruction to analyticity was a simple pole at $z=a$. You may wonder if it's always possible to ``go around'' such singularities and analytically continue a function outside of the radius of convergence. This has to be decided case-by-case. For example, \eqref{eq:1-za} can be clearly continued to the whole complex plane minus the point $z=a$. To see this isn't always possible, it's a fun exercise to consider the function $\sum_{k=0}^{\infty} z^{2^k}$. It has a unit radius of convergence, but also a dense set of singularities for every $2^k$-th root of unity, which prevents it from being analytically continued past the unit disk. Such accumulations of singularities are called \emph{natural boundaries} and they're known to arise in scattering amplitudes.

We can make analogous statements for the Legendre, or partial-wave, expansion
\be
g(z) = \sum_{j=0}^{\infty} d_j P_j(z)\, .
\ee
Let's say that $g(z)$ is already regular on the interval $[-1,1]$ and we know the behavior of the coefficients $d_j$. One example is
\begin{subequations}
	\begin{align}
		\frac{1}{\sqrt{1-2z/a + 1/a^2}} &= P_0(z) + \frac{P_1(z)}{a} + \frac{P_2(z)}{a^2} + \ldots\\
		&= \sum_{j=0}^{\infty} a^{-j} P_j(z)\, .
	\end{align}
\end{subequations}
The coefficients decay as $\sim a^{-j}$ and the left-hand side tells us that the function is singular at $z = \frac{1+a^2}{2a}$. If we plot this singular locus $a = |a| \e^{i\phi}$ for all phases $\phi \in [0,2\pi]$, we get the following red curve in the $z$-plane:
\be
\includegraphics[scale=1,valign=c]{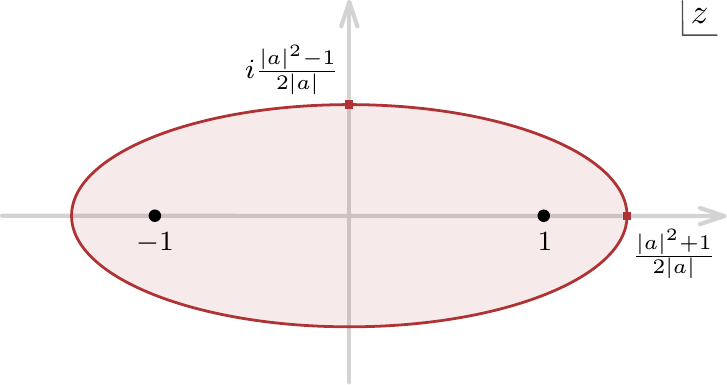}
\ee
It's an ellipse with foci $z = \pm 1$ and semi-major/minor axes (the largest and smallest distance to the center) $\frac{|a|^2 \pm 1}{2|a|}$. 

The Cauchy--Hadamard theorem generalizes in a natural way: as long as for large $k$, the absolute values of the coefficients of partial-wave expansion decay as $a^{-k}$, the function $g(z)$ is analytic everywhere inside the ellipse. This fact is known as the Neumann theorem.

\subsubsection{Analyticity in the angles}

The above theorem is very useful when applied to the partial-wave expansion of $f(p,\theta)$, because we already know something about the asymptotic behavior of partial-wave amplitudes $f_j(p)$ at large $j$. Namely, for finite-range scatterers we found that they decay \emph{faster} than exponentially with the spin $j$. Recall that this was a consequence of locality of interactions, or more precisely, that the scatterer was confined entirely within some finite radius $R$. Therefore, Neumann's theorem proves that $f(p,\theta)$ has to be analytic in the whole $\cos \theta$ complex plane!

Notice that the above arguments were entirely independent of causality, which instead was connected with analyticity in the momentum variable $p$. To summarize, analyticity of scattering amplitudes is underpinned by two physical principles: causality and locality. In Sec.~\ref{sec:Mandelstam-representation}, we'll combine these two statements about analyticity into a single equation.

\subsection{Exercises}

In this lecture, we've seen a lot of cool analytic functions. Go ahead and plot them using \texttt{ComplexPlot} to see how the poles move around and bound states appear in the upper half-plane as you tune the values of $j,p,U,R$. For example, to implement the function $\mathtt{S[j,p,U]}$ from \eqref{eq:S-j-spherical}, you'll need the functions \texttt{SphericalBesselJ}, \texttt{SphericalHankelH1}, and \texttt{SphericalHankelH2}. The goal of this set of exercises is to learn about how one would infer an existence of bound states and count them using only the data on the real axis.

\subsubsection{Counting the number of zeros and poles}

We start with a simpler question first. Let's say we're given a function $f(z)$ on the boundary of a unit disk, $|z|=1$, and asked to compute the number of poles $n_p$ and zeros $n_z$ in its interior. (From now on, we count poles and zeros with multiplicity, so a double pole counts as two etc.) We're going to claim that
\be\label{eq:nz-np-identity}
n_z - n_p = \frac{1}{2\pi i} \oint_{|z|=1} \d \log f(z)\, .
\ee
This identity is not so difficult to prove. For every zero of $f(z)$, say at $f(z) \sim z - a$, the integrand looks like $\d \log f(z) \sim \frac{\d z}{z - a}$ and it contributes $+1$ to the result by Cauchy's theorem whenever $|a| < 1$. Similarly, for every simple pole the logarithm picks up an additional minus sign and it contributes a $-1$ to the result if it's located within the unit radius.

Note that another way to think about the formula \eqref{eq:nz-np-identity} is using Stokes' theorem. Since the integrand is a total derivative, we can localize it at the endpoints. However, we have to be more careful since every zero and pole $f(z)$ give a logarithmic branch cut of $\log f(z)$ passing through the unit circle at some point. In order to use Stokes' theorem, we have to stop right before and right after the cut. As a result, the integral picks up discontinuities across every branch cut, which contribute $+2\pi i$ for every simple zero and $-2\pi i$ for every simple pole of $f(z)$.

If we wanted to count the number of zeros and poles individually, we would need more information. For example, if we knew the function can be represented as
\be\label{eq:ratio}
f(z) = \frac{g(z)}{h(z)}\, ,
\ee
with both $g$ and $h$ analytic in the unit disk, then
\be\label{eq:n-poles}
n_z  = \frac{1}{2\pi i} \oint_{|z|=1} \d \log g(z)\, , \qquad n_p  = \frac{1}{2\pi i} \oint_{|z|=1} \d \log h(z)\, .
\ee
Therefore, if we want to count the number of poles of a scattering amplitude, we'll first have to learn how to canonically separate it into a ratio like \eqref{eq:ratio}.

\subsubsection{Jost functions}

For simplicity, we're going to focus on the zero angular momentum case, $S_0(p)$. Recall that it can be read off from the asymptotic behavior of the total wave function:
\be\label{eq:asymptotic-total}
\psi \sim \frac{1}{r} \left[ \e^{-ipr} - S_0(p)\, \e^{ipr} \right]\, .
\ee
The idea is to write a clever ansatz for the solutions of the Schr\"odinger equation and massage it to the above form. It's a second-order differential equation, so there must exist two independent solutions. We're going to call them $\frac{1}{r} J(p,r)$ and $\frac{1}{r} J(-p,r)$, such that asymptotically we have
\be
J(p,r) \sim \e^{-ipr}\, .
\ee
You can go ahead and show that the two putative solutions are indeed independent by computing the Wronskian. Moreover, since the Schr\"odinger equation is real and the two solutions are asymptotically complex conjugates, it must be that $J^\ast(p,r) = J(-p,r)$ for all radii $r$ and real $p$.

Once we've established this property, the total wave function $\psi$ must be expressible as a linear combination of the form
\be
\psi = \frac{1}{r} \Big[ A_1 J(p,r) + A_2 J(-p,r) \Big]\, . 
\ee
Imposing regularity at the origin $r = 0$ implies that
\be\label{eq:S-Jost}
S_0(p) = -\frac{A_2}{A_1} = \frac{J(p)}{J(-p)}\, .
\ee
The quantity $J(p) = J(p,0)$ is called the \emph{Jost function}. By matching with \eqref{eq:asymptotic-total}, we recognized the ratio $-\frac{A_2}{A_1}$ as the scattering amplitude. Using $J^\ast(p) = J(-p)$, you can verify the unitarity and symmetry relations of $S_0(p)$ for real $p$.

A lot of results mentioned throughout the lectures are derived using analyzing analyticity properties of Jost functions.
For us, only two properties will be important: that $J(p)$ is analytic in the lower half-plane including the real axis (except possibly at $p=0$), and that it approaches $J(p) \to 1$ as $|p| \to \infty$ in the lower half-plane. Moreover, all the bound state \emph{poles} of $S_0(p)$ arise as \emph{zeros} of $J(p)$ on the negative imaginary axis.

\subsubsection{Levinson's theorem}

To combine all the above findings, let's consider the derivative of the logarithm of the Jost function, $\log J(p)$, in the lower half-plane of $p$. We had another name for this quantity: it's the time delay (up to constants),
\be\label{eq:dlog-J}
\d \log J(p) = \d \left[ i \eta_0(p) \right] = i \eta_0'(p)\, \d p\, .
\ee
We can now consider the following contour $\Gamma$:
\be
\includegraphics[scale=1,valign=c]{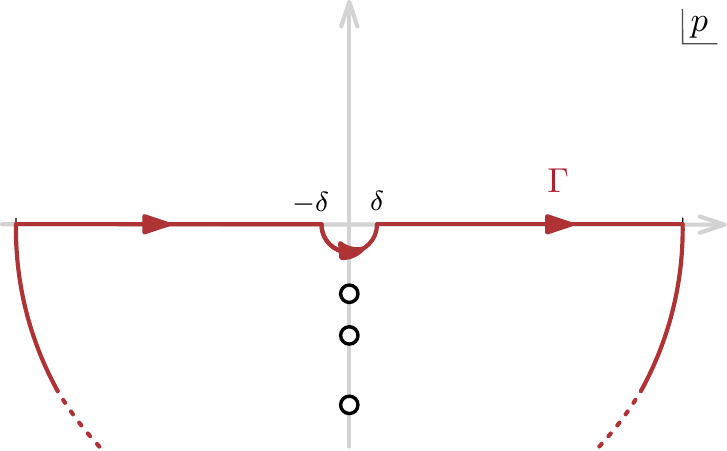}
\ee
We want to count the number of zeros (circles in the above picture) of $J(p)$ enclosed by it, or equivalently, the number of bound states $n_{\mathrm{bound}}$ of the amplitude. Here, for later purposes, we made a detour around the origin in a small arc of radius $\delta$. Using the second identity in \eqref{eq:n-poles}, we have
\be\label{eq:n-bound}
n_{\mathrm{bound}} = -\frac{1}{2\pi} \oint_{\Gamma} \d \eta_0(p)\, .
\ee
The overall minus sign comes from the fact $\Gamma$ has a negative (clockwise) orientation.
At this stage, we can start analyzing which parts of the contour can be dropped.

The fact that $J(p) \to 1$ along the large arc in the lower half-plane, translates to $\eta_0(p) \to 0$. We can therefore forget about it in the formula \eqref{eq:n-bound}. Let's for the time being also assume that $J(0) \neq 0$, since it will need to be considered as a special case. This allows us to drop the semi-circle near the origin. We are left with
\be
n_{\mathrm{bound}} = -\frac{1}{2\pi} \left( \int_{-\infty}^{-\delta} + \int_{\delta}^{\infty} \right) \d \eta_0(p)\, .
\ee
Further, using the fact that $\eta_0(p)$ is odd and localizing the integrals on the boundary we have
\be
n_{\mathrm{bound}} = \frac{\eta_{0}(0) - \eta_{0}(\infty)}{\pi}\, ,
\ee
where we could safely take the limit $\delta \to 0$. By \eqref{eq:dlog-J}, you can also think about it as time delay integrated between low and high momenta, up to a constant.

Go ahead and extend the proof to the case in which $J(0) = 0$. Here, we can no longer drop the contribution from the small semi-circle around the origin. Instead, prove that the additional term gives
\be
\frac{1}{2\pi i} \int_{\mathrm{\rotatebox[origin=c]{180}{$\curvearrowleft$}}} \d \log (p) = -\frac{1}{2}\, .
\ee
Whether we count a possible pole at $p=0$ as a bound state is pretty much a matter of definition. We're going to not count it. Since $n_{\mathrm{bound}}$ is an integer, we can remove any possible factor of $-\frac{1}{2}$ by always rounding down: 
\begin{empheq}[box=\graybox]{equation}
	n_{\mathrm{bound}} = \left\lfloor \frac{\eta_{0}(0) - \eta_{0}(\infty)}{\pi} \right\rfloor\, .
\end{empheq}
This result is known as the \emph{Levinson's theorem}.

\subsubsection{Winding phase}

It's time to get some hands-on experience with Levinson's theorem. Let's first illustrate how \emph{not} to go about using it. We're going to work with the example of the spherical well amplitude $S_0^{\mathrm{well}}(p)$ from Sec.~\ref{sec:solving-Schrodinger}. First, let's implement it: 
\begin{minted}[firstnumber=1]{mathematica}
Tanc[x_] := Tan[x]/x;
S[0, p_, U_] := Exp[-2I*p]*(1 + I*p*Tanc[Sqrt[p^2 - U]])/
						   (1 - I*p*Tanc[Sqrt[p^2 - U]]);
\end{minted}
It's going to be simple to set the unit $R=1$ everywhere. Recalling that $S_0(p) = \e^{2i \eta_0(p)}$, let's first try to naively plot the phase shift $\eta_0^{\mathrm{well}} (p)$ in the complex plane, say for $U=-50$:
\begin{minted}{mathematica}
ComplexPlot[Log[S[0, p, -50]]/(2I), {p, -10 - 10I, 10 + 10I},
			PlotPoints -> 300]
\end{minted}
As a result, you should see something like this:
\be
\includegraphics[valign=c,scale=0.6]{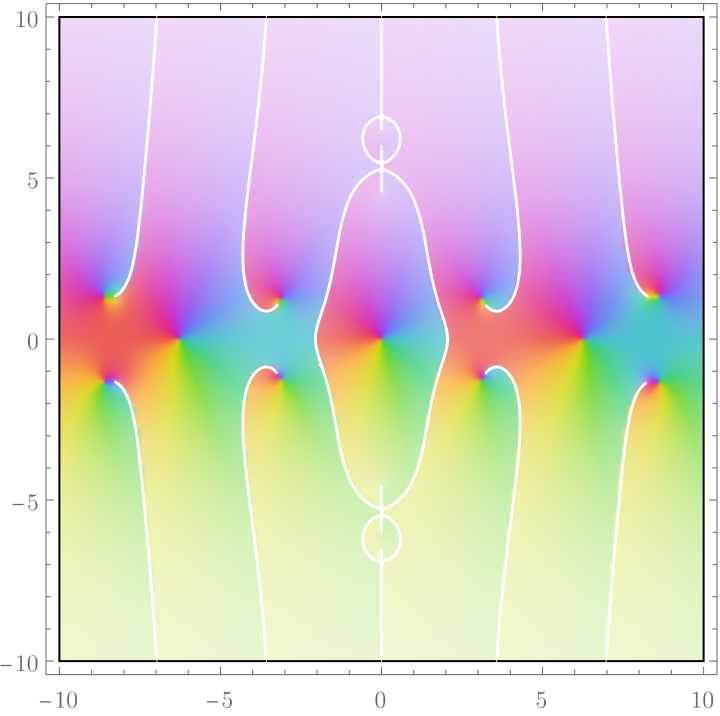}
\ee
The white lines are branch cuts. As promised earlier, they originate from the poles and zeros of $S_0(p)$. The fact that some of them pass through the real axis prevents us from accessing $\eta_0(p)$ directly. Instead, what we'd have to do is to rotate some of the cuts so that the positive real axis lies on the same sheet. In other words, we'd like to keep track of $\eta_0(p)$ as a continuous function between $p=0$ and $\infty$. We can certainly attempt to do it, but it turns out that there's a slicker approach we turn to now.

The other strategy will be to simply plot the result of $S_0^{\mathrm{well}}(p)$ for increasing values of $p$ and keep track of how its phase rotates. To make it more realistic, we're going to pretend we got the data from a noisy experiment. For example, we can define:
\begin{minted}{mathematica}
Noise[x_] := 1 + RandomVariate[NormalDistribution[]]*x;
data = Table[Join[ReIm[S[0, p, -50]*Noise[0.05]], {Log[p]}],
	         {p, 0.01, 100.01}];
\end{minted}
Here, the function $\mathtt{Noise}$ adds a certain amount of random fluctuations, say $5\%$. We then sample the amplitude at larger and larger values of $p$, starting from $p=0.01$ and stopping at $p=100.01$. The resulting list $\mathtt{data}$ consists of triples $(\Re \hat{S}_0^{\mathrm{well}}(p),\, \Im \hat{S}_0^{\mathrm{well}}(p),\, \log p)$, where the hat denotes added noise.

The idea is to make a three-dimensional plot of $\mathtt{data}$, so that the horizontal plane is the complex plane recording the values of $\hat{S}_0^{\mathrm{well}}(p)$ (the so-called Argand diagram) and the vertical axis keeps track of which $p$ is visualized. This procedure can be achieved with the command
\begin{minted}{mathematica}
ListLinePlot3D[data, PlotRange -> Full, Filling -> Bottom]
\end{minted}
You should be able to see a spiral like this:
\be
\includegraphics[valign=c,scale=0.8]{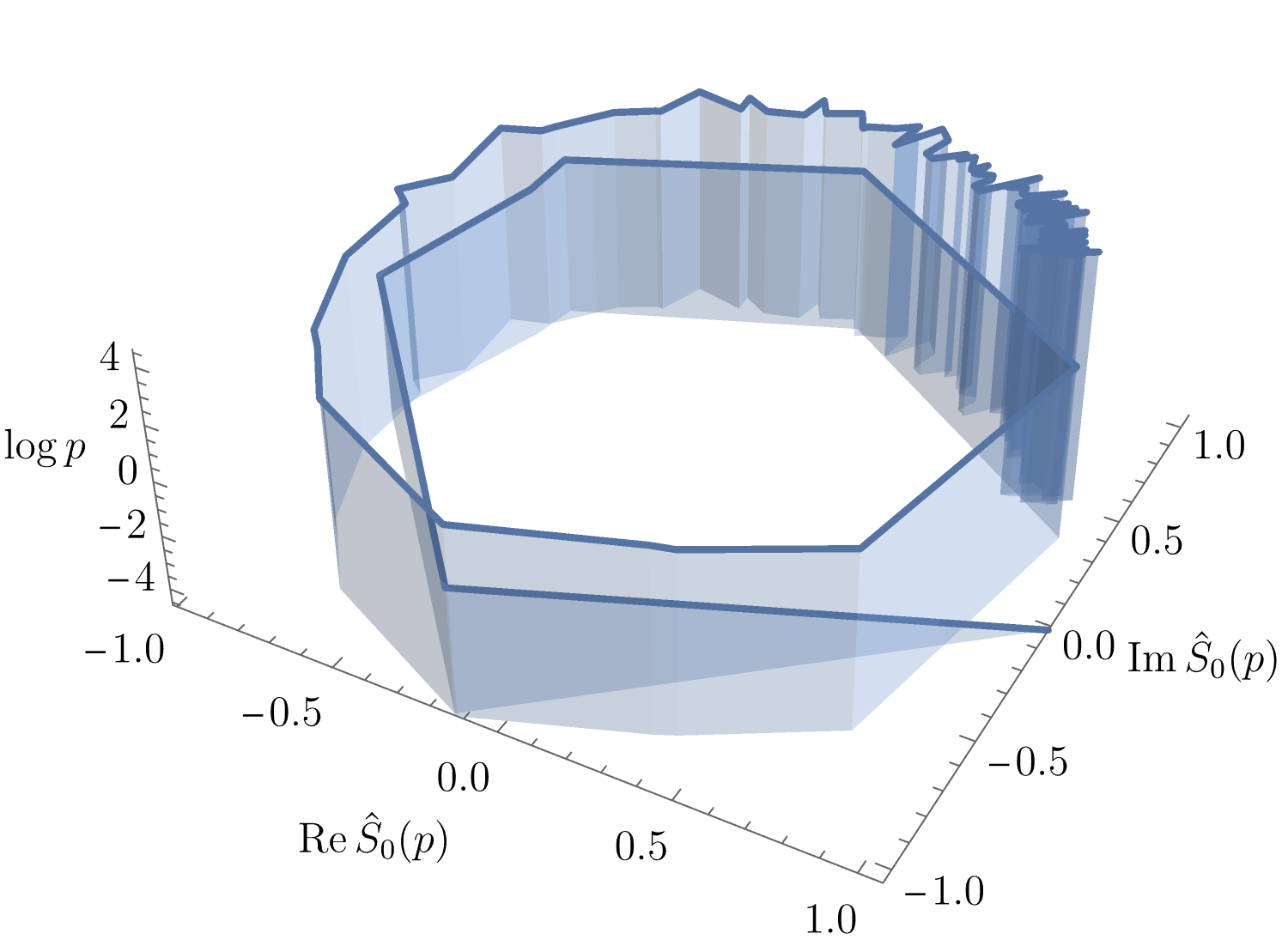}
\ee
The reason to include the third axis is just so that it's easier to count how many windings there are. In the above plot, we can see approximately $2$ windings. In other words, the difference of $2\eta_0(p)$ between small and large $p$ is roughly $4\pi$. According to Levinson's theorem, this fact indicates $2$ bound states. As you can read off from \eqref{eq:U-plots}, this is indeed the correct answer for $U=-50$.

Experiment with different values of $U$, levels of noise, and the number of samples.  Consider the function $\mathtt{S[j,p,U]}$ based on the higher-$j$ amplitudes \eqref{eq:S-j-spherical}. Generate ``data'' on the real axis and apply Levinson's theorem to estimate the number of bound states for a few values of $j$ and $U$. Compare this prediction with the exact counting obtained by plotting $\mathtt{S[j,p,U]}$ on the positive imaginary axis.  In particular, what should happen when we take $U>0$ for which there are no bound states? For consistency with the Levinson's theorem, you should be able to find that the phase winds and unwinds in such a way that the net number of windings in zero.
\section{Lecture III: Potential scattering}

In the previous lecture we've learned about scattering off finite-range potentials. We now wish to explore the more interesting cases of Coulomb or Yukawa interactions and learn how their infinite-range nature reflects itself in the analytic properties.

\medskip
\noindent\rule{\textwidth}{.4pt}
\vspace{-2em}
\localtableofcontents
\noindent\rule{\textwidth}{.4pt}

\pagebreak

\subsection{\label{sec:Yukawa-potential}Yukawa potential}

Our next goal is to understand what happens when the scattering can't be isolated to take place within a finite region of space.
To this end, we'll consider the classic case of ``screened'' Coulomb, or Yukawa, scattering. It's a model for scattering of nuclei mediated by pions and many other physical phenomena. As you might know, full scattering dynamics in this case becomes extremely complicated and we usually resort ourselves to perturbation theory in the weak coupling. We'll be able to learn some general lessons about analyticity from this approach, almost all of which will survive to the non-perturbative setting.

\subsubsection{Lippmann--Schwinger equation}

The strategy will be to first solve for the wave function $\psi_0$ in the absence of the potential and then perturb around it. The unperturbed time-independent Schr\"odinger equation is
\be\label{eq:Schrodinger-unperturbed}
\left(\nabla^2 + p^2 \right) \psi_0(\p,\x) = 0\, .
\ee
The solutions are of course the plane waves $\psi_0 = \e^{\pm i \p \cdot \x}$. On the other hand, the full wave function $\psi$ satisfies
\be\label{eq:Schrodinger-perturbed}
\left(\nabla^2 + p^2 \right) \psi(\p,\x) = U(\x)\, \psi(\p,\x)\, ,
\ee
where we switched on the (rescaled) potential $U = \frac{\hbar^2}{2m} V$. We can think of $m$ as either the mass of the particle scattering off the potential $V$, or as the reduced mass of two particles interacting through the same potential.
The idea is to invert this equation by hitting both sides with a putative inverse operator $\left(\nabla^2 + p^2\right)^{-1}$.

There's not just one but two ambiguities hidden in this procedure. The first has to do with the zero mode: we could've added an arbitrary multiple of \eqref{eq:Schrodinger-unperturbed} to \eqref{eq:Schrodinger-perturbed} and after inverting the resulting equation, $\psi$ would've shifted by a multiple of $\psi_0$. We fix this ambiguity in the following way:
\be\label{eq:psi-expansion}
\psi(\p,\x) = \psi_0(\p,\x) + \left(\nabla^2 + p^2\right)^{-1} U(\x)\, \psi(\p,\x)\, .
\ee
In other words, we require that the right-hand side goes back smoothly to the unperturbed solution $\psi_0$ as the potential is switched off, $U \to 0$.

The second ambiguity has to do with the definition of the inverse operator itself. The way to see it is to realize that, as usual, the inverse operator is the Green's function $G$ for the problem, satisfying
\be\label{eq:G-definition}
(\nabla^2 + p^2)\, G(\p,\x-\x') = \delta^3(\x - \x')\, ,
\ee
where the Laplacian $\nabla^2$ acts on the variable $\x-\x'$.
The are two possible solutions given by
\be\label{eq:Green}
G_\pm(\p,\x-\x') = -\frac{\e^{\pm i p |\x - \x'|}}{4\pi |\x - \x'|}\, .
\ee
To verify them, it's the simplest to go to the momentum space and then Fourier transform the result back to position space. Let us postpone going through this exercise for a while. It will turn out that $G_+$ is the choice consistent with causality and hence we'll stick with it for now.

In this way, \eqref{eq:psi-expansion} leads to a reformulation of the Schr\"odinger equation:
\begin{empheq}[box=\graybox]{equation}\label{eq:Lippmann-Schwinger}
	\psi(\p,\x) = \e^{i \p \cdot \x} - \frac{1}{4\pi} \int \d^3 \x'\, \frac{\e^{ip|\x - \x'|}}{|\x - \x'|}\, U(\x')\, \psi(\p,\x')\, .
\end{empheq}
It's called the \emph{Lippmann--Schwinger equation}. It's not quite a solution yet, because both sides still depend on the wave function $\psi$. However, we'll be able to use it recursively to compute approximate solutions.

\subsubsection{Amplitude from wave function}

Before doing so, let's see how we can extract the total scattering amplitude from \eqref{eq:Lippmann-Schwinger}. It's contained within the asymptotic limit of the wave function, so all we have to do is to expand the exponential for large $\x$ using
\be
|\x - \x'| = r \sqrt{ 1 - \frac{2\x \cdot \x'}{r^2} + \frac{r'^2}{r^2} } \sim r - \frac{\x \cdot \x'}{r}\, ,
\ee
where $r = |\x|$ and $r' = |\x'|$. Hence the Green's function behaves asymptotically as a spherical wave with a slightly shifted phase:
\be
G_+(\p,\x-\x') \sim - \frac{\e^{i p r - i \p'\cdot \x'}}{4\pi r}\, .
\ee
Here, we defined the momentum $\p' = p \frac{\x}{r}$ to have the same magnitude as $\p$, but pointing in the direction of $\x$.
Comparing \eqref{eq:Lippmann-Schwinger} with the definition of the total scattering amplitude \eqref{eq:psi-out}, which is the coefficient of the $\frac{\e^{i p r}}{r}$ term, we find
\begin{empheq}[box=\graybox]{equation}
	\label{eq:T-LS}
	f(p,\theta) = - \frac{1}{4\pi} \int \d^3 \x'\, \e^{- i \p' \cdot \x'}\, U(\x')\, \psi(\p,\x')\, .
\end{empheq}
This formula gives a direct way of extracting the amplitude from the wave function. 
Recall that $\theta$ was the angle between $\p$ and $\x$. We can therefore equivalently treat the amplitude as a function of two momenta, $\p$ and $\p'$, such that
\be
\p \cdot \p' = p^2 \cos \theta\, .
\ee
To manifest the fact that it really depends on only two degrees of freedom, we can also parameterize it by $p$ and the modulus $q = |\mathbf{q}|$ of the momentum transfer  $\q = \p - \p'$. It's related to the remaining variables through
\be
q^2 = 2p^2(1 - \cos \theta) = 4p^2 \sin^2 \tfrac{\theta}{2}\, ,
\ee
such that $\p \cdot \p' = p^2 - q^2/2$. Physical kinematics corresponds to real momenta and angles, which translates to the requirement $0 \leq q^2 \leq 4p^2$.
We'll use the notations $f(p,\theta)$, $f(\p,\p')$, and $f(p,q)$ interchangeably, depending on which variables are the most convenient for a given purpose.

Note that in the derivation we made a tacit assumption that the potential decays sufficiently fast at infinity, so that expanding in large $\x$ under the integral sign was valid and \eqref{eq:T-LS} converges. This assumption will come to haunt us soon enough.

\subsubsection{Born approximation}

The idea now is to plugging in \eqref{eq:Lippmann-Schwinger} into \eqref{eq:T-LS} and keep doing it over and over every time we encounter $\psi$. This leads to an expansion:
\begin{empheq}[box=\graybox]{align}\label{eq:Born-series}
	f(\p, \p') = &- \frac{1}{4\pi} \int \d^3 \x'\, \e^{i (\p-\p') \cdot \x'}\, U(\x') \\
	&- \frac{1}{4\pi} \int \d^3 \x'\, \d^3 \x'' \e^{i (\p-\p') \cdot \x'}\, U(\x')\, G_+(\p, \x'- \x'')\, U(\x'') + \ldots . \nn
\end{empheq}
This is known as the \emph{Born series} and the first term as the \emph{Born approximation} of the amplitude. In general, the $k$-th term will involve $k$ integrals with exponentials and $k$ powers of the potential. Hence, you can think about it as a weak-coupling expansion of the amplitude $f(\p,\p')$.

It's around the time things become more convoluted, so let's go through an example. We take the Yukawa potential with a coupling constant $\lambda$ and an effective range of interactions $r \sim 1/\mu$:
\be
U(\x) = \lambda \frac{\e^{-\mu r}}{r}\, .
\ee
The Born approximation is essentially the Fourier transform of this potential:
\be
f_\Born(\p,\p') = -\frac{\lambda}{4\pi} \int \d^3 \x'\, \e^{i \q \cdot \x'} \frac{\e^{-\mu r'}}{r'}\, ,
\ee
where $\q = \p - \p'$ is the momentum transfer.
This integral can be computed using fairly standard techniques. We first go to spherical coordinates with $\d^3 \x' = r'^2\, \d r'\, \d (\cos \theta')\, \d \phi$ and write $\q \cdot \x' = q r \cos \theta'$. Integrating out the angular components we get
\begin{subequations}
	\begin{align}
		f_\Born(\p, \p') &= -\frac{\lambda}{2} \int_0^{\infty} r'^2\, \d r' \int_{-1}^{1} \d(\cos \theta') \e^{i q r'\! \cos \theta'} \frac{\e^{-\mu r'}}{r'} \\
		&= -\frac{\lambda}{q}\int_0^{\infty} \d r' \sin (qr') \e^{-\mu r'}\, .
	\end{align}
\end{subequations}
Performing this integration gives the final answer:
\begin{empheq}[box=\graybox]{equation}
	\label{eq:Yukawa-Born}
	f_\Born(\p, \p') = -\frac{\lambda}{q^2 + \mu^2}\, .
\end{empheq}
The fact that the Born approximation is a Fourier transform is another manifestation of the uncertainty principle: if a potential has some features at range $r \sim 1/\mu$, they give significant contributions only at the momentum transfer $q \sim \mu$. Note that, once again, locality is reflected in analyticity in the momentum transfer $q$.

\subsubsection{Feynman \texorpdfstring{$i\varepsilon$}{i epsilon} and causality}

The fact that the Born approximation of the Yukawa scattering \eqref{eq:Yukawa-Born} turned out to be so simple might inspire us to study the whole Born expansion purely in the momentum space. Diagrammatically, it looks as follows:
\be
\includegraphics[valign=c,scale=1]{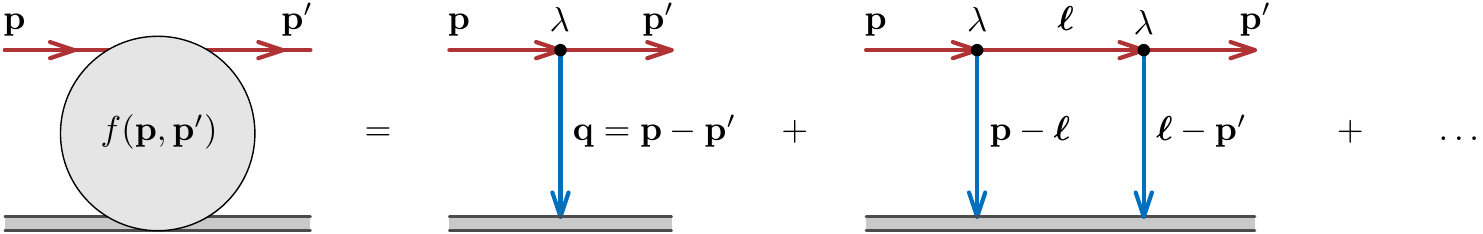}
\ee
The heavy line downstairs represents the source of the potential at the origin, while the red line upstairs is the particle we scatter, starting with momentum $\p$ and ending up with $\p'$. The blue lines represent the interaction with the potential. The series continues with more and more lines and vertices added up. Each picture is nothing but a non-relativistic version of a Feynman diagram. Note that each order in the expansion, there's only a single term.

The rules for writing an expression associated to each diagram are as follows. Momentum at each vertex is conserved and we integrate over all unconstrained (loop) momenta. The heavy line represents the source and hence doesn't itself contribute any terms. You can think of it as having an infinite mass. Each blue line with a momentum $\mathbf{k}$ has a propagator
\be
\tilde{U}(\mathbf{k}) = \frac{4\pi}{k^2 + \mu^2}\,,
\ee
coming from the Fourier transform of the potential, as computed in \eqref{eq:Yukawa-Born}. Interactions can only happen at red-red-blue vertices and each one of them comes with a factor of $\lambda$. Finally, a red line with momentum $\mathbf{k}$ evolves with the free propagator in momentum space $\tilde{G}_+(\p,\mathbf{k})$ and we'll turn to defining it more properly now.

We can now go back to the ambiguity with constructing $G_\pm$ by studying it in momentum space. Naively inverting \eqref{eq:G-definition} would give us $\frac{1}{-k^2 + p^2}$, where $\mathbf{k}$ is the Fourier dual of $\x - \x'$ and $k = |\mathbf{k}|$. However, this definition would be ambiguous when $k = p$, which would make inverting the Fourier transform ill-defined. Instead, we consider two possibilities of slightly deformed Green's functions:
\be
\tilde{G}_\pm(\p, \mathbf{k}) = \frac{1}{-k^2 + p^2 \pm i\eps}
\ee
and treat $\eps$ as an infinitesimal regulator that we send to zero at the end of the computation.

The choice of $\pm i\eps$ is related to causality. To see why this is the case, let us perform an inverse transform of $\tilde{G}_\pm(\p, \mathbf{k})$ and try to recover \eqref{eq:Green}. We have
\be
G_\pm(\p, \x - \x') = \int \frac{\d^3 \mathbf{k}}{(2\pi)^3} \frac{\e^{i \mathbf{k} \cdot (\x - \x')}}{-k^2 + p^2 \pm i\eps}\, .
\ee
Proceeding as before, we integrate out the angular components and are left with a single integral over the modulus $k$:
\begin{subequations}
	\begin{align}
		G_\pm(\p, \x - \x') &= -\frac{i}{(2\pi)^2 |\x - \x'|} \int_{0}^{\infty} \frac{k\, \d k}{-k^2 + p^2 \pm i\eps} \left( \e^{i k |\x - \x'|} - \e^{-i k |\x - \x'|} \right)\\
		&= \frac{i}{(2\pi)^2 |\x - \x'|} \int_{-\infty}^{\infty} \frac{k\, \e^{i k |\x - \x'|}\, \d k}{(k - p  \pm i\eps)(k + p \mp i\eps)} \, .
	\end{align}
\end{subequations}
In the second transition we combined both exponentials in the bracket at a cost of extending the integration domain to the whole real axis. The final integral can be done by residues. Since $|\x - \x'| > 0$, we can close the contour in the upper half-plane of $k$. It only encloses a single pole, but which one depends on the choice of $\pm i \eps$. For the minus, the relevant pole is at $k \approx -p + i\eps$, while for the plus it's $k \approx p + i\eps$ (recall that $p \geq 0$).

Going through the above derivation lands us on the two solutions we quoted in \eqref{eq:Green}, which only differ by the sign in the exponent $\e^{\pm i p |\x - \x'|}$.
The minus choice corresponds to an \emph{ingoing} wave. In other words, if we used it in the Lippmann--Schwinger equation \eqref{eq:Lippmann-Schwinger}, it would propagate signals back in time. On the contrary, the choice of the plus sign gives back the \emph{outgoing} Green's function $G_+$, which is why we selected it as the physically relevant solution. We conclude that the choice of $+ i\eps$ selects the correct causality condition.

To sum everything up, the diagrammatic rules are
\be
\includegraphics[scale=1.2,valign=c]{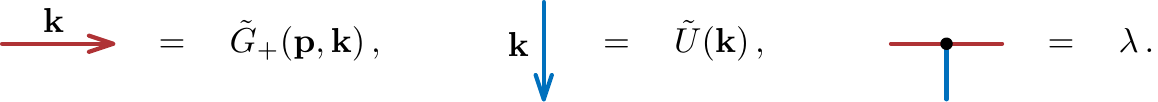}
\ee
The propagators are amputated for external legs. For every undetermined loop momentum $\bm{\ell}$ we integrate over $\frac{\d^3 \bm{\ell}}{(2\pi)^3}$. The final result is normalized by a factor of $-\frac{1}{4\pi}$.

\subsection{\label{sec:anomalous}Anomalous thresholds}

The leading Born approximation \eqref{eq:Yukawa-Born} turned out to be quite simple. However, as you might expect from the above discussion, subleading correction start involving rather complicated integrals rather quickly.
Therefore, what we want to learn right now is how to be able to predict analytic properties of the answer without evaluating the integrals explicitly.

We're going to work with the next-to-leading order (NLO) Born approximation for the Yukawa scattering, i.e., the part proportional to $\lambda^2$. Using the above diagrammatic rules, we can write down the integral
\begin{empheq}[box=\graybox]{equation}
	\label{eq:f-NLO}
	f_{\mathrm{NLO}}(\p,\p') = -\frac{\lambda^2}{2\pi^2} \int \frac{\d^3 \bm{\ell}}{ [(\bm{\ell} - \p)^2 + \mu^2] [-\ell^2 + p^2 + i\eps] [(\p' - \bm{\ell})^2 + \mu^2]}\, .
\end{empheq}
Let us hold off with computing it and instead think about how a  singularity can potentially arise. We've already seen that the integral is naively singular close to $\ell^2 = p^2$, but the $i\eps$ factor saves it. More generally, if the integrand has a pole, but a contour can be deformed to avoid it, the resulting integral is non-singular. Thus, the relevant question to ask is when the integrand has a divergence and no contour deformation can rescue it.

\subsubsection{\label{sec:pinch-endpoint}Pinch and endpoint singularities}

There are a few mechanisms for this to happen. Let's briefly abstract away from the integral \eqref{eq:f-NLO} and instead consider a toy-model one-dimensional example to understand the way singularities can arise better. One simple integral we can consider is
\be\label{eq:Ip}
I(p) = \int_0^{\infty} \frac{\d z}{(z - p)(z + 1)} = -\frac{\log(-p)}{p + 1}\, .
\ee
Since the integral is so simple, we can immediately write down the answer, but it's going to be useful to see how much we can learn without looking at it. We're going to treat $I(p)$ as a contour integral in the $z$-plane. The integrand has two simple poles at $z = p$ and $z = -1$, out of which only the first one depends on the external kinematics $p$.

Recall that our task is to determine what happens to $I(p)$ as a function of $p$. As we move it around in the $p$-plane, also the position of the pole the $z$-plane shifts around. For example, starting with $p$ in the upper half-plane, we can ask what happens as it approaches $-1$. There are two distinct ways to make this happen, either by moving anti-clockwise or clockwise around the origin, respectively:
\be\label{eq:z-plane}
\includegraphics[scale=1,valign=c]{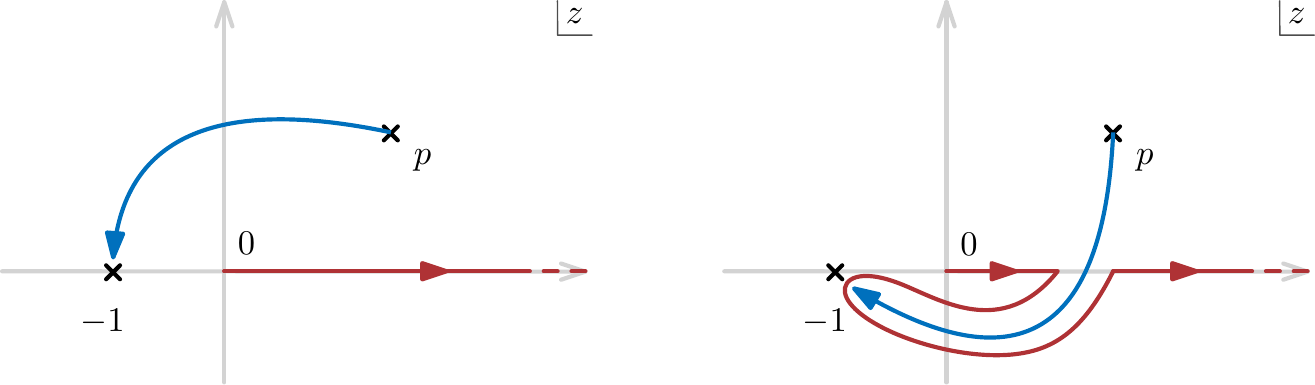}
\ee
The integration contour is illustrated in red and the trajectory of the pole in blue. These two paths lead to two distinct behaviors. Along the first one, the two poles indeed collide, but it happens far away from the integration domain, so the integral remains finite at all times.

On the other hand, going clockwise around the origin is more dangerous because the pole approaches the integration contour. We can of course keep deforming it to avoid the singularity as much as possible. However, as $p \to -1$, the contour gets trapped between the two roots and a divergence is unavoidable. It's called a \emph{pinch} singularity. How does it look like in terms of $p$? 
The simplest way to find out is to expand the integrand, which close to $p=-1$ is a total derivative of a simple pole $\sim -\frac{\partial}{\partial z}\left( \frac{1}{z+1}\right)$. Therefore, this specific pinch singularity gives a simple pole of $I(p)$ at $p=-1$ once we follow the blue path on the right of \eqref{eq:z-plane}.

We can keep going with this exercise and ask what happens as we wind around the origin in the $p$-space multiple times. Just drawing the counterparts of the blue paths, we find that only the zero-winding case (left figure in \eqref{eq:z-plane}) is non-singular. We conclude that $I(p)$ must be a multi-sheeted function, with a pole on every sheet but one.

What causes this multi-valuedness in the first place? There's another natural place to look for singularities, which happen when $p$ goes to one of the endpoints of integration, either $p \to 0$ or $p \to \infty$. In such a case, no contour deformation can save us because the boundaries of the contour stay fixed. This is another class called \emph{endpoint} singularities. When either $p\to0$ or $\infty$, the integrand looks locally like $\sim \frac{\partial}{\partial z} \log z$ and leads to a logarithmic divergence of $I(p)$. This is precisely the logarithm causing the multi-valuedness.

The information above pretty much nails down the answer to be that given in \eqref{eq:Ip} up to a constant (we're going to fix it in a second anyway). The pole at $p=-1$ is absent on the principal sheet of the logarithm because its residue is proportional to $\log(1) = 0$. Recall that on other sheets we have $\log(\e^{2\pi k i}) = 2\pi k i$, where the integer $k$ labels which sheet we are on.

Once again, notice that the multitude of sheets arose only because we're trying to pack too much information in a single complex plane. We could've easily unfolded it by, say, changing variables to $p = \e^{w}$. The function $I(p)$ in the complex $p$- and $w$-planes looks as follows:
\be\label{eq:I-p}
\includegraphics[scale=0.53,valign=c]{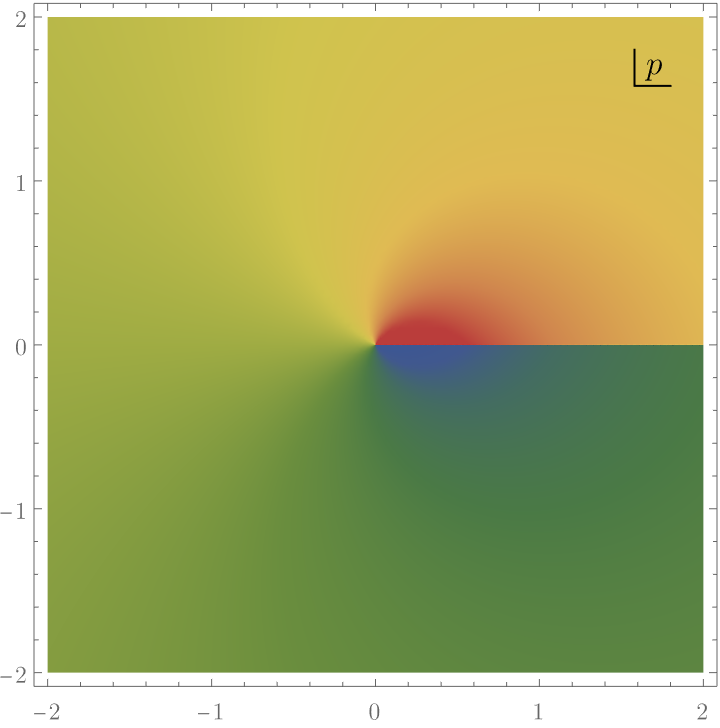}\quad
\includegraphics[scale=0.53,valign=c]{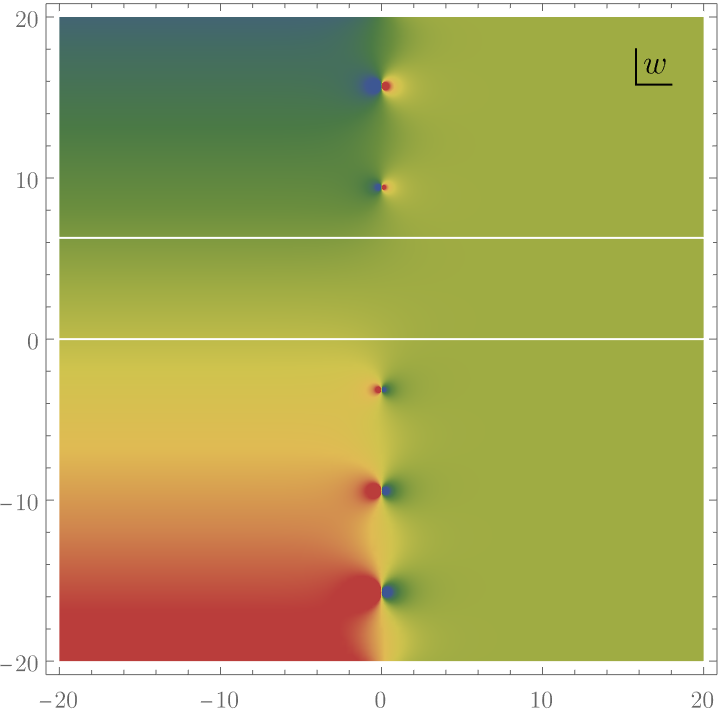}
\ee
Here, we plotted $\Im\, I(p)$ so that branch cuts are more visible. The $p$-plane maps to the strip $0 < \Im\, w < 2\pi$ delineated by the white lines on the right-hand plot (colors don't match just because the scale is different). The poles appear for $w = (2k+1) \pi i$ for all integers $k$ except $0$. Note that because we plot the imaginary part in \eqref{eq:I-p}, in a neighborhood of a simple pole we see large positive and negative values, which look like red-blue dipoles above.

There are two matters we should settle to complete our understanding of $I(p)$. First, what exactly selected that we write down $\log(-p)$ with the branch cut running to the right in \eqref{eq:Ip}? As emphasized in the previous lecture, the way we draw a cut is completely arbitrary. However, in writing a contour integral such as \eqref{eq:f-NLO}, we implicitly committed to a specific choice by specifying the integration contour. For example, if $p$ was on the real positive axis, we'd have to make a choice whether to run the contour right above or right below it, just as in the discussion of the Feynman $i\eps$ before. This choice is precisely what selects if we probe the function $I(p)$ below or above the branch cut. Integrating along a deformed contour in the $z$-plane would shift the branch ambiguity to different values of $p$ and hence affect the placement of the branch cut in the $p$-plane.

To make this more intuitive, consider the following rerouting of the contour in the $z$-plane:
\be
\includegraphics[scale=1,valign=c]{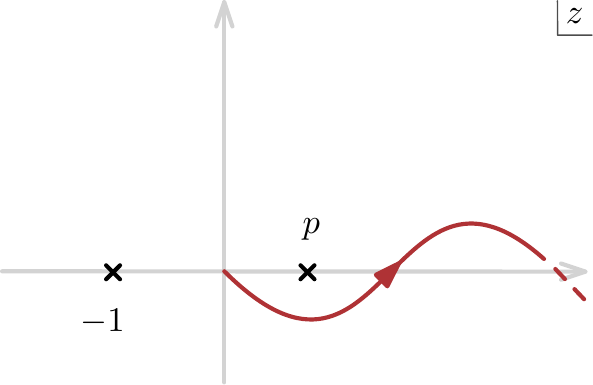}\qquad
\includegraphics[scale=0.53,valign=c]{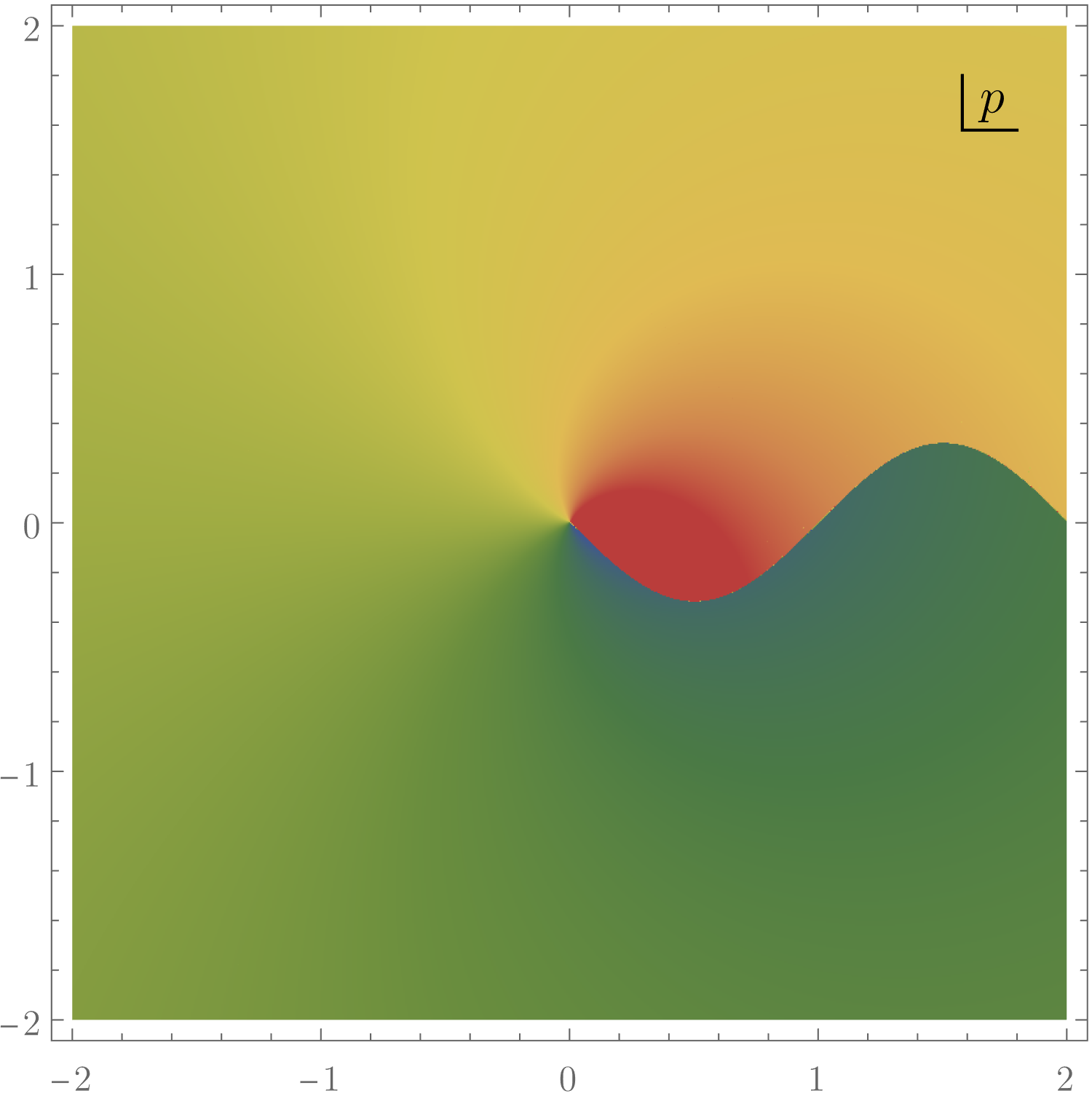}
\ee
The right-hand side illustrates the result of integrating over this contour for any given complex $p$. As you can tell, the wavy branch cut appears only where the ambiguity in defining the integrand arises. Some points that used to be on the second sheet on the left plot of \eqref{eq:I-p} are now on the first and vice versa. This teaches us an important lesson that the arbitrariness in choosing how to place branch cuts of integrals is a direct consequence of the specific choice of the integration contour.

The second, related, question is how to define the analytic continuation on the other sheets of $I(p)$. To this end, we can trace what happens to the integration contour as we wind around the origin once:
\be
\includegraphics[scale=1,valign=c]{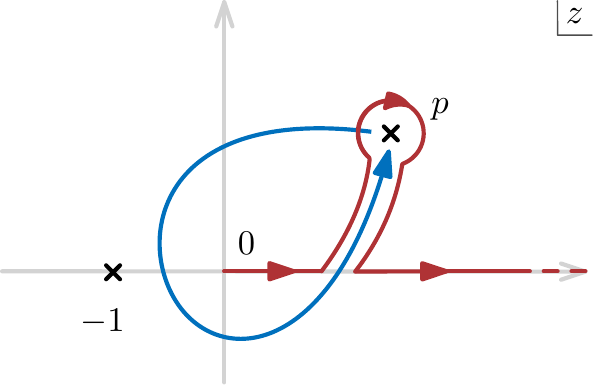}
\ee
The difference between the resulting contour and the original one is just a small clockwise loop around $z=p$. We call this difference the discontinuity across the branch cut:
\be\label{eq:Disc-Ip}
\Disc_{p=0} I(p) = -\oint_{z = p} \frac{\d z}{(z - p)(z + 1)} = -\frac{2\pi i}{p+1}\, .
\ee
Often, the term ``discontinuity'' is used to mean the jump of a function across a branch cut. Here, we're using it in an analytically-continued sense, meaning the difference between the value of a function at the end and beginning of the blue path above. Notice that \eqref{eq:Disc-Ip} has fewer singularities than $I(p)$ by itself, which is a general feature.
Going on the $k$-th sheet amounts to picking up more and more residues, and hence adding \eqref{eq:Disc-Ip} $k$ times to the answer. Computing the discontinuity this way would've also helped with determining the overall constant in the final expression \eqref{eq:Ip} that we couldn't fix by just knowing positions of singularities alone. 

One simple aspect of \eqref{eq:Ip} was that the position of the pole in the $z$-plane depended linearly on $p$. In more complicated situations, the poles themselves might be interesting functions of the external variables. Likewise, in more than one complex variable, the geometry of pinches can become arbitrarily complicated.

\subsubsection{Landau equations}

Back to the NLO computation \eqref{eq:f-NLO}, the essential intuition remains unchanged, though details become more involved. First of all, we'll need to express everything algebraically since it's difficult to draw pictures in three complex dimensions. Each of the three propagator factors
\be\label{eq:propagators}
\mathsf{D}_1 = (\bm{\ell} - \p)^2 + \mu^2\, , \qquad \mathsf{D}_2 = \ell^2 - p^2\, , \qquad \mathsf{D}_3 = (\p' - \bm{\ell})^2 + \mu^2\, .
\ee
defines a quadratic surface $\mathsf{D}_a = 0$ in the $\bm{\ell}$-space. Note that poles are determined by one complex condition, so each $\mathsf{D}_a = 0$ is really a two-complex-dimensional (or four-real-dimensional) surface in $\mathbb{C}^{3}$. This, quite literally, adds a new dimension to the problem, because unlike in the $I(p)$ example, where poles were just points, poles in higher dimensions can have an intricate geometry on their own.

It can happen that either of them becomes singular by itself (we can call it a self-pinch), or two or more surfaces pinch for a specific value of the external kinematics. Just as in our toy model, if this happens along the original integration domain $\bm{\ell} \in \mathbb{R}^{3}$ and the measure $\d^3\bm{\ell}$ doesn't cancel the pole of the integrand, the integral is ambiguous and develops a branch cut.
A contour deformation is needed and the specific choice of this deformation translates to the choice of the branch of $f_{\mathrm{NLO}}(\p,\p')$. Whenever such a deformation can't be done, the integral is genuinely singular, meaning that it develops a branch point or a pole.

Let's first determine the condition for when a contour deformation can't be done. We can parameterize it by a simple change of variables:
\be
\bm{\ell} \to \bm{\ell} + \delta \bm{\ell} 
\ee
with an infinitesimal complex vector $\delta \bm{\ell}$, which typically adds imaginary contributions just like the Feynman $i\varepsilon$.
The three propagator surfaces \eqref{eq:propagators} get deformed by the amounts
\be\label{eq:delta-D}
\delta \mathsf{D}_1 = 2(\bm{\ell} - \p)\cdot \delta \bm{\ell}\, ,\qquad
\delta \mathsf{D}_2 = 2 \bm{\ell} \cdot \delta \bm{\ell}\, ,\qquad
\delta \mathsf{D}_3 = 2(\bm{\ell} - \p')\cdot \delta \bm{\ell}\, .
\ee
We now have many options for which possible singularity we want to study. For example, let's consider the first and second propagator surfaces pinching.

The algebraic condition we want to impose is that both propagators diverge simultaneously, $\mathsf{D}_1 = \mathsf{D}_2 = 0$, but also that the contour deformation can't save us. In other words, that the displacements $\delta \mathsf{D}_1$ and $\delta \mathsf{D}_2$ are aligned along the same direction, i.e., are linearly dependent. While it might be difficult to visualize this situation in three complex dimensions, the following cartoon illustrates the basic point:
\be
\includegraphics[scale=1,valign=c]{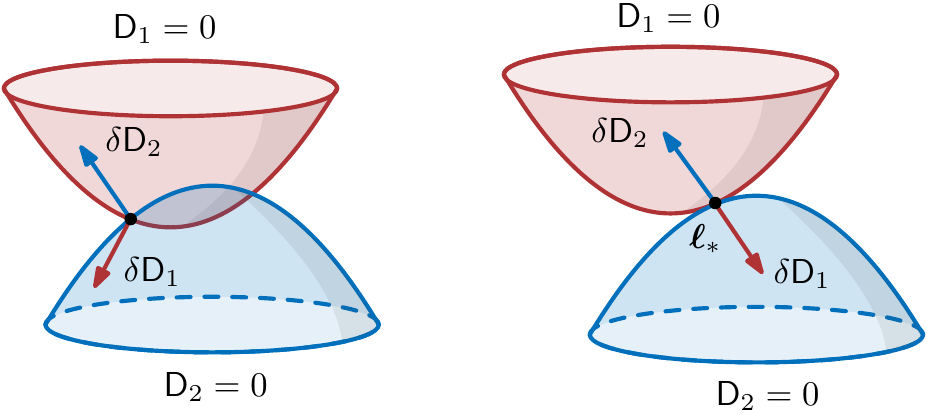}
\ee
In both cases the surfaces intersect, so there are points in the complexified loop momentum space that satisfy the first two conditions \eqref{eq:delta-D}. The vectors represent each $\delta \mathsf{D}_a$, which are tangent to their respective surface. In the first case, the vectors at any of the intersection points are aligned in different directions and the contours can be pulled apart (into the remaining complex directions). In the second, there's an intersection point $\bm{\ell}_\ast$ at which the displacement vectors are aligned, which leads to a pinch singularity. Stated more mathematically, we're looking for conditions for the union of propagator surfaces to change topology. For instance, the self-pinch comes from a sphere (the mass shell) degenerating into a point. As you can imagine, geometry of these pinches can become extremely complicated, which is why it pays off to formulate the conditions algebraically.

Anyway, one way to summarize this is to ask for a solution to the system of equations:
\be
(\bm{\ell} - \p)^2 + \mu^2 = 0\, , \qquad \ell^2 - p^2 = 0\, , \qquad \alpha_1 (\bm{\ell} - \p) + \alpha_2 \bm{\ell} = 0\, .
\ee
for some $\alpha_1$ and $\alpha_2$ that aren't simultaneously zero. This is one of the simplest instances of \emph{Landau equations}, whose solutions determine singularities of scattering amplitudes.
Note that the equations are unchanged under rescaling the $\alpha$'s so only the ratio $\frac{\alpha_2}{\alpha_1}$ matters. Since the last equality is a vector equation, we have the total of $5$ constraints, but the number of unknowns is only $4$: the ratio $\frac{\alpha_2}{\alpha_1}$ and the $3$ components of $\bm{\ell}$. Hence we expect one net constraint on the external kinematics. It's going to determine the position of a potential singularity. Let's just go ahead and solve the system to see how it comes about.
Solving the third equation determines the value of the loop momentum:
\be
\bm{\ell}_\ast = \frac{\alpha_1}{\alpha_1 + \alpha_2} \p\, .
\ee
Plugging it into the second equation tells us that either $\frac{\alpha_2}{\alpha_1} = -2$ or $0$ (equivalently, $\bm{\ell}_\ast = -\p$ or $\p$). Finally, using them on the first equation yields two solutions, respectively:
\be\label{eq:D1-D2}
p = \pm \frac{i\mu}{2}\, , \qquad \mu = 0\, .
\ee
The first one is associated to possible singularities in the $p$-plane. The second one happens in the limit as the Yukawa potential becomes Coulomb and it's related to infrared (IR) divergences we'll analyze more closely in Sec.~\ref{sec:Coulomb}. By symmetry, identical solutions are obtained by pinching $\mathsf{D}_2$ and $\mathsf{D}_3$.

Recall that just like in the toy model integral, pinching is only a \emph{necessary} condition for a singularity but not a sufficient one. It often happens that the contour doesn't get trapped. It typically takes much more work to determine if it does actually happen and on which sheet. We'll come back to one criterion in a second.

Using similar reasoning we can write down analogous Landau equations for other subsets of propagator surfaces pinching. We always impose the $\mathsf{D}_a = 0$ conditions and linear dependence of contour deformations. I'm going to spare you the details and just quote the final answers. From pinching $\mathsf{D}_1$ and $\mathsf{D}_3$ we have:
\be\label{eq:D1-D3}
\bm{\ell}_\ast = \frac{\p + \p'}{2}\,, \qquad \frac{\alpha_3}{\alpha_1} = 1\,, \qquad q = \pm 2i\mu
\ee
and
\be\label{eq:second-type}
\bm{\ell}_\ast = \infty\,, \qquad \frac{\alpha_3}{\alpha_1} = -1\,, \qquad q = 0\, .
\ee
The self-pinch singularities of either $\mathsf{D}_1$ or $\mathsf{D}_3$ naively happen for $\mu = 0$ with $\bm{\ell}_\ast = \p$ and $\bm{\ell}_\ast = \p'$ respectively. Likewise, the self-pinch for $\mathsf{D}_2$ appears when $p=0$ with $\bm{\ell}_\ast = 0$. However, none of the three is singular because of a cancellation with the measure $\d^3 \bm{\ell}$.
Finally, we can pinch all the three surfaces together, where we impose $\mathsf{D}_a=0$ for $a=1,2,3$ and $\sum_{a=1}^{3} \alpha_a \,\delta\mathsf{D}_a = 0$. This is often referred to as the \emph{leading singularity}, in contrast with all the previous ones called \emph{subleading}. The solution is given by
\be\label{eq:leading}
\bm{\ell}_\ast = \frac{p^2}{2p^2 + \mu^2}(\p + \p'),\,\qquad
\frac{\alpha_2}{\alpha_1} = \frac{\mu^2}{p^2},\,\qquad \frac{\alpha_3}{\alpha_1} = 1,\,\qquad
p = \pm \frac{i\mu^2}{\sqrt{q^2 + 4\mu^2}}\, ,
\ee
as well as \eqref{eq:second-type} with $\frac{\alpha_2}{\alpha_1} = 0$.

Another way to streamline the above discussion would be to introduce Schwinger (or Feynman) parameters directly in \eqref{eq:f-NLO} using the identity
\be
\frac{1}{\mathsf{D}_1 \mathsf{D}_2 \cdots \mathsf{D}_m} = (m{-}1)! \int_{0}^{1}  \frac{\d^m \alpha\, \delta(\alpha_1 {+} \alpha_2 {+} \ldots {+} \alpha_m - 1)}{(\alpha_1 \mathsf{D}_1 + \alpha_2 \mathsf{D}_2 + \ldots + \alpha_m \mathsf{D}_m)^m}\, 
\ee
with $m=3$. This way, we obtain a representation of the integral \eqref{eq:f-NLO} with only a single surface to pinch, though now in higher dimension:
\be
f_{\mathrm{NLO}}(\p,\p') = -\frac{\lambda^2}{\pi^2} \int \frac{\d^3 \bm{\ell} \, \d^3 \alpha \, \delta(\alpha_1 + \alpha_2 + \alpha_3 - 1) }{ \left\{\alpha_1 [(\bm{\ell} - \p)^2 + \mu^2] + \alpha_2 [-\ell^2 + p^2 + i\eps] + \alpha_3 [(\p' - \bm{\ell})^2 + \mu^2] \right\}^3 }\, .
\ee
Repeating the singularity analysis from above, we'd arrive at the same Landau equations. Contour deformations in $\bm{\ell}$ give rise to the equation $\sum_{a=1}^{3} \alpha_a\, \delta\mathsf{D}_a = 0$. For every $a$, it can either happen that $\alpha_a = 0$, or $\alpha_a \neq 0$ in which case we need to impose that the derivative of $\sum_{a=1}^{3} \alpha_a \mathsf{D}_a$ with respect to $\alpha_a$ vanishes, or both at the same time. The $\alpha_a$-derivative condition is the same as $\mathsf{D}_a = 0$ we encountered before. You can convince yourself that the curly bracket automatically vanishes on the support of these new conditions, and going through all the possibilities lands us on the same sets of Landau equations we enumerated above.
The delta function only sets the overall scale for the $\alpha_a$'s, so it plays the same role as taking the ratios above. 

This integral representation might make it more intuitive why solutions with $\alpha_a \geq 0$ are the ones that are the most relevant: they occur on the undeformed integration contour in the new variables. For this reason, if there's no additional branching, i.e., we work in the quadratic variables $p^2$ and $q^2$, the $\alpha_a \geq 0$ condition determines singularities on the physical sheet.

Landau equations turn out to be much more difficult to formulate for higher-order contributions with more than one loop, both in quantum mechanics and quantum field theory. The reason is that one needs to be extra careful with the rates at which surfaces pinch. For example, if three poles collide, it can happen in six inequivalent ways, depending on the relative ``speeds'' at which they approach each other. Mathematically, resolving such ambiguities is addressed by \emph{compactification} of the integration domain by \emph{blow-ups}. The situation gets even more complicated in the presence of genuine UV and IR divergences in quantum field theory. We won't have time to discuss any of it in detail.

\subsubsection{Direct integration}

It's interesting to compare the above predictions with the result of integrating \eqref{eq:f-NLO} directly. Let's focus on the case $\mu>0$. The integration itself is not entirely illuminating, so let's just quote the final result:
\begin{empheq}[box=\graybox]{equation}\label{eq:f-NLO-final}
	f_{\mathrm{NLO}}(\p,\p') = \frac{i \lambda^2}{2q \sqrt{\Delta}} \left[ \log(\frac{\sqrt{\Delta} - i \mu q/2}{\sqrt{\Delta} + i \mu q/2}) + \log(\frac{\sqrt{\Delta} + p q}{\sqrt{\Delta} - p q}) \right]
\end{empheq}
with
\be\label{eq:Delta}
\Delta = p^2 q^2 + 4\mu^2 p^2 + \mu^4\, .
\ee
The branch point at $\Delta = 0$ is the leading singularity \eqref{eq:leading}. The first logarithm has branch points starting from subleading singularities \eqref{eq:D1-D2} and \eqref{eq:D1-D3} at $p^2 = -\mu^2/4$ and $q^2 = -4\mu^2$ respectively. The second logarithm has branch points at $p^2 = -\mu^2/4$ as well. One has to be very careful with analyzing nested branch points, especially that there are a few cancellations. As usual, it's much easier to just plot the result in the complex plane (with $\mu=\lambda=1$), for a couple of values of $q$:
\be\label{eq:f-NLO-plots}
\includegraphics[scale=0.5,valign=c]{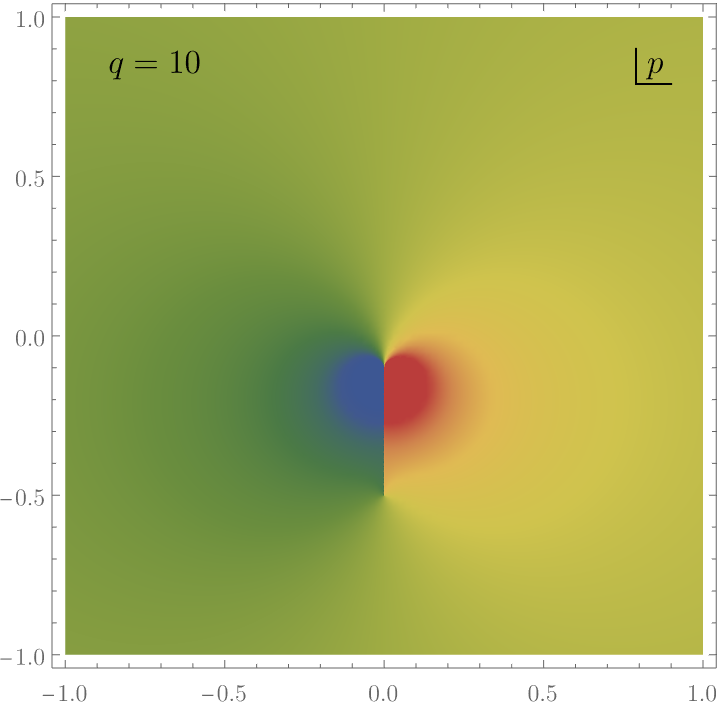}\quad
\includegraphics[scale=0.5,valign=c]{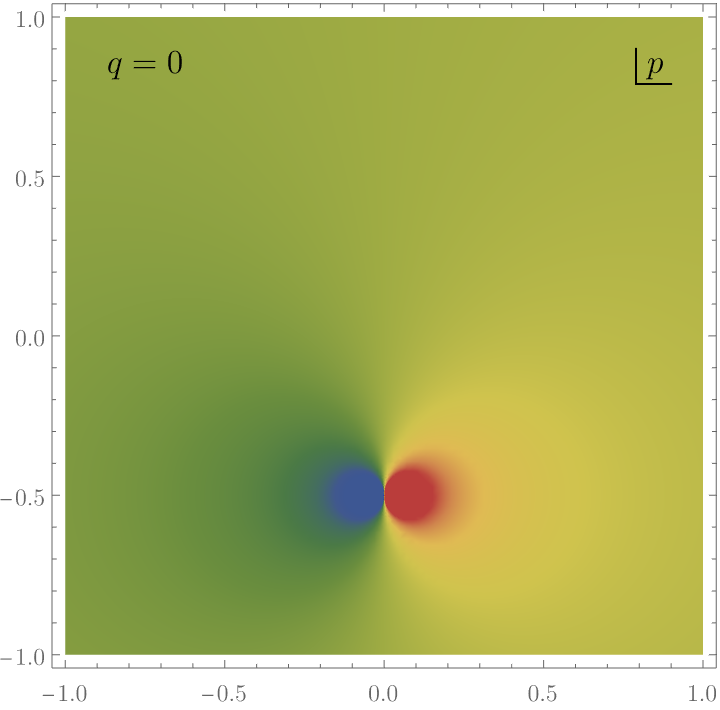}
\ee
Here, we once again plotted the imaginary part to see branch cuts more easily. At finite $q$, the branch cuts in the upper half-plane cancel out and $f_{\mathrm{NLO}}(\p,\p')$ is analytic there. However, it still has singularities in the lower half-plane. In the forward limit, the situation simplifies rather drastically since we have
\be\label{eq:f-NLO-forward}
f_{\mathrm{NLO}}(\p,\p) = \frac{\lambda^2}{2\mu^2 (\mu - 2 i p)} \, ,
\ee
which only has a simple pole at $p = -i\mu/2$.

It's a general result in elimination theory that kinematics solving Landau equations can be phrased as vanishing of polynomials in the variables $p^2$ and $q^2$ (but not $p$ and $q$), for example $\Delta = 0$ from \eqref{eq:Delta}. This is one of the reasons why later we'll switch to working with the quadratic combinations, $p^2$ and $q^2$, entirely.

\subsubsection{Physical interpretation}

In the previous lecture we've seen that singularities of scattering amplitudes can often be given physical interpretation. This begs the question of what's the meaning of the singularities we found using Landau equations.

The conditions $\mathsf{D}_a = 0$ are simple to interpret. They determine when the ``virtual'' particle becomes a physical one, i.e., it goes on the mass shell. For example, $\mathsf{D}_1$ and $\mathsf{D}_3$ become a particle with mass $\mu$ and $\mathsf{D}_2$ is just the original particle with mass $m$ deflected. Moreover, it turns out we can also interpret each $\alpha_a$ variable as the Schwinger proper time elapsed between two collision events in the units of the mass. As pointed out by Coleman and Norton, denoting the momentum flowing through a given propagator with $\mathbf{k}_i$, combinations of the type $\alpha_i \mathbf{k}_i$ are simply the spatial displacement between two interaction vertices, say at $\mathbf{x}_{i-1}$ and $\mathbf{x}_{i}$ that would happen if they were classical particles:
\be
\x_{i-1} - \x_{i} = \alpha_i \mathbf{k}_i\, .
\ee
Therefore, we can think of Landau equations as determining when a quantum process becomes classical.
For instance, the leading singularity can be interpreted as the following series of classical particle collisions:
\be\label{eq:triangle}
\includegraphics[scale=1.15,valign=c]{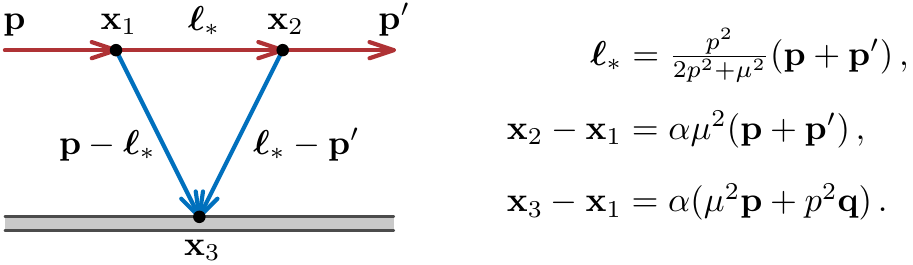}
\ee
Note that the positions of vertices $\mathbf{x}_i$ are only determined up to an overall rescaling we called $\alpha$. Here, $\mathbf{x}_3 = 0$ is fixed to the origin. Such singularities are called \emph{anomalous thresholds} (the term ``anomalous'' is stuck with us due to history and shouldn't be taken too seriously). It turns out that they are a consequence of unitarity and, as we'll see below, in fact an everyday occurrence.

There's only one catch. As we determined in \eqref{eq:leading}, the above triangle singularity requires the kinematic constraint $p^2 = -\frac{\mu^4}{q^2 + 4\mu^2}$ to be satisfied. This is the place where a new branch cut opens up. However, it can't be realized in real kinematics, as we've already seen in \eqref{eq:f-NLO-plots}, where the corresponding singularity was located in the lower half-plane of $p$. Essentially, the problem is caused by the interactions at the $\mathbf{x}_1$ and $\mathbf{x}_2$ vertices, which look like a decay of single particle into two (or its time inverse). Real kinematics for such an interaction can't be realized classically. Hence, when we talk about \eqref{eq:triangle}, we really talk about an analytically-continued classical process with complex external momenta.

This shouldn't make you think that anomalous thresholds are something exotic. In fact, they contribute very often in multi-particle scattering processes. This is well-known to billiard aficionados as the following trick shot:
\be\label{eq:triangle-threshold}
\begin{gathered}
	\begin{tikzpicture}[rotate=180]
		
		\foreach \t in {0,1,...,5} {\filldraw[color=black, fill=white, fill opacity=(\t)*0.1, draw opacity=(\t)*0.1, thick] (\t/5,-\t/5) circle (0.2);};
		\foreach \t in {1,...,10} {\filldraw[color=black, fill=white, semitransparent, thick] (1+\t/5,-1+\t/5) circle (0.2);};
		\foreach \t in {1,...,10} {\filldraw[color=black, fill=white, semitransparent, thick] (3+\t/5,1-\t/5) circle (0.2);};
		\foreach \t in {0,1,...,5} {\filldraw[color=black, fill=white, fill opacity=(5-\t)*0.1, draw opacity=(5-\t)*0.1, thick] (5+\t/5,-1+\t/5) circle (0.2);};
		
		\foreach \t in {0,1,...,5} {\filldraw[color=Maroon, fill=Maroon, fill opacity=(\t)*0.05, draw opacity=(\t)*0.1, thick] (\t/5,-1.65+\t/20) circle (0.2);};
		\foreach \t in {1,...,19} {\filldraw[color=Maroon, fill=Maroon, fill opacity=0.3, draw opacity=0.5, thick] (1+\t/5,-1.4) circle (0.2);};
		\foreach \t in {0,1,...,5} {\filldraw[color=Maroon, fill=Maroon, fill opacity=(5-\t)*0.05, draw opacity=(5-\t)*0.1, thick] (5+\t/5,-1.4-\t/20) circle (0.2);};
		
		\foreach \t in {0,1,...,15} {\filldraw[color=RoyalBlue, fill=RoyalBlue, fill opacity=(\t)*0.05, draw opacity=(\t)*0.05, thick] (\t/5,1.8-\t/40) circle (0.2);};
		\foreach \t in {1,...,15} {\filldraw[color=RoyalBlue, fill=RoyalBlue, fill opacity=(15-\t)*0.05, draw opacity=(15-\t)*0.05, thick] (3+\t/5,1.4+\t/40) circle (0.2);};
		
	\end{tikzpicture}
\end{gathered}
\ee
with time flowing to the right. The cue (white) ball scatters elastically three times off the red and blue balls. The fact you can observe it with naked eyes means the process is very classical and with real momenta, giving us a practical example of an anomalous threshold.

The interpretation of some of the anomalous thresholds will later change in quantum field theory. In fact they foreshadow an appearance of a general principle called \emph{crossing symmetry}, which says that a given scattering amplitudes can be reinterpreted as a different process in which some particles are exchanged for anti-particles traveling back in time. For example, the singularity at $q^2 = -4\mu^2$ will be understood in terms of particle production in a different scattering channel. We'll return to such questions later in Sec.~\ref{sec:crossing}.

\subsection{\label{sec:unitarity-cuts}Unitarity cuts and analyticity}

One can of course continue with higher orders in the perturbative expansion, but we've pretty much learned everything we needed for our purposes. In this section, we're going to briefly review how the analyticity statements we previously made for finite-range potentials generalize to Yukawa-like potentials.

\subsubsection{Fixing the angle or momentum transfer}

It turns out one can analyze convergence of the Born expansion \eqref{eq:Born-series} for any potential that decays asymptotically as $\e^{-\mu r}/r$, just like in the Yukawa case. Let's keep the momentum transfer $q$ fixed and look at the complex plane of $p$. While the details are rather complicated, the result reaffirms our observations from the previous section: singularities, now also including branch points, can happen in the lower half-plane of $p$. The total amplitude is also meromorphic in the upper half-plane, i.e., analytic except for a number of bound-state poles.

Correspondingly, the amplitude is also meromorphic on the first sheet of the $E$-plane minus the energy branch cut on the positive real axis. We can summarize these findings schematically as
\be
\includegraphics[scale=1.05,valign=c]{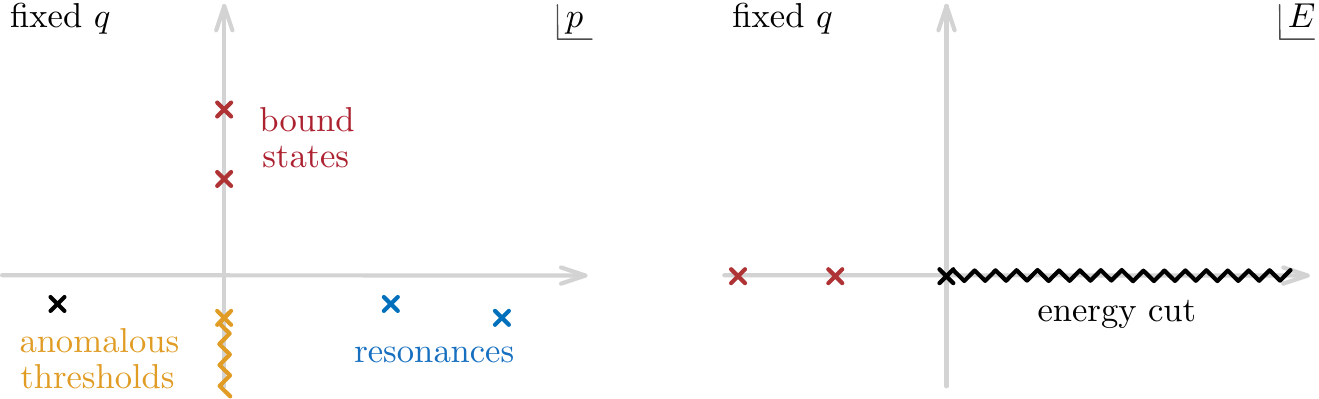}
\ee
Let me emphasize that bound states are not visible at any order in perturbation theory. They instead come from the fact that the Born series diverges for specific momenta.

So far we mostly looked into the fixed-$q$ scenario. A very important comment is that the above discussion strongly depends on which variable is held fixed when looking at the $p$-plane: the scattering angle $\theta$ or the momentum transfer $q$. For example, the Born approximation $-\frac{\lambda}{q^2 + \mu^2}$ from \eqref{eq:Yukawa-Born} is constant in $p$ at fixed $q$, but after plugging in $q^2 = 4p^2 \sin^2 \tfrac{\theta}{2}$ and keeping $\theta$ fixed, the amplitude has two simple poles at
\be
p = \pm \frac{i\mu}{2 \sin \tfrac{\theta}{2}}\, .
\ee
Likewise, the subleading Born correction \eqref{eq:f-NLO-final} also features singularities in the upper half-plane. For example, for fixed $\theta = \pi/3$, it looks like:
\be
\includegraphics[scale=0.55,valign=c]{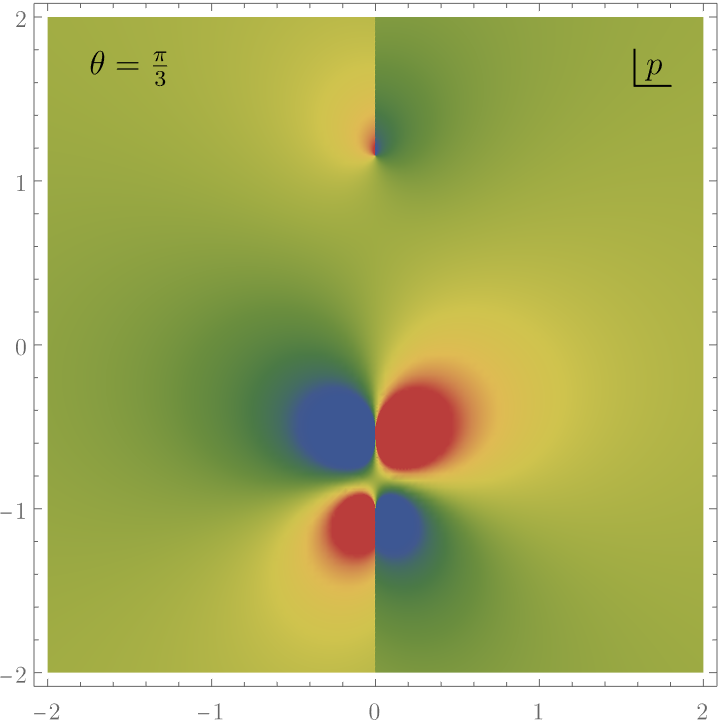}
\ee

This discussion highlights that what we're really supposed to be doing is looking at the joint analyticity in two variables, say $(p,\cos \theta)$ or $(p^2,q^2)$. Determining those properties is a more complicated task, which we'll study in the next lecture.

\subsubsection{To partial amplitudes and back again}

Another thing we can ask about is the behavior of the partial wave amplitudes $f_j(p)$. One can compute them with a modified version of the Born expansion, but once we evaluated the total amplitude, it's actually easier to determine $f_j(p)$ directly using the formula \eqref{eq:total-to-partial}.

For example, in the scalar case $j=0$, we only have to compute
\be
f_0(p) = \frac{p}{2} \int_{-1}^{1} f(p,\theta)\, \d(\cos \theta)\,.
\ee
Plugging in the Born approximation \eqref{eq:Yukawa-Born} with $q^2 = 2p^2 (1 - \cos\theta)$, we find
\begin{subequations}\label{eq:f0-Born}
	\begin{align}
		f_0^{\mathrm{Born}}(p) &= -\frac{\lambda p}{2} \int_{-1}^{1} \frac{\d(\cos \theta)}{2p^2(1- \cos \theta) + \mu^2}\\
		&= -\frac{\lambda}{4p} \log \left( 1 + \frac{4p^2}{\mu^2} \right)\, .
	\end{align}
\end{subequations}
We see that even the leading Born term has branch cuts starting at $p = \pm i\mu/2$. It arises as an endpoint singularity at $\cos \theta = -1$, i.e., backwards scattering (forward scattering $\cos \theta = 1$ would give a singularity at $\mu=0$). One can show that this property is shared among all $j$ and to all orders in perturbation theory. The resulting analytic properties in the $p$ and $E$-plane look like this:
\be
\includegraphics[scale=1.05,valign=c]{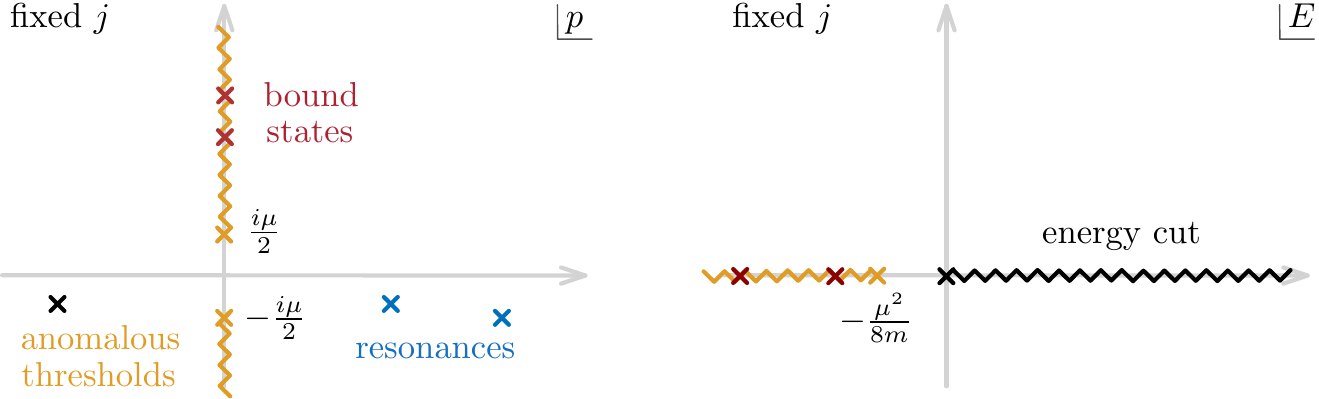}
\ee

There are two interesting questions we can ask in relation to previous results. The first one is: if all partial amplitudes $f_j(p)$ really have a branch cut in the upper half-plane, and the total amplitude $f(p,\theta)$ can be written as a sum over $f_j$'s, how come it doesn't inherit the cut? The resolution of this puzzle is that, in general, one cannot read off analytic properties of a function if its series expansion diverges. For example, take a look at the function
\be
\sum_{j=1}^{\infty} z^{-j} = \frac{1}{z-1}\, .
\ee
Every coefficient in the Taylor expansion has a pole at $z=0$. But the series doesn't converge in the unit disk $|z|\leq 1$ (it only does on the outside of it, $|z|>1$), which means we can't use it directly to determine analytic properties in $|z|\leq 1$. As a matter of fact, the result on the right-hand side is non-singular at $z=0$.

Another puzzle is about compatibility with the discussion of finite-range potentials in the previous lecture, where we claimed that the total amplitude is always meromorhpic and in particular doesn't have any branch cuts. You might think that since the Yukawa potential has an effective range of $\sim 1/\mu$ and drops off exponentially thereafter, we might've cut it off at some larger radius, say using
\be
U(r) = \lambda \frac{\e^{-\mu r}}{r}\, \theta\! \left(10^{10}/\mu - r\right)\, 
\ee
and the physics would remain the same. After all, the mistake we're making is of order $\sim \e^{-10^{10}}$. The only way all these statements can be compatible is if the branch cut effectively appears in the limit of infinite cutoff. The way this happens is that a large number of poles start appearing as we cut off the potential at farther and farther distances.
In the limit, they all ``condense'' into a branch cut. A lesson we learn from this discussion is that branch cuts in the $p$-plane are direct consequence of the tail of the potential extending to $r\to \infty$. Note that the energy cut in the $E$-plane is always there, even for finite-range scattering.

\subsubsection{\label{sec:cutting}Unitarity cuts}

In the previous lectures we talked a lot about unitarity constraints of partial amplitudes. For example, in \eqref{eq:Im-fj} and \eqref{eq:Im-fj-bound} we found that $0 \leq \Im f_j(p) = |f_j(p)|^2 \leq 1$ for all angular momenta $j$ and momenta $p$. It's interesting to ask how unitarity works for total scattering amplitudes.

Before starting this investigation, let me emphasize an important point about the interplay between unitarity and perturbative expansion. To illustrate it, it's enough to take the Born approximation we computed in \eqref{eq:f0-Born} and plot its absolute value for a few values of the Yukawa scale $\mu$:
\be
\includegraphics[valign=c,scale=1.1]{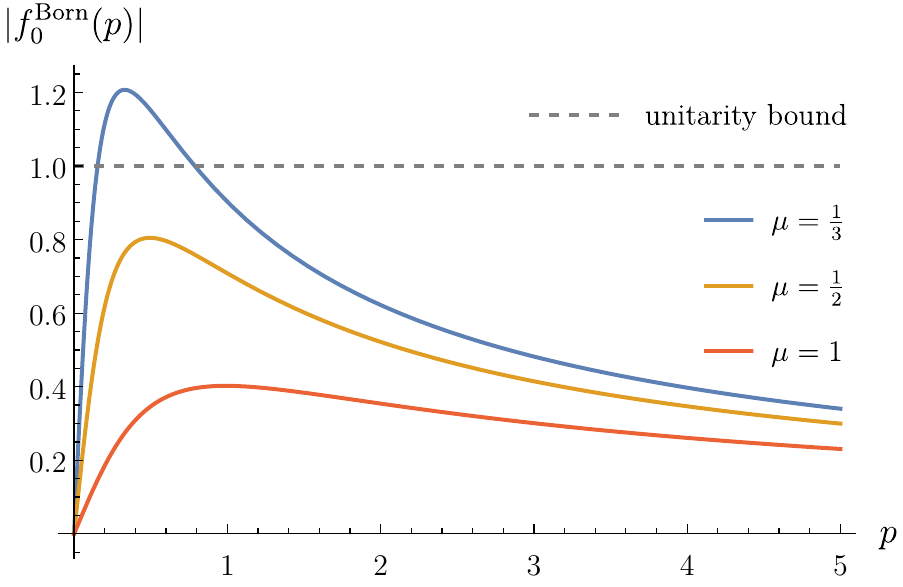}
\ee
Working in the units of $\lambda=1$, we see that for sufficiently small $\mu$, the bound $|f_0^{\mathrm{Born}}| \leq 1$ no longer works. This is not a pitfall of unitarity, but that of perturbative expansion. We can only trust it in the regime where $\lambda$ is much smaller than the other mass-scales in the problem, $\mu$ and $p$. You can think of it as putting a bound on the coupling constant beyond which the effective description in terms of Born approximation breaks down. As we'll see in Sec.~\ref{sec:EFT-bounds}, bounds of this type can give a lot of information.

With this caveat, we're now ready to ask about unitarity for the total amplitude. To emulate what happens for partial amplitudes, we're going to ask how to constrain $\Im f(\p,\p')$. In the case of the leading Born approximation, this is quite boring because $\Im f_{\mathrm{Born}}(\p,\p') = 0$, so we move on to the subleading correction \eqref{eq:f-NLO}. We can certainly compute the imaginary part directly from the answer \eqref{eq:f-NLO-final}, but it's going to be more instructive to determine it at the level of the loop integral.

Since the imaginary part of \eqref{eq:f-NLO} enters only through the propagator $\frac{1}{-\ell^2 + p^2 + i\eps}$, we can simply use the distributional identity
\begin{subequations}
	\begin{align}
		\Im\, \frac{1}{-\ell^2 + p^2 + i\eps} &= \frac{1}{2i} \left( \frac{1}{-\ell^2 + p^2 + i\eps} - \frac{1}{-\ell^2 + p^2 - i\eps} \right)\\
		&= -\pi \delta(-\ell^2 + p^2)\, .
	\end{align}
\end{subequations}
Similarly to the Sokhotski--Plemelj identity we've encountered in the first lecture, the above formula can be represented diagrammatically as 
\be
\includegraphics[scale=1,valign=c]{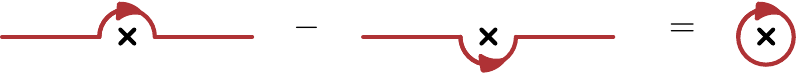}
\ee
Here, it was important that both $\ell$ and $p$ remain real. The result is that $\bm{\ell}$ is put on a similar footing to the physical momenta $\mathbf{p}$ or $\mathbf{p}'$ and it becomes on-shell. We say that there's a \emph{unitarity cut} slicing the corresponding propagator. Diagrammatically, we denote it by
\be
\includegraphics[scale=1,valign=c]{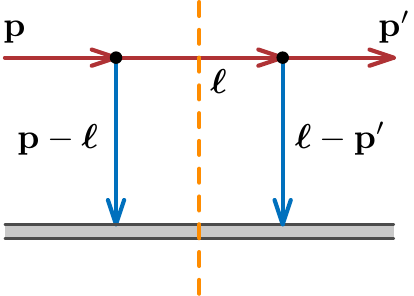}
\ee
where technically speaking also the gray line is cut, but it doesn't affect the computation, so we'll leave it like that. From the picture, both sides of the cut look like the diagram we drew for the leading Born approximation. Indeed, at the level of the equations the two uncut propagators in \eqref{eq:f-NLO} are respectively
\be
\frac{1}{(\bm{\ell} - \p)^2 + \mu^2} = -\frac{1}{\lambda} f_\Born(\p, \bm{\ell})\, ,\qquad \frac{1}{(\p' - \bm{\ell})^2 + \mu^2} = -\frac{1}{\lambda} f_\Born(\bm{\ell}, \p')\, ,
\ee
since $\bm{\ell}$ satisfies $\ell^2 = p^2$ and hence it's indistinguishable from an incoming/outgoing state.

Putting everything together, we're left with
\begin{subequations}\label{eq:unitarity-f-NLO}
	\begin{align}
		\Im f_{\mathrm{NLO}}(\p,\p') &= \frac{1}{2\pi} \int \d^3 \bm{\ell}\, \delta(\ell^2 - p^2)\, f_\Born(\p, \bm{\ell}) f_\Born(\bm{\ell}, \p')\\
		&= \frac{p}{4\pi} \int \d \Omega_{\bm{\ell}}\, f_\Born(\p, \bm{\ell}) f_\Born(\bm{\ell}, \p')\, ,
	\end{align}
\end{subequations}
where in the final transition we decomposed the measure $\d^3 \bm{\ell}$ in the spherical coordinates $\ell^2\, \d \ell \, \d \Omega_{\bm{\ell}}$ and performed the integration over $\d \ell$ using $\delta(\ell^2 - p^2) = \frac{1}{2|\ell|}[\delta (\ell - p) + \delta (\ell + p)]$, where only the first delta function has support.

We see that unlike in the fixed-$j$ case, after multiplying the two amplitudes on the right-hand side, we still have to integrate over the intermediate phase space comprised of the angular directions $\d \Omega_{\bm{\ell}}$. This result almost generalizes to the full non-perturbative expansion, except for a small subtlety that we'll get to now.

\subsubsection{\label{sec:optical}Optical theorem}

Let's now turn to a different perspective on unitarity for the full, non-perturbative amplitude, which is going to be a generalization of \eqref{eq:unitarity-f-NLO}. There's a small computation ahead, but if you're only interested in the final result, you can skip to \eqref{eq:f-unitarity}.

To this end, we're going to imagine that we're evolving a wave function $\psi_{\p}$ with incoming momentum $\p$ and measuring its amplitude along the direction of $\bm{\ell}$ (such that $|\bm{\ell}|=p$). On the other hand, we're going to consider a ``backward'' wave function $\psi_{\p'}^\ast$ with a similar property, except we replace $\p \to \p'$ and take a complex conjugate. Asymptotically, the two wave functions can be written as
\begin{subequations}
	\begin{align}
		\psi_{\p} &\sim \e^{i \p \cdot \x} + f(\p,\bm{\ell}) \frac{\e^{ip r}}{r}\, ,\\
		\psi_{\p'}^\ast &\sim \e^{-i \p' \cdot \x} + f^\ast (\p',\bm{\ell}) \frac{\e^{-ip r}}{r}\, .
	\end{align}
\end{subequations}
In order to relate the two quantities, let us introduce the following combination:
\be\label{eq:psi-p-pp}
\psi_{\p} \nabla^2 \psi_{\p'}^\ast - \psi_{\p'}^\ast \nabla^2 \psi_{\p} = 0\, ,
\ee
which is zero after using the fact that both wave functions satisfy the Schr\"odinger equation.

One thing we can do with \eqref{eq:psi-p-pp} is to integrate it inside a ball of radius $r$. By means of the second Green's identity, equivalently we can rephrase it as the following integral over the sphere $S_r$:
\be\label{eq:unitarity-psi-p-pp}
\int_{S_r} \left[ \psi_{\p} \frac{\partial \psi_{\p'}^\ast}{\partial r}  - \psi_{\p'}^\ast \frac{\partial \psi_{\p}}{\partial r} \right] r^2 \d \Omega_{\x} = 0	\, .
\ee
Here, we have to integrate over all the solid angles and we might as well use $\bm{\ell}$ as a reference vector, which amounts to setting $\d \Omega_{\x} = \d \Omega_{\bm{\ell}}$.
Now, the goal is to evaluate it in the limit of large radius $r$. We need to deal with $r$-derivatives hitting plane waves such as $\e^{i \p \cdot \x}$. Here, it's going to be useful to use the following identity:
\be
\e^{i \p \cdot \x} \sim \frac{2\pi}{i p r} \left[ \delta(\Omega_\x - \Omega_\p)\, \e^{ipr} - \delta(\Omega_\x + \Omega_\p)\, \e^{-ipr} \right]\, ,
\ee
which holds up to $\frac{1}{r^2}$ corrections. You can show it using the expansion of plane waves in spherical harmonics. Equipped with this trick, we have
\begin{subequations}
	\begin{align}
		\psi_{\p} &\sim \frac{2\pi}{ipr} \left[ \delta(\Omega_{\bm{\ell}} - \Omega_\p)\, \e^{ipr} - \delta(\Omega_{\bm{\ell}} + \Omega_{\p})\, \e^{-ipr} + \frac{ip}{2\pi} f(\p, \bm{\ell})\, \e^{i p r} \right] ,\\
		\frac{\partial \psi_{\p'}^\ast}{\partial r} &\sim \frac{2\pi}{r} \left[ \delta(\Omega_{\bm{\ell}} - \Omega_{\p'})\, \e^{-ipr} + \delta(\Omega_{\bm{\ell}} + \Omega_{\p'})\, \e^{ipr}  - \frac{ip}{2\pi} f^\ast(\p', \bm{\ell})\, \e^{-i p r}\right] .
	\end{align}
\end{subequations}
We dropped all the subleading terms. Let's use it to compute the first term in \eqref{eq:unitarity-psi-p-pp}. The result is
\begin{align}\label{eq:psi-p-der}
	\psi_{\p} \frac{\partial \psi_{\p'}^\ast}{\partial r} r^2 \sim &\; 2\pi \Big[ \delta(\Omega_{\bm{\ell}} + \Omega_{\p'}) f (\p,\bm{\ell})\, \e^{2 i p r} + \delta(\Omega_{\bm{\ell}} + \Omega_{\p}) f^\ast (\p',\bm{\ell})\, \e^{-2ipr} \Big] \\
	&+ 2\pi \Big[  \delta(\Omega_{\bm{\ell}} - \Omega_{\p'}) f(\p,\bm{\ell})  - \delta(\Omega_{\bm{\ell}} - \Omega_{\p}) f^\ast(\p',\bm{\ell}) \Big]  -ip f(\p, \bm{\ell}) f^\ast(\p', \bm{\ell})\, ,\nn
\end{align}
where we ignored terms with two delta functions multiplying each other.
After subtracting the second term in \eqref{eq:unitarity-psi-p-pp}, which amounts to replacing $\p \leftrightarrow \p'$ and complex conjugating, both terms in the first line cancel out and all those in the second line double up. Using symmetry of the problem we have $f(\p,\p') = f(\p',\p)$. 

Localizing the angular integral \eqref{eq:unitarity-psi-p-pp} using delta function we get our final answer:
\begin{empheq}[box=\graybox]{equation}
	\label{eq:f-unitarity}
	\Im f(\p, \p') = \frac{p}{4\pi} \int f(\p, \bm{\ell}) f^{\ast} (\p', \bm{\ell})\, \d \Omega_{\bm{\ell}}\, .
\end{empheq}
This is the unitarity condition on the total scattering amplitude. Diagrammatically, it translates to the unitarity cut:
\be\label{eq:optical-theorem}
\includegraphics[scale=1,valign=c]{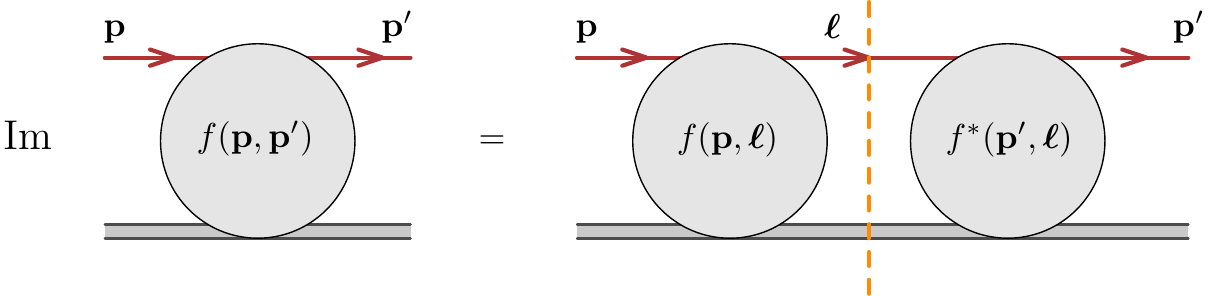}
\ee
Note that the amplitude to the right of the unitarity cut is complex conjugated. This equation is referred to as the \emph{generalized optical theorem}. In relativistic theory, diagrammatics of replacing propagators with unitarity cuts leads to the so-called Cutkosky rules.

We can no longer conclude that the imaginary part of the total amplitude has to be positive. This only happens in the forward limit, $\p = \p'$, where the integrand becomes $|f(\p,\bm{\ell})|^2$, thus implying that $\Im f(\p,\p) \geq 0$. For example, using \eqref{eq:f-NLO-forward}
\be
\Im f_{\mathrm{NLO}}(\p,\p) = \frac{p \lambda^2}{\mu^2 (4p^2 + \mu^2) }\, ,
\ee
which remains positive for all $p>0$. The forward-limit amplitude is also proportional to the total cross section, $\sigma_{\mathrm{tot}} = \frac{4\pi}{p}\, \Im f(\p,\p)$, in which case the application of the above unitarity equation is \emph{the} optical theorem, guaranteeing that $\sigma_{\mathrm{tot}}$ stays non-negative.

The unitarity equation \eqref{eq:f-unitarity} can be also applied term by term in the Born expansion. Comaparing with \eqref{eq:unitarity-f-NLO}, there's an additional complex conjugate, but it doesn't matter since the amplitudes on both sides of the cut are purely real at NLO. The first time complex conjugate starts mattering is NNLO, where it takes a little more work to show directly that \eqref{eq:f-unitarity} holds at the level of loop integration.

\subsection{\label{sec:Coulomb}Coulomb potential}

Let us now turn to Rutherford scattering. It's mediated by the Coulomb potential,
\be\label{eq:Coulomb}
U(\x) = \frac{\lambda}{r}\, .
\ee
Naively, it seems like it's simply the $\mu \to 0$ limit of the Yukawa scattering problems we just encountered above. However, \eqref{eq:Coulomb} comes with a crucial difference, in that the potential is no longer exponentially suppressed at infinity and hence can make particles interact even at large distances. We'll see that there's something fundamentally different about such potentials.

\subsubsection{IR divergences}

Let's first see what goes wrong doing a naive computation. The leading Born approximation is
\be\label{eq:Coulomb-Born}
f_{\Born}^{\mathrm{Coulomb}}(\p,\p') = -\frac{\lambda}{q^2}\, .
\ee
This amplitude gives rise to infinite cross-section at small angles, but otherwise there's nothing suspicious about it. The trouble starts with the first correction, which is now given by
\be\label{eq:f-NLO-Coulomb}
f_{\mathrm{NLO}}^{\mathrm{Coulomb}}(\p,\p') = -\frac{\lambda^2}{2\pi^2} \int \frac{\d^3 \bm{\ell}}{(2\pi)^3} \frac{\lambda}{(\bm{\ell} - \p)^2} \frac{1}{-\ell^2 + p^2 + i\eps} \frac{\lambda}{(\p' - \bm{\ell})^2}\, .
\ee
This integral is divergent no matter what value of the external momenta $\p$ and $\p'$ we choose. As we've seen from the analysis based on Landau equations, the divergence comes from the ``soft'' region where $\bm{\ell} \propto \p$ and $\bm{\ell} \propto \p'$, which implies $\ell^2 = p^2$. Unlike in the case of kinematic singularities, there's no way we can deform away from it and the integral \eqref{eq:f-NLO-Coulomb} plainly diverges. These type of problems are called \emph{IR divergences}.

One can in principle make sense of it through a trick called dimensional regularization, which instructs us to first compute \eqref{eq:f-NLO-Coulomb} in $\D$ dimensions and then expand around $\D = 3 - 2 \epsilon$ with $\epsilon \to 0$ (remember, this is a separate $\epsilon$ from that in the $i\eps$ prescription). The result is proportional to $\frac{1}{\epsilon}$.
Alternatively, we could've mass-regulated the integral, or equivalently, taken the $\mu \to 0$ limit of the Yukawa answer \eqref{eq:f-NLO-final}:
\be\label{eq:Yukawa-small-mu}
f_{\mathrm{NLO}}(\p,\p') \big|_{\mu \to 0} = -\frac{i \lambda^2}{p q^2} \log \left( \frac{\mu}{q}\right) + \ldots
\ee
One can then hope that the divergent terms proportional to $\frac{1}{\epsilon}$ or $\log \mu$ disappear for one reason or another from physical observables such as the cross-section.\footnote{Another method sometimes taken in the literature on IR divergences is to introduce an energy cutoff. Especially in the relativistic context, this approach is so disappointing, its mention was relegated to a footnote.} Physically, however, neither of these tricks is the correct way to think about IR divergences.

\subsubsection{Revisiting asymptotic states}

The core of the problem is that we really made a mistake in \eqref{eq:psi-out} by claiming that the scattering amplitude can be read off as the coefficient of $\frac{\e^{ipr}}{r}$ of the outgoing wave function. This form tacitly assumes that the waves behaving as $\sim \frac{1}{r}$ have already interacted with the scatterer and evolve freely at infinity. But in the Coulomb case, the potential itself behaves as $\sim \frac{1}{r}$. Hence our assumption wasn't correct and we need to revisit the definition of the scattering amplitude. All the trouble with dimensional or mass regularization had to do with forcing the computation into a framework that shouldn't have been applied in the first place.

Luckily for us, scattering off a Coulomb potential is a problem that can be solved \emph{exactly}. We start, as usual, with the Schr\"odinger equation:
\be\label{eq:Schrodinger-Coulomb}
\left(\nabla^2 + p^2 - \frac{\lambda}{r}\right) \psi(\p,\x) = 0\, .
\ee
To understand the nature of the asymptotic state, let's make an ansatz for the new behavior at infinity. First, we consider a correction to the outgoing wave function with an undetermined function $g(r,z)$:
\be
\psi_{\mathrm{out}}(\p,\x) \sim \frac{\e^{i p r + g(r,z)}}{r}\, ,
\ee
where $g(r,z) = 0$ corresponds to free asymptotic propagation. Recall that $z = r \cos \theta$. To simplify the manipulations, it'll be useful to use parabolic coordinates $\xi_{\pm} = r \pm z$, in terms of which the Laplacian equals
\be
\nabla^2 = \frac{4}{\xi_+ + \xi_-} \left( \frac{\partial}{\partial \xi_+} \xi_+ \frac{\partial}{\partial \xi_+} + \frac{\partial}{\partial \xi_-} \xi_- \frac{\partial}{\partial \xi_-} \right) + \frac{1}{\xi_+ \xi_-} \frac{\partial^2}{\partial \phi^2}\, .
\ee
Plugging everything back into the Schr\"odinger equation \eqref{eq:Schrodinger-Coulomb} and expanding to leading order in $\xi_{\pm}$, we get the constraint
\be
\left(\xi_+ \frac{\partial}{\partial \xi_+} + \xi_- \frac{\partial}{\partial \xi_-} + 1 + \frac{i\lambda}{2p}\right)  g(\xi_+, \xi_-) = 0\, .
\ee
The simplest non-zero solution would be if $g$ had only dependence on one of the variables, say $\xi_-$. We can then drop the $\xi_+$ derivatives and are left with
\begin{subequations}\label{eq:g}
	\begin{align}
		g(\xi_-) &= C - \frac{i\lambda}{2p} \log(\xi_-) \\
		&= -\frac{i\lambda}{2p} \log[p (r-z)]\, ,
	\end{align}
\end{subequations}
where we fixed the constant $C = -\frac{i\lambda}{2p} \log p$, which doesn't really matter for the asymptotic behavior and we include it to make the argument of the logarithm dimensionless.

We can repeat a similar exercise for incoming wave functions, which results in the refined asymptotic behavior
\be\label{eq:Coulomb-incoming}
\psi_{\mathrm{in}}(\p, \x) \sim \e^{i \p \cdot \x + \frac{i\lambda}{2p} \log[p(r-z)]}\, .
\ee
These are the corrected incoming and outgoing asymptotic wave functions for Coulomb scattering. Notice that this correction becomes particularly important for soft momenta ($p \to 0$) and in the collinear limit ($r \to z$). Let me also draw attention to the fact that the relative phase of the incoming and outgoing wave functions was chosen somewhat arbitrarily and will eventually trickle into the choice of phase for the scattering amplitude, which is inherently non-unique.

\subsubsection{Exact scattering amplitude}

The Coulomb interaction is actually a rare case for which the Schr\"odinger equation \eqref{eq:Schrodinger-Coulomb} can be solved exactly. This is in contrast with the Yukawa potential, for which such a solution isn't known and we have to resort to perturbation theory. 

Let's try the ansatz $\psi(\p,\x) = \e^{i \p \cdot \x} h(\xi_-)$ and once again demand that the additional correction $h(\xi_-)$ is $\xi_+$- and $\phi$-independent. It translates to the following differential equation for $h(\xi_-)$:
\be 
\left[ \xi_- \frac{\partial^2}{\partial \xi_-^2} + (1 - i p \xi_-) \frac{\partial}{\partial \xi_-} - \frac{\lambda}{2}\right] h(\xi_-) = 0\, .
\ee
This happens to be a famous differential for the confluent hypergeometric function ${}_1 F_1$. Imposing regularity at $r=0$, the solution gives:
\be
\psi(\p, \x) = A\, \e^{i \p \cdot \x}\, {}_1 F_1 \left(\tfrac{\lambda}{2ip},\, 1 ;\, ip(r-z) \right)\, .
\ee
The overall normalization $A$ doesn't matter for our purposes.
Recognizing the fact it's a function a German mathematician named a couple of centuries ago doesn't immediately solve our problems, but it means its properties are probably well-understood by now. For example, we can look up how it behaves asymptotically in the large argument $r-z$, giving:
\be
\psi(\p, \x) \sim \frac{A}{\Gamma(1+ \frac{i\lambda}{2p})} \left[ \e^{i \p \cdot \x + \frac{i\lambda }{2 p}\log[p(r-z)]} + \left\{- \frac{\lambda}{2p^2}\frac{\Gamma(1+\frac{i\lambda}{2p})}{\Gamma(1-\frac{i\lambda}{2p})} \frac{r}{r-z} \right\} \frac{\e^{i p r - \frac{i\lambda}{2p} \log [p(r-z)]}}{r}  \right] .
\ee
Here, $\Gamma$ is the gamma function. Everything looks in order: the incoming and outgoing waves received corrections consistent with \eqref{eq:g} and \eqref{eq:Coulomb-incoming}. The coefficient of the latter, in curly brackets, is what we \emph{define} to be the scattering amplitude. Using some trigonometry we can simplify $\frac{r}{r-z} = \frac{1}{2 \sin^2(\theta/2)} = \frac{2p^2}{q^2}$, which gives 
\begin{empheq}[box=\graybox]{equation}
	\label{eq:Coulomb-amplitude}
	f_{\text{Coulomb}}(\p,\p') = - \frac{\lambda}{q^2}\frac{\Gamma(1+\frac{i\lambda}{2p})}{\Gamma(1-\frac{i\lambda}{2p})}\, .
\end{empheq}
This is a remarkable formula and deserves to be boxed. Analogous formulae exist for partial wave amplitudes.

We can attempt to expand \eqref{eq:Coulomb-amplitude} in perturbation theory. Since $\Gamma(1)=1$, at leading order we find agreement with the Born approximation \eqref{eq:Coulomb-Born}. At NLO, we find a result proportional to $\frac{1}{p q^2}$, which is the same kinematic dependence as in \eqref{eq:Yukawa-small-mu}. Effectively, the extra naive logarithm $\log (\mu/q)$ came from the fact we forgot to include $\log[p(r-z)]$ in the definition of asymptotic wave function.

\subsubsection{Spectrum of the hydrogen atom}

It would be a shame not to plot \eqref{eq:Coulomb-amplitude}. Setting $-\lambda = q^2 = 1$, the result in the $p$-plane looks as follows:
\be
\includegraphics[scale=0.6,valign=c]{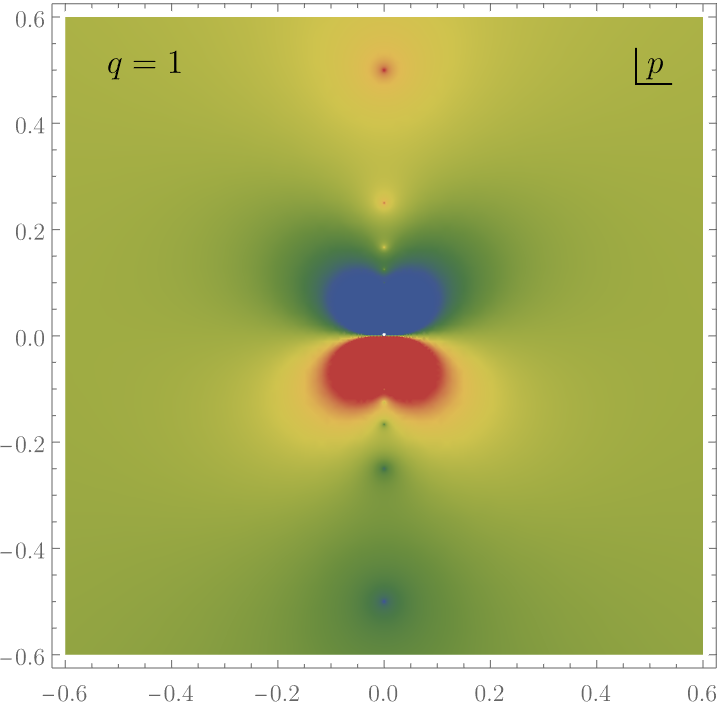}
\ee
The choice of $\lambda < 0$ corresponds to attractive Coulomb potential, so as expected, there are bound state poles in the upper half-plane. In fact, there's an infinite number of them.

As a matter of fact, these bound states are nothing but the energy levels of the hydrogen atom.  It's fun to see how they arise from the scattering amplitude \eqref{eq:Coulomb-amplitude}. In this case, the coupling constant is $\lambda = - \frac{2m_e}{\hbar^2} e^2 k_e$, where $m_e$ is the mass of the electron and $k_e = \frac{1}{4\pi \varepsilon_0}$ is the Coulomb constant. The gamma function $\Gamma$ has simple poles whenever its argument is a non-positive integer. In other words, the amplitude has a simple pole for
\be
1 + \frac{i \lambda}{2p} = 1-n
\ee
for $n = 1,2,3,\ldots$. Recall that the energy is given by $E_n = \frac{\hbar^2 p^2}{2m_e}$ after reinstating the $\hbar$, and hence we find
\be
E_n = - \frac{ k_e^2 m_e e^4 }{2 \hbar^2} \frac{1}{n^2}\, ,
\ee
which is the correct spectrum of the hydrogen atom.

\subsection{Exercises}

In this set of exercises, we're going to combine what we've learned about numerical methods for reconstructing analytic structure from data to rediscover QCD resonances.

\subsubsection{BESIII data}

We're going to use data from the BESIII (Beijing Spectrometer III) experiment, which studied radiative decays of the $J/\psi$ meson into pairs of neutral pions $\pi^0 \pi^0$ and a photon $\gamma$. While the physics of this process is slightly more complicated than what we studied, the basic feature of unstable particles remains the same. Recall that the place to look for them is the lower half-plane of $p$.

The first goal is to download the experimental data. We're going to need the file \texttt{nominal\_results\_1.txt} from
\be
{\small
\href{https://journals.aps.org/prd/supplemental/10.1103/PhysRevD.92.052003}{\texttt{https://journals.aps.org/prd/supplemental/10.1103/PhysRevD.92.052003}}\, .}
\ee
If you're reading these notes in a distant future and the link doesn't work, just get hold of some other experimental data with peaks from the internet. If you're reading it in a post-apocalyptic future and the internet doesn't exist, you probably should be doing other things anyway.

The data has many columns in it, but for us we're interested in just the first three, representing the momentum $p$ of the $\pi^0 \pi^0$ system, the intensity $I_0(p)$ of the $0^{++}$ amplitude (equivalent of our $j=0$), and its error $E_0(p)$. The mass is in the units of $\mathrm{GeV}/c^2$ and the others in some arbitrary normalization that doesn't matter for us. You can extract the data as follows:
\begin{minted}[firstnumber=1]{mathematica}
data = Import["path/to/nominal_results_1.txt", "Data"];
\end{minted}
There are $182$ data points. We're only interested in its first three columns.

To have a first look at the data, we can simply plot it using
\begin{minted}{mathematica}
ErrorBars[expr_] := Map[Function[{#[[1]], Around[#[[2]], #[[3]]]}], expr];
ListPlot[ErrorBars[data], AspectRatio -> 1/3, PlotRange -> Full]
\end{minted}
The function \texttt{ErrorBars[expr]} in the first line just creates the data points with error bars. The result should look more or less like this:
\be
\includegraphics[scale=0.95,valign=c]{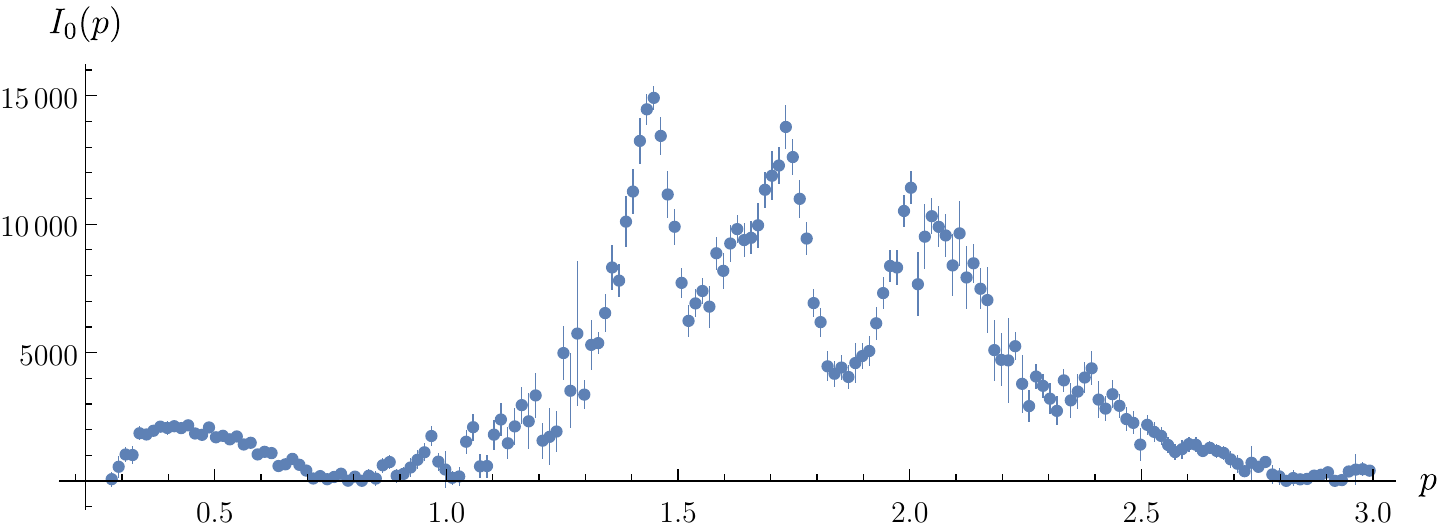}
\ee
Even by eye, we can observe three distinct peaks. They come from resonances called
\be
f_0(1500), \quad f_0(1710), \quad f_0(2020)
\ee
respectively. The names include their approximate mass $m$ in $\mathrm{MeV}/c^2$ (please don't confuse them for the partial amplitude $f_0(p)$). We can try to read off a more accurate measurement of the mass and width by fitting to a Breit--Wigner profile. You're free to try your hand at it. But this is a course on analyticity, so we're going to follow a different, though equivalent, path.

\subsubsection{Schlessinger point method}

First of all, intensity $I_0(p)$ is not really an analytic function. It equals
\be
I_0(p) = \sqrt{1 - 4m_{\pi}^2/p^2}\, |f_0(p)|^2\, ,
\ee
where $m_\pi \approx 135\, \mathrm{MeV}/c^2$ is the pion mass and $f_0(p)$ is the partial amplitude. However, if we approximate $f_0(p)$ as a rational function with poles in the lower half-plane (which is a valid assumption in this case), then $I_0(p)$ also shares the same poles and their complex conjugates. Therefore, we're going to look at $I_0(p)$ in the lower half-plane and assume it's rational.
This might naively seem like a great real-life application for the Pad\'e approximants we've been playing with in the previous set of exercises. However, applying them directly doesn't give great results, essentially because the data is quite noisy. Try it.

Instead, we're going to explore another option, which is a statistical version of Pad\'e approximants called the \emph{Schlessinger point method}. The basic idea is to select a random sample of $2N-1$ data points, compute the Pad\'e approximant with $(m,n) = (N-1,N)$ exactly (instead of optimizing for it), and record the positions of poles of the result. We then repeat the same procedure for a large number of random samples. Positions of poles will tend to cluster around the ``true'' poles, giving us a reasonable estimate for the masses and widths of the resonances.

To account for the error bars, for every sample we're going to wiggle the data point within the error bars. This way, we're going to get more accurate results in the limit of large number of samples. Getting a single sample with a specific size can be achieved with the code:
\begin{minted}{mathematica}
getSample[data_, size_] := Block[{sample,sampleNoise},
	sample = RandomSample[data, size];
	sampleNoise = Table[{s[[1]], s[[2]] +
				   		 RandomVariate[NormalDistribution[]]*s[[3]]},
				  {s, sample}];
	Return[sampleNoise];
];
\end{minted}
Here, \texttt{sample} is a random sample of \texttt{size} data points from \texttt{data}. For every data point \texttt{s} in \texttt{sample}, we add random noise consistent with the error bars. You can play around with different ways of selecting samples. For example, it would make sense to preferentially keep the data points that have the smallest error bars, since they're the most reliable. Plot a few random samples and make sure they look reasonable.

Now it's time to compute Pad\'e approximants. You can of course do it by writing an ansatz, just like we did in \eqref{eq:Pade-approximant} and then solving numerically for the coefficients. However, there's a better way. Let's rewrite the Pad\'e ansatz as a continued fraction:
\be\label{eq:g-m}
g_{m}(p) = \frac{f(p_1)}{1 + \frac{c_1(p - p_1)}{1 + \frac{c_2(p - p_2)}{1 + \cdots}}}\, ,
\ee
which is equivalent to $g_{\lfloor \frac{m-1}{2} \rfloor, \lfloor \frac{m}{2} \rfloor}(p)$ if we stop the recursion at the term $c_{m-1}(p - p_{m-1})$. In this form, the solution for fitting $m$ data points by imposing $g_m(p_i) = f(p_i)$ for all $i=1,2,\ldots,m$. It takes a simple recursive form:
\be
c_1 = \frac{\frac{f(p_1)}{f(p_2)}-1}{p_2 - p_1}
\ee
and
\be
c_k = \frac{1}{p_k - p_{k+1}}\left( 1 + \frac{c_{k-1}(p_{k+1} - p_{k-1})}{1 + \frac{c_{k-2}(p_{k+1} - p_{k-2})}{1 + \cdots}} \right)
\ee
The last term is $\frac{c_1 (p_{k+1} - p_1)}{1 - f(p_1)/f(p_{k+1})}$. The proof isn't entirely illuminating, but to make sure you understand the formula, it's instructive to test it out for a couple of values of $m$.

Anyway, this way we can quickly get the numerical solution, which is important to speed up the process when the number of samples is large.
One way to implement this solution is as follows:
\begin{minted}{mathematica}
Continued[terms_] := Fold[#2/(1 + #1) &, Reverse@terms];
SPM[dataSample_] := Block[{n, pi, f, c},
    n = Length[dataSample];
    pi = dataSample[[All, 1]];
    f = dataSample[[All, 2]];
    c[1] := c[1] = (f[[1]]/f[[2]] - 1)/(pi[[2]] - pi[[1]]);
    c[k_] := c[k] = 1/(pi[[k]] - pi[[k + 1]])*(1 +
     				Continued[Join[Table[c[i]*(pi[[k + 1]] - pi[[i]]),
     				{i, k - 1, 1, -1}], {-(f[[1]]/f[[k + 1]])}]]);
    Return[Together[Continued[
    	   Join[{f[[1]]},Table[c[i]*(p - pi[[i]]), {i, n - 1}]]]]];
];
\end{minted}
It might look slightly intimidating, but it only implements the solution \eqref{eq:g-m}.
Here \texttt{SPM[dataSample]} is a function that takes the \texttt{dataSample} and spits out the fitted function. \texttt{Continued[terms]} is just an auxiliary functions implementing continued fractions for a list of \texttt{terms}. Recall that the notation \texttt{c[k\_] := c[k] =} saves the result in the memory so that one doesn't have to recompute \texttt{c[k]} over and over.

Make sure that your \texttt{SPM} routine gives correct results in simple examples. Also test it out on some individual samples from the BESIII data. Notice that the size of the sample can't be too large, because at some point the numerics breaks down.

\subsubsection{Analyticity from experiment} 

We can now put everything together by generating many many random samples, fitting them using the Schlessinger formula, and saving the positions of the poles. One way to implement it is
\begin{minted}{mathematica}
noSamples = 10^5;
pointsPerSample = 25;
poles = Flatten[ParallelTable[Select[p /.
	    Solve[Denominator[SPM[getSample[data, pointsPerSample]]] == 0],
	    -0.25 < Im[#] < 0 && 0 < Re[#] < 3 &], {i, noSamples}]];
\end{minted}
The two parameters \texttt{noSamples} and \texttt{pointsPerSample} are self-explanatory. Within the loop, we solved for positions of the poles of each $g_m(p)$ and saved only those ones in the kinematic region we expect to see interesting features.

It remains to plot the results in the complex $p$-plane. We should probably put \texttt{poles} in bins and plot their histogram, but there's a hacky solution that gets us results faster. Let's just plot all the data points with a larger point size and some transparency:
\vspace{-1em}
\begin{minted}{mathematica}
ListPlot[ReIm[poles], PlotRange -> {{1, 2.5}, {-0.25, 0}},
	     PlotStyle -> {Opacity[0.017], PointSize[0.015]}]
\end{minted}
This way, clusters of poles will be more visible. After running the code, I get the following result:
\be
\includegraphics[scale=0.8,valign=c]{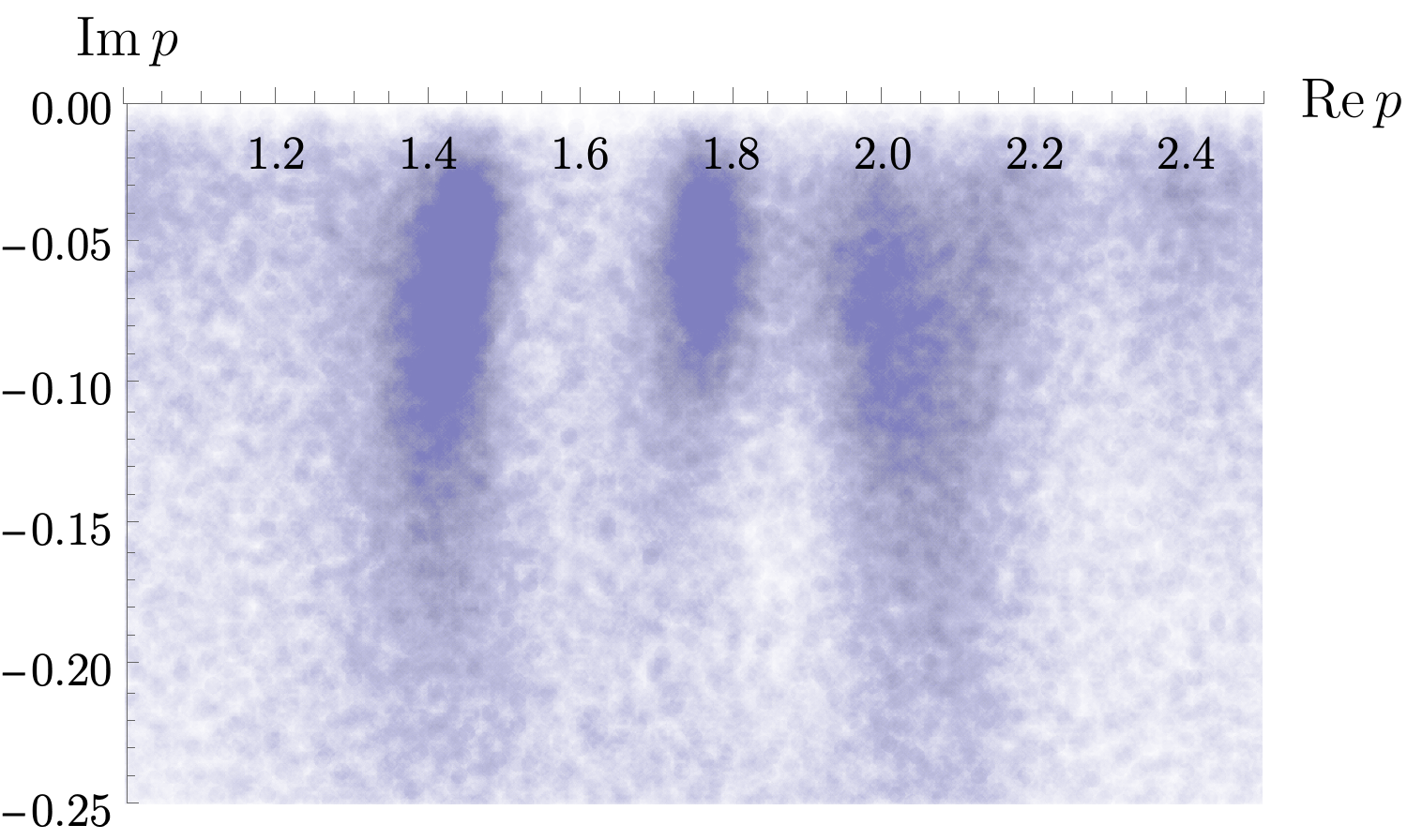}
\ee
This is the first (and the last) time we see a lower half-plane coming from a real experiment! We can clearly see three resonances mentioned before. Recall that if their position is $p = m - i\Gamma/2$, then $m$ is the mass and $\Gamma$ is the decay width. Can you measure them from the plot? Compare the result with the current estimates by the JPAC collaboration, which as of 2021 are:
\begin{subequations}
	\begin{align}
		m_{f_0(1500)} = 1450 \pm 10\, \text{MeV}/c^2, \qquad &\Gamma_{f_0(1500)} = 106 \pm 16\, \text{MeV}/c^2\\
		m_{f_0(1710)} = 1769 \pm 8\, \text{MeV}/c^2, \;\;\qquad &\Gamma_{f_0(1710)} = 156 \pm 12\, \text{MeV}/c^2\\
		m_{f_0(2020)} = 2038 \pm 48\, \text{MeV}/c^2, \qquad &\Gamma_{f_0(2020)} = 312 \pm 82\, \text{MeV}/c^2
	\end{align}
\end{subequations}
Are the results consistent?
Experiment with adjusting the parameters, different ways of selecting the samples, reading off the position of the poles and their uncertainties, etc. Ask your favorite artificial intelligence assistant to do the analysis for you and compare how your results stack against it.

\section{Lecture IV: Maximal analyticity}

So far, we primarily focused on studying properties of scattering amplitudes in terms of the complexified momentum following from causality. In this lecture, we'll look more closely at the analytic properties in the momentum transfer, or equivalently the scattering angle, and their connection to locality. Finally, we'll wrap everything together by making statements about joint analyticity in both variables simultaneously, and see how in a sense it allows us to replace the Schr\"odinger equation.

\medskip
\noindent\rule{\textwidth}{.4pt}
\vspace{-2em}
\localtableofcontents
\noindent\rule{\textwidth}{.4pt}

\pagebreak

\subsection{\label{sec:complex-angles}Complex angles}

Our first goal in this lecture will be to look at the analytic properties following from locality for Yukawa-like potentials. As we've seen in the previous lectures, there's a certain amount of freedom in which specific kinematic invariants we choose to work with. From now on, we're going to commit to the following \emph{Mandelstam invariants}:
\begin{empheq}[box=\graybox]{equation}\label{eq:Mandelstam-invariants}
s = \p^2 = p^2, \qquad t = -(\p - \p')^2 = -q^2\, .
\end{empheq}
Recall that $t = -2s (1 - \cos\theta)$, so the physically admissible values are
\be\label{eq:t-physical}
0 \leq -t \leq 4s\, ,
\ee
which amounts to requiring that energies are positive and angles are real.
We already know a fair bit about analyticity properties in $s$-plane and now we want to look at those in the $t$-plane. Since we already have some theoretical data at this stage, it seems prudent to analyze it first.

The Born approximation $\frac{\lambda}{t - \mu^2}$ had a simple pole at $t =  \mu^2$. On the other hand, the NLO term has two branch cuts: a logarithmic one starting at
\be
t = 4\mu^2
\ee
and a square-root one starting at 
\be
t = \frac{\mu^2}{s}(\mu^2 + 4s) > 4\mu^2\, .
\ee
We can read-off this information either from the Landau equation analysis in Sec.~\ref{sec:anomalous} or the explicit expression \eqref{eq:f-NLO-final}. The naive NLO singularity at $t = 0$ cancels out on the principal sheet. 

Based on this observation, we might expect that also the full non-perturbative amplitude has singularities in those places and perhaps more. In Sec.~\ref{sec:locality}, we've seen that analyticity in the angle or momentum transfer follows from locality of interactions. Back then, it was the fact that the scatterer was contained within some finite radius that guaranteed higher-$j$ partial amplitudes were extremely suppressed. By Neumann's theorem, this fact translated to analyticity within an ellipse in the complex $\cos \theta$-plane. For cutoff potentials supported inside a given radius, the size of this ellipse is infinite, which guarantees analyticity for any complex value of $\cos \theta$.

We can repeat a similar analysis for the Yukawa case. Even before doing any computations, we can notice that the two leading terms in the Born expansion are analytic within some regions near the origin in the $t$-plane, respectively: 
\be\label{eq:q2-singularities}
\includegraphics[scale=0.53,valign=c]{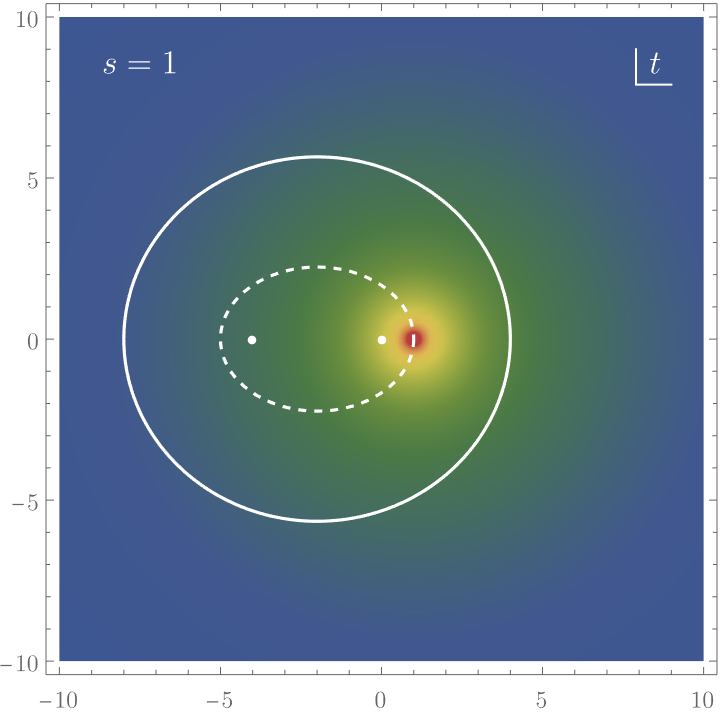}\quad
\includegraphics[scale=0.53,valign=c]{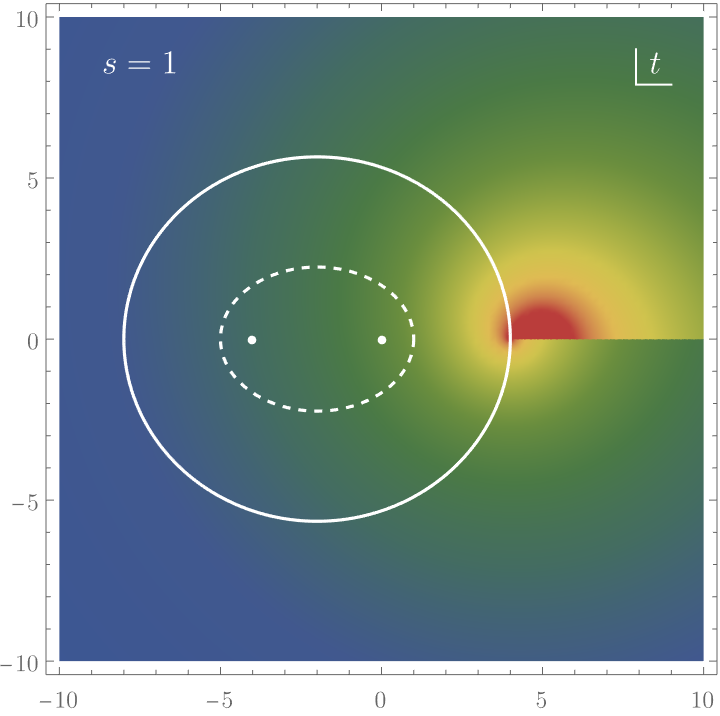}
\ee
For the purpose of the plot, we set $\mu=s=1$. The Born term doesn't have singularities within the smaller (dashed) ellipse, and the NLO term is analytic in the larger (solid) one. Recall that the focii of the ellipses are at $\cos \theta = \pm 1$, which translates to $t = -4s$ and $0$, indicated with dots above. The physical region \eqref{eq:t-physical} is the interval stretching between those two points. Sizes of the ellipses are dictated by the distance to the nearest singularity, $t=\mu^2$ and $t = 4\mu^2$ respectively, which grows with larger $\mu$.

This is indeed what we expect from the general arguments: Yukawa potentials have a range $\sim 1/\mu$, so the smaller the range, the more analyticity in $t$ we expect. It's time to make these statements more quantitative.

\subsubsection{Probing locality with wave packets}

The idea will be to set up a scattering process that probes locality of interactions. The most intuitive way to do it is to send a wave packet that comes within some distance $b$ of the origin, called the \emph{impact parameter}, and ask how the amplitude changes as we increase $b$. After some computations, we're going to find an upper bound on the outgoing wave function. If you're just interested in the final result, feel free to skip directly to Sec.~\ref{sec:Lehmann-ellipses}.

In momentum space, we can use an incoming wave packet whose momentum is centered around some mean momentum $\p$:
\be
\psi_{\mathrm{in}}(\p') = \frac{\gamma^{3/2}}{\pi^{3/4}} \e^{ - \gamma^2 (\p' - \p)^2/2  - i \p' \cdot \mathbf{b}}\, ,
\ee
where $\mathbf{b}$ is the impact vector with $b = |\mathbf{b}|$ and $\gamma$ measures its width in momentum space. The prefactors are just the canonical normalization of the wave function. It might be simpler to understand it in the position space, which after a Fourier transform gives the probability density
\begin{subequations}\label{eq:psi-in-packet}
\begin{align}
|\psi_{\mathrm{in}}(\x, \t)|^2 &= \left| \int  \psi_{\text{in}}(\p')\, \e^{-\frac{i {\p'}^2 t}{2m}} \, \d^3 \p' \right|^2  \\
&= \frac{1}{\pi^{3/2} \Gamma^{3}(\t)} \e^{ - (\mathbf{x} - \hat{\mathbf{x}})^2 /\Gamma^2 }\, .
\end{align}
\end{subequations}
The quickest way to perform this integral is to complete the square.
The result is written in terms of the mean position of the peak $\hat{\mathbf{x}}$ and the width of the wave packet $\Gamma$ in positions space, which are given by
\be
\hat{\mathbf{x}}(\t) = \mathbf{b} + \p\, \t/m\, , \qquad \Gamma(\t) = \sqrt{\gamma^2 + \frac{\t^2}{m^2 \gamma^2}}\, .
\ee
They both depend on the time $\mathrm{t}$ (not to be confused with the Mandelstam invariant $t$).
To simplify our lives, we can take $\p \cdot \mathbf{b} = 0$, so that the closest the wave packet comes to the origin happens at $\t=0$.
\be
\includegraphics[valign=c,scale=1.1]{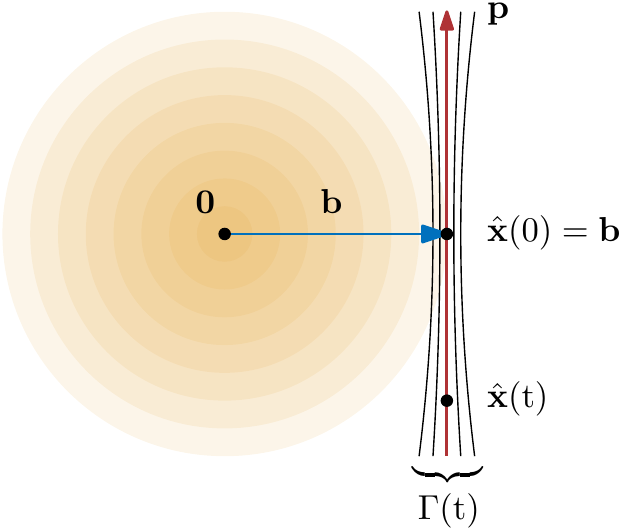}
\ee

Notice how the width of the wave packet $\Gamma(\t)$ depends on time. When $\t\sim 0$, the width equals roughly $\gamma$, but as we cross $\t \sim m \gamma $, it starts \emph{spreading} more or less linearly in time. So if we kept $\gamma$ and $\p$ fixed, the asymptotic behavior for small and large $\t$ is
\be\label{eq:psi-in-asymptotic}
\left| \psi_{\mathrm{in}}(\x, 0) \right|^2 \sim \e^{- (\x - \mathbf{b})^2/ \gamma^2 }\, ,\qquad
\left|\psi_{\mathrm{in}}(\x, \pm \infty) \right|^2 \sim \e^{- p^2 \gamma^2}\, .
\ee
respectively, with $p = |\p|$.
We see that the dependence on the impact parameter $b$ is entirely washed out at late and early times $\t \to \pm \infty$. This is a rather undesirable effect if want to probe locality, because the system essentially forgets how close the wave function made it to the scatterer. For this purpose, we need to design more special wave packets. 

The degree of freedom we have at our disposal is how to tune $\gamma$ as a function of the impact parameter $b$. The optimal choice would be to set $\gamma^2 \propto b$, which makes $\psi_{\mathrm{in}}$ exponentially suppressed with the impact parameter $b$ at finite and asymptotic times. In fact, if we choose
\be
\gamma^2 = \frac{b}{p}\, ,
\ee
then both contributions \eqref{eq:psi-in-asymptotic} go as $\sim \e^{-pb}$ as $b \to \infty$. It will turn out that this is the optimal choice of wave packets probing locality.

\subsubsection{Cook's inequality}

Intuitively, we expect that an incoming wave function $\psi_{\mathrm{in}}$ can't produce an arbitrarily large outgoing $\psi_{\mathrm{out}}$ if the potential $V$ is well-behaved. Therefore, we expect that there's a bound on the norm of $\psi_{\mathrm{out}}$ in terms of the other two quantities. We'll derive such a bound now.

The total wave function $\psi = \psi_{\mathrm{in}} + \psi_{\mathrm{out}}$ satisfies the Schr\"odinger equation, which we can write as
\be
i\hbar \frac{\partial (\psi_{\mathrm{in}} + \psi_{\mathrm{out}})}{\partial \t} = (H_0 + V) (\psi_{\mathrm{in}} + \psi_{\mathrm{out}})\, ,
\ee
where $H_0$ is the free Hamiltonian and $V$ is the potential. Subtracting away the Schr\"odinger equation for the incoming free wave ($i\hbar \frac{\partial \psi_{\mathrm{in}}}{\partial \t} = H_0 \psi_{\mathrm{in}}$), we get
\be\label{eq:psi-out-Schrodinger}
i\hbar \frac{\partial  \psi_{\mathrm{out}}}{\partial \t} = H_0 \psi_{\mathrm{out}} + V (\psi_{\mathrm{in}} + \psi_{\mathrm{out}})\, .
\ee
Let's now compute how the norm of the outgoing wave packet
\be
\norm{\psi_{\mathrm{out}}(\t)}^2 = \int |\psi_{\mathrm{out}}(\x,\t) |^2\, \d^3 \x
\ee
changes with time. Using \eqref{eq:psi-out-Schrodinger} and setting $\hbar=1$, its time derivative becomes
\begin{subequations}
	\begin{align}
		\frac{\partial \norm{\psi_{\mathrm{out}}}^2 }{\partial \t} &= -i \int \left( V \psi_{\mathrm{in}} \psi^\ast_{\mathrm{out}} - \psi_{\mathrm{out}} V \psi_{\mathrm{in}}^\ast \right) \d^3 \x \\
		&= 2\, \Im \int \psi_{\mathrm{out}}^\ast\, V \psi_{\mathrm{in}}\, \d^3 \x\, .
	\end{align}
\end{subequations}
Here, the terms quadratic in $\psi_{\mathrm{out}}$ all canceled out and we're left with only cross-terms. We used the fact that the potential is real. The imaginary part is certainly no greater than the absolute value, so we find
\begin{subequations}
	\begin{align}
		\frac{\partial \norm{\psi_{\mathrm{out}}}^2 }{\partial \t} &\leq 2 \left| \int \psi_{\mathrm{out}}^\ast\, V \psi_{\mathrm{in}}\, \d^3 \x \right|\\
		&\leq 2 \norm{\psi_{\mathrm{out}}} \norm{ V \psi_{\mathrm{in}}}\, ,
	\end{align}
\end{subequations}
where in the second line we used the Cauchy--Schwarz inequality, treating $V \psi_{\mathrm{in}}$ together. This tells us $\frac{\partial}{\partial \t} \norm{\psi_{\mathrm{out}}} \leq \norm{V \psi_{\mathrm{in}}}$. Assuming that the incoming and outgoing wave packets decay in the infinite past and future, we can integrate this inequality over time, resulting in
\begin{empheq}[box=\graybox]{equation}\label{eq:Cooks-inequality}
	\norm{\psi_{\mathrm{out}}(\t)} \leq \int_{-\infty}^{\t} \norm{ V \psi_{\mathrm{in}}(\t')}\, \d \t'\, .
\end{empheq}
This statement is called the \emph{Cook's inequality}. Since $\norm{\psi_{\mathrm{out}}(\t)}$ in the limit $\t \to \infty$ measures scattering probability, \eqref{eq:Cooks-inequality} sets an upper bound on the amount of scattering once $\psi_{\mathrm{in}}$ and $V$ are provided.

\subsubsection{Bound on the wave function}

We're now ready to put a bound on $\norm{\psi_{\mathrm{out}}}$. Basically, we want to make the quantity
\be\label{eq:V-psi-in}
\norm{V \psi_{\mathrm{in}}(\t) }^2 = \int |V(r)|^2\, |\psi_{\mathrm{in}}(\x,\t)|^2\, \d^3 \x
\ee
as small as possible for large impact parameters $b$ across all times $\t$. Assuming that the potential is Yukawa-like amounts to imposing that it's asymptotically bounded by $|V(r)| \leq \lambda \frac{\e^{-\mu r}}{r}$ for some positive $\lambda$ and $\mu$. We take it to be time-independent.

To find out what's the optimal bound, we're going to split the integral \eqref{eq:V-psi-in} into two pieces: inside of a sphere of some radius $R < b$ and outside of it. The potential is the largest on the inside and the wave packet is localized on the outside, so we have two competing effects. The strategy is to put a bound on both and optimize for the value of $R$ at the end.

To get an estimate on the integral outside of the sphere, we're going to use the crude bound $|\psi_{\mathrm{in}}|^2 < \frac{1}{\pi^{3/2} \Gamma^3}$ coming from \eqref{eq:psi-in-packet}, which leads to
\begin{subequations}\label{eq:inside-bound}
\begin{align}
\int_{r > R} |V(r)|^2\, |\psi_{\mathrm{in}}(\x,\t)|^2\, \d^3 \x &< \frac{4\lambda^2}{\sqrt{\pi} \Gamma^3} \int_{R}^{\infty} \e^{-2\mu r}\, \d r\\
&= \frac{2\lambda^2}{\sqrt{\pi}\mu \Gamma^3} \e^{-2\mu R}\, .
\end{align}
\end{subequations}
In the first line we used spherical symmetry so that only the radial part of the integral needs to be done. Note that the result depends on time only through $\Gamma(\t)$.

Estimating the inside contribution is slightly more involved. First, we need to determine how close to the center of the wave packet we can get within the sphere, or equivalently what's the smallest possible $(\x - \hat{\x})^2$. This can be answered with some elementary geometry:
\be
\includegraphics[valign=c,scale=1.1]{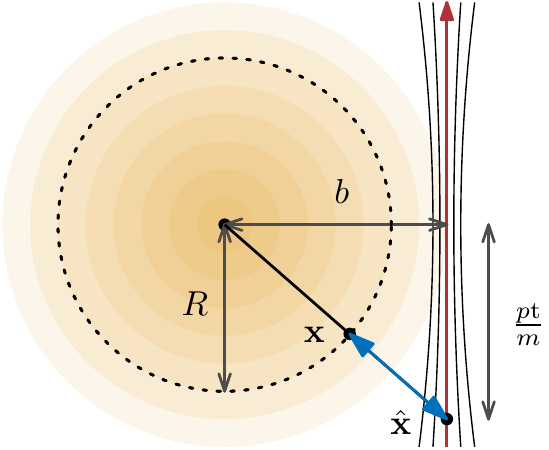}
\ee
For a given point $\hat\x$, the closest $\x$ is on the straight line from the origin at distance $R$. We're interested in the norm of the blue segment, which satisfies
\be
|\x - \hat\x| = \sqrt{b^2 + \frac{p^2 \t^2}{m^2}} - R\, > b - R\, .
\ee
This gives us the upper bound on $|\psi_{\mathrm{in}}|^2 < \frac{1}{\pi^{3/2} \Gamma^3} \e^{-p(b-R)^2/b}$ since $\Gamma^2 > \gamma^2 = b/p$. Repeating the same steps as before, we find that the contribution from the inside of the sphere of radius $R$ is
\be
\int_{r < R} |V(r)|^2\, |\psi_{\mathrm{in}}(\x,\t)|^2\, \d^3 \x < \frac{2\lambda^2}{\sqrt{\pi}\mu \Gamma^3} \e^{- p (b - R)^2/b}\, ,
\ee
which has the same prefactor as \eqref{eq:inside-bound}, but a different exponent.

At this stage we can choose the optimal $R$ that balances out the two contributions. We simply solve for the exponentials to be equally damped: $2 \mu R = p(b-R)^2/b$ for $R$. Out of the two solution of this quadratic equation, we pick the one with $R < b$:
\be
R_\ast = \frac{b}{p} \left( \mu + p - \sqrt{\mu(\mu + 2p)} \right)\, .
\ee
This peculiar dependence on $p$ will dictate analyticity properties of the corresponding scattering amplitude.

All in all, the bound on the norm $\norm{ V \psi_{\mathrm{in}}}$ becomes:
\be
\norm{ V \psi_{\mathrm{in}}} < \frac{2\lambda}{\pi^{1/4} \mu^{1/2} \Gamma^{3/2}} \e^{- \mu R_\ast}\, .
\ee
In the final step, we use Cook's inequality \eqref{eq:Cooks-inequality} with $\t \to \infty$ to get a bound on the norm of the outgoing wave function. Since the only time dependence enters through $\Gamma(\t)$, it will result in a constant $C$ which isn't important for our purposes. We obtain
\be
\norm{\psi_{\mathrm{out}}} < C \e^{-\mu R_\ast}\, .
\ee
Let's make a couple of cross-checks. In the massless limit, $\mu \to 0$, the exponential disappears, which means the outgoing wave packet is no longer suppressed at large impact parameters. When the momentum of the incoming wave packet is large, $p \to \infty$, the right-hand side becomes proportional to $\e^{-\mu b}$, which is the naive rate of decrease we would expect classically.

\subsubsection{\label{sec:Lehmann-ellipses}Lehmann ellipses}

In order to connect what we just found to the behavior of partial amplitudes $f_j(s)$ at large angular momentum $j$, we have to relate them to $\norm{\psi_\mathrm{out}}$. This can be certainly done, but it turns out that a more intuitive approach will yield the same result, so let's run with it.

Just as in Sec.~\ref{sec:low-spin}, we're going to use the fact that a wave packet localized at the momentum $p = \sqrt{s}$ and the impact parameter $b$ contributes vast majority of its energy to inducing the angular momentum
\be
j \sim b \sqrt{s}\, .
\ee
The identification is valid when $\sqrt{s} \gg \mu$, i.e., when the characteristic wave length is much smaller than the range of the potential.
Moreover, the amplitude $f_j(s)$ is linear in the norm $\norm{\psi_\mathrm{out}}$, so it should satisfy the same asymptotic bound
\begin{empheq}[box=\graybox]{equation}\label{eq:f-j-bound}
|f_j(s)| \lesssim \e^{-j \mu/\sqrt{s}}\, ,
\end{empheq}
now translating to the large-$j$ behavior. A more rigorous derivation would result in the same bound. Note that we couldn't have just guessed it by dimensional analysis, because the exponents might've had logarithmic corrections, like in the potential well/barrier case from Sec.~\ref{sec:low-spin}.

This is precisely what we needed to prove analyticity in the $t$-plane. Recall that exponential decrease of $f_j(s)$ of the form $a^{-j}$ guarantees analyticity in the ellipse with foci at $\cos \theta = \pm 1$ and semi-major and -minor axes $\frac{|a|^2 \pm 1}{2|a|}$. From \eqref{eq:f-j-bound} we have
\be
|a| = \e^{\mu/\sqrt{s}} \approx 1 + \frac{\mu}{\sqrt{s}}\, ,
\ee
where we expanded the right-hand side to be consistent with the assumption $\sqrt{s} \gg \mu$. The resulting ellipse coincides with the one drawn with dashed lines in \eqref{eq:q2-singularities}. It's called the \emph{small Lehmann ellipse}. On the real axis, it stretches between $\cos \theta \approx \pm (1 + \frac{\mu^2}{2s})$. Equivalently, in the $t$-plane, it covers the real axis from $t = \mu^2$ to $t = -4s - \mu^2$. What limits expanding the ellipse is precisely the pole at $t = \mu^2$ due to the Born term.

We'll have to put more work into understanding analyticity properties of the total amplitude beyond the small Lehmann ellipse. However, there's a trick we can do to remove the obstruction from the Born simple pole, which amounts to considering the imaginary part of the amplitude, $\Im f(s,t)$, whose convergence is governed by the large-$j$ properties of $\Im f_j(s)$ instead of $f_j(s)$ themselves. Since unitarity tells us that $\Im f_j = |f_j|^2$, using \eqref{eq:f-j-bound} we have
\be
|\Im f_j(s)| \lesssim \e^{- 2j \mu/\sqrt{s}}\, ,
\ee
which is twice as fast of a decay compared to $f_j(s)$ and hence leads to a larger ellipse called, inventively, the \emph{large Lehmann ellipse}. It amounts to replacing $\mu \to 2\mu$ in the above discussion. For example, its endpoints on the real $t$-axis are $t = 4\mu^2$ and $t = -4s -4\mu^2$. It's illustrated with solid line in \eqref{eq:q2-singularities}. As you might've expected, the size of the large ellipse is now limited by the NLO singularity, the anomalous threshold at $t = 4\mu^2$.

\subsection{\label{sec:Regge-theory}Regge theory}

Our next goal will be to further extend the domain of analyticity in $t$. 
One useful tool at our disposal will be treating the angular momentum $j$ as a complex variable, instead of just an integer multiple of $\hbar$. At first, it might seem rather silly to do so, but remember that we already complexified momenta and angles and got quite a lot of mileage out of it. As a matter of fact, if we write down the Schr\"odinger equation for higher-$j$ wave functions \eqref{eq:j-Schrodinger-equation},
\be\label{eq:Schrodinger-Regge}
\left[ -\nabla^2 + \frac{j(j+1)}{r^2} + U - s \right] \psi_j(s,r) = 0\, ,
\ee
the variable $j$ is more or less put on the same footing as $s$, so there's no excuse to not analytically continue it to the complex plane.

After its pioneer, the area of studying scattering amplitudes with complex $j$ is known as \emph{Regge theory}. In this section, we briefly review its basics that are needed for our purposes.

\subsubsection{Watson transform}

The starting point of our discussion will be the following formula for the total scattering amplitude $f(s,t)$ called the \emph{Watson transform} (sometimes referred to as the Sommerfeld--Watson transform):
\begin{empheq}[box=\graybox]{equation}\label{eq:Watson-transform}
f(s,t) = -\frac{1}{2i\sqrt{s}} \oint_{C} \frac{(2j+1) f_j(s) P_j(-\cos \theta)}{\sin (\pi j)}\, \d j\, .
\end{empheq}
In the integrand, $f_j(s)$ denotes an analytic continuation of the partial amplitudes for any complex $j$. Note the minus sign in the argument of the Legendre polynomial $P_j$. The contour $C$ encloses the positive real axis as follows:
\be\label{eq:C-contour}
\includegraphics[scale=1.2,valign=c]{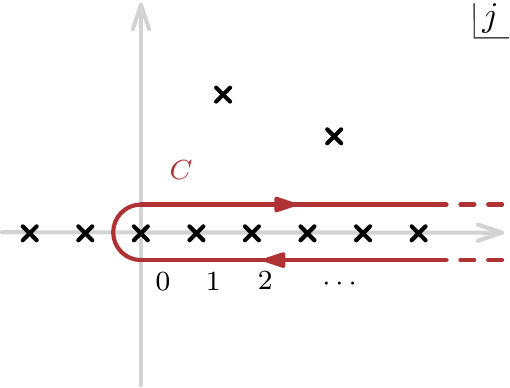}
\ee
First, let's make sure that \eqref{eq:Watson-transform} is in fact correct. We can simply deform the contour to enclose the poles at non-negative integers $j=0,1,2,\ldots$ individually and compute them by residues. Here, we get a factor of $\frac{(-1)^j}{\pi}$ from the sine function. In addition, we need to use the symmetry relation of Lengendre polynomials
\be
P_j(- \cos \theta) = (-1)^j P_j(\cos \theta)\, .
\ee
Each circle also gives rise to a factor of $-2\pi i$ due to clockwise orientation, which in the end gives
\be
f(s,t) = \frac{1}{\sqrt{s}} \sum_{j=0}^{\infty} (2j+1) f_j(s) P_j(\cos \theta)\, .
\ee
This is indeed the partial-wave expansion of the total amplitude we previously encountered in \eqref{eq:partial-wave-expansion}. Convergence of the Watson transform is therefore guaranteed by the arguments given in the previous section. However, it turns out we can do better if we deform the contour $C$ in a different way. The idea is that such a deformation can extend the domain of analyticity in $t$, if we manage to prove that the resulting integral converges in a larger domain.

\subsubsection{\label{sec:Regge-trajectories}Regge trajectories}

As we preemptively indicated in \eqref{eq:C-contour}, $f_j(s)$ might have poles in the complex $j$-plane. These are in fact the bound states we studied in the previous lectures, except now their position is viewed as a function of $j$. Treated from this perspective, they're called \emph{Regge poles}. Note that a pole of the scattering amplitude is described by a single complex condition, so in the two-dimensional complex space, say parametrized by $(s,j)$, a bound state corresponds to a one-dimensional complex surface (equivalently, two real dimensional). What we used to do is to study the intersection of this surface with constant $j$ in the $s$-plane, where a bound state looked like a point. Now we do it the other way around. 

What's interesting about this picture, is that we can trace what happens to a bound state for a given angular momentum, say at $j=0$, as we start increasing its energy $s < 0$ towards zero. Recall that the energy of a bound state is always negative. In other words, we can look at the quantity $\frac{\d j}{\d s}$. As a starting point, we can take the $s$-derivative of the Schr\"odinger equation \eqref{eq:Schrodinger-Regge}:
\be\label{eq:ds-Schrodinger}
\left[ -\nabla^2 + \frac{j(j+1)}{r^2} + U - s \right] \frac{\partial \psi_j(s,r)}{\partial s} + \left( \frac{2j+1}{r^2} \frac{\d j}{\d s} - 1 \right) \psi_j(s,r) = 0\, ,
\ee
where $\psi_j$ is the wave function associated to the specific bound state under question. The next step is to multiply \eqref{eq:ds-Schrodinger} by $r^2 \psi_j$ and subtract from it the Schr\"odinger equation multiplied by $r^2 \frac{\partial \psi_j}{\partial s}$. The result is
\begin{subequations}
\begin{align}
\left( \frac{2j+1}{r^2} \frac{\d j}{\d s} - 1 \right) r^2 \psi_j^2 &= r^2 \left(  \psi_j \nabla^2 \frac{\partial \psi_j}{\partial s} - \frac{\partial \psi_j}{\partial s} \nabla^2 \psi_j \right) \\
&= \frac{\partial}{\partial r} \left[ r^2 \left( \psi_j \frac{\partial^2 \psi_j}{\partial r \partial s} - \frac{\partial \psi_j}{\partial r} \frac{\partial \psi_j}{\partial s}\right) \right] .
\end{align}
\end{subequations}
In the second line we recognized that the combination on the right-hand side is in fact a total derivative with respect to $r$. For wave functions decaying sufficiently fast at infinity, it integrates to zero and we get the identity
\be\label{eq:dj-ds}
(2j+1)\frac{\d j}{\d s} = \frac{\int_0^{\infty} r^2\, \psi^2_j(s,r)\, \d r }{\int_0^{\infty} \psi^2_j(s,r) \, \d r}\, .
\ee
This equation should be intuitive if you paid attention in classical mechanics classes: the quantity on the right-hand side measures the moment of inertia, which for a particle with mass $m$ spinning with angular velocity $\omega$ at radius $r$ is $I = mr^2$. On the left-hand side, we compute the change of the angular momentum $j = I \omega$ with respect to the energy $s = \frac{1}{2}I\omega^2$, and we find agreement in the classical limit $j \gg \frac{1}{\hbar}$.

In any case, for a bound state, the wave function $\psi_j$ is real, so the right-hand side of \eqref{eq:dj-ds} is positive. Hence, we find
\begin{empheq}[box=\graybox]{equation}\label{eq:Regge-slope}
\frac{\d j}{\d s} > 0
\end{empheq}
as long as $s<0$. This inequality should not be surprising given the above classical intuition: as we keep pumping more energy into the system, its angular momentum increases. What happens to a Regge pole in the $j$-plane can be illustrated as follows:
\be
\label{eq:Regge-trajectory}
\includegraphics[scale=1.2,valign=c]{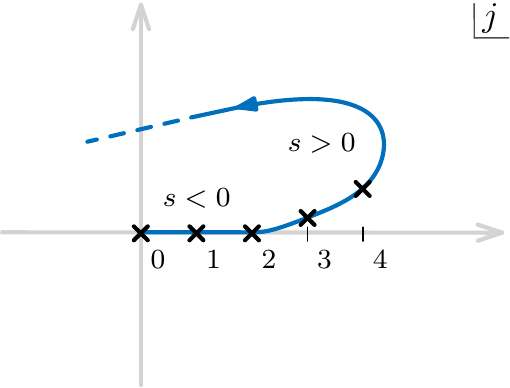}
\ee
For a certain $s<0$, we have a bound state at $j=0$. As we keep increasing $s$, according to \eqref{eq:Regge-slope} the angular momentum increases, giving us bound states for higher and higher $j$. At some point, we cross $s=0$, in which case \eqref{eq:Regge-slope} is no longer valid and the pole can detach from the real axis. What happens for Yukawa potentials afterwards is illustrated in the diagram above. Before the Regge pole completely shoots off to the complex plane, it can be still felt as a resonance if it's close enough to the real axis, even though it doesn't lie directly on it. The path a given pole follows as we increase $s$ is called its \emph{Regge trajectory}.

This story tells us that the bound states we observed in Sec.~\ref{sec:higher-angular-momenta} for individual $j$ are in a sense different aspects of the same phenomenon. Here, we managed to track how a given bound state continuously transforms into another one by treating $j$ as a complex variable. In general, because of the detaching phenomenon illustrated in \eqref{eq:Regge-trajectory}, one finds that the number of bound states for angular momentum $j+1$ can't exceed that for $j$.

\subsubsection{Analyticity in the cut plane}

Now that the origin of the poles in the first quadrant of \eqref{eq:C-contour} is explained, we can start deforming the integration contour. Specifically, we deform $C$ by wrapping it out as follows:
\be\label{eq:C-prime-contour}
\includegraphics[scale=1.2,valign=c]{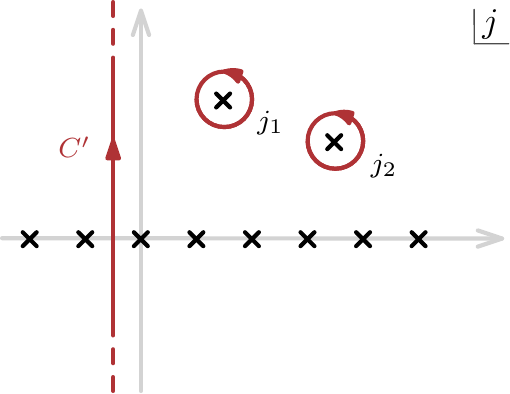}
\ee
Here, we stopped deforming at the line $C'$ parallel to the imaginary axis whose real part is between $-1$ and $0$, say $C' = -\frac{1}{2} + i\R$. On the way, we picked up the Regge poles at some $j = j_1, j_2,\ldots$. Moreover, using techniques similar to those from previous sections, one can show that $f_j(s)$ decays sufficiently fast as $|j| \to \infty$ in the complex directions, which allows us to drop large arcs in the upper and lower half-planes. We arrive at the formula
\be\label{eq:Watson-transform-improved}
f(s, t) = -\frac{1}{2i\sqrt{s}} \int_{C'} \frac{(2j+1) f_j(s) P_j(-\cos \theta)}{\sin (\pi j)}\, \d j + \sum_{\substack{\text{Regge}\\ \text{poles }j_a}} R_a\, ,
\ee
where $R_a$ denotes the result of performing the residue around each $j_a$. We're going to fix physical $s>0$, which means the Regge poles are only above the positive real axis.
The advantage of using this form of the Watson transform is that its convergence properties are much better compared to \eqref{eq:Watson-transform}, which allows us to enlarge the domain of analyticity in $t$.

As a matter of fact, the only dependence on $t$ in \eqref{eq:Watson-transform-improved} enters through the scattering angle in the Legendre polynomial $P_j(-\cos \theta)$. Recall that $\cos \theta = 1 + \frac{t}{2s}$. Once $j$ is made complex, $P_j$ develops a branch cut for $\cos \theta \geq 1$, translating to $t\geq 0$ in the $t$-plane. One can show that everywhere away from this cut, the integrand in \eqref{eq:Watson-transform-improved} is suppressed as $C'$ approaches infinity. Therefore, \eqref{eq:Watson-transform-improved} defines the analytic continuation of the total amplitude $f(s,t)$ for all $t$ except for $t \geq 0$. Moreover, we also know that the amplitude is analytic in the small Lehmann ellipse, so we can extend the domain of analyticity to everywhere except for $t \geq \mu^2$. Finally, we can even subtract away the Born pole at $t = \mu^2$ by considering the subtracted amplitude
\be
\label{eq:subtracted-amplitude}
f(s,t) - f_{\Born}(s,t)\, ,
\ee
which has an improved domain of analyticity extending to the large Lehmann ellipse.

We can summarize everything with the following diagram:
\be\label{eq:t-analyticity}
\includegraphics[scale=1,valign=c]{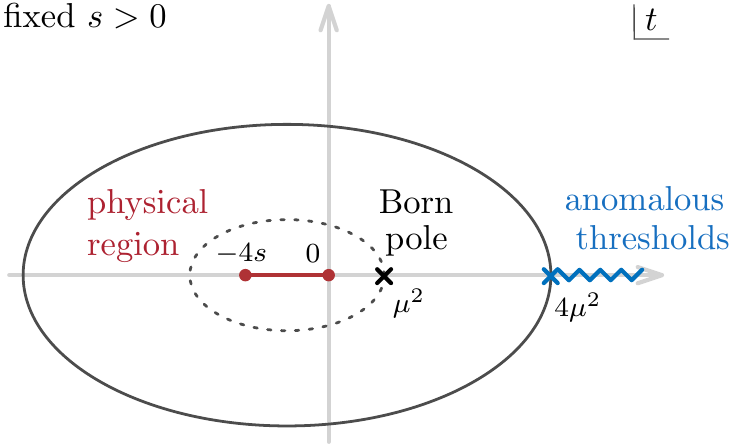}
\ee
The dashed and solid lines indicate the small and large Lehmann ellipses respectively. The physical region is $-4s \leq t \leq 0$. Watson transform arguments enabled us to further prove analyticity in the whole $t$-plane except for the Born pole at $t = \mu^2$ and branch cuts for $t \geq 4\mu^2$.

\subsubsection{Large momentum transfer}

One of the remarkable properties of the Watson transform is that it allows us to also study the $|t| \to \infty$ limit. Here, once again, it suffices to know the behavior of the Legendre polynomial, which for $|z| \to \infty$ behaves as
\be
P_j(-z) \sim (-z)^{j}\, .
\ee
Recall that $z = \cos \theta = 1 + \frac{t}{2s}$.
This limit follows directly from definition of $P_j$ we've seen in \eqref{eq:Legendre-Rodrigues} and is also valid for non-integer $j$.
Since along the integration contour $C'$ we have $\Re j = -\frac{1}{2}$, the integral is suppressed with the power $\sim|\!\cos \theta|^{-1/2}$, or equivalently $\sim t^{-1/2}$.

Clearly, what matters the most in the large-$t$ limit are the contributions with the largest $\Re j$. Therefore, the asymptotics of the total amplitude at large momentum transfer is controlled by the Regge pole with the largest $\Re j$. Let's call it $J(s)$. It can in principle depend on $s$. In the $|t| \to \infty$ limit we hence find
\begin{empheq}[box=\graybox]{equation}\label{eq:large-t}
f(s,t) \sim (-t)^{J(s)}\, .
\end{empheq}
In particular, if there are no bound states, the amplitude decays to zero in the $t \to \infty$ limit (this statement changes once we scatter particles with spin). Notice that if $J(s)$ is non-integer, the amplitude has a branch cut in $t$ going to the right, which is consistent with the picture \eqref{eq:t-analyticity}.

Although we won't discuss it more than in passing, also the imaginary parts $\Im j$ have a physical interpretation. They turn out to measure the angular decay of the corresponding bound-state wave function.

\subsection{\label{sec:dispersion-once-again}Dispersion relations once again}

Causality and locality, supplemented by other physical properties such as unitary evolution of wave functions, allowed us to prove various analytic properties of scattering amplitudes in the $s$ and $t$ invariants. The next logical step is to combine them. The way we're going to proceed is to exploit dispersion relations. More concretely, let's analyze the amplitude $f(s',t)$ in the $s'$-plane, which we want to integrate over the following contour:
\be\label{eq:s-dispersion}
\includegraphics[valign=c,scale=1]{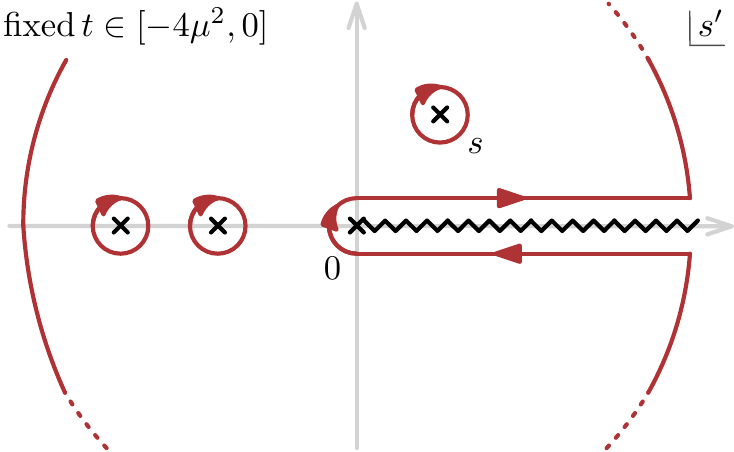}
\ee
It's equivalent to the contour \eqref{fig:dispersion-contour} we previously encountered studying dispersion relation for the Green's function. We're going to hold off specifying the precise function we want to integrate, before dissecting different parts of \eqref{eq:s-dispersion} one by one.

\subsubsection{High-energy bound}

First, we need to understand the high-energy, or $|s| \to \infty$, limit of the amplitude, which will tell us how many subtractions we need to make contributions from the arc at infinity disappear. I should point out straight away that the high-energy limit is of course where the non-relativistic approximation breaks down physically. However, it still makes sense to ask about the $|s| \to \infty$ limit staying within the realm of quantum mechanics. We'll revisit this question in relavistic quantum field theory later in Sec.~\ref{sec:Froissart}.

As before, we're going to use perturbation-theory examples to motivate the general statement. The Born approximation itself was independent of $s$. On the other hand, using the explicit result for the NLO correction, we find that it behaves as
\be
\lim_{|s| \to \infty} f_{\mathrm{NLO}}(s,t) = - \frac{i}{2\sqrt{s}\sqrt{-t}\sqrt{4\mu^2 - t}} \log \left( \frac{\sqrt{4\mu^2 - t} + \sqrt{-t}}{\sqrt{4\mu^2 - t} - \sqrt{-t}} \right)
\ee
and hence it's subleading to the Born term. This pattern continues and one can show that the full non-perturbative amplitude approaches the Born term at high energies:
\be
\lim_{|s| \to \infty} f(s,t) = f_{\mathrm{Born}}(s,t) = \frac{\lambda}{t - \mu^2}\, .
\ee
In the language of Sec.~\ref{sec:subtractions}, this behavior tells us that we need one subtraction. Moreover, we already have an excellent candidate for the thing to subtract: it's the Born approximation itself. In other words, if we choose to integrate
\be\label{eq:dispersion-integrand}
\frac{1}{2\pi i}\frac{f(s',t) - f_{\Born}(s',t)}{s' - s}
\ee
on the contour \eqref{eq:s-dispersion} for some constant $s$, then we can drop the arcs at infinity entirely.

\subsubsection{Schwarz reflection principle}

Next up, the let's look at the part of the contour lying right above and below the positive real axis in \eqref{eq:s-dispersion}. If we call its displacement into the complex plane $\pm i\delta$, we have to deal with the quantities of the form
\be\label{eq:Disc-s}
\Disc_{s=0}f(s,t) = \lim_{\delta \to 0} \Big[ f(s + i\delta, t) - f(s - i\delta, t) \Big] ,
\ee
for $s>0$, where the minus sign arises because of the difference in orientations between the two parts of the contour, right above and right below the positive real axis. This quantity is called the \emph{total discontinuity} across the branch cut starting at $s=0$.
Ultimately, we want to tighten the contour by sending $\delta \to 0$. The first term on the right-hand side is precisely what becomes the physical scattering amplitude. Recall that in terms of the momentum $p$, this is the part that corresponded to $p>0$, while the second term gives $p<0$.

We can make a further simplification by relating \eqref{eq:Disc-s} to the imaginary part of the amplitude. To see it, let me invoke one of the elementary results in complex analysis called the \emph{Schwarz reflection principle}. It says that if a function $g(z)$ is analytic in the upper half-plane and also real on an interval on the real $z$-axis, it admits a unique analytic continuation into the lower half-plane such that $g(z^\ast) = g^\ast(z)$. For instance, $g(z)=z$ is an example of such a function, but $g(z) = i z$ isn't because it isn't real for real $z$.

Scattering amplitudes $f(s,t)$ are real for almost all $s<0$, except for a finite number of bound-state poles. It's the simplest to see this in perturbation theory, where the only source of imaginary part for any Feynman diagram comes from the propagator $\tilde{G}_+(\p,\mathbf{k}) = \frac{1}{-k^2 + p^2 + i\eps}$ we encountered in Sec.~\ref{sec:Yukawa-potential}. Since $s = p^2 <0$, the propagator can never go on shell and hence we can discard the $i\eps$, making the result real. With slightly more work, it's also possible to prove the same statement non-perturbatively. This allows us to use the Schwarz reflection principle, and in particular write
\be
f(s - i\delta, t) = f^\ast(s + i\delta, t)\, .
\ee
Note similarity to the symmetry relation $S_j(p) = S_j^\ast(-p)$ we encountered previously.
Hence, we find that \eqref{eq:Disc-s} is the imaginary part of the amplitude, up to a constant:
\be
\Disc_{s=0} f(s,t) = 2i\, \Im\, f(s,t)\, .
\ee
This result is interesting, because it connects back to unitarity, as we've seen in Sec.~\ref{sec:unitarity-cuts}.

\subsubsection{Putting the pieces together}

We can finally put all the pieces together. Integrating the quantity \eqref{eq:dispersion-integrand} along the contour \eqref{eq:s-dispersion} gives us the dispersion relation:
\begin{empheq}[box=\graybox]{equation}\label{eq:f-dispersion}
f(s,t) = f_{\Born}(s,t) + \sum_{a=1}^{k} \frac{\Res_{s=s_a}f(s, t)}{s - s_a} + \frac{1}{\pi} \int_{0}^{\infty} \frac{\Im f(s',t)}{s' - s}\, \d s'\, .
\end{empheq}
Here, $s$ is taken to be any complex variable away from the positive real axis.
The first term on the right-hand side is simply the Born term. The second contribution is a sum over bound-state poles and the third is the integral over the imaginary part of the total amplitude. Note that the Born approximation doesn't enter either of the latter two contributions since it's real and constant in $s$. The relation is valid for $-4\mu^2 \leq t \leq 0$.

A particularly interesting special case is the forward limit $t \to 0$. In this case, the integrand in \eqref{eq:f-dispersion} depends on the total cross-section in the physical region since $\Im f(s',t) = \frac{\sqrt{s'}}{4\pi} \sigma(s')$. If we tune $s$ to be physical as well, the left-hand side computes the total amplitude. The Born and bound-state terms can be either calculated directly or inserted as free parameters. In this way, the dispersion relation provides a powerful consistency check that directly probes physically-measurable quantities.

The non-forward version loses some of this appeal for two reasons. First of all, the imaginary part is no longer directly related to the cross section. Secondly, the integral over $s > 0$ involves the chunk $0 < s < -\frac{t}{4}$, which is not in the physical region. This motivates us to improve upon the dispersion relation and understand how it can be extended to other values of $t$.

\subsection{\label{sec:Mandelstam-representation}Mandelstam representation}

In this section we'll combine everything we've learned about analyticity of the total amplitude into a single formula.

\subsubsection{Maximal analyticity}

For simplicity, let's treat the case with no bound states first. The idea is to apply dispersion relations, this time in the $t$ variable, for the quantity $\Im f(s', t)$ in the integrand of \eqref{eq:f-dispersion}. Strictly speaking, the imaginary part is not an analytic function, but we can think of it as a difference between $\frac{1}{2i}\left[ f(s' + i\delta, t) - f(s' - i\delta, t) \right]$ in the precise sense explained below. Remembering that $s' > 0$, we can now exploit the analyticity domain we found in \eqref{eq:t-analyticity} to integrate each
\be\label{eq:t-integrand}
\frac{1}{2\pi i} \frac{f(s' \pm i\delta, t')}{t' - t}
\ee
over the following contour:
\be\label{eq:t-dispersion}
\includegraphics[valign=c,scale=1]{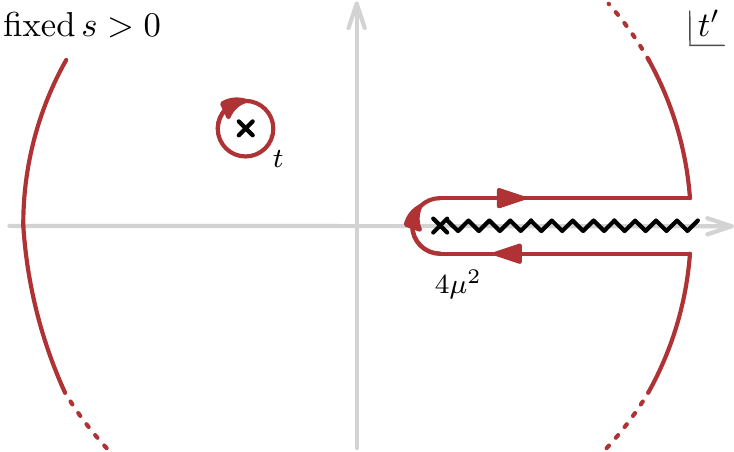}
\ee
Since we assume there are no bound states, the integrand \eqref{eq:t-integrand} decays at infinity and we can drop the arcs. We're left with
\be
f(s' \pm i \delta,t) = \frac{1}{2\pi i} \int_{4\mu^2}^{\infty} \frac{f(s' \pm i\delta, t' + i\delta') - f(s' \pm i\delta, t' - i\delta')}{t' - t} \,\d t'\,,
\ee
where we introduced an infinitesimal parameter $\delta'$ (separate from $\delta$), which measures from which side to approach the branch cut on the real axis. In the next step, we're going to plug this expression into the third term in the dispersion relation \eqref{eq:f-dispersion}. We need the following combination, known as the \emph{double spectral function}, or sometimes \emph{double discontinuity}:
\begin{subequations}
\begin{align}
\rho(s,t) &= \Disc_{t=4\mu^2} \Disc_{s=0} f(s,t)\\
&= \lim_{\delta,\delta' \to 0} \Big[ f(s + i\delta, t + i\delta') - f(s + i\delta, t - i\delta') \\
&\qquad\qquad - f(s - i\delta, t + i\delta') + f(s - i\delta, t - i\delta') \Big],\nonumber
\end{align}
\end{subequations}
where both $s >0$ and $t > 4\mu^2$. Its computation can be rather subtle, but we'll see an example shortly. In terms of $\rho(s,t)$, the double dispersion relation becomes
\begin{empheq}[box=\graybox]{equation}\label{eq:Mandelstam-representation}
f(s,t) = f_{\Born}(s,t) - \frac{1}{4\pi^2} \int_{0}^{\infty} \int_{4\mu^2}^{\infty} \frac{\rho(s',t')}{(s' - s)(t' - t)}\, \d s'\, \d t'\, .
\end{empheq}
This form of writing the amplitude is called the \emph{Mandelstam representation}. Notice that it only depends on the values of the double spectral function for uphysical values of the kinematics. The fact that $\rho(s,t)$ is zero in the physical region is one of the simplest examples of \emph{Steinmann relations}, which more generally say that double discontinuities in overlapping channels vanish for physical kinematics.

As a matter of fact, the equation \eqref{eq:Mandelstam-representation} also defines the analytic continuation of $f(s,t)$ to a larger region in the $(s,t)$ complex space than we've seen before. Since the only way $s$ and $t$ enter \eqref{eq:Mandelstam-representation} is through the Born term or as poles of the integrand, the only possible singularities can happen are when
\be\label{eq:st-cuts}
s > 0 \quad\mathrm{or}\quad t > 4\mu^2 \quad \mathrm{or}\quad t = \mu^2\, .
\ee
This allows us to conclude that the total amplitude is in fact analytic in the entire two-dimensional complex space minus the branch cuts and the Born pole in \eqref{eq:st-cuts}! This is called \emph{maximal analyticity} of the total amplitude.

\subsubsection{Including bound states}

Analogous expressions can be of course derived in the presence of $n_{\mathrm{bound}}$ bound states. The number of subtractions needed is controlled by the leading Regge trajectory $J(s)$. Recall from \eqref{eq:large-t} that the amplitude goes as $(-t)^{J(s)}$ in the $t$-plane. Hence, we need $\lceil J\rceil$ subtractions, which can be implemented by taking the function
\be
\frac{1}{2\pi i} \frac{f(s' \pm i\delta, t')}{(t' - t)\prod_{a=1}^{\lceil J\rceil} (t' - t_a)}
\ee
instead of \eqref{eq:t-integrand} to derive dispersion relations in $t'$. Here, $t_a$ are some extra kinematic points at which we measure the amplitude. The resulting formula for the Mandelstam representation gets a few additional terms compared to \eqref{eq:Mandelstam-representation}. It's not extremely illuminating, so we'll not print it here.

\subsubsection{Support of the double spectral function}

A little more work allows us to improve on the region where the double spectral density $\rho(s,t)$ needs to be integrated in \eqref{eq:Mandelstam-representation}. The reason for this is that $\rho(s,t)$ is a double discontinuity. As we've learned before, it means it can have better analyticity properties than the amplitude itself. For example, in Sec.~\ref{sec:pinch-endpoint} we've learned that the discontinuity of the toy model $-\frac{\log(-p)}{p+1}$ across the $p>0$ branch cut equals to $- \frac{2\pi i}{p+1}$, which doesn't have a singularity at $p=0$ itself. If you're impatient and want to know the support of $\rho(s,t)$ straight away, it's summarized in \eqref{eq:support-rho}.

As a matter of fact, we can see it explicitly when computing $\rho(s,t)$ in the Born expansion to NLO using the expression for $f_{\mathrm{NLO}}$ found in \eqref{eq:f-NLO}. Since we already established that computing the discontinuity in $s$ is the same as taking the imaginary part, we get
\be\label{eq:Disc-f-NLO}
\Disc_{s=0} f_{\mathrm{NLO}}(s,t) = 2i\, \Im f_{\mathrm{NLO}}(s,t) = -\frac{\lambda^2}{\sqrt{t} \sqrt{\Delta}} \log\left( \frac{\sqrt{\Delta} + \sqrt{-s}\sqrt{t}}{\sqrt{\Delta} - \sqrt{-s} \sqrt{t}} \right) \theta(s)
\ee
in the region $0 \leq -t \leq 4s$.
Compared to \eqref{eq:f-NLO}, we've set $p = i\sqrt{-s}$ and $q = -i\sqrt{t}$.
Recall that
\be
\Delta = \mu^4 + 4\mu^2 s - s t
\ee
is the anomalous threshold curve responsible for the triangle Landau singularity.

While the discontinuity \eqref{eq:Disc-f-NLO} was originally computed for $s$ and $t$ in the physical region, we'll treat it now as an analytic function of both variables. This amounts to stripping off the step function $\theta(s)$. At first, you might be concerned whether this uplift in unique. For example, could we have added another analytic function in such a way that its restriction to the physical region $0 \leq -t \leq 4s$ gives zero? One of the fundamental results in analysis, called the \emph{identity theorem}, says that any such function has to be zero identically (more generally, if the set of zeros of an analytic function accumulates to a point, the function vanishes identically). It means that the analytic continuation we're making is, in fact, unique. Any further doubt should be dissipated if we use the definition \eqref{eq:Disc-s} directly: $\Disc_{s=0}$ is initially defined on the real axis by the $\pm i\delta$ prescription. Its analytic continuation is defined by deforming both terms in \eqref{eq:Disc-s} smoothly. For example, continuing $s$ to in the upper half-plane, we compute the difference between $f_{\mathrm{NLO}}(s,t)$ on the first ($p = i\sqrt{-s}$) and second ($p = -i\sqrt{-s}$) sheet. The result agrees with \eqref{eq:Disc-f-NLO} after scratching off $\theta(s)$.

At any rate, we can now analytically continue to $t > 4\mu^2$ and compute the second discontinuity. Before doing so, however, notice that the singularity at $t = 4\mu^2$, which was present in the full amplitude, is now absent from \eqref{eq:Disc-f-NLO}. To convince you of this fact, let's plot the discontinuity for $s=\mu=1$:
\be
\includegraphics[scale=0.53,valign=c]{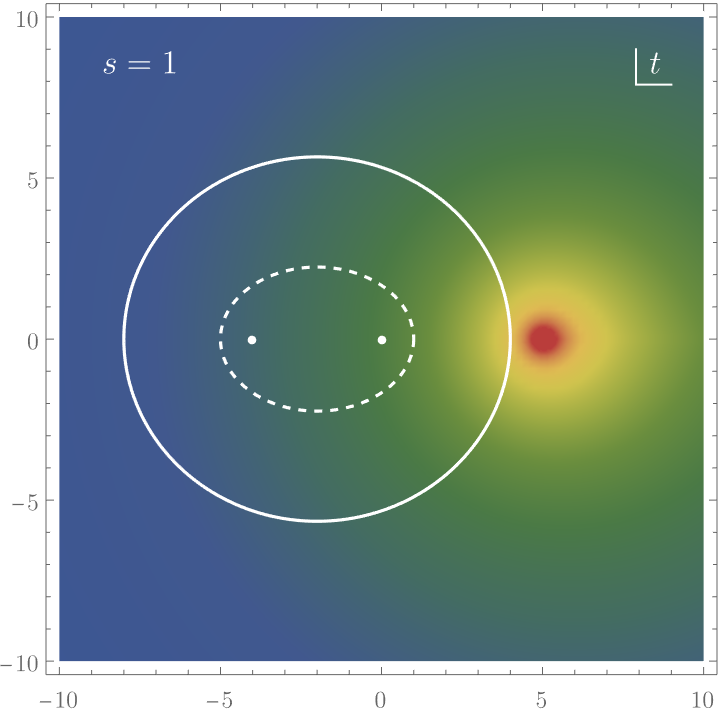}
\ee
To remind you, the small ellipse was limited by the Born pole $t = \mu^2$, the large one by the singularity at $t = 4\mu^2$. We now find that the nearest singularity of the discontinuity \eqref{eq:Disc-f-NLO} occurs at the triangle threshold $\Delta = 0$ located even farther (in the above plot, at $t = 5$). It's a square-root branch point.

In order to complete the computation of $\rho(s,t)$, we need to calculate the second discontinuity across the branch cut at $\Delta < 0$. We can do it, for example, by computing a difference between the original expression \eqref{eq:Disc-f-NLO} and that with $\sqrt{\Delta} \to -\sqrt{\Delta}$. The result is
\be
\rho_{\mathrm{NLO}}(s,t) = \Disc_{t = 4\mu^2} \Disc_{s=0} f_{\mathrm{NLO}}(s,t) =\frac{\lambda^2\, }{\sqrt{-t\Delta}}\, \theta(-\Delta)\theta(s)\, .
\ee
As expected, the result has only support in the region $s>0$ and $t>4\mu^2$, but we also get a stronger requirement that $\Delta < 0$. It means that in the Mandelstam representation, only the smaller region is needed.

To sum up, we can plot everything in the real $s$ and $t$ space as follows:
\be\label{eq:support-rho}
\includegraphics[valign=c,scale=1.2]{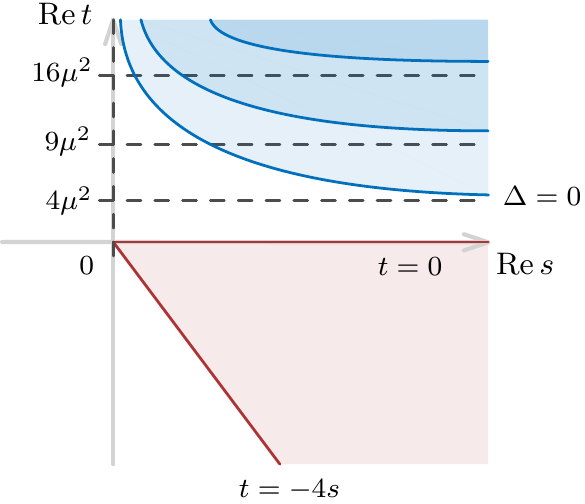}
\ee
Shaded in red is the physical region. The blue region is the domain of support of the double spectral density $\rho(s,t)$. Note that this region asymptotes to $s \to 0$ and $t \to 4\mu^2$ as $t$ and $s$ are large and positive, respectively. One can show that contributions from more subleading terms in the Born expansion start having their support farther and farther in the $s,t>0$ quadrant. They open up at the Landau curves illustrated in blue. For the $k$-th term in the Born expansion, they asymptote to $s\to 0$ and $t \to k^2\mu^2$. The same structure survives at the non-perturbative level.

Roughly speaking, the double spectral density admits the following diagrammatic representation in terms of unitarity cuts:
\be\label{eq:generalized-unitarity}
\includegraphics[valign=c,scale=1.1]{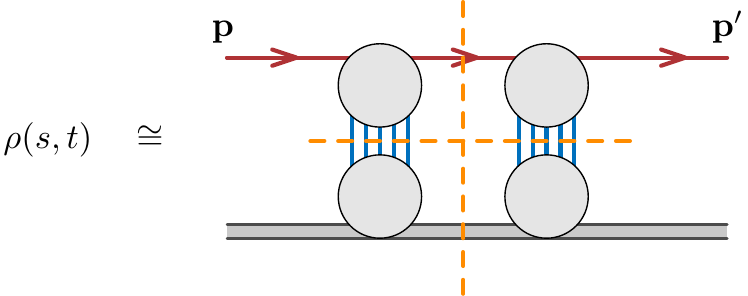}
\ee
However, one needs to be careful with how to interpret this picture. The first one (vertical) is the Cutkosky cut we studied in Sec.~\ref{sec:cutting}. The second cut (horizontal), is however never admissible physically since it would correspond to complex angles. It should be understood in terms of the analytic continuation we performed above on the NLO example. One can show that the mathematics of performing the cut in $s,t>0$ is the same, even though the physics becomes a bit more obscured. In quantum field theory, computing such analytically-continued cuts (forgetting about imposing positivity of energies) is a commonplace and closely related to \emph{generalized unitarity}.

\subsubsection{The dream}

The appeal of the Mandelstam representation comes from the following observation. Let's say we didn't know about the Schr\"odinger equation, but wanted to be able to compute the scattering amplitude $f(s,t)$ from a minimal amount of data. As an input, we're given $f_{\Born}(s,t)$, which is the same as specifying the potential. By the generalized optical theorem \eqref{eq:optical-theorem}, it allows us to compute $\Im f(s,t)$ exactly all the way up to $t < 4\mu^2$ and consequently also the double spectral density for $t < 9\mu^2$. In fact, the answer is equal to \eqref{eq:f-NLO-final} we derived before. Using the Mandelstam representation \eqref{eq:Mandelstam-representation}, this allows us to compute the total amplitude to the NLO (if there are bound states, they should be taken as input as well).

To summarize, we traced the following steps:
\be\label{eq:Mandelstam-steps}
f_{\Born}(s,t) \;\xrightarrow[s\text{-channel cut}]{\text{gen. optical thm}}\, \Im f_{\mathrm{NLO}}(s,t) \;\xrightarrow[t\text{-channel cut}]{\text{discontinuity}}\, \rho_{\mathrm{NLO}}(s,t) \;\xrightarrow[\text{integrate}]{\text{Mandelstam rep.}}\, f_{\mathrm{NLO}}(s,t)\, .
\ee
We can now use this result as a new input in the whole machinery again and go to higher orders in perturbation theory and in $t$. This algorithm allows us to compute $f(s,t)$ in principle, though not always in practice, for arbitrary values of $s$ and $t$. Equivalently, you can think of the problem as specifying the Born amplitude, or even just its generalized unitarity cuts, and gluing them together by iterating the picture \eqref{eq:generalized-unitarity}.

Either way, the bottom line is that the Mandelstam representation can effectively \emph{replace} the Schr\"odinger equation. The reason why this is possible in the first place is that the Mandelstam representation implements all the physical constraints we've learned about causality, locality, and unitarity into a single equation. The dream that has driven, and continues to drive, this line of research, is the prospect of formulating a Mandelstam representation for relativistic scattering amplitudes, in which the analogue of the Schr\"odinger equation isn't known.

\subsection{Exercises}

In this set of exercises, we're going to look at the largely mysterious subject of quantum chaos in scattering amplitudes.  

\subsubsection{Leaky torus}

One of a few explicit examples of a chaotic scattering amplitude is that of the leaky torus. The setup is as follows. Consider the hyperbolic surface with constant negative curvature, which can be described by the upper half-plane parameterized by $z = x+ i y$ with $y > 0$ and endowed with the metric $\d s^2 = \frac{1}{y^2}(\d x^2 + \d y^2)$. The Schr\"odinger equation for a particle with mass $m$ reads
\be
i \hbar \frac{\partial \Psi}{\partial t} = -\frac{\hbar^2}{2m} \left(\nabla^2 + \frac{1}{4}\right) \Psi\, ,
\ee
where the Laplacian is
\be
\nabla^2 = y^2 \left( \frac{\partial^2}{\partial x^2} + \frac{\partial^2}{\partial y^2} \right) .
\ee
The stationary solutions can be described by $\Psi = \psi\, \e^{-iEt/\hbar}$, where in the simplest case
\be
\psi(p,y) = y^{1/2 \pm i p}
\ee
with $E = \frac{\hbar^2 p^2}{2m}$ as usual. These are the analogues of incoming and outgoing plane waves, which emanate from $y \to \infty$. The factor of $y^{1/2}$ plays more or less the same role as $\frac{1}{r}$ in flat space, and is there to ensure that the probability flux remains constant.

Gutzwiller considered a setup in which we carve out the following shaded shape from the upper half-plane:
\be
\includegraphics[scale=1.2,valign=c]{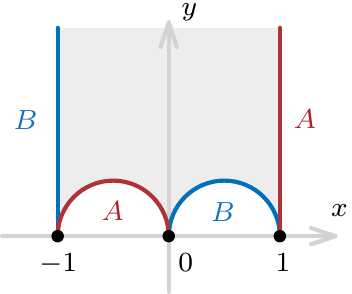}
\ee
The opposite boundaries $A$ and $B$ are identified as indicated by the colors, so that walking through one side we emerge on the other. The result would be a torus if it wasn't for the four points at $x=-1,0,1$ on the real axis and $y=\infty$. They are identified with a single point on the torus and create a singularity, indicated by the black dot:
\be
\includegraphics[scale=1,valign=c]{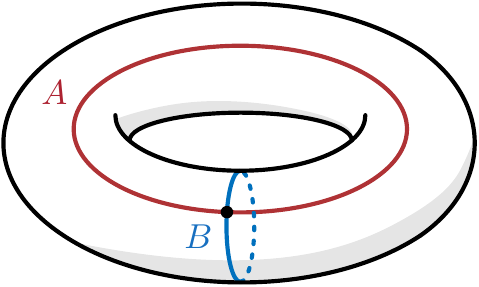}
\ee
This is the place where plane waves can come in and escape from, at least quantum-mechanically, which is why the space is called the \emph{leaky torus}. We won't need to add any scatterer: the curvature of the space ensures non-trivial scattering on its own. The waves entering the torus through the singularity will in general bounce of the geometry a given number of times before escaping through the hole at the end. It turns out that the system is chaotic, meaning that arbitarily small changes to the initial momentum $p$ can alter the scattering amplitude unpredictably.

Solving the Schr\"odinger equation becomes rather involved. While I'm not going to review it, the important part for us is that there exists an exact solution. After integrating over the $x$ direction, the wave function behaves asymptotically as
\be\label{eq:leaky-torus}
\psi(p,y) \sim y^{1/2 - i p} + S_{\mathrm{G}}(p)\, y^{1/2 + i p}\, ,
\ee
which allows us to define the scattering amplitude as the coefficient of the second term, which explicitly reads
\begin{empheq}[box=\graybox]{equation}
S_{\mathrm{G}}(p) = \pi^{-2ip} \frac{\Gamma(\tfrac{1}{2} + ip)\, \zeta(1+ 2ip)}{\Gamma(\tfrac{1}{2} - ip)\, \zeta(1-2ip)}\, .
\end{empheq}
This is the Gutzwiller's S-matrix. Here, $\zeta$ is the Riemann zeta function. Although defined in a slightly different setup, $S_{\mathrm{G}}(p)$ is a quantum-mechanical amplitude that satisfies the same properties of those encountered throughout these lectures. In particular, we'd like to study its analyticity in the $p$-plane.

\subsubsection{Riemann zeta function}

Before analyzing the amplitude \eqref{eq:leaky-torus} further, let us review a few facts about the Riemann zeta function that are going to be relevant. The traditional definition
\be
\zeta(s) = \sum_{n=1}^{\infty} \frac{1}{n^s}\, ,
\ee
converges for $\Re s > 1$. Nevertheless, we can define its values in the remainder of the complex $s$-plane by analytic continuation. For example, $\zeta(-1) = -\frac{1}{12}$.

Convergence of the series representation guarantees analyticity of $\zeta(s)$ in $\Re s > 1$. In fact, there's a singularity at $s=1$. In order to extend it to other values $s$, one can use the reflection identity:
\be\label{eq:Riemann-reflection}
\zeta(s) = 2^s \pi^{s-1} \sin \left(\tfrac{\pi s}{2} \right) \Gamma(1-s) \zeta(1-s)\, ,
\ee
which is valid for any $s$. Since we already identified the only singularity of $\zeta(s)$, to understand the analytic structure of the scattering amplitude \eqref{eq:leaky-torus}, it remains to study positions of the \emph{zeros} of the Riemann zeta function. This is somewhat of a famous math problem and we won't be able to solve it today. Let's instead review what's known about the zeros.

First, the formula \eqref{eq:Riemann-reflection} guarantees that $\zeta(s)$ has a simple zero for every even negative integer $s = -2, -4, -6, \ldots$, coming from the sine function. They are known as the \emph{trivial zeros}. For positive even $s$, the zeros of the sine are canceled by the poles of the gamma function. The reflection formula doesn't allow us to say anything concrete about the possible zeros in the \emph{critical strip} $0 < \Re s < 1$.

The Riemann hypothesis states that all the non-trivial zeros of $\zeta(s)$ are located on the \emph{critical line}
\be
\Re s = \tfrac{1}{2}\, .
\ee
There's a lot of interesting literature on this question that we won't have time to review. For example, it's known that there are infinitely many zeros in the critical strip and that at least $40\%$ of them lie on the critical line.%
\footnote{
	Among other fascinating properties of the Riemann zeta function is its ability to mimic other functions. For example, let's imagine a function $f(s)$ that paints the face of Bernhard Riemann himself, analytic in a disk of radius $\frac{1}{2}$. You can show that as you go higher and higher in the critical strip of $\zeta(s)$, you'll keep finding better and better approximations to $f(s)$. It's a feature similar to $\pi$ having an arbitrary sequence of numbers embedded in its decimal expansion, in fact, infinitely often.
}
For our purposes, we'll simply assume the Riemann hypothesis is true.

We can now translate the above statements to the analytic properties of the leaky torus scattering amplitude \eqref{eq:leaky-torus}. The singularity of the Riemann zeta function at $s=1$ translates to $p=0$ and hence cancels out after taking the ratio in \eqref{eq:leaky-torus}. The trivial zeros would have been located at $p = \frac{2n+1}{2i}$ for positive integers $n$. But for such $p$ the denominator of \eqref{eq:leaky-torus} reads $\Gamma(-n)\zeta(-2n)$, which give a pole-zero cancellation for all such $n$. The only source of divergence is $n=0$, translating to a zero of \eqref{eq:leaky-torus} at $p = -\tfrac{i}{2}$. The same analysis for the numerator gives a simple pole of the amplitude at $p = \frac{i}{2}$.

Finally, there are the non-trivial zeros of the zeta function. Let us denote their location with $s = \frac{1}{2} - 2 i z_j$ for $j$-th zero. These do not cancel out in the amplitude and give rise to simple poles at $p = z_j - \frac{i}{4}$. Similarly, there are simple zeros at $p = z_j + \frac{i}{4}$. We can summarize all this information by means of a plot. First, we define $S_{\mathrm{G}}(p)$ with
\begin{minted}[firstnumber=1]{mathematica}
S[p_] := π^(-2I*p)*(Gamma[1/2 + I*p]*Zeta[1 + 2I*p])/
				   (Gamma[1/2 - I*p]*Zeta[1 - 2I*p]);
\end{minted}
The amplitude can be plotted in the complex plane using the command
\begin{minted}{mathematica}
ComplexPlot3D[S[p], {p, -12 - 12I, 12 + 12I},
			  PlotPoints -> 100, Mesh -> Automatic]
\end{minted}
You should see a plot looking like this:
\be
\includegraphics[scale=0.8,valign=c]{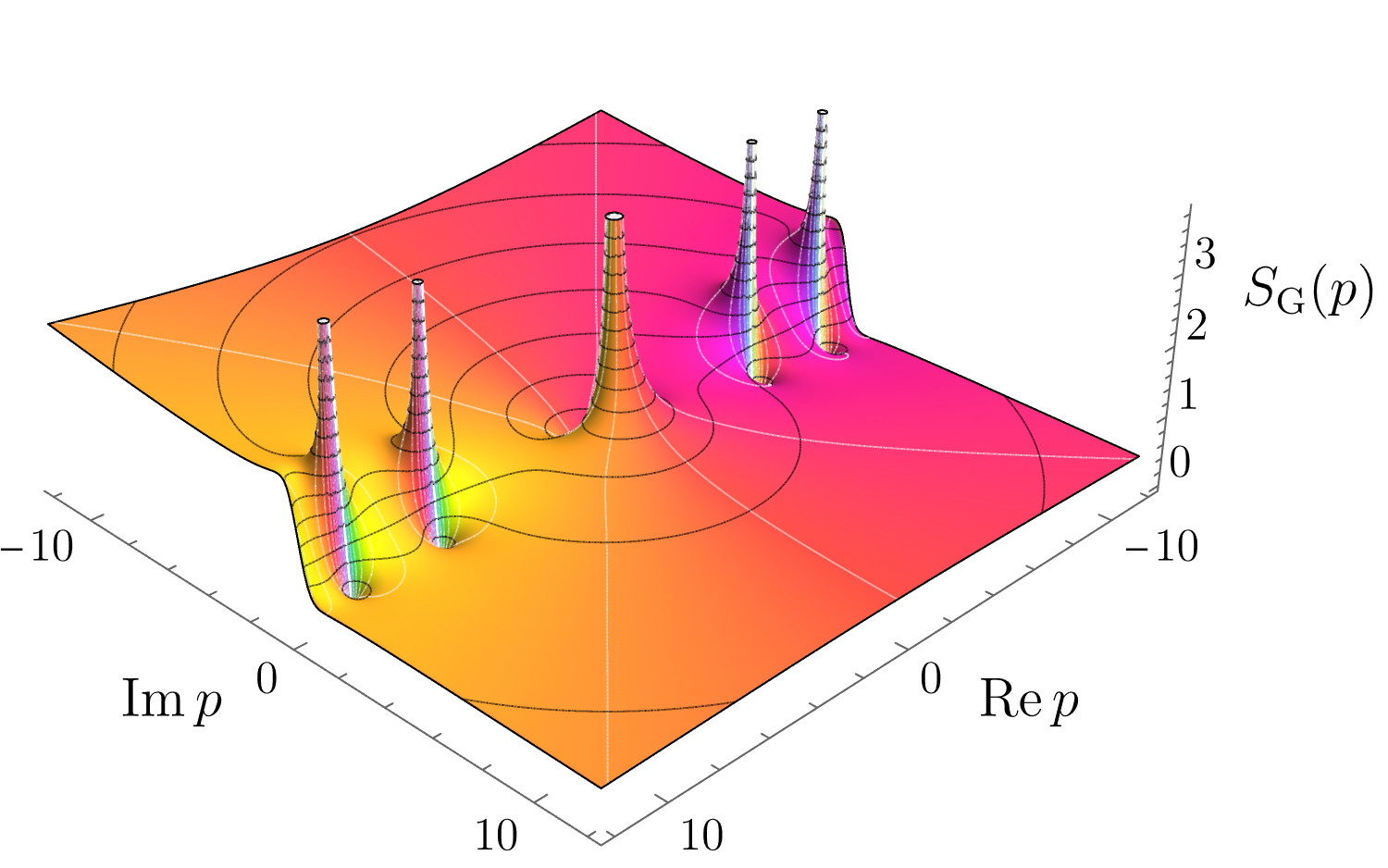}
\ee
Recall that the black spots are the zeros and white are the poles. The picture is entirely consistent with the above discussion. For example, the smallest non-trivial zero of $\zeta(s)$ occurs for $z_1 \approx 7.07$, which translates to poles and zeros at $p \approx \pm \frac{i}{4} \pm 7.07$. The Riemann zeros are conveniently provided with the command $\mathtt{ZetaZero[j]}$.

In the scattering amplitudes language, the pole at $p = \frac{i}{2}$ corresponds to a bound state and all those for $p = z_j -\frac{i}{4}$ are resonances.

\subsubsection{Signs of chaos}

We're going to probe the behavior of the amplitude by looking at time delays. Recall that $S_{\mathrm{G}}(p)$ can be written as an exponential of the phase shift:
\be
S_{\mathrm{G}}(p) = \e^{2 i \eta(p)}	\, .
\ee
Let's separate the phase into two parts, the background and the fluctuations, such that $\eta = \eta_{\mathrm{bg}} + \eta_{\mathrm{fluct}}$ with
\begin{subequations}
\begin{align}
\eta_{\mathrm{bg}}(p) &= -p \log \pi + \arg \Gamma(\tfrac{1}{2} + i p)\, ,\\
\eta_{\mathrm{fluct}}(p) &= \arg \zeta(1+ 2i p)\, .
\end{align}
\end{subequations}
The origin of their names becomes clear when looking at their asymptotics. Using the Stirling's approximation we have $\log \Gamma(x) \sim x \log x - x$, so the background contribution grows as
\be
\eta_{\mathrm{bg}}(p) \sim p \log(\frac{p}{\pi}) - p\, .
\ee
On the other hand the second contribution grows much slower (one can show that for sufficiently large $p$, $\eta_{\mathrm{fluct}}$ is bounded by a multiple of $\log(2p)$). In fact, let's just plot it to see ourselves. It can be done with the commands
\begin{minted}{mathematica}
Plot[{p*Log[p/π] - p,
	  p*Log[p/π] - p + Arg[Zeta[1 + 2I*p]]}, {p, 20, 70}]
\end{minted}
The first line plots both the background and the second one the full phase shift $\eta$.
The result should look more or less like this:
\be
\includegraphics[scale=1.1,valign=c]{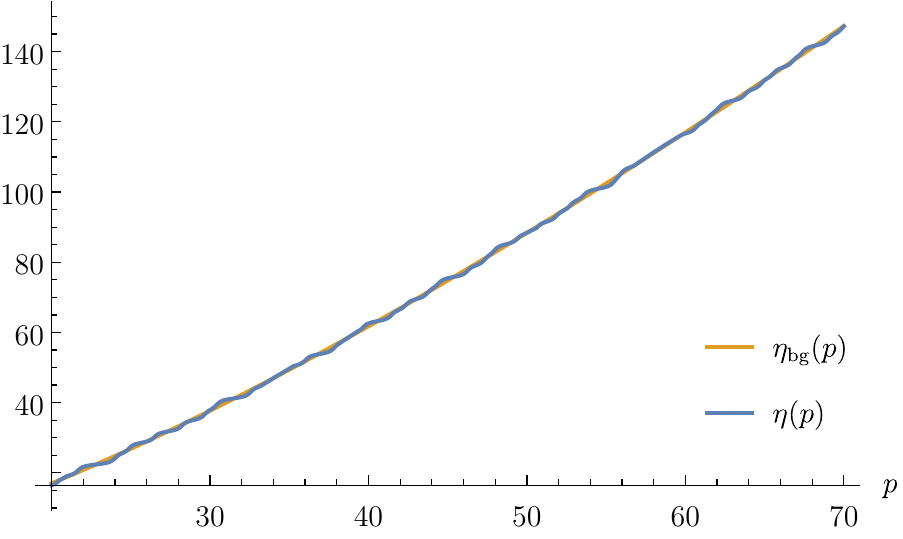}
\ee
We (barely) see that $\eta_{\mathrm{fluct}}$ adds tiny fluctuations above the smooth background contribution $\eta_{\mathrm{bg}}(p)$. In fact, this is easier seen by plotting the ratio:
\be
\includegraphics[scale=1.10,valign=c]{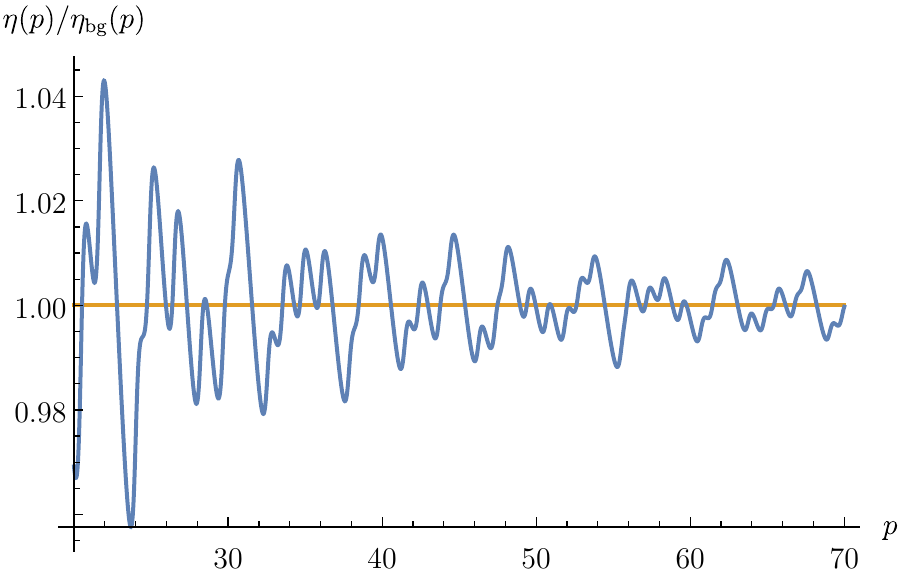}
\ee
This also illustrates that the zeta function is highly erratic.

A more physical way of probing chaotic behavior is going to be to study the time delay as a function of the momentum $p$. Recall from Sec.~\ref{sec:time-delays} that time delay is proportional to $\eta'(p)$. Once again, we can separate the background and fluctuation components. The former achieves its Stirling's asymptotics extremely fast, so we can efficiently approximate it as
\be\label{eq:eta-prime-bg}
\eta_{\mathrm{bg}}'(p) = \Re \frac{\Gamma'(\tfrac{1}{2}+ip)}{\Gamma(\tfrac{1}{2}+ip)} - \log \pi \approx \log \left( \frac{p}{\pi} \right)\, .
\ee
You can verify this by plotting the exact and approximate answers. You should find that the two are virtually indistinguishable for $p \gtrsim 10$.

The fluctuation term needs some massaging to bring it into a form we can study efficiently at large $p$. The starting point is
\be\label{eq:eta-prime-fluct}
\eta_{\mathrm{fluct}}'(p) = 2 \Re \frac{\zeta'(1 + 2ip)}{\zeta(1+2ip)}\, .
\ee
Here, it's going to be very convenient to use the following representation of the logarithmic derivative of the zeta function:
\be
\frac{\zeta'(s)}{\zeta(s)} = b - \frac{1}{s-1} - \frac{1}{2} \frac{\Gamma'(1+s/2)}{\Gamma(1+s/2)} + \!\!\!\!\sum_{\substack{\text{Riemann}\\ \text{zeros }\Im\, s_j<0}} \!\!\!\! \left( \frac{1}{s_j} + \frac{1}{s_j^\ast} + \frac{1}{s - s_j} + \frac{1}{s - s_j^\ast} \right),
\ee
where $b = \log(2\pi) - 1 - \gamma_{\mathrm{E}}/2$ is a constant.
The sum goes over all zeros $s_j = \frac{1}{2} - 2 i z_j$ with negative imaginary part, or $z_j > 0$. Show that after substituting into \eqref{eq:eta-prime-fluct} we get
\be\label{eq:eta-fluct-sum}
\eta_{\mathrm{fluct}}'(p) = b -\Re \frac{\Gamma'(\tfrac{3}{2}+ip)}{\Gamma(\tfrac{3}{2}+ip)} + \frac{1}{4}\!\! \sum_{\substack{\text{Riemann}\\ \text{zeros }z_j>0}} \!\!\! \left( \frac{1}{(p{-}z_j)^2 + (\tfrac{1}{4})^2} + \frac{1}{(p{+}z_j)^2 + (\tfrac{1}{4})^2} + \frac{2}{z_j^2 + (\tfrac{1}{4})^2}\right)\!.
\ee
The contribution from the gamma functions almost entirely cancels out with a similar contribution from the background time delay \eqref{eq:eta-prime-bg}:
\be
\Re \left[ \frac{\Gamma'(\tfrac{1}{2}+ip)}{\Gamma(\tfrac{1}{2}+ip)} - \frac{\Gamma'(\tfrac{3}{2}+ip)}{\Gamma(\tfrac{3}{2}+ip)} \right] = -\frac{1}{2} \frac{1}{p^2 + (\tfrac{1}{2})^2}\, .
\ee
Therefore, the only terms that depend on $p$ are of the Breit--Wigner form. The above term is a consequence of the pole right above the origin in the $p$-plane and looks like a bound state with width $\tfrac{1}{2}$. Similarly, the first two terms in the sum from \eqref{eq:eta-fluct-sum} look like resonances with width $\frac{1}{4}$. Their positions are determined by the Riemann zeros. Their unpredictability is precisely what gives rise to the chaotic nature of the time delay.

Let's go ahead and plot the time delay. It can be now computed much more efficiently in terms of the Riemann zeros, especially if we notice that only a finite number of terms are relevant in a given window of momenta $p$. For example, let's find out how $\eta'(p)$ looks like close to $p \approx 10^6$. We can use the function $\mathtt{ZetaZero[j,k]}$, which gives the $j$-th whose imaginary part is greater than $k$:
\begin{minted}{mathematica}
z[j_, k_] := N[Im[ZetaZero[j, 2*k]]/2];
ηp = 1/4*Sum[1/((p - z[j, 10^6 - 50])^2 + (1/4)^2), {j, 600}];
\end{minted}
The function $\mathtt{z[j, k]}$ gives the numerical values of $z_j$ larger than $k$. 
We're going to take only terms relevant around the value of $p \approx 10^6$. It turns out that $600$ zeros are sufficient for that purpose. Let's also drop all the constant shifts of the time delay. Now it only remains to plot the function with
\begin{minted}{mathematica}
Plot[ηp, {p, 10^6, 10^6 + 50}]
\end{minted}
The result, indeed, looks rather chaotic:
\be
\includegraphics[scale=1.2,valign=c]{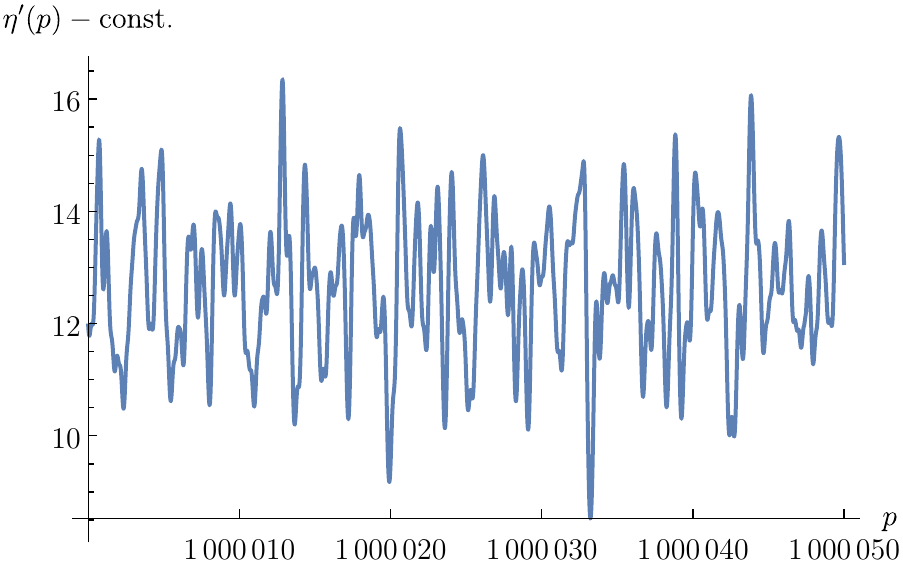}
\ee
By increasing precision, make sure that the jaggedness of the above plot is not a numerical artifact. Experiment with different ranges of $p$. Include the other terms in the time delay $\eta'(p)$ and confirm that they are indeed small.

\section{\label{sec:lecture5}Lecture V: Quantum field theory}

In this lecture we'll highlight a few examples of analyticity being used to learn new things about quantum field theory. After emphasizing the main differences to quantum-mechanical scattering, we're going to study crossing symmetry, dispersive bounds on effective field theories, and the S-matrix bootstrap.

\medskip
\noindent\rule{\textwidth}{.4pt}
\vspace{-2em}
\localtableofcontents
\noindent\rule{\textwidth}{.4pt}

\pagebreak

\subsection{What changes in quantum field theory?}

In the previous lectures we've seen that analytic properties of scattering amplitudes in quantum mechanics can be understood essentially entirely as a consequence of physical principles. A telegraphic summary is that causality leads to analyticity in the energies, or the center of mass energy squared $s$, while locality guarantees analyticity in the angles, or the momentum transfer squared $t$. Unitarity puts non-linear constraints on the analytic structure and implies existence of singularities such as anomalous thresholds. Throughout the lectures, we emphasized that branch cuts are very much properties of physicists (who choose the variables such as $s$ and $t$), not the physics.

As highlighted from the outset, much less is known in relativistic quantum field theory. As a case in point, even basic aspects, such as the upper half-plane analyticity in the energy, which in quantum mechanics was a straightforward consequence of causality, is not fully understood. In view of these difficulties, most physicists are instead comfortable with taking analyticity as an \emph{axiom}, since it has to be connected with causality and locality, at least at a hand-wavy level. As we'll see, this approach can be quite fruitful. We should always remember, though, that it's important to establish connections between analyticity and physics concretely, since ultimately axioms of physical theories should be rooted in physics.

From this point of view, what makes relativistic quantum field theory so much more difficult than quantum mechanics? At a conceptual level, it's just trickier to probe causality and locality relativistically (unitarity remains rather clear). One of the goals of the previous lectures was to familiarize ourselves with these notions in the non-relativistic and time-independent setup, so that the connection might not seem that far-fetched in quantum field theory. At the technical level, most of the complications can be traced to particle production: it opens up a slew of scattering channels, new production thresholds, singularities, and so on.

Another underlying problem is the absence of the counterpart of the Schr\"odinger equation. If we trace back where most of the analyticity statements came from in the previous lectures, one way or another, they followed from looking at regularity properties of the Schr\"odinger equation, possibly with complex kinematics. In quantum field theory, the closest starting points would be either the Lehmann--Symanzik--Zimmermann (LSZ) reduction formula, or the perturbative expansion in terms of Feynman diagrams. In the recent years, we've also seen hints that a holographic dictionary might be useful as yet another, completely different, starting point.

The purpose of this last lecture is to give you an idea of the type of physics we can probe with analyticity in quantum field theory. It's a vast topic spanning a large chunk of the recent literature on scattering amplitudes. Due to time constraints, I allowed myself to pick three topics that are particularly exciting to me and that give a cross-section of the type of questions studied nowadays. These are: crossing symmetry, dispersive bounds on effective field theories, and the S-matrix bootstrap. 

In order to adhere to the conventions employed in the literature on quantum field theory, we're going to denote the $S$-matrix operator by, well, $S$, and split it into the non-interacting part $\boldone$ and the rest denoted by $T$, so that
\be
S = \boldone + i T\,.
\ee
We'll use the same symbol $T$ to denote the interacting scattering amplitude itself, once the momentum conservation delta function has been stripped of. For example, $T(s,t)$ means essentially the same as the non-relativistic $f(s,t)$, perhaps up to some irrelevant normalization factors. It's important to remember that for multi-particle scattering, $T$ can still have disconnected components. 

\subsection{\label{sec:crossing}Crossing symmetry}

The first topic we're going to study is \emph{crossing symmetry}, which is a mysterious property relating the dynamics of particles to that of anti-particles moving back in time. It can be stated in the most crisp way as an analyticity property, saying that scattering amplitudes for different processes are actually described by \emph{the same} function.

Let me illustrate the insanity of this claim on a simple example. We can consider the process of an apple falling on the Earth, which is described by an elastic $2 \to 2$ scattering amplitude similar to the Coulomb case we studied before. It's actually described very well by the Born approximation, which recovers Newton's laws of gravity.

We can now ask what happens if we take an apple from the ingoing state and replace it with an anti-apple (you know about anti-apples, right?) moving back in time, i.e., as an outgoing state. This would result in a $1 \to 3$ amplitude for the Earth decaying to an apple-anti-apple pair and an Earth. Such a process isn't allowed kinematically, because the Earth is known to be stable, at least on the scale of a few billion years. What we have to do instead is to also replace the outgoing Earth with an incoming anti-Earth. This $2 \to 2$ process now describes an Earth-anti-Earth system decaying to gravitational radiation and producing an apple-anti-apple pair:
\be\label{eq:apple-Earth}
\begin{gathered}
\begin{tikzpicture}
\transparent{0.8};
\foreach \t in {0,1,...,5} {\node at (\t/4,-\t/10) {
		\includegraphics[scale=0.2]{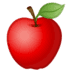}};};
\foreach \t in {0,1,...,5} {\node at (3+\t/4,\t/10-0.5) {
		\includegraphics[scale=0.2]{apple}};};
\foreach \t in {0,1,...,5} {\node at (\t/4,\t/10-2.5) {
		\includegraphics[scale=0.2]{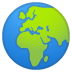}};};
\foreach \t in {0,1,...,5} {\node at (3+\t/4,-\t/10-2) {
		\includegraphics[scale=0.2]{Earth}};};
\transparent{1};
\filldraw[color=black, fill=lightgray!50] (2.1,-1.25) circle (1) node {$T(s,t)$};
\end{tikzpicture}
\end{gathered}
\quad\xleftrightarrow[\text{continuation}]{\text{analytic}}\quad
\begin{gathered}
\begin{tikzpicture}
	\transparent{0.8};
	\foreach \t in {0,1,...,5} {\node at (\t/4,-\t/10) {
			\includegraphics[scale=0.2]{Earth}};
			\draw[thick] (\t/4-0.22,-\t/10+0.27) -- (\t/4+0.22,-\t/10+0.27);	
	};
	\foreach \t in {0,1,...,5} {\node at (3+\t/4,\t/10-0.5) {
			\includegraphics[scale=0.2]{apple}};};
	\foreach \t in {0,1,...,5} {\node at (\t/4,\t/10-2.5) {
			\includegraphics[scale=0.2]{Earth}};};
	\foreach \t in {0,1,...,5} {\node at (3+\t/4,-\t/10-2) {
			\includegraphics[scale=0.2]{apple}};
			\draw[thick] (3+\t/4-0.22,-\t/10+0.27-2) -- (3+\t/4+0.22,-\t/10+0.27-2);	
	};
	\transparent{1};
	\filldraw[color=black, fill=lightgray!50] (2.1,-1.25) circle (1) node {$T(s,t)$};
\end{tikzpicture}
\end{gathered}
\ee
Crossing symmetry states that the two amplitudes are in fact connected by analytic continuation, or equivalently, are boundary values of the same analytic function $T(s,t)$ in different kinematical regimes. The only difference between the left- and right-hand sides in \eqref{eq:apple-Earth} is the range of $s$ and $t$ for which the amplitude $T(s,t)$ is computed.

This thought experiment can be also stated in a more tangible way at a particle-physics level as the analytic continuation of the Coulomb scattering $e^- \gamma \to e^- \gamma$ to electron-positron annihilation $e^+ e^- \to \gamma \gamma$, but \eqref{eq:apple-Earth} gives crossing symmetry the level of glamour it really deserves.\footnote{I'd like to thank Aaron Hillman for suggesting to use this fruitful example.}

\subsubsection{Crossing channels}

Let's put some math behind this cartoon picture. It's conventional to treat all the external momenta uniformly and assign to them the Lorentz-momenta $p_i^\mu$, such that they satisfy the momentum conservation $\sum_{i=1}^{4} p_i^\mu = 0$. In mostly-minus signature, the incoming particles have positive energies, $p_i^0 > 0$, and the outgoing ones have negative, $p_i^0 < 0$. The Mandelstam invariants are
\be
s = (p_1 + p_2)^2\,, \qquad t = (p_2 + p_3)^2\, \qquad u = (p_1 + p_3)^2\, .
\ee
To simplify all the manipulations, we're going to assume elastic scattering of two particles, meaning that the external masses are $M_1 = M_3$ and $M_2 = M_4$. Recall that $M_i^2 = p_i^2$.
Momentum conservation then translates to $s + t + u = 2M_1^2 + 2M_2^2$, so we can eliminate $u$ and use the notation $T(s,t)$ to denote the $2\to 2$ amplitude.

Let us first establish which regions of the kinematic invariants correspond to physical scattering, i.e., real momenta and angles. The slickest way to answer this question is to impose that the determinant of the Gram matrix $\mathbf{G}_{ij} = 2p_i \cdot p_j$ is positive. For the specific choice of masses above, it gives
\be
\det \medmath{\begin{pmatrix}
	2 M_1^2 & s-M_1^2 - M_2^2 &  u - 2M_1^2 \\
	s-M_1^2 - M_2^2 & 2M_2^2 & t - M_1^2 - M_2^2 \\
	u - 2M_1^2 & t - M_1^2 - M_2^2 & 2M_1^2 
\end{pmatrix}}
= 2 u \left[s t  - (M_1 - M_2)^2 \right] > 0\, .
\ee
Depending on the specific masses, this inequality can carve out $3$ or $4$ disjoint regions. They are responsible for the $3$ scattering channels: $12 \to 34$, $13 \to 24$, and $14 \to 23$, as well as a possible decay channel such as $1 \to 234$ if it's kinematically allowed. The first three are usually called the $s$-, $u$-, and $t$-channels respectively. If we take $M_1 \gg M_2$ (having in mind that an apple is $\sim 10^{26}$ times lighter than the Earth), these regions look more or less like so:
\be
\includegraphics[scale=1.1,valign=c]{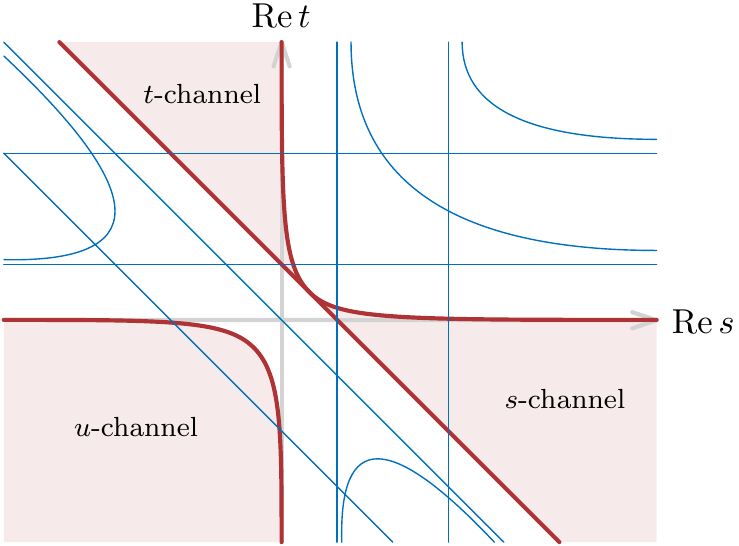}
\ee
With blue lines we illustrated some examples of possible singularities.
Since there are in principle branch cuts within the red regions, we should be more precise about how to approach them. Using either the LSZ formula or Feynman integrals, one can show that the correct choices are
\begin{subequations}\label{eq:f-approaches}
\begin{gather}
T_s(s,t) = \lim_{\delta \to 0} T(s + i\delta, t)\, ,\qquad T_t(s,t) = \lim_{\delta \to 0} T(s, t + i\delta)\, ,\\
T_u(s,t) = \lim_{\delta \to 0} T(s - i\delta, t) = \lim_{\delta \to 0} T(s, t - i\delta)\, .
\end{gather}
\end{subequations}
Here, $T_s$ denotes the scattering amplitude in the $s$-channel and so on. Notice that the $u$-channel case has to be approached from the direction $u \to u + i\delta$, which by momentum conservation translates to either of the choices above. Validity of \eqref{eq:f-approaches} crucially depends on the assumption that all the external particles are stable.

\subsubsection{Bros--Epstein--Glaser analyticity}

We can now ask about analyticity properties in the complex plane of $s$ (we'll come back to analyticity in the $t$ variable in Sec.~\ref{sec:Froissart}). Proving anything in this regard turns out to be a fiendishly complicated task and involves intricate theorems from analysis in several complex variables. In contrast with the non-relativistic case, we need to make a few more assumptions:
\begin{itemize}[leftmargin=*]
\item {\bf Unitarity:} the $S$-matrix operator satisfies $S S^\dag = \boldone$,
\item {\bf Locality:} existence of local operators $\phi(x)$,
\item {\bf Causality:} commutators $[\phi(x), \phi(y)]$ vanish at space-like separations of $x$ and $y$,
\item {\bf Poincar\'e invariance}: the amplitude $T(s,t)$ depends only on $s$ and $t$,
\item {\bf Mass gap:} the lightest particle in the spectrum has a positive mass $m>0$,
\item {\bf Stability:} all the external states are stable.
\end{itemize}	
Notice that these results won't directly apply to our apple-Earth scattering problem, since gravitons are massless and hence there's no mass gap. Also, the term ``locality'' is now used in a different sense to what we meant by it in the previous lectures.

Using these ingredients, Bros, Epstein, and Glaser showed that for any fixed $t<0$, the amplitude has to be analytic in the upper half-plane of $s$ for sufficiently large $|s|$:
\be\label{eq:BEG}
\includegraphics[scale=1.1,valign=c]{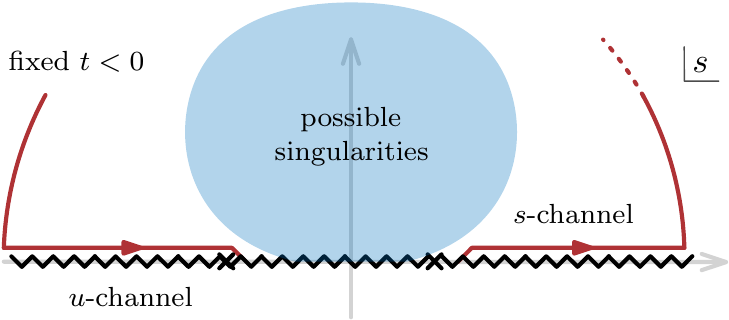}
\ee
If there were any additional anomalous thresholds, they would have to be contained within the blue shaded region. I should mention that no example of such a singularity has ever been found in perturbation theory (on the other hand, for fixed angle $\cos \theta$ or angular momentum $j$, such examples are easy to find). 
Anyway, the picture \eqref{eq:BEG} establishes that there is an analytic continuation between the $s$- and $u$-channels following the red path. Unfortunately, compared with \eqref{eq:f-approaches}, the $u$-channel is approached from the wrong side, $T(s + i\delta,t)$ instead of $T(s - i\delta, t)$. In other words, we land on the complex-conjugated amplitude in the $u$-channel.

This problem can be quickly remedied by repeating the same derivations with $s \leftrightarrow t$, which allows us to analytically continue between the $t$-channel and $u$-channel from the wrong side. Composition of these two paths of analytic continuation gives a link between $T_s$ to $T_t$. This is the simplest instance of crossing symmetry.%
\footnote{A word of warning that the term ``crossing symmetry'' is also very often used for a \emph{different} concept of permutation invariance in the labels. For example, shuffling around the external momenta of the process $\pi \pi \to \pi \pi$ permutes the set $\{ s,t,u\}$ under which the amplitude stays invariant. This notion of crossing symmetry has nothing to do with the one described in the lecture, and they should not be confused with each other.}

Perhaps this is a good place to mention that a little more is known in the case where $t$ is sufficiently small compared to the mass gap $m^2$. More precisely, one can show that the amplitude $T(s,t)$ is real and analytic in the region below all \emph{normal thresholds}:
\be\label{eq:Euclidean-region}
s < 4m^2, \qquad t < 4m^2, \qquad u = -s-t+ 2M_1^2 + 2M_2^2 < 4m^2\, ,
\ee
except for possible simple poles, just like in the Born term. The normal thresholds in the $s$-channel occur at $s = (km)^2$, where the energy is high enough that a possibility of $k$-particle production opens up, and similarly in the other channels (in quantum mechanics, the $s$-channel normal thresholds didn't exist and the $t$- and $u$- ones were classified as anomalous thresholds). The kinematics determined by \eqref{eq:Euclidean-region} is called the \emph{Euclidean region}. In terms of the momentum transfer squared $t$, it exists in the little window when $t$ satisfies
\be
2M_1^2 + 2M_2^2 - 8m^2 = t_0 < t < 4m^2\, .
\ee
Presence of this region allows us to prove with relatively little effort analyticity in the whole $s$ complex plane minus the normal-threshold branch cuts and poles. Here, crossing symmetry can be obtained in a much simpler way by simply continuing through the Euclidean region:
\be\label{eq:Euclidean-path}
\includegraphics[scale=1.1,valign=c]{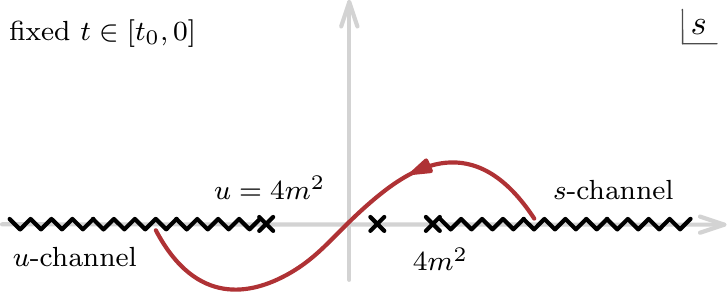}
\ee
One can show that similar paths of analytic continuation exist for $n$-particle scattering if the mass gap isn't too small compared with the external masses. These arguments, once again, technically speaking can't say anything about crossing symmetry in the presence of massless fields. The only concrete result with massless particles currently is that crossing symmetry holds for all \emph{planar} Feynman diagrams.

\subsubsection{Analytic continuation of energies}

Only limited results are known about analyticity beyond the $2\to2$ case. However, even before sitting down to prove it rigorously, we should ask what's the best possible crossing statement we could hope for.

Recall that crossing symmetry relates particles to anti-particles traveling back in time. We couldn't have just flipped the sign of the energy of a single particle, because this would violate momentum conservation. The simplest move we can make is to simultaneously cross two particles. The way this can be achieved is to take two particles, say incoming $2$ and outgoing $3$, and deform their momenta with a complex parameter $z$ as follows:
\begin{subequations}\label{eq:p23-deformation}
\begin{align}
p_2^\mu(z) = (z p_2^+, \tfrac{1}{z} p_2^-, p_2^\perp)\, ,\\
p_3^\mu(z) = (z p_3^+, \tfrac{1}{z} p_3^-, p_3^\perp)\, .
\end{align}
\end{subequations}
Here, we used lightcone coordinates $p^\mu = (p^+, p^-, p^\perp)$ such that $p^2 = p^+ p^- - (p^\perp)^2$. This deformation achieves two things at the same time. First of all, the $z$ cancels out when computing the masses of the external particles, so all of them remain on shell. Secondly, if we go to a Lorentz frame in which $p_2^\pm + p_3^\pm = 0$, momentum conservation remains preserved for any complex value of $z$.

Analytic continuation can be performed along a path in the upper half-plane of $z$ as follows:
\be\label{eq:z-path}
\includegraphics[scale=1.1,valign=c]{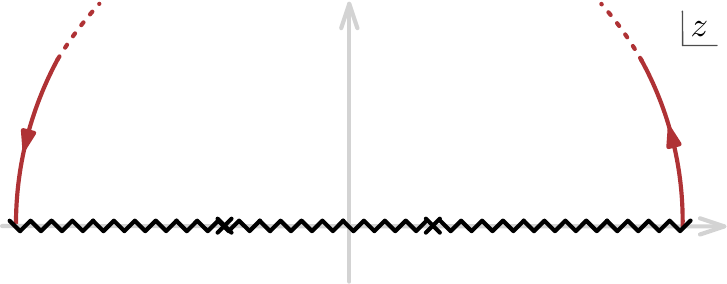}
\ee
The starting point is at some large $z \gg 1$ and the endpoint is at $z \ll -1$. The result is that we flipped the energies of particles $2$ and $3$, as required. Notice that the Mandelstam invariant $t = (p_2 + p_3)^2$ remains $z$-independent and hence we can keep it fixed. On the other hand, at large $z$:
\be
s(z) = [p_1 + p_2(z)]^2 \sim z p_2^+ p_1^-\, ,
\ee
explaining why we took $|z|$ to be large in \eqref{eq:z-path}. This way, we can stay within the region of analyticity indicated in \eqref{eq:BEG}. As we've seen above, the result of this continuation is the $u$-channel amplitude, except complex conjugated.

The immediate guess would be that the same story carries over to all higher-multiplicity scattering amplitudes and the result of the above continuation is the amplitude in a different channel, up to a complex conjugate. This prediction turns out to be wrong. In fact, it's so wrong, we can already see it at tree level.

Tree-level diagrams give a great playground for crossing symmetry. Here, we don't have to worry about any unknown singularities, since they only come from the poles explicitly written down as propagators. Nevertheless, we can still meaningfully probe questions about the result of analytic continuation in different channels. Studying it carefully will give us an idea of what we can reasonably expect from crossing symmetry beyond tree level.

Let's see the simplest example in action. Consider the $u$-channel diagram, proportional to $\frac{1}{u(z) - m^2}$, but evaluated in the $s$-channel kinematics. We're supposed to perform the analytic continuation along the arc \eqref{eq:z-path}, which means that $u(z)$ is deformed in the \emph{lower} half-plane, since the original kinematic invariant $u = (p_1 + p_3)^2 < 0$ starts off negative and we rotate anti-clockwise. When $z$ is very close to the real axis at the beginning of the continuation in \eqref{eq:z-path}, we have
\be
u(z) \approx u - i\delta\,,
\ee
where $\delta$ parameterizes how far we go into the complex plane. Hence, at the starting point we have: 
\be
T_s(s,t) = \lim_{\delta \to 0} \frac{1}{u - m^2 - i\delta} = \includegraphics[valign=c,scale=1]{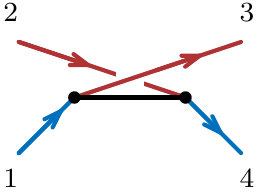}\, .
\ee
In the diagram, we color-coded the particles we're crossing in red and the arrows denote the flow of energy. Notice that the $-i\delta$ approach is opposite to the Feynman $+i\eps$ prescription we'd normally follow. However, this difference doesn't matter in the $s$-channel since the propagator can never go on shell there. On the other hand, at the endpoint of the continuation \eqref{eq:z-path} we find
\be\label{eq:f-u-conjugate}
T_u^\ast(s,t) = \lim_{\delta \to 0} \frac{1}{u - m^2 - i\delta} = \left( \includegraphics[valign=c,scale=1]{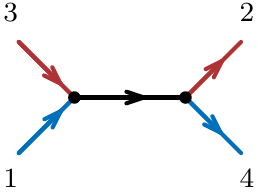}\right)^{\!\!\!*}\, .
\ee
Here, the $-i\delta$ does matter and we find that the limit is the complex conjugated amplitude in the $u$-channel. We're still using the same variable $u$, but according to \eqref{eq:z-path}, it is now rotated to be positive. This is precisely what we described earlier.

Since this example was so trivial, we should ask what could've possibly gone wrong? Essentially, the only other possibility is that the result of analytic continuation could've been $T_u(s,t)$ without the complex conjugate. The difference between that and \eqref{eq:f-u-conjugate} is
\begin{subequations}\label{eq:crossing-cut}
\begin{align}
T_{u}(s,t) - T_{u}^\ast(s,t) &= \lim_{\delta \to 0} \left( \frac{1}{u - m^2 + i\delta} - \frac{1}{u - m^2 - i\delta} \right)\\
&= -2\pi i \,\delta(u - m^2)\, .
\end{align}
\end{subequations}
It's yet another application of the Sokhotski--Plemelj identity. The second line is what we'll refer to as $\Cut_u T(s,t)$. It's simply the unitarity cut of the tree-level diagram, up to a normalization factor.

This manipulation teaches us an important lesson. Crossing symmetry at tree level is all about keeping track of the delta functions. It's essentially a book-keeping device that tells us when a unitarity cut needs to be included in an expression, which is a general lesson extending to loop-level diagrams, up to subtleties we'll discuss below.

\subsubsection{S-matrix theory isn't just about scattering amplitudes}

Equipped with the above set of tricks, we can now ask about crossing symmetry for $5$-point amplitudes. For example, we can consider the following diagram:
\be
T_{12 \to 345}(s_{ij}) = \frac{1}{(s_{13} - m^2)(s_{45} - m^2 + i\delta)} = \includegraphics[valign=c,scale=1]{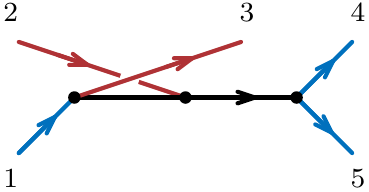}\, .
\ee
We drop $\lim_{\delta \to 0}$ from the notation for readability.
Let's cross the particles $2$ and $3$ according to the deformation \eqref{eq:p23-deformation}.
The above diagram only depends on two Mandelstam invariants $s_{13}$ and $s_{45}$. The first one starts off space-like, $s_{13} < 0$, in the $12 \to 345$ channel and hence it doesn't need any $i\delta$ prescription. The invariant $s_{45}>0$ doesn't change under the analytic continuation, but $s_{13}$ does:
\be
s_{13}(z) = [p_1 + p_3(z)]^2 \sim z p_3^+ p_1^-\, .
\ee
Therefore, $s_{13}$ rotates along a large arc in its lower half-plane and approaches the kinematic region with a $-i\delta$. The result of the continuation is therefore
\be
[T_{12 \to 345}(s_{ij})]_{2 \leftrightarrow 3} = \frac{1}{(s_{13} - m^2 - i\delta)(s_{45} - m^2 + i\delta)}
\ee
with $s_{13}, s_{45} > 0$.
This is \emph{not} the scattering amplitude in the $13 \to 245$ channel, nor its complex conjugate. In fact, after using the same identity as in \eqref{eq:crossing-cut}, we find that the continued amplitude can be written as
\begin{subequations}\label{eq:13-245-channel}
\begin{align}
[T_{12 \to 345}(s_{ij})]_{2 \leftrightarrow 3} &= \frac{1}{(s_{13} - m^2 - i\delta)(s_{45} - m^2 - i\delta)} - \frac{2\pi i \delta(s_{45} - m^2)}{s_{13} - m^2 - i\delta}\\
&= \left(\includegraphics[scale=1,valign=c]{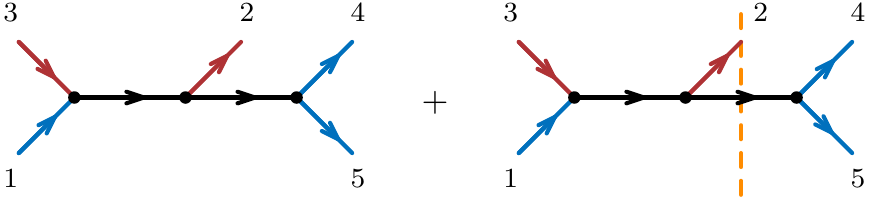} \right)^{\!\!\ast} .
\end{align}
\end{subequations}
We can identify it as the complex-conjugated amplitude in the $13 \to 245$ channel \emph{plus} a unitarity cut in the $s_{45}$ invariant:
\begin{empheq}[box=\graybox]{equation}
[T_{12 \to 345}(s_{ij})]_{2 \leftrightarrow 3} = T_{13 \to 245}^\ast(s_{ij}) + \Cut_{s_{45}} T_{13 \to 245}^\ast(s_{ij})\, .
\end{empheq}
It turns out that the right-hand side is the tree-level approximation not of a scattering amplitude, but of a different observable! In terms of the blob picture, it can be written as
\be\label{eq:observable}
\includegraphics[scale=1.2,valign=c]{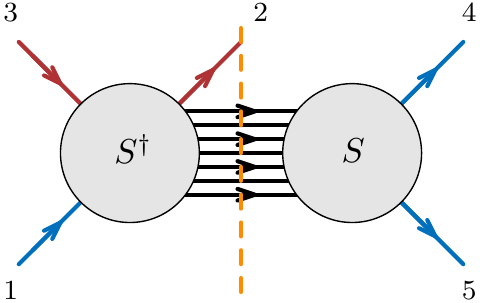}
\ee
Here, both blobs are really $S = \boldone + i T$ and they can have disconnected components. The dagger denotes complex conjugation and we sum over all possible states crossing the unitarity cut in orange. Note that the particle $2$ doesn't participate in the $S$ interaction after scattering off $S^\dag$. The two terms in \eqref{eq:13-245-channel} are the only two ways you can distribute the interaction vertices to fit into the picture \eqref{eq:observable}. In the first one, we see the non-interacting part of $S$: particles $4$ and $5$ simply pass through the $S$ and only interact within the $S^\dagger$ blob, giving the complex-conjugated diagram. In the second term, $4$ and $5$ interact within $S$ and are connected by a unitarity cut of a single propagator to the $S^\dagger$ blob containing the remaining two interaction vertices. You can convince yourself that these are the only two terms contributing.

The above observable has a simple physical interpretation. Imagine we want to compute the expectation value of gravitational radiation emitted in the background of two heavy objects, such as black holes, inspiraling into each other. Such a situation is described by the picture in \eqref{eq:observable}, where $1$ and $5$ are one black hole, $3$ and $4$ are the other, and $2$ is a graviton. In the above language, we obtained it as a crossing of a regular $2\to 3$ scattering amplitude:
\be
\begin{gathered}
	\begin{tikzpicture}[scale=0.85]
		\transparent{0.6};
		\foreach \t in {0,1,...,5} {\node at (3+\t/4,\t/10-0.5) {
				\includegraphics[scale=0.2]{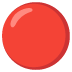}};
				\draw[thick] (3+\t/4-0.25,\t/10-0.17) -- (3+\t/4+0.25,\t/10-0.17);	
		};
		\foreach \t in {0,1,...,5} {\node at (3+\t/3.5,-1.25) {
					\includegraphics[scale=0.2]{red-black-hole}};};
		\foreach \t in {0,1,...,5} {\node at (\t/4,\t/10-2.5) {
				\includegraphics[scale=0.2]{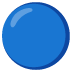}};};
		\foreach \t in {0,1,...,5} {\node at (3+\t/4,-\t/10-2) {
				\includegraphics[scale=0.2]{blue-black-hole}};};
		\transparent{1};
		\draw[photon,very thick] (-0.2,0.2) -- (2,-1);
		\filldraw[color=black, fill=lightgray!50] (2.1,-1.25) circle (1) node {$S$};
	\end{tikzpicture}
\end{gathered}
\quad\xleftrightarrow[\text{continuation}]{\text{analytic}}
\begin{gathered}
	\begin{tikzpicture}[scale=0.85]
	\transparent{0.6};
	\foreach \t in {0,1,...,5} {\node at (6+\t/5,\t/10-0.5) {
			\includegraphics[scale=0.2]{red-black-hole}};	
	};
	\foreach \t in {0,1,...,5} {\node at (0.2+\t/5,-\t/10) {
			\includegraphics[scale=0.2]{red-black-hole}};	
	};
	\foreach \t in {0,1,...,5} {\node at (0.2+\t/5,\t/10-2.5) {
			\includegraphics[scale=0.2]{blue-black-hole}};};
	\foreach \t in {0,1,...,5} {\node at (6+\t/5,-\t/10-2) {
			\includegraphics[scale=0.2]{blue-black-hole}};};
	\transparent{1};
	\draw[photon,very thick] (2.5-0.2,-0.8) -- (3.7,0.4);
	\draw[draw=none,fill=black!50] (2,-1.65) rectangle ++(3,0.8);
	\filldraw[color=black, fill=lightgray!50] (2.1,-1.25) circle (1) node {$S^\dag$};
	\filldraw[color=black, fill=lightgray!50] (5.1,-1.25) circle (1) node {$S$};
	\draw[very thick, dashed, orange] (3.7,0.5) -- (3.7,-3.2);
	\end{tikzpicture}
\end{gathered}
\ee
Recently, there has been a lot of excitement about using the right-hand side to predict parts of the waveform detected by gravitational-wave interferometers such as LIGO. Crossing symmetry tells us that, in an appropriate sense, it can be obtained by analytic continuation of the left-hand side: a process in which a (blue) black hole-graviton system scattering into a black hole and a (red) black hole-anti-black hole pair.

\subsubsection{Crossing equation}

The above example is of course just the tip of an iceberg. The most obvious generalization is to take a $AB \to CD$ scattering process, where $A$, $B$, $C$, and $D$ denote non-empty sets of external particles. Together with Caron-Huot, Giroux, and Hannesdottir, we proved that at tree level, crossing of all particles in the set $B$ with those from $C$ gives the following result:
\be\label{eq:crossing-equation}
\includegraphics[scale=1.2,valign=c]{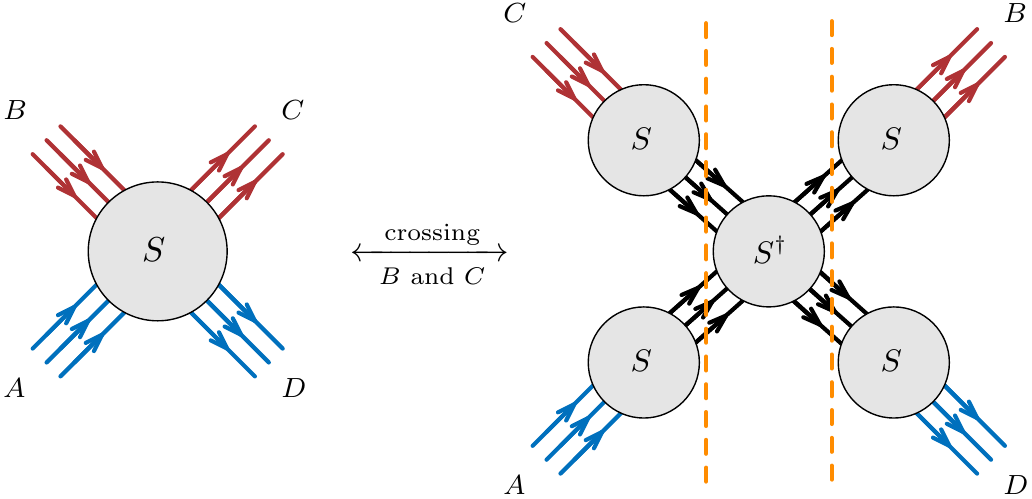}
\ee
where we used the same notation as before. It's called the \emph{crossing equation}. This time, we have to sum over four families of intermediate particles put on unitarity cuts. The four $S$ blobs only talk to each other through the $S^\dag$ in the middle. To see that \eqref{eq:observable} can be recovered as a special case, it's enough to notice that whenever $A,B,C,$ or $D$ is a single stable particle, the blob it attaches to has to be $S = \boldone$, since a stable particle can't decay.

The right-hand side of \eqref{eq:crossing-equation} produces a whole zoo of new kinds of observables, out of which the \eqref{eq:observable} was just one example. Surprisingly, all of them are given by the same analytic function as the original amplitude on the left-hand side, except continued to different kinematics.

However, beyond $4$-point scattering, it's no longer guaranteed that an amplitude can be written as a boundary value of a \emph{single} analytic function as in \eqref{eq:f-approaches}, beyond tree level. For example, it may happen that two coincident thresholds (say, normal and anomalous) have to be approached from the opposite $\pm i\delta$ directions, causing a clash at the endpoints of the analytic continuation. The above prescription assumes that kinematics is such that these local analyticity problems don't appear. Hence crossing symmetry is as subtle as it is fascinating.
Even at tree level, proof of the crossing equation uses a fair deal of combinatorics together with unitarity. It can be also verified on a number of loop-level examples. It's tempting to propose that it holds non-perturbatively. 

\subsection{\label{sec:EFT-bounds}Bounds on effective field theories}

The second topic I wanted to discuss is the application of dispersion relations to put constraints on physically-admissible effective field theories. As the first step, we need to reconsider the high-energy behavior of scattering amplitudes, which will determine how many subtractions to make. Since large energy is precisely when the non-relativistic approximation breaks down, this aspects changes somewhat to what we discussed before.

\subsubsection{Unitarity and inelasticity}

Before starting, let's review a few things about unitarity, which embodies the physical principle of probability conservation and simply says that $S$ is a unitary operator:
\be
S S^\dagger = \boldone\,.
\ee
After plugging in $S = \boldone + i T$ and reaorganizing the terms, it's equivalent to the constraint
\be\label{eq:pre-unitarity}
\tfrac{1}{2i}\left( T - T^\dag \right) = \tfrac{1}{2} T T^\dag\,. 
\ee
In this section, we'll consider $2 \to 2$ scattering of identical particles, in which case the left-hand side becomes the imaginary part of the amplitude $T(s,t)$. After inserting a complete basis of states between $T$ and $T^\dag$ on the right-hand side, we get the generalized optical theorem, which diagrammatically says
\be\label{eq:unitarity}
\includegraphics[scale=1.1,valign=c]{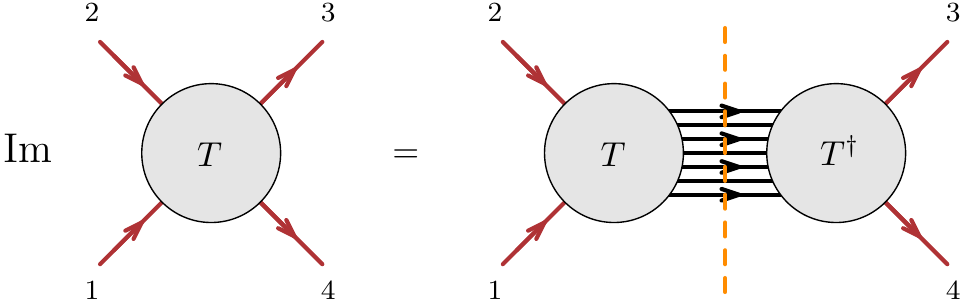}
\ee
It's essentially the same equation as that described in Sec.~\ref{sec:optical}, except for two differences. The unitarity cut involves a sum over all the intermediate states and an integration over their phase space. Most importantly, they can be $k$-particle states if the external energy $s$ is large enough to produce them. For example, if each intermediate particle has mass $m$, we'd need $s \geq (km)^2$. These are the normal thresholds. In Sec.~\ref{sec:optical}, only $k=2$ was allowed. The contributions from $k > 2$ are called \emph{inelastic}, because they require the knowledge of $2 \to k$ and $k \to 2$ amplitudes in the intermediate steps. Recall that unitarity in the form \eqref{eq:unitarity} holds for physical kinematics and it takes more work to continue it the complex values.

One can iterate \eqref{eq:pre-unitarity} by inserting $T^\dagger$ from the left-hand side to the right-hand side over and over to obtain other, equivalent, forms of the unitarity equation. In particular, remembering that $T$ can contain disconnected contributions, we find that consistency with unitarity implies a collection of possible thresholds opening up. These are the anomalous thresholds. For $2 \to 2$ scattering of stable particles, however, only the normal thresholds are kinematically allowed in the physical region. As a simple example, iterating this procedure twice, we arrive at the following form of the unitarity equation:
\be
\tfrac{1}{2i} (T - T^\dagger) = \tfrac{1}{2} \left( T^2 - i T^3 - T^3 T^\dagger\right)\, .
\ee
The second term on the right-hand side involves contributions with two unitarity cuts, e.g.,
\be
\includegraphics[scale=1.1,valign=c]{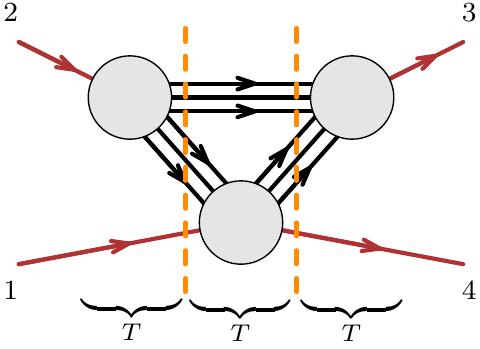}
\ee
This anomalous threshold is disallowed kinematically, because the left- and right-most interactions would involve a decay process. It means that the corresponding threshold can only contribute in complexified kinematics, possibly on different sheets. However, if we considered $3 \to 3$ scattering, just like we did in \eqref{eq:triangle-threshold}, the triangle anomalous threshold would contribute in the physical region.

Anyway, one simplification is to stay within the elastic $k=2$ world and parameterize any inelasticity. Concretely, \eqref{eq:unitarity} gives us the formula
\be
\Im\, T(s,t) = \frac{\sqrt{s-4m^2}}{128\pi^2 \sqrt{s}}	\!\!\int\!
\d \Omega_{\ell}\,
T(s,t_{\mathrm{L}})\,
T^{\ast}(s,t_{\mathrm{R}})
\ee
for $s < 9m^2$. We set $t_{\mathrm{L}} = (p_2 - \ell)^2$ and $t_{\mathrm{R}} = (\ell + p_3)^2$ for the intermediate loop momentum $\ell^\mu$.
Here, we see the second difference to the non-relativistic case in the overall normalization. It originates from the fact that we integrate over the Lorentz-invariant phase space of the $2$-particle intermediate state. The contributions from inelastic terms start appearing when $s \geq 9m^2$.

It will be important to understand the unitarity constraint on partial-wave amplitudes as well. We're going to use the partial-wave decomposition
\be\label{eq:T-partial-wave}
T(s,t) = \frac{\sqrt{s}}{4\sqrt{s-4m^2}} \sum_{j=0}^{\infty} (2j+1) f_j(s) P_j(\cos \theta)\, ,
\ee
which, to follow conventions, differs from the one encountered before by the normalization factor.
After plugging it into the unitarity constraint and some math gymnastics, we obtain the condition
\be\label{eq:f-unitarity-inelastic}
\Im f_j(s) = |f_j(s)|^2 + \Delta
\ee
for every $j$.
Here, $\Delta \geq 0$ represents the contributions from inelastic scattering that turn on for $s \geq 9m^2$. Note that this condition can be satisfied only if $\Delta \leq \frac{1}{4}$. The philosophy will be to forget about $\Delta$, which is generally harder to compute, and instead get a weaker condition:
\begin{empheq}[box=\graybox]{equation}\label{eq:unitarity-inequalities}
0  \leq |f_j(s)|^2 \leq \Im f_j(s) \leq 1
\end{empheq}
The first inequality is obvious and the second follows once we drop inelastic contributions. The third one is yet another consequence of \eqref{eq:f-unitarity-inelastic} once we notice that $f_j = \frac{1}{2i}\left( \e^{2i \eta_j } \sqrt{1-4\Delta} - 1\right)$ for real $\eta_j$, so $\Im f_j \leq \frac{1}{2}(1 + \sqrt{1 - 4\Delta}) \leq 1$.

\subsubsection{\label{sec:Froissart}Froissart bound}

The unitarity constraints \eqref{eq:unitarity-inequalities} can be very inefficient. As a matter of fact, we've seen in Sec.~\ref{sec:locality} and \ref{sec:complex-angles} that locality (understood in terms of bounded range of interactions, not necessarily local fields) guaranteed that $f_j(s)$ has to decay exponentially at large $j$. In fact, one can set up the same calculation with wave packets probing locality in quantum field theory and this result remains unchanged. Let's now combine it with unitarity to get a more efficient bound on partial-wave amplitudes. Through the expansion \eqref{eq:T-partial-wave}, it will ultimately allow us to put an upper bound on the large-$s$ behavior. 

In general, we might expect that $|f_j(s)|$ behaves as follows:
\be
\includegraphics[valign=c,scale=1.1]{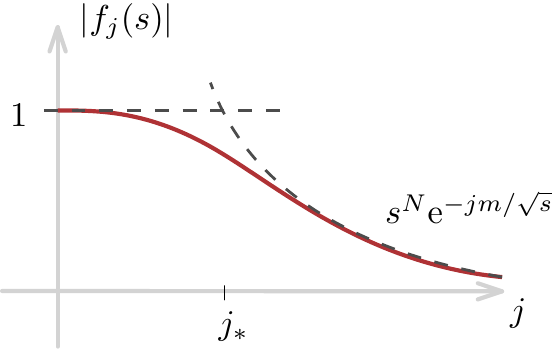}
\ee
For high angular momentum, we get an exponential suppression, but as we lower $j$, there's some critical $j_\ast$ at which the unitarity bound is stronger. To get an estimate on $j_\ast$, let's also include possible polynomial corrections to the exponential bound by saying that
\be
|f_j(s)| < C s^{N} \e^{-j m / \sqrt{s}}	
\ee
for large $s$ and some constants $C$ and $N$. The bound becomes weaker than $|f_j(s)| \leq 1$ at the crossover point $j_\ast$ when
\be
\log C + N \log s \sim j_\ast \frac{m}{\sqrt{s}}\, .
\ee
Hence the cutoff $j_\ast$ grows at high energy as
\be
j_\ast \sim \sqrt{s} \log s\, .
\ee
Since we're only interested in the behavior at large $s$, the precise proportionality constant will not be important.

At first, let's focus on the physical region with the real scattering angle $\theta$, in which case we can use the fact that the Legendre polynomials are bounded:
\be
| P_j(\cos \theta) | \leq 1\, .
\ee
Hence, \eqref{eq:T-partial-wave} tells us that for large $s$ the amplitude is bounded by
\be
|T(s,t)| \leq \frac{1}{2} \sum_{j=0}^{\infty} (2j+1) |f_j(s)|\, .
\ee
We can now split the terms depending on whether $j \lessgtr j_\ast$. Those smaller than $j_\ast$ have $|f_j(s)| \leq 1$ and simply give contributions of order $\sim j_\ast^2 \sim s \log^2 s$. Those with $j$ at least $j_\ast$ are more suppressed because
\be
\sum_{j = j_\ast}^{\infty} (2j + 1) \e^{-jm/\sqrt{s}} \sim \frac{\sqrt{s}}{m} j_\ast\, \e^{- j_\ast m/\sqrt{s}}\, ,
\ee
which is much smaller than $\sim s \log^2 s$. Therefore, the behavior is dominated by low partial-wave spins and we find the following high-energy bound:
\begin{empheq}[box=\graybox]{equation}\label{eq:T-bound5}
|T(s,t)| \lesssim s \log^2 s\, .
\end{empheq}
Note that we assumed presence of a mass gap $m>0$. A similar derivation can be made for the imaginary part $\Im\, T$ and it would lead to the same bound. In terms of the total cross-section, it translates to $\sigma_{\mathrm{tot}} \lesssim \log^2 s$. This is known as the \emph{Froissart bound}.

So far, we worked in the physical region, meaning that the angle $\theta$ was real, or equivalently $0 \leq -t \leq s$. But notice that the arguments only depended on the exponential suppression of the terms in the partial-wave expansion \eqref{eq:T-partial-wave}. As we've seen in Sec.~\ref{sec:complex-angles}, the exponential suppression can be extended to the ellipses in the $t$-plane using locality of interactions. This shows that the bound \eqref{eq:T-bound5} holds also within the small Lehmann ellipse, and a similar one applies for $|\Im\, T(s,t)|$ in the large Lehmann ellipse.

\subsubsection{Dispersive bounds}

We have now all the ingredients needed to write down dispersion relations. The setup is going to be different from that employed in the previous lectures. We're going to assume that we have access to a low-energy (IR) effective theory, through its scattering amplitude for small $s$ and $t$ below some mass scale $\Lambda^2$. We're going to be agnostic about the large-energy (UV) completion of our model and only assume that it satisfies the general physical principles such as causality, locality, and unitarity. The way this assumption manifests itself is through analyticity. Combination of these ingredients will allow us to put bounds on Wilson coefficients in the low-energy expansion. 

Following Adams, Arkani-Hamed, Dubovsky, Nicolis, and Rattazzi, who first came up with this way of constraining effective field theories (EFTs), we'll consider the example of a massless scalar field $\pi$ with a shift symmetry $\pi \to \pi + \mathrm{const}$. In such a theory, the leading terms in the effective Lagrangian we can write down are
\be\label{eq:L-EFT}
\mathcal{L}_{\mathrm{EFT}} = \frac{1}{2} \partial_\mu \pi \partial^\mu \pi + \frac{a}{\Lambda^4} (\partial_\mu \pi \partial^\mu \pi)^2 + \ldots,
\ee
where $\Lambda$ a mass scale. Here, $a$ is the Wilson coefficient which a priori can be any real number. At the leading order in perturbation theory, the $\pi \pi \to \pi \pi$ scattering amplitude is given by
\be\label{eq:T-EFT}
T(s,t) = \frac{a}{\Lambda^4} (s^2 + t^2 + u^2) + \ldots\, .
\ee
We treat it as a low-energy description of the full amplitude, which satisfies all the physical requirements we've been discussing throughout the lectures. In particular, we'll assume that it satisfies the Froissart bound $\lesssim s \log^2 s$.

Let's now write down dispersion relations in the forward limit, $t=0$. In order to dampen the high-energy behavior, we need two subtractions, which we'll implement by considering
\be
\frac{1}{2\pi i} \frac{T(s',0)}{s'^3}\, .
\ee
The contour of integration is almost the same as in Sec.~\ref{sec:dispersion-once-again}, except in relativistic theory we also have to take into account branch cuts coming from the $u$-channel normal thresholds. Let's assume that they kick in only at some scale $\sim \Lambda^2$ of new physics beyond the low-energy description. We have
\be
\includegraphics[scale=1.1,valign=c]{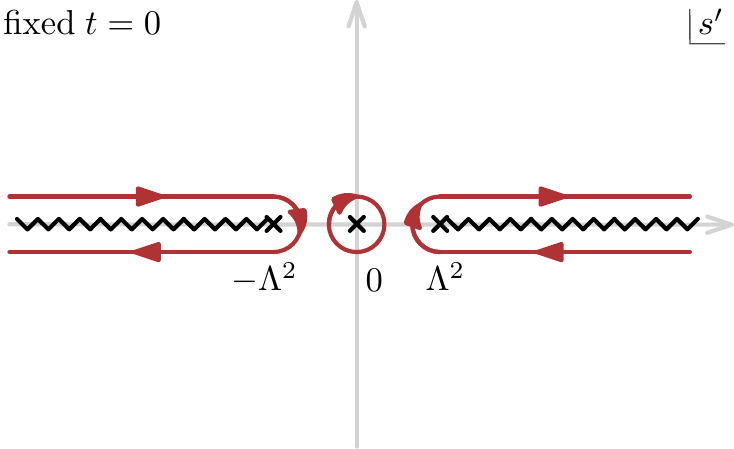}
\ee
The arcs at infinity don't contribute because of the subtractions. Additionally, we can use permutation symmetry of the problem to relate the $s$-channel amplitude to $u$-channel through the relabelling $s \leftrightarrow u = -s-t$, which in the forward limit says that $T(s,0) = T(-s,0)$. Evaluating all the pieces of the above contour, we end up with
\be
\frac{1}{2} T''(s,0) \Big|_{s=0} = \frac{2}{\pi} \int_{\Lambda^2}^{\infty} \frac{\Im\, T(s,0)}{s^3}\, \d s\, ,
\ee
where the residue at $s=0$ translates to taking the second derivative of $T(s,0)$ with respect to $s$.
The imaginary part of the amplitude is related to the total cross-section through $\Im\, T(s,0) = s \sigma_{\mathrm{tot}}(s)$. Once we plug in the EFT expansion, \eqref{eq:T-EFT}, we have $T''(s,0) = 2 a / \Lambda^4$. Positivity of the cross-section tells us that
\begin{empheq}[box=\graybox]{equation}
a = \frac{2 \Lambda^4}{\pi} \int_{\Lambda^2}^{\infty} \frac{ \sigma_{\mathrm{tot}}(s,0)}{s^2}\, \d s > 0\, . 
\end{empheq}
We found that $a$ has to be positive!
This is a very surprising constraint from the point of view of effective field theory, which in principle doesn't have anything to say about the specific value of $a$.

What we've seen above is that consistency of the model at high energies, implemented through analyticity and the Froissart bound, means that only a certain class of EFTs is admissible as its low-energy description. As you might expect, this simple toy model only begins to scratch the surface of possible EFT bounds. One immediate observation is that instead of making two subtractions, we could've used $\frac{T(s',0)}{s'^{2k+1}}$ for arbitrary $k \geq 1$, which leads to an \emph{infinite} number of positivity constraints for deeper and deeper Wilson coefficients in the EFT expansion. One can also go beyond the forward limit and to gauge and gravity theories. There has been a lot of recent work on using more interesting inequalities to carve out the space of admissible EFTs within the space of Wilson coefficients.

\subsubsection{Wilsonian intuition in the presence of time}

Before closing our discussion of this topic, let's ask how is the constraint $a>0$ compatible with the usual Wilsonian intuition on EFTs. Recall that it says we can organize our way of thinking about physics in terms of a hierarchy of energy scales. For example, the result of integrating out high-frequency modes in the UV can be phrased as inducing irrelevant operators in the IR. From this perspective, the coupling $a$ reflects the unknown UV physics and hence there's no reason to expect that it should be constrained in the IR, especially that it's not protected by any symmetry.
Crucially, the above intuition is entirely \emph{Euclidean}, while the analyticity statements in $s$ have to do with causality, which is a uniquely \emph{Lorentzian} concept.

Consider the above example. Here, $\pi$ is a massless field, which means its excitations travel exactly along the light-cone. In non-flat backgrounds, it might easily happen that higher-derivative operators might push these excitations outside of the light-cone. This is precisely what's at odds with the Euclidean intuition of decoupling of short and long wavelengths. In Lorentzian signature, even UV effects can survive to low energies, if they happen in the neighborhood of the light-cone.

To see this explicitly, consider expanding \eqref{eq:L-EFT} around the solution $\pi = \pi_0 + \varphi$, where $\partial_\mu \pi_0 = V_\mu$ is a constant light-like vector with $V^2 = 0$. The fluctuations $\varphi$ over this background satisfy the linearlized equation of motion
\be
\left( \eta^{\mu\nu} + \frac{8 a}{\Lambda^4} V^\mu V^\nu + \ldots \right) \partial_\mu \partial_\nu \varphi = 0\, .
\ee
We can obtain the dispersion relation (in the sense of speed of propagation, not contour integrals) in momentum space by plugging in a plane wave solution $\varphi = \e^{i p \cdot x}$, giving:
\be
p^2 + \frac{8 a}{\Lambda^4} (p \cdot V)^2 + \ldots = 0\, .
\ee
Since $(p\cdot V)^2 \geq 0$, absence of superluminal propagation requires $a \geq 0$ (which, note, is ever so slightly weaker than $a>0$ we got before). It pushes the excitations $\varphi$ \emph{into} the light-cone a tiny bit.

At first glance, it might seem rather puzzling why a small effect associated to a higher-derivative operator can be seen at low energies at all. The reason for it is that, while the correction is indeed small, it can build up over long distances. This is entirely analogous to the discussion of time delays from Sec.~\ref{sec:causal-transforms-angles}. There, we've seen that time delay can be made arbitrarily large if we simply increase the size of the scattering region $R$.

\subsection{S-matrix bootstrap}

The last topic I'd like to highlight is the S-matrix bootstrap. It's, in fact, one of the oldest ideas in the S-matrix theory, championed by Geoffrey Chew in the 1960's. The hope was that imposing consistency conditions on scattering amplitudes, mostly through its analytic properties, would allow us to reconstruct the S-matrices for strong interactions ``out of thin air'' without Feynman-diagrammatic computations. While the idea thrives and reverberates throughout large chunks of theoretical high-energy physics, the original dream of discovering the Standard Model from analyticity turned out to be far too ambitious.

The main point I wanted to emphasize is the change in perspective in the modern incarnation of the S-matrix bootstrap. We are no longer searching for a unique theory, but instead for special corners in the appropriately-defined theory space. This approach turned out to be a great success story in the neighboring problem of bootstrapping conformal field theories, leaving us with hope it can carry over to scattering amplitudes as well.

\subsubsection{\label{sec:2d-amplitudes}Two-dimensional amplitudes}

In order to illustrate the principles of the S-matrix bootstrap on the simplest possible example, we'll consider scattering of a single scalar in $1+1$ space-time dimensions. It turns out that in this case, the bootstrap program can be completed entirely analytically without resorting to numerics. Although some special cases were known for longer, viability of this approach to bootstrap scattering amplitudes was demonstrated in a series of papers by Paulos, Penedones, Toledo, van Rees, and Vieira, and we closely follow their simplest toy-model example.

We'll consider the $2\to 2$ scattering amplitude of a scalar with mass $m$ and a single bound state with mass $M < 2m$. A special fact about two-dimensional scattering is that momentum conservation forces the  kinematics to be either forward ($t=0$) or backward ($u=0$). Let's pick the second choice without loss of generality. Since $s+t = 4m^2$, the scattering amplitude depends only on the single independent Mandelstam invariant $s$ and we'll denote it by $S(s)$. Another special property is that, just like in the case of spherical scattering from the first lecture, the $2\to 2$ amplitude is just a single function. There's no angular momentum. Compared to the four-dimensional case, the restricted kinematics gives rise to an additional Jacobian ${\cal J}_s = 1 / (2\sqrt{s}\sqrt{s - 4m^2})$ that we have to tack on to $S(s)$.

The main technical simplification in two dimensions comes from the fact that unitarity amounts to imposing
\be\label{eq:unitarity-2d}
|S(s)|^2 \leq 1
\ee
in the physical region $s > 4m^2$. We have an inequality since we're not taking into account possible inelastic contributions. The amplitude has an $s$-channel branch cut when $s \geq 4m^2$ and a $t$-channel one when $s \leq 0$ (translating to $t \geq 4m^2$). Moreover, the presence of the bound state is seen as two simple poles at $s=M^2$ and $s = 4m^2 - M^2$. The function $S(s)$ is analytic everywhere in the $s$-plane outside of these cuts and poles.

Close to the bound state at $s=M^2$, the amplitude behaves as
\be\label{eq:bound-state-2d}
S(s) =  \frac{g^2 m^4}{s - M^2}\, {\cal J}_{\small M^2} + \ldots\, .
\ee
Here, $g$ denotes the coupling of two external scalars to the bound state, and the normalization by powers of $m$ is chosen so that $g$ is dimensionless. Note that we're treating $S(s)$ as the exact amplitude, so $g$, $m$, and $M$ are the renormalized coupling constant and masses.

We'll phrase the bootstrap problem as asking what's the maximum value of $g$ attainable for a given spectrum of the masses, $m$ and $M$. It's intuitively clear that $g$ can't be too large, since otherwise new bound states could be produced. The challenge is to turn this intuition into an inequality. As ingredients, we're going to use analyticity in the cut $s$-plane, the unitarity constraint \eqref{eq:unitarity-2d}, and the behavior near the bound state \eqref{eq:bound-state-2d}.

\subsubsection{Maximum modulus principle}

The idea will be to argue that the unitarity constraint \eqref{eq:unitarity-2d}, which holds only in the physical regions, can be in fact extended to all values of $s$ in an appropriate sense. The strategy will be similar to the one we've encountered in Sec.~\ref{sec:high-energy-bound}.

As the first step, we're going to change variables from $s$ to the cunningly-chosen $z$:
\be\label{eq:z-2d}
z(s) = \frac{\sqrt{s(4m^2 - s)} - M \sqrt{4m^2 - M^2}}{\sqrt{s(4m^2 - s)} + M \sqrt{4m^2 - M^2}}\,.
\ee
The following plots illustrate how this change of variables maps the part of the complex plane with $\Re s > 2m^2$ into a unit disk in the $z$-plane:
\be
\includegraphics[scale=0.78,valign=c]{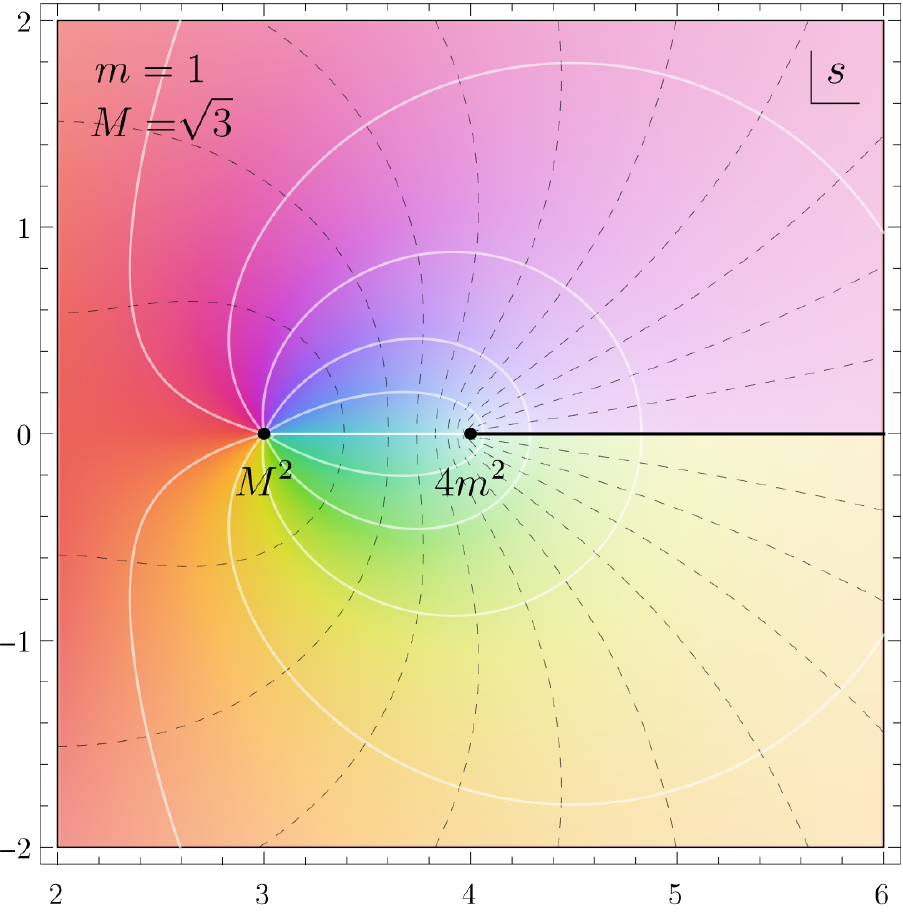}\quad
\includegraphics[scale=0.8,valign=c]{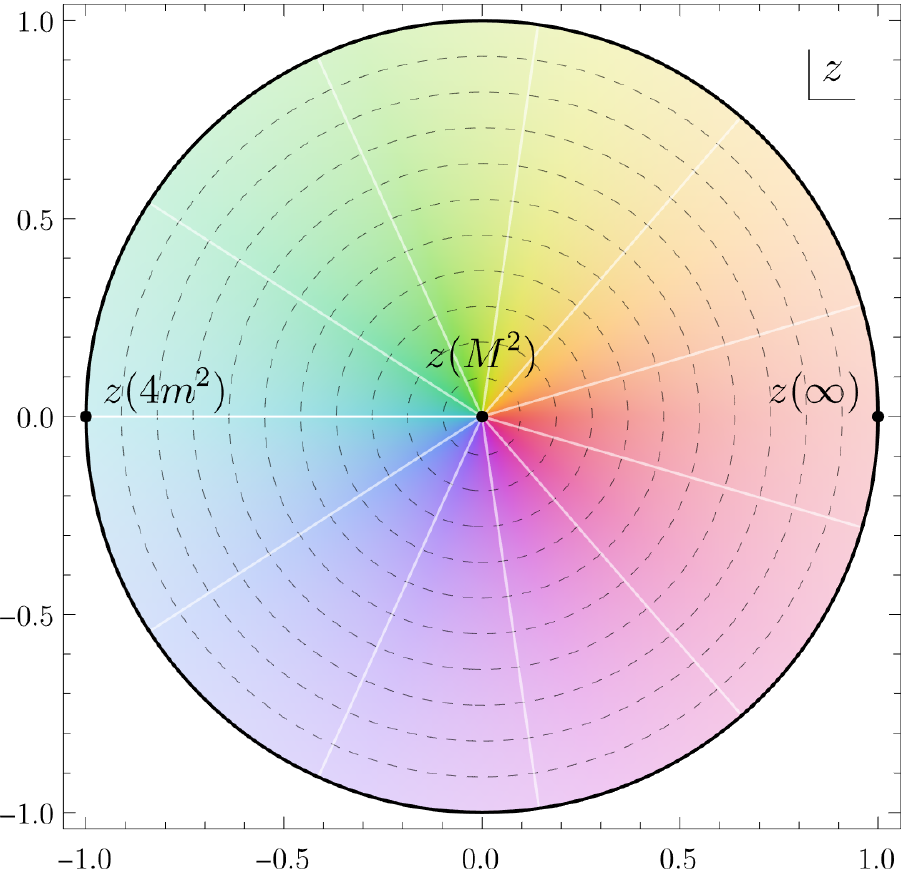}
\ee
For the purpose of the plot, we chose $m=1$ and $M = \sqrt{3}$. What happened is that the physical region $s \geq 4m^2$ above the cut (pink-ish) got mapped into the semi-circle in the lower half-plane of $z$. Likewise, the unphysical side of $s \geq 4m^2$ (yellow-ish) is mapped into the semi-circle in the upper half-plane of $z$. As you can check using \eqref{eq:z-2d}, the branch points at $s=4m^2$ and $s\to \infty$ are mapped to $z = -1$ and $z = 1$, respectively. The axis $\Re s = 2m^2$ (red-ish) is folded such that its points with opposite imaginary parts are identified. Finally, the bound state pole at $s=M^2$ is mapped to the origin of $z$. For simplicity, we're going to assume that $M > \sqrt{2}m$ so that the bound-state pole is located within $\Re s > 2m^2$.

The amplitude $S(z)$ is now analytic in the unit disk except at for the bound-state pole at $z = 0$. We can write its residue as
\be
\Res_{z = 0} S(z) = \frac{1}{2\pi i} \oint_{|z'| = 1} S(z')\, \d z' = \frac{1}{2\pi} \int_{0}^{2\pi} S(\e^{i\phi})\, \e^{i\phi}\, \d \phi\, ,
\ee
where the second inequality is just a trivial change of variables $z' = \e^{i\phi}$.
This is of course just the dispersion relation is disguise. We can now use the fact that the absolute value of an integral is upper-bounded by the integral of the absolute value:
\be\label{eq:Res-inequality}
|\!\Res_{z = 0} S(z)| \leq \frac{1}{2\pi} \int_{0}^{2\pi} \d\phi  = 1\, .
\ee
After the first inequality we used $|\e^{i \phi}| = 1$ and the unitarity constraint $|S(\e^{i\phi})| \leq 1$. We're going to use this bound in a short while.

We can actually do much better than just bounding the residue by using one of the fundamental results in complex analysis called the \emph{maximum modulus principle}. It says that a function $f(z)$ analytic in a unit disk can only attain its maximum modulus $|f(z)|$ at the boundary $|z|=1$. Moreover, the bound is saturated only for the constant function. The proof of this statement is quite similar to the contour manipulations that lead to \eqref{eq:Res-inequality}.

We can't directly apply the maximum modulus principle to the amplitude $S(z)$, because it's not analytic in the unit disk due to the pole at the origin. However, we can take $f(z) = z S(z)$, which tells us that
\begin{empheq}[box=\graybox]{equation}\label{eq:S-bound-2d}
|z S(z)| \leq 1
\end{empheq}
for any $z$ in the unit disk. This inequality can be viewed as the extension of the unitarity condition \eqref{eq:unitarity-2d} to complex kinematics. We could've obtained a similar result directly in the cut $s$-plane, but its derivation would be slightly more technical since one has to employ the Phragm\'en--Lindel\"of principle we encountered in Sec.~\ref{sec:high-energy-bound}, which is an extension of the maximum modulus principle to unbounded domains.

\subsubsection{Saturating the bound}

It only remains to translate all the statements from $z$ back to $s$. The residues $\Res_{s = M^2} = \frac{\partial s}{\partial z} \Res_{z = 0}$ are connected through a Jacobian factor, which reads
\be
\frac{\partial s}{\partial z} \bigg|_{s = M^2} = 2M^2 \frac{M^2 - 4m^2}{M^2 - 2m^2}\, .
\ee
The inequality \eqref{eq:Res-inequality} together with \eqref{eq:bound-state-2d} immediately allows us to place an upper bound on the bound-state coupling $g$:
\begin{empheq}[box=\graybox]{equation}\label{eq:g2-bound}
g^2 \leq \frac{4 M^3}{m^4} \frac{(4m^2 - M^2)^{3/2}}{|M^2 - 2m^2|}\, .
\end{empheq}
Note that the absolute value in the denominator is not strictly necessary because so far we took $M > \sqrt{2} m$. Let's now relax this condition.

Recall that there are two bound-state poles: at $s = M^2$ and $s = 4m^2 - M^2$ coming from the $s$- and $u$-channel exchanges respectively. Close to the second one, the amplitude behaves as
\begin{subequations}
\begin{align}
S(4m^2 - M^2) &=  \frac{g^2 m^4}{u - M^2}\, {\cal J}_{4m^2 - M^2} + \ldots \\
&= -\frac{g^2 m^4}{s - 4m^2 + M^2}\, {\cal J}_{4m^2 - M^2} + \ldots.
\end{align}
\end{subequations}
Hence its residue reverses the sign compared to \eqref{eq:bound-state-2d}. Once we take $M < \sqrt{2} m$, this is the pole that crosses $\Re s > 2m^2$ and enters the unit disk in $z$. The only thing that changes is the overall minus sign of the residue, but since we're only bounding the modulus, the result \eqref{eq:g2-bound} stays intact. Recall that we need $0 < M^2 < 4m^2$ so that the bound state is below the thresholds in both channels.

It's the simplest to understand the bound \eqref{eq:g2-bound} by plotting it as a function of $M/m$:
\be\label{eq:g2-plot}
\includegraphics[scale=1.1,valign=c]{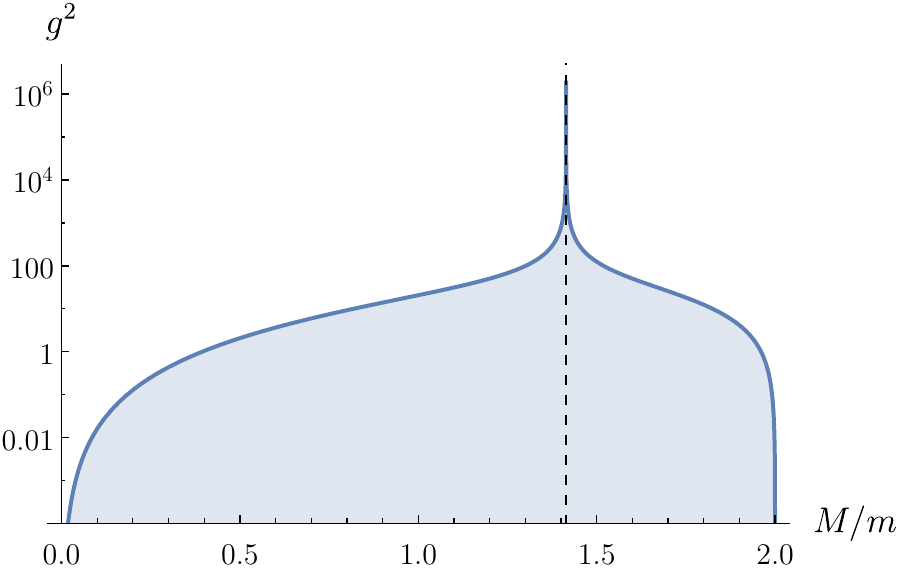}
\ee
Note the scale is logarithmic. All the couplings outside of the shaded region are physically disallowed. The features of the bound can be easily explained. It diverges when $M = \sqrt{2}m$ (dashed line), which corresponds to the situation in which two bound states collide and cancel out. Here, we obviously lose control over bounding $g$. On the other hand, very close to $M = 0$ and $M=2m$, the bound state can be interpreted as a $2$-particle state made of the scalar with mass $m$, which requires weak coupling.

It's interesting to ask what saturates the bound. We've already seen around \eqref{eq:S-bound-2d} that the maximum modulus principle said the scattering amplitude saturating the bound has to satisfy $|z S(z)| = 1$ for all $z$ in the disk. The solution of this equation (up to a phase) turns out to be precisely the amplitude in a famous two-dimensional theory called the \emph{sine--Gordon model}:
\begin{empheq}[box=\graybox]{equation}
S_{\mathrm{SG}}(s) = \frac{1}{z(s)} = \frac{\sqrt{s(4m^2 - s)} + M \sqrt{4m^2 - M^2}}{\sqrt{s(4m^2 - s)} - M \sqrt{4m^2 - M^2}}\, .
\end{empheq}
Excitations in this theory are known as \emph{breathers} (named this way because they look like they breathe) and the above amplitude corresponds to the $2\to 2$ scattering of the lightest one with mass $m$, involving a bound state with mass $M > \sqrt{2} m$. The amplitude saturating the same bound for $M < \sqrt{2} m$ is $- S_{\mathrm{SG}}(s)$, but its physical interpretation isn't known. Similar analysis can be carried out in the presence of multiple bound states.

\subsubsection{Numerical strategy}

It would be too much to expect that we can pull off the same analysis entirely analytically beyond the two-dimensional cases. In this final section, let me give a glimpse of how the bootstrap algorithm is implemented numerically. For simplicity, we remain in $1+1$ dimensions, which illustrates the general principle but with fewer variables.

The first step is to pretend the amplitude can be continued off-shell and treat it as a function $S(s,t)$ of independent variables $s$ and $t$ (in higher dimensions, we would treat $s$, $t$, and $u$ as independent). Ultimately, we're interested in the on-shell subspace $s+ t = 4m^2$ on which $S(s)$ is defined. We have branch cuts when $s> 4m^2$ and $t>4m^2$, as well as a number of possible bound-state poles.

As before, we can map the cut $s$- and $t$-planes to unit disks by changing variables to
\be
\rho_s(s) = \frac{\sqrt{2}m - \sqrt{4m^2 - s}}{\sqrt{2}m + \sqrt{4m^2 - s}}, \qquad \rho_t(t) = \frac{\sqrt{2}m - \sqrt{4m^2 - t}}{\sqrt{2}m + \sqrt{4m^2 - t}}\, .
\ee
This allows us to write down an ansatz for the amplitude as a Taylor expansion around the origin in $\rho_s$ and $\rho_t$:
\be\label{eq:bootstrap-ansatz}
S(\rho_s,\rho_t) = - \frac{\hat{g}_s}{s - M^2} - \frac{\hat{g}_t}{t - M^2} + \sum_{a,b = 0}^{\infty} c_{ab}\, \rho_s^a \rho_t^b\, .
\ee
Here, we used a pair of bound states and assumed that neither is mapped to the origin of the disks for simplicity.
Since the external particles are identical, we should impose invariance under $s \leftrightarrow t$, which amounts to equating the (rescaled) coupling constants $\hat{g}_s = \hat{g}_t$ and setting $c_{ab} = c_{ba}$.

In principle, the on-shell condition $s + t = 4m^2$ imposes the constraint
\be
\rho_s^2 \rho_t + \rho_t^2 \rho_s + 4 \rho_s \rho_t + \rho_s + \rho_t = 0\, ,
\ee
which means there are a lot of dependencies among the coefficients in the expansion \eqref{eq:bootstrap-ansatz}. These can be used to reduce the size of the ansatz by setting many $c_{ab}$'s to zero. You might be worried that going off-shell could've introduced new singularities. However, a result in complex convexity theory known as the \emph{Cartan's extension theorem} says that analyticity of $S(\rho_s)$ implies that there exists a function $S(\rho_s,\rho_t)$ analytic in the product of two disks, though in general it might not be unique.

The bootstrap strategy can be implemented as follows. We truncate the sums in \eqref{eq:bootstrap-ansatz} at some large cutoff for the exponents $a$ and $b$. We then sample points on the semi-circle in the upper half-plane of $\rho_s$, which correspond to $s \geq 4m^2$. For each of them, we can impose the unitarity constraint. This gives a large system of inequalities for the coefficients $\hat{g}_s$ and $c_{ab}$ of the ansatz, which we can use to maximize $\hat{g}_s$. This procedure can be done using similar numerical tools to those we used in the first set of exercises. When applied to the same problem as in Sec.~\ref{sec:2d-amplitudes}, the outcome becomes a numerical version of the plot \eqref{eq:g2-plot}. Once can of course experiment with different ansatze than \eqref{eq:bootstrap-ansatz}, for example, by using Zernike polynomials or a neural network.

As you might expect by now, the above example is just the simplest of toy models for the larger S-matrix bootstrap program. The extra technical difficulty arising in higher dimensions becomes the fact that unitarity has to imposed for each partial-wave amplitude and hence involves bounding integrals. Nevertheless, many variations of the bootstrap problem have been successfully implemented in the recent years. One can also connect it with the EFT approach we discussed in Sec.~\ref{sec:EFT-bounds}. Hopefully, in the future, the S-matrix bootstrap can be also used to constrain other observables beyond scattering amplitudes using the crossing equation.

\subsection{Exercises}

In this final set of exercises, we'll get a glimpse of the topic of resurgence, which applied to quantum theory says that perturbative expansions around different vacua are all interconnected. Although this topics hasn't yet played a significant role in the S-matrix theory, it's really a part of the same story, and learning about it might give you a different perspective on perturbation theory.

\subsubsection{Asymptotic series}

We're going to study the follow toy model integral:
\be\label{eq:Z-lambda}
\mathcal{Z}(\lambda) = \frac{1}{\sqrt{2\pi}} \int_{-\infty}^{\infty} \e^{-\frac{1}{2} \phi^2 - \frac{\lambda}{4} \phi^4} \, \d \phi\, .
\ee
We first focus on the case $\lambda > 0$. In a fancy language, you can think of it as the partition function of the $\phi^4$ theory in zero space-time dimensions, where \eqref{eq:Z-lambda} is the path integral over the value of the field $\phi$ at a single point.  

Anyway, $\mathcal{Z}(\lambda)$ is a function of only the coupling $\lambda$ and it seems natural to expand it in perturbation theory as
\be\label{eq:Z-as}
\mathcal{Z}_{\mathrm{pert}}(\lambda) = \sum_{k=0}^{\infty} a_k \lambda^{k}\, .
\ee
The coefficients $a_k$ of this series can be computed by expanding the exponential $\e^{-\frac{\lambda}{4}\phi^4}$ and using the identity
\be
\frac{1}{\sqrt{2\pi}} \int_{-\infty}^{\infty} \e^{-\frac{1}{2}\phi^2} \phi^{4k}\, \d \phi = (4k-1)!!\, .
\ee
Recall that the double factorial equals $m!! = m(m-2)(m-4) \cdots 1$ for odd $m$. This gives
\be
a_k = \frac{(4k-1)!!}{(-4)^k k!}\, .
\ee
You can in principle reproduce $a_k$ by expanding $\mathcal{Z}(\lambda)$ in terms of Feynman diagrams, being careful about the symmetry factors. The next step is to plot partial sums of \eqref{eq:Z-as} to see how it converges, which can be done with the snippet
\begin{minted}[firstline=1]{mathematica}
ListPlot[Accumulate[
	Table[(4k - 1)!!/((-4)^k*k!) λ^k, {k, 0, 15}]
] /. λ -> 1/20, DataRange -> {0, 15}, PlotRange -> Full]
\end{minted}
Here, we set $\lambda = \frac{1}{20}$ and considered partial sums up to $k \leq k_\ast = 15$. The result looks as follows:
\be\label{eq:partial-sums}
\includegraphics[scale=1.1,valign=c]{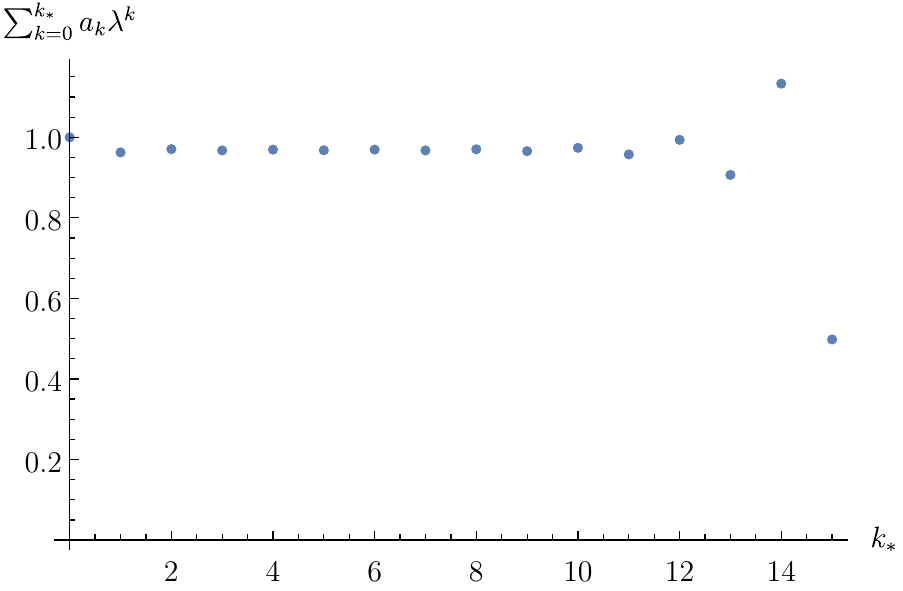}
\ee
What happened? It looks like the series converged for a while, but then it went crazy. You can check that numerical precision is not the issue here. The culprits are the coefficients of the series \eqref{eq:Z-as}, which become increasing large for large $k$. In fact, they go as $a_k \sim (-4)^k k!$. Since they don't decay exponentially (in fact, they grow), \eqref{eq:Z-as} has a zero radius of convergence. It means that for large enough $k$, adding more terms in the sum leads to \emph{worse} approximation of the true answer. This is an example of an \emph{asymptotic series} and the reason we called it $\mathcal{Z}_{\mathrm{pert}}$ instead of $\mathcal{Z}$. 

It's useful to estimate what's the cutoff $k_\ast$ on the sum that gives the most accurate results. Asymptotically for large $k$, the series coefficients grow as $|a_k \lambda^k| \sim |4\lambda|^k k!$. Using the Stirling's approximation, we get $\sim \e^{k [\log k - 1 + \log (4\lambda)]}$. Minimizing for $k$, we find that its critical value is
\be
k_\ast = \frac{1}{4\lambda}\, .
\ee
For example, for $\lambda = \frac{1}{20}$, the optimal truncation would've been at $k \lesssim k_\ast = 5$. You can read off from \eqref{eq:partial-sums} that this would indeed give a result that seem to have stabilized. You can go ahead and try repeating the same exercise for other values of $\lambda$.

We can estimate the error we make in truncating the series by asking what's the size of the terms around where we've put the cutoff. They are of order
\be
\sim \e^{- k_\ast} \, .
\ee
Hence, the result of the truncation is accurate more or less to this order. For us, it's $\e^{-5} \approx 0.007$, so the asymptotic series is useful to make predictions at around a percent level in this case. Any corrections are non-perturbative in $\lambda$. Indeed, you can expand $\e^{-\frac{1}{4\lambda}}$ in $\lambda$ and convince yourself that every term in the Taylor expansion vanishes, so they are not visible at any order in perturbation theory.

\subsubsection{Dyson's argument}

Asymptotic series might seem rather exotic, but in fact almost every expansion we make in quantum field theory is of this kind. The classic physical argument making it plausible is due to Dyson, who pointed out that Feynman-diagrammatic expansion of observables in QED has to be asymptotic. Consider the perturbative series in the charge $e^2$. If it really had a non-zero radius of convergence around the origin, it would mean that we can formulate the theory for some small $e^2<0$. Such a theory would correspond to like charges attracting, instead of repelling, each other. We know that it can't exist because it would give rise to pathologies such as unstable vacuum: an infinite number of repelling positron-electron pairs could be created. Therefore, perturbative expansion of scattering amplitudes and observables in QED has to be asymptotic. Can you estimate at which loop order the perturbation theory will start giving inaccurate answers?

This discussion is of course general and extends to other theories. Another quick argument is to notice that the number of Feynman diagrams in a garden-variety quantum field theory grows much faster than exponentially with the number of loops. Unless there are huge cancellations, they have to lead to an asymptotic series.

\subsubsection{Borel resummation}

One possible way of making sense of asymptotic series such as \eqref{eq:Z-as} is the \emph{Borel transform}, which is defined by
\be\label{eq:BZ}
\mathcal{B}\mathcal{Z}_{\mathrm{pert}}(\lambda) = \sum_{k=0}^{\infty} \frac{a_k}{k!} \lambda^k\, .
\ee
In our case, the terms now go as $|\frac{a_k}{k!}| \sim |4\lambda|^{k}$ and the series converges when $|\lambda| < \frac{1}{4}$. As a matter of fact, it can be written down in closed form:
\be\label{eq:BZ2}
\mathcal{B}\mathcal{Z}_{\mathrm{pert}}(\lambda) = \frac{2}{\pi (1 + 4\lambda)^{1/4}}
K\left(\sqrt{\tfrac{1}{2} - \tfrac{1}{2\sqrt{1+4\lambda}}} \right)\, ,
\ee
where $K$ is the complete elliptic integral of the first kind. Convergence is now limited by singularities in the $\lambda$-plane, also known as the Borel plane. Let's inspect it visually by running
\begin{minted}{mathematica}
ComplexPlot3D[
	2*EllipticK[1/2 - 1/(2*Sqrt[1 + 4λ])]/(π(1 + 4λ)^(1/4)),
{λ, -1 - I, 1 + I}]
\end{minted}
Note the convention $\mathtt{EllipticK[k^2]} = K(k)$. The result looks like this:
\be
\includegraphics[scale=1.1,valign=c]{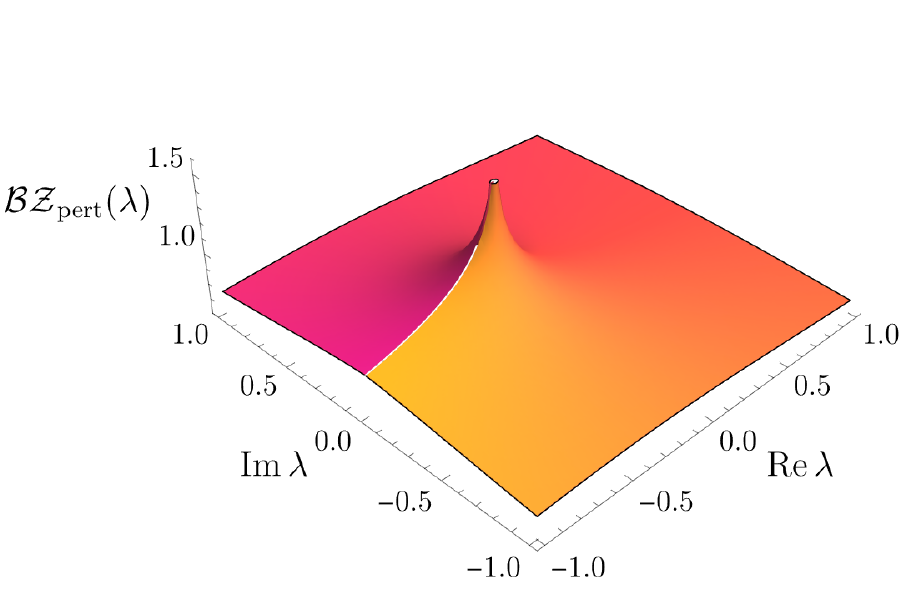}
\ee
As you can also work out from \eqref{eq:BZ2}, the answer has only a logarithmic branch cut starting from $\lambda = -\frac{1}{4}$.

Expansion around this singularity turns out to involve somewhat familiar ingredients:
\be\label{eq:BZ-log}
\mathcal{B}\mathcal{Z}_{\mathrm{pert}}(\lambda) = -\frac{1}{\sqrt{2} \pi} \log(\lambda + \tfrac{1}{4})\, \mathcal{B}\mathcal{Z}_{\mathrm{inst}}(\lambda + \tfrac{1}{4}) + \mathrm{regular}\, .
\ee
where ``regular'' denotes terms analytic around $\lambda = -\tfrac{1}{4}$. The coefficient of the logarithm gives yet another Taylor series:
\be
\mathcal{B}\mathcal{Z}_{\mathrm{inst}}(\lambda) = \sum_{k=0}^{\infty} (-1)^k \frac{a_k}{k!} \lambda^k\, ,
\ee
where the constants $a_k$ are the same as before. The meaning of the subscript ${}_{\mathrm{inst}}$ will become clear shortly. Coefficients of this expansion are almost exactly the same as in \eqref{eq:BZ}, except for the additional sign $(-1)^k$. As a matter of fact, we can also go back and reconstruct the corresponding asymptotic series: 
\begin{subequations}\label{eq:Z-inst}
\begin{align}
\mathcal{Z}_{\mathrm{inst}}(\lambda) &= \sum_{k=0}^{\infty} (-1)^k a_k \lambda^k \\
&= 1 + \frac{3}{4} \lambda + \frac{105 }{32} \lambda^2 + \frac{3465}{128} \lambda^3 + \frac{675675}{2048}\lambda^4 + \frac{43648605}{8192} \lambda^5 + \ldots \, .
\end{align}
\end{subequations}
The conclusion is that singularities of the Borel transform allowed us to find a new asymptotic series. We will see its importance in a second.

Without assuming the specific form of the Borel transform \eqref{eq:BZ}, but only the fact it's polynomially-bounded at infinity and has a logarithmic branch cut starting at $\lambda = -\frac{1}{4}$, derive dispersion relations for $\mathcal{BZ}_{\mathrm{pert}}(\lambda)$. Explain how they link the coefficients of the perturbative expansion $\mathcal{B}\mathcal{Z}_{\mathrm{pert}}(\lambda)$ around $\lambda = 0$ to the coefficient of $\mathcal{B}\mathcal{Z}_{\mathrm{inst}}(\lambda)$. You should be able to find that low-$k$ terms of the former are related to the high-$k$ terms of the latter. This is one the simplest forms of \emph{resurgence}: the idea that perturbative expansions around different values of $\lambda$ communicate with one another.

The original function $\mathcal{Z}(\lambda)$ can be reconstructed from its Borel transform using
\be
\mathcal{Z}_B(\lambda) = \int_{0}^{\infty} \d t\, \e^{-t}\, \mathcal{B}\mathcal{Z}_{\mathrm{pert}}(t\lambda)\, .
\ee
By employing the series expansion of $\mathcal{B}\mathcal{Z}_{\mathrm{pert}}(t\lambda)$ from \eqref{eq:BZ}, you can convince yourself that the result agrees with the asymptotic expansion \eqref{eq:Z-as}. This doesn't yet mean that $\mathcal{Z}_B(\lambda) = \mathcal{Z}(\lambda)$, because the two results could've differed by non-perturbative terms. Notice that the singularity at $t\lambda = -\frac{1}{4}$ is precisely what limits convergence of the integral. If $\lambda < 0$, there's a branch cut on the integration domain and the integrand goes as $\sim \e^{\frac{1}{4\lambda}}$. This ambiguity translates to the appearance of non-perturbative corrections. For any other complex value of $\lambda$, however, we have $\mathcal{Z}_B(\lambda) = \mathcal{Z}(\lambda)$ and we say that the series is Borel resummable. In fact, we can integrate it directly, which gives
\be
\mathcal{Z}(\lambda) = \frac{\e^{\frac{1}{8\lambda}}}{2\sqrt{\pi \lambda}}\, K_{1/4}\! \left( \tfrac{1}{8\lambda} \right)\, ,
\ee
where $K_a$ is the modified Bessel function.

\subsubsection{Saddle points and transseries}

To tie everything together, we're going to evaluate the integral \eqref{eq:Z-lambda} through the method of steepest descents. By setting the first derivative of $S(\phi) = \frac{1}{2} \phi^2 + \frac{\lambda}{4}\phi^4$ to zero, you should find three saddle points at\vspace{-0.5em}
\be
\phi_0^\ast = 0, \qquad \phi_\pm^\ast = \pm \frac{i}{\sqrt{\lambda}}\, .\vspace{-0.5em}
\ee
On these saddles, the action takes the values
\be
S(\phi_0^\ast) = 0, \qquad S(\phi_\pm^\ast) = -\frac{1}{4\lambda}\, .
\ee
Contributions from each of them behave as $\e^{-S(\phi^\ast)}$ times a power series in $\lambda$. For example, expanding around $\phi = \phi_0^\ast$ is going to give us the asymptotic series $\mathcal{Z}_{\mathrm{pert}}(\lambda)$. This is a toy model for the perturbative vacuum, while $\phi_\pm^\ast$ would be called \emph{instantons}.

Let's evaluate a contribution from the instanton saddle points. Due to symmetry $\phi \leftrightarrow -\phi$ symmetry, it doesn't matter which of the two saddles we pick, so let's choose $\phi_+^\ast$ for concreteness. Expanding around it can be done by plugging in $\phi = \phi_+^\ast + i \delta \phi$, around which
\be
S(\phi) = S(\phi_+^\ast) + (\delta \phi)^2 \left( 1 + \tfrac{1}{2} \sqrt{\lambda}\, \delta \phi \right)^2\, .
\ee
The fact of $i$ in front of $\delta \phi$ is important to make it a Gaussian.
Either with pen and paper, or using \texttt{Series} and \texttt{Integrate}, show that the saddle-point expansion around it gives
\be
\sqrt{\pi} \e^{\frac{1}{4\lambda}} \left( 1 + \frac{3}{4} \lambda + \frac{105 }{32} \lambda^2 + \frac{3465}{128} \lambda^3 + \frac{675675}{2048}\lambda^4 + \frac{43648605}{8192} \lambda^5 + \ldots \right) ,
\ee
which organizes itself into $\sqrt{\pi} \e^{\frac{1}{4\lambda}}$ times the asymptotic series $\mathcal{Z}_{\mathrm{inst}}(\lambda)$ we encountered in \eqref{eq:Z-inst}. It's non-perturbative in $\lambda$.

Approximating the integral is more complicated than just expanding around the saddles, because one needs to decide which of them are relevant for a given contour in the first place. This can be done by writing the original integration contour, which for us was the real line, as a linear combination of steepest-descent path from each saddle, also known as Lefschetz thimbles. Systematic way of decomposing a given integration contour in such a basis is called Picard--Lefschetz theory (or complex Morse theory). In our simple setup, however, this can be done more straightforwardly by the means of a plot. We can just plot the values along which the phase $\Im\, S(\phi)$ is stationary. For example, we can use
\begin{minted}{mathematica}
S[φ_, λ_] := φ^2/2 + λ*φ^4/4;
With[{λ = 1 + I},
gradient = ContourPlot[Re[S[x + I*y, λ]], {x, -5, 5}, {y, -5, 5}, 
					   PlotRange -> Full, ContourLines -> False
					   Contours -> 20, ColorFunction -> "RedBlueTones"];
thimbles = ContourPlot[{Im[S[x + I*y, λ]] == 0, 
						Im[S[x + I*y, λ]] == Im[S[I/Sqrt[λ], λ]]},
					   {x, -5, 5}, {y, -5, 5}, ContourStyle -> Black];
points = ListPlot[{ReIm[0], ReIm[I/Sqrt[λ]], ReIm[-I/Sqrt[λ]]}, 
				  PlotStyle -> Red];
]
Show[gradient, thimbles, points]
\end{minted}
Here, $\mathtt{S}[\phi,\lambda]$ defines the exponent, \texttt{gradient} plots its real part and \texttt{thimbles} (black) are the constant-phase curves emanating from the saddle points \texttt{points} (red). An example plot for $\lambda = 1 + i$ looks as follows:
\be
\includegraphics[scale=0.63,valign=c]{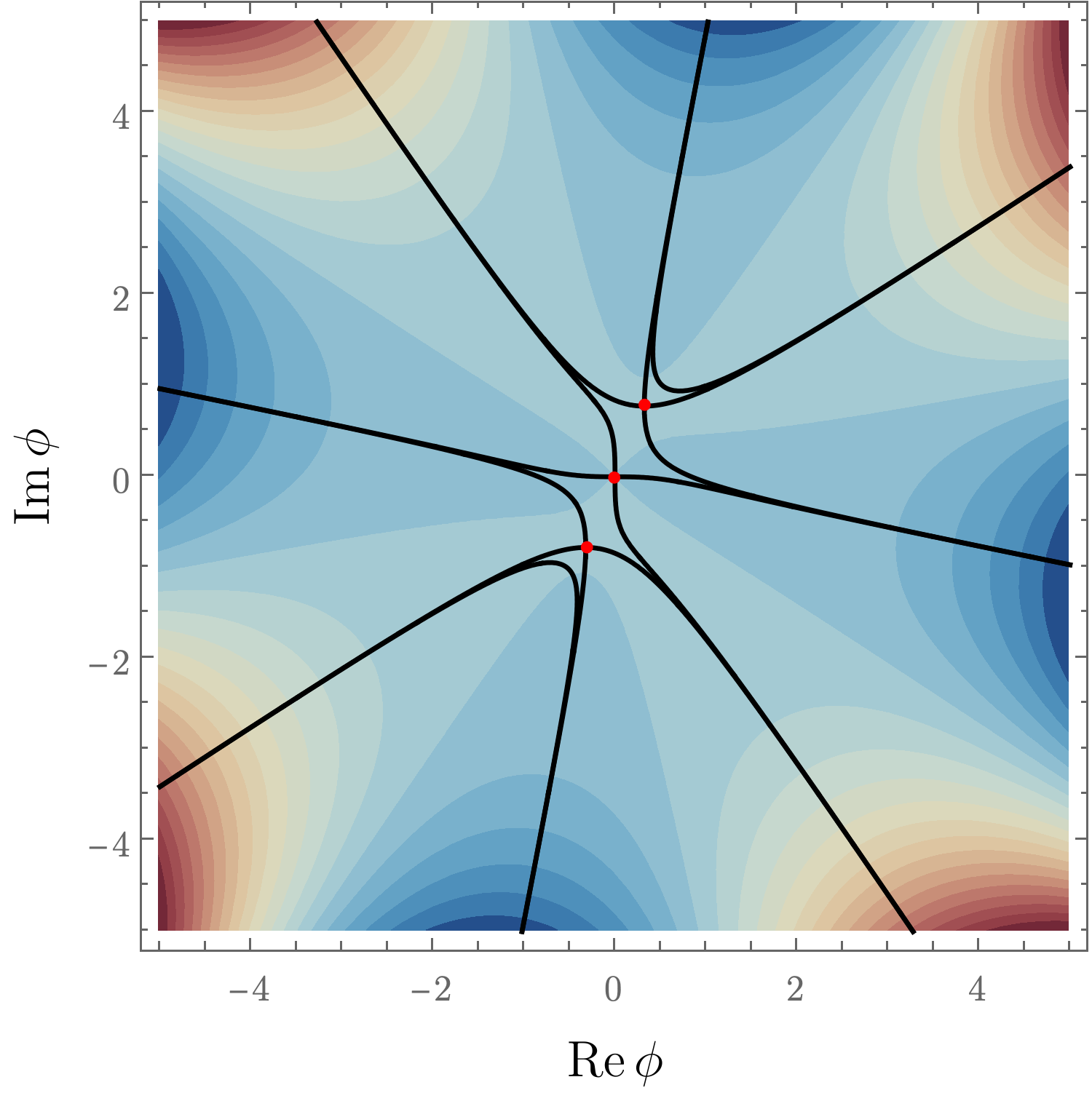}\qquad\;
\ee
The bluer colors denote higher values of $S$, which are the places where the integrand is more suppressed. From each of the three saddle points, there are two stationary-phase curves: one is the steepest ascent path asymptoting to the red regions, and the other one is the steepest descent reaching the blue ones. Our original contour, which was just the real axis, can be deformed into the steepest descent path passing through the perturbative saddle at $\phi = 0$. It means it gives the leading contribution to the integral. However, as we adjust $\lambda$, the saddle points and thimbles move around and the situation might change drastically.

Play around with different values of $\lambda$ and classify for which ranges the other two saddles play a role. Which linear combinations of steepest descent paths do we need? Note that for certain values of $\lambda$ there's a ``phase transition'' in which saddles are being picked up. 
The fact that a function can have different asymptotic behavior depending on the parameter $\lambda$ is related to \emph{Stokes phenomena}.

One can also compute subleading corrections to the saddle-point approximation by successively include terms from adjacent saddles to the dominant one. This is what Berry and Howls called \emph{hyperasymptotics} and it allows us to consistently improve on the poor convergence of the perturbative expansion $\mathcal{Z}_{\mathrm{pert}}$. For example, for $\lambda>0$, corrections are of order $\sim \e^{-[S(\phi_0^\ast) - S(\phi_\pm^\ast)]}$, which are the non-perturbative terms we were missing before. This is yet another way resurgence creeps in: improving perturbative expansion involves knowing something about the non-perturbative vacua.

More generally, observables in quantum field theory admit a \emph{transseries}, which is a more refined expansion than just a Taylor series. In simple cases, say expanding in $\hbar \to 0$, it looks like:
\be
\sum_{p,q,r} c_{p,q,r}\, \hbar^{p} \, \log^q \hbar\, \e^{-\frac{S_r}{\hbar}}\, ,
\ee
which involves powers, powers of logarithms, and exponentials.
			
\end{document}